\documentclass[%
reprint,
 amsmath,amssymb,
 aps,
floatfix,
unsortedaddress,
]{revtex4-2}

\usepackage{graphicx}% Include figure files
\usepackage{dcolumn}% Align table columns on decimal point
\usepackage{bm}% bold math
\usepackage{siunitx} %SI unit
\usepackage{blindtext}
\usepackage{tikz}
\usetikzlibrary{}
\usepackage{caption}
\usepackage{subcaption}
\usepackage{url} 

\usepackage{textcomp}
\usetikzlibrary{trees}
\usetikzlibrary{shadings}
\tikzstyle{every node}=[draw=black,thin,anchor=west, minimum height=2.5em]

\newcommand{\bmat}[1]{{\ensuremath{\bm {#1}}}}
\newcommand{\bvec}[1]{{\ensuremath{\bm {#1}}}}

\begin{document}

\preprint{APS/123-QED}

\title{2022 Review of Data-Driven Plasma Science}
% Force line breaks with \\
%\affiliation{% Authors' institution and/or address\\ This line break forced with \textbackslash\textbackslash}%
%\affiliation{ Second institution and/or address\\ This line break forced% with \\}%
%\affiliation{ Third institution, the second for Charlie Author}%
%\affiliation{% Authors' institution and/or address\\  This line break forced with \textbackslash\textbackslash }%

\author{Rushil Anirudh}
%\email{anirudh1@llnl.gov}
\affiliation{Lawrence Livermore National Laboratory, Livermore, CA 94550, USA}

\author{Rick Archibald}
%\email{archibaldrk@ornl.gov}
\affiliation{Oak Ridge National Laboratory, Oak Ridge, TN 37830, USA}

\author{M. Salman Asif}
%\email{sasif@ece.ucr.edu}
\affiliation{University of California, Riverside, Riverside, CA 92521, USA}

\author{Markus M. Becker}
%\email{markus.becker@inp-greifswald.de}
\affiliation{Leibniz Institute for Plasma Science and Technology (INP), Felix-Hausdorff-Str. 2, 17489 Greifswald, Germany}

\author{Sadruddin Benkadda}
%\email{sadruddin.benkadda@univ-amu.fr}
\affiliation{Aix Marseille University, CNRS, PIIM UMR 7345, Marseille, France}

\author{Peer-Timo Bremer}
%\email{bremer5@llnl.gov}
\affiliation{Lawrence Livermore National Laboratory, Livermore, CA 94550, USA}

\author{Rick H.S. Bud{\'e}}
%\email{r.h.s.bude@tue.nl}
\affiliation{Department of Applied Physics, Eindhoven University of Technology, The Netherlands}

\author{C.S. Chang}
%\email{cschang@pppl.gov}
\affiliation{Princeton Plasma Physics Laboratory, Princeton, NJ 08540, USA}

\author{Lei Chen}
%\email{Lei.Chen@iter.org}
\affiliation{Institute of Ion Physics and Applied Physics University of Innsbruck, Technikerstrasse 25, 6020 Innsbruck, Austria
}

\author{R. M. Churchill}
%\email{rchurchi@pppl.gov}
\affiliation{Princeton Plasma Physics Laboratory, Princeton, NJ 08540, USA}

\author{Jonathan Citrin}
%\email{J.Citrin@differ.nl}
\affiliation{DIFFER - Dutch Institute for Fundamental Energy Research, Eindhoven, the Netherlands}
\affiliation{Science and Technology of Nuclear Fusion Group, Eindhoven University of Technology, Eindhoven, Netherlands}

\author{Jim A Gaffney}
%\email{gaffney3@llnl.gov}
\affiliation{Lawrence Livermore National Laboratory, Livermore, CA 94550, USA}

\author{Ana Gainaru}
%\email{gainarua@ornl.gov}
\affiliation{Oak Ridge National Laboratory, Oak Ridge, TN 37830, USA}

\author{Walter Gekelman}
%email{gekelman@physics.ucla.edu}
\affiliation{Department of Physics and Astronomy,
University of California, Los Angeles, CA 90095, USA
}

\author{Tom Gibbs}
%\email{tgibbs@nvidia.com}
\affiliation{NVIDIA, 2788 San Tomas Expressway
Santa Clara, CA 95051, USA}

\author{Satoshi Hamaguchi}
%\email{hamaguch@ppl.eng.osaka-u.ac.jp}
\affiliation{Center for Atomic and Molecular Technologies, Graduate School of Engineering, Osaka University, 2-1 Yamadaoka Suita, Osaka, Japan}

\author{Christian Hill}
%\email{ Ch.Hill@iaea.org}
\affiliation{Department of Nuclear Sciences and Applications, International Atomic Energy Agency, Wagramer Strasse 5, PO Box 100, 1400 Vienna, Austria}

\author{Kelli Humbird}
%\email{humbird1@llnl.gov}
\affiliation{Lawrence Livermore National Laboratory, Livermore, CA 94550, USA}

\author{S\"oren Jalas}
%\email{soeren.jalas@desy.de}
\affiliation{Center for Free-Electron Laser Science and Department of Physics Universit\"at Hamburg, Luruper Chaussee 149, 22761 Hamburg, Germany}

\author{Satoru Kawaguchi}
%\email{skawaguchi@mmm.muroran-it.ac.jp}
\affiliation{Division of Information and Electronic Engineering, Graduate School of Engineering,
Muroran Institute of Technology, Muroran, Hokkaido, 050-8585, Japan}

\author{Gon-Ho Kim}
%\email{ghkim@snu.ac.kr}
\affiliation{Seoul National University, Seoul 151-741, Republic of Korea}

\author{Manuel Kirchen}
%\email{manuel.kirchen@desy.de}
\affiliation{Deutsches Elektronen-Synchrotron DESY, Notkestr. 85, 22607 Hamburg, Germany}

\author{Scott Klasky}
%\email{klasky@ornl.gov}
\affiliation{Oak Ridge National Laboratory, Oak Ridge, TN 37830, USA}

\author{John L. Kline}
%\email{jkline@lanl.gov}
\affiliation{Los Alamos National Laboratory, Los Alamos, NM 87545, USA}

\author{Karl Krushelnick}
%\email{kmkr@umich.edu}
\affiliation{Center for Ultrafast Optical
Science, University of Michigan, Ann Arbor, MI 48109-2099, USA}

\author{Bogdan Kustowski}
%\email{kustowski1@llnl.gov}
\affiliation{Lawrence Livermore National Laboratory, Livermore, CA 94550, USA}

\author{Giovanni Lapenta}
%\email{giovanni.lapenta@kuleuven.be}
\affiliation{Department of Mathematics, KULeuven, University of Leuven, Belgium}

\author{Wenting Li}
%\email{wenting@lanl.gov}
\affiliation{Los Alamos National Laboratory, Los Alamos, NM 87545, USA}

\author{Tammy Ma}
%\email{ma8@llnl.gov}
\affiliation{Lawrence Livermore National Laboratory, Livermore, CA 94550, USA}

\author{Nigel J. Mason}
%\email{n.j.mason@kent.ac.uk}
\affiliation{Department of Physical Sciences, The University of Kent, Canterbury, CT2 7NH, UK}

\author{Ali Mesbah}
%\email{mesbah@berkeley.edu}
\affiliation{Department of Chemical and Biomolecular Engineering, University of California Berkeley, USA}

\author{Craig Michoski}
%\email{michoski@gmail.com}
\affiliation{The Oden Institute for Computational Engineering \& Sciences, University of Texas at Austin, Austin, Texas 78712-1229 USA}

\author{Todd Munson}
%\email{tmunson@mcs.anl.gov}
\affiliation{Mathematics and Computer Science Division, Argonne National Laboratory, 9700 S. Cass Ave., Lemont, IL 60439, USA}

\author{Izumi Murakami}
%\email{murakami.izumi@nifs.ac.jp}
\affiliation{National Institute for Fusion Science, 
National Institutes of Natural Sciences, Toki, Gifu 509-5292, Japan \\
Department of Fusion Sciences, The Graduate University for Advanced Studies, 
SOKENDAI, Toki, Gifu, 509-5292, Japan}

\author{Habib N. Najm}
%\email{hnnajm@sandia.gov}
\affiliation{Sandia National Laboratories, Albuquerque, NM 87185, USA}

\author{K. Erik J. Olofsson}
%\email{olofsson@fusion.gat.com}
\affiliation{General Atomics, PO Box 85608, San Diego, California 92186-5608 USA}

\author{Seolhye Park}
%\email{druyvesteyndf@gmail.com}
\affiliation{Samsung Display Co., Ltd.,Asan-si, Chungcheongnam-do 31454, Republic of Korea}

\author{J. Luc Peterson}
%\email{peterson76@llnl.gov}
\affiliation{Lawrence Livermore National Laboratory, Livermore, CA 94550, USA}

\author{Michael Probst}
%\email{Michael.Probst@uibk.ac.at}
\affiliation{Institute of Ion Physics and Applied Physics University of Innsbruck, Technikerstrasse 25, 6020 Innsbruck, Austria
}
\affiliation{School of Molecular Science and Engineering
Vidyasirimedhi Institute of Science and Technology, Rayong 21210, Thailand}

\author{David Pugmire}
%\email{pugmire@ornl.gov}
\affiliation{Oak Ridge National Laboratory, Oak Ridge, TN 37830, USA}

\author{Brian Sammuli}
%\email{sammuli@fusion.gat.com}
\affiliation{General Atomics, PO Box 85608, San Diego, California 92186-5608 USA}

\author{Kapil Sawlani}
%\email{Kapil.Sawlani@lamresearch.com}
\affiliation{Lam Research Corporation, Fremont, CA 94538, USA}

\author{Alexander Scheinker}
%\email{ascheink@lanl.gov}
\affiliation{Los Alamos National Laboratory, Los Alamos, NM 87545, USA}

\author{David P. Schissel}
%\email{schissel@fusion.gat.com}
\affiliation{General Atomics, PO Box 85608, San Diego, California 92186-5608 USA}

\author{Rob J. Shalloo}
%\email{rob.shalloo@desy.de}
\affiliation{Deutsches Elektronen-Synchrotron DESY, Notkestr. 85, 22607 Hamburg, Germany}

\author{Jun Shinagawa}
%\email{Jun.Shinagawa@us.tel.com}
\affiliation{Tokyo Electron America, Inc., 2400 Grove Blvd., Austin, TX 78741 USA}

\author{Jaegu Seong}
%\email{seongjaegu@gmail.com}
\affiliation{Samsung Display Co., Ltd.,Asan-si, Chungcheongnam-do 31454, Republic of Korea}

\author{Brian K. Spears}
%\email{spears9@llnl.gov}
\affiliation{Lawrence Livermore National Laboratory, Livermore, CA 94550, USA}

\author{Jonathan Tennyson}
%\email{j.tennyson@ucl.ac.uk}
\affiliation{Department of Physics and Astronomy, University College London, London WC1E 6BT, UK}

\author{Jayaraman Thiagarajan}
%\email{jayaramanthi1@llnl.gov}
\affiliation{Lawrence Livermore National Laboratory, Livermore, CA 94550, USA}

\author{Catalin M. Tico\c{s}}
%\email{catalin.ticos@eli-np.ro}
\affiliation{National Institute for Laser, Plasma and Radiation Physics, M\u{a}gurele 077125, Romania}
%\email{catalin.ticos@eli-np.ro}

\author{Jan Trieschmann}
%\email{jt@tf.uni-kiel.de}
\affiliation{Theoretical Electrical Engineering, Faculty of Engineering, Kiel University, Kaiserstraße 2, 24143 Kiel, Germany}

\author{Jan van Dijk}
%\email{J.v.Dijk@tue.nl}
\affiliation{Department of Applied Physics, Eindhoven University of Technology, The Netherlands}

\author{Brian Van Essen}
%\email{vanessen1@llnl.gov}
\affiliation{Lawrence Livermore National Laboratory, Livermore, CA 94550, USA}

\author{Peter Ventzek}
%\email{peter.ventzek@us.tel.com}
\affiliation{Tokyo Electron America, Inc., 2400 Grove Blvd., Austin, TX 78741 USA}

\author{Haimin Wang}
%\email{haimin.wang@njit.edu}
\affiliation{Institute for Space Weather Sciences,
New Jersey Institute of Technology, Newark, NJ 07102, USA}

\author{Jason T. L. Wang}
%\email{jason.t.wang@njit.edu}
\affiliation{Institute for Space Weather Sciences,
New Jersey Institute of Technology, Newark, NJ 07102, USA}

\author{Zhehui Wang}
%\email{zwang@lanl.gov}
\affiliation{Los Alamos National Laboratory, Los Alamos, NM 87545, USA}

\author{Kristian Wende}
%\email{kristian.wende@inp-greifswald.de}
\affiliation{Leibniz Institute for Plasma Science and Technology (INP), Felix-Hausdorff-Straße 2, 17489 Greifswald, Germany}

\author{Xueqiao Xu}
%\email{xu2@llnl.gov}
\affiliation{Lawrence Livermore National Laboratory, Livermore, CA 94550, USA}

\author{Hiroshi Yamada}
%\email{yamada.hiroshi@edu.k.u-tokyo.ac.jp}
\affiliation{Graduate School of Frontier Sciences, The University of Tokyo, Kashiwa, Chiba 277-8568 Japan}

\author{Tatsuya Yokoyama}
%\email{yokoyama.tatsuya17@ae.k.u-tokyo.ac.jp}
\affiliation{Graduate School of Frontier Sciences, The University of Tokyo, Kashiwa, Chiba 277-8568 Japan}

\author{Xinhua Zhang}
%\email{xinhuazhang@lanl.gov}
\affiliation{Los Alamos National Laboratory, Los Alamos, NM 87545, USA}

%**************************

\date{\today}% It is always \today, today,

\begin{abstract}
Data science and technology offer transformative tools and methods to science. This review article highlights latest development and progress in the interdisciplinary field of data-driven plasma science (DDPS). A large amount of data and machine learning algorithms go hand in hand. Most plasma data, whether experimental, observational or computational, are generated or collected by machines today. It is now becoming impractical for humans to analyze all the data manually. Therefore, it is imperative to train machines to analyze and interpret (eventually) such data as intelligently as humans but far more efficiently in quantity. Despite the recent impressive progress in applications of data science to plasma science and technology, the emerging field of DDPS is still in its infancy. Fueled by some of the most challenging problems such as fusion energy, plasma-processing of materials, and fundamental understanding of the universe through observable plasma phenomena, it is expected that DDPS continues to benefit significantly from the interdisciplinary marriage between plasma science and data science into the foreseeable future.

%Data science and technology offer transformative tools and methods to science. This community-based review highlights some exciting development and progress in data-driven plasma science. Data science and machine learning algorithms go hand in hand. Most of the plasma data, whether experimental or computational, are generated or collected by machines today, and it is imperative to train machines to analyze the data similar to or better than humans since it is impractical to analyze the vast amount of data manually. Despite of the impressive progress in applications of machine learning to plasma science and technology, we are still in the early phase of the growth. Fueled by some of the most challenging problems such as fusion energy, plasma-processing of materials, and fundamental understanding of the universe through the observable plasma phenomena, the interdisciplinary marriage of plasma science with data science will certainly be productive in the foreseeable future.

%\begin{description}
%\end{description}
%https://ja.overleaf.com/project/60860fbda10c77acd62bf9bc
\end{abstract}

\maketitle

%\tableofcontents
\clearpage
\tableofcontents
\clearpage

\section{\label{sec:level1} Introduction}
%\red{JT: I have been cleaning up the referencing. The citations contain two different papers called Chen:2014. I don't know which or both are cited below.}
%{\color{blue} ZW: Which .bib files contain Chen:2014? Sec.~III, VI do not seem to have it.}
%\red{The problem has been solved.}

 Plasma science, like other branches of natural science such as particle and high-energy physics, condensed matter physics, physics of fluids, nuclear physics, atomic and optical physics, astrophysics, cosmology, material science, biology and chemistry, is founded on experiments and observations. Experimental data-driven activities through collection of experimental data, analysis of the data, reduction of the data to knowledge and comparison of experimental data with theory, computational and statistical models play a central role in plasma science and technology. In recent years, data-driven plasma science and technology is going through a renaissance, picking up new meanings, and revealing unexplored directions because of the advances in data science and technology both inside and outside the domain of plasma research and applications. 

One of the most widely known, and possibly most successful, examples of data-driven science applied to conventional scientific disciplines is the Materials Gnome Initiative (MGI)~\cite{MLGreen_APR2017}.%, initiated by President Barak Obama of the USA in 2011.
~Similar projects also took place around the world around the same time.  In this project, the search for new functional materials was assisted by data-driven approaches, rather than the experience and intuition of engineers in materials science,  and the efficiency of discovery of new materials is said to have been significantly improved.  In the project, not only the existing material data were fully exploited by newly developed machine learning (ML) techniques and artificial intelligence (AI), but also efficient methods to collect a large amount of data in relatively short periods, which is called ``high-throughput screening (HTS),’’  were also developed. In general, in materials science,   shortage or the lack of data is often the problem for efficient material discovery, so the development of HTS techniques, especially those fit to the latest ML and AI techniques, played a key role in the success of the MGI project. 

Similarly, the search for the best plasma conditions for specific applications, such as nuclear fusion and semiconductor device manufacturing, is often one of the most essential research-and-development (R\&D) activities in plasma science and technologies. Therefore, similar approaches developed in the MGI may also be useful in this field. Especially, systematic collection, classification, and improved accessibility of data for reuse may also be crucial in promoting data-driven approaches to problem-solving in plasma science and technologies.  

Data-driven science is sometimes called the fourth paradigm of discovery~\cite{HTT:2009}. The previous three paradigms are empirical or experimental (Galileo Galilei), theoretical (Issac Newton), and computational (It may be hard to credit a single person for this) according to a classification by Jim Gray in his talk to NRC-CSTB in 2007. Data-driven science is fundamentally different from the previous three paradigms and thus transformational. Most notably, it could take human intelligence out of the discovery process, and make fully automated scientific discovery possible through artificial intelligence. It has been predicted by Frank Wilczek that such a transition could take about 100 years~\cite{Wilczek:2015,You:2015}. 

We may recognize several pillars in data-driven plasma science: availability of big data (come in different forms), availability of a large number of advanced algorithms and methods including theoretical-driven algorithms such as a finite element solver, statistical driven algorithms, and availability of inexpensive computational platform. We have summarized the current status of data-driven research activities in plasma science and technologies in this review article. The article is organized in the following manner: In Sec. II, fundamental data science is briefly reviewed, especially in the light of applications to plasma science and technologies in general.  In Sec. III, examples of data-driven approaches for the analyses of basic plasma physics and laboratory experiments are discussed. In Sec. IV, an overview of data-driven analyses in magnetic confinement fusion (MCF) research is presented. 
Another large field of high-power/high-energy plasma physics is inertial confinement fusion (ICF), 
whose latest data-driven analyses are presented in Sec. V. 
In space and astronomical plasmas, a large amount of observational data has been accumulated over many decades and data-analytic techniques have been extensively studied. The latest development of such research activities are summarized in Sec. VI.   
Plasma technologies are also widely used in industries and cost-effective development is always of interest to the industries. Some latest development of data-driven approaches to R\&D in industries and related academic problems are highlighted in Sec. VII.  Sec. VIII discusses the current status of various databases that may be of interest to the plasma community. A final summary is given in section IX.

\noindent
[Zhehui Wang and Satoshi Hamaguchi]

\section{Fundamental Data Science}
\label{Section2}
\subsection{Introduction}
%all -> C.S.Chang (we need to list the person's name who actually and essentially wrote this; cannot be all for this short paragraph) 
%
%
This section provides a brief overview of the present status and future direction of the fundamental analysis methods for scientific data, on which the present and future plasma data science rely.  We discuss data reduction and compression methods that operate on diverse architectures and also on streaming data, capable of high compression rates while preserving targeted quantities-of-interest  (QoIs).  We describe dimensional reduction and sparse modeling techniques that promote scientific understanding of high dimensional data and reduces analysis and storage cost of the data. We cover machine learning (ML) enhancements to modeling and simulation that can be utilized to accelerate simulations and to provide accurate and robust closure models.  In addition we discuss other fundamental ML methods throughout this section. Intrinsic to the analysis methods and tools presented in this section is the hardware used to execute these methods.  We cover the techniques used to integrate hardware capabilities with the numerical methods presented in this section.  We focus on advancements in workflow automation, which is necessary to store, move, and process the complex scientific data produced at leadership experimental and computing facilities.  We overview the explosion of theory, algorithms, and tools that have been developed over recent decades in uncertainty quantification (UQ).  Advancements in visualization and data understanding are described that can be used by domain experts to facilitate knowledge and discovery from scientific data.  This section ends with ML control theories that are applicable to highly nonlinear and multivariable plasma dynamics and that can take into account of the safety-critical plasma applications.

\noindent
[C.S. Chang]

\subsection{Data Reduction/Compression}
%Klasky

Experimental, Observational, and Computational facilities are facing a crisis because of the large increase in data being produced at these facilities.
New  technologies allow more data to be captured at higher rates, which increases data volumes and velocities, and necessitates the need for streaming reduction techniques. Hence, there is a crucial need for fast reduction techniques that must work on diverse
architectures and stream data across processes in complex workflows, ensuring that short and long term events can be captured and analyzed in the reduction process. 

There are several cross-cutting challenges which are not specialized to a particular application and can be thought of as reduction motifs that can work for a variety of applications and can be further customized and tuned for different scientific instruments.
The first motif is for reducing ``noisy'' data, 
when the signal to noise ratio is low and where computational signatures are often needed to extract the signal.
The second motif is for high-dimensional data, 
often illustrated in plasma physics applications, which often simulate six dimensional physics.
The third motif is for non-uniform and unstructured data, often produced by Magneto Hydro Dynamics (MHD) codes.
The fourth motif is for reducing data as it streams, which can be from a live experiment or from an exascale simulation. 
Finally, the fifth motif is to ensure that simple and complex QoIs (derived quantities) from downstream processing has a user-specified uncertainty to ensure trustworthy data used for later post processing.
Many of these motifs can be put together to illustrate new scientific challenges; for example when 
large simulations that search for features, events, and
anomalies and produce QoIs that could be combined with in-situ machine learning and artificial intelligence workflows to produce reduced order models in addition to a complete data model repository. 
In all of these cases, there is a crucial need for fast reduction techniques that must work on diverse architectures and stream data across processes in complex experimental workflows, ensuring that short-term events can be captured and analyzed in the reduction process.

Both compression and analysis share  a common goal: to extract science from the raw data which involves extracting  the essential structure and key features of the phenomenon under study while ignoring or discarding the noise and  data that have little or no impact on the quantities of interest.
It is important to understand how reduction methods affect the specific quantities of interest used in the analysis so that reduction does not alter the results of the analysis. 
Lossless methods have unfortunately been generally unable to achieve the high compression ratios needed to handle the large quantities of data generated by facilities, often reducing
data by less than 15\%. This means that
we have to look at lossy methods, which bring the fidelity of
the reduced data into question. Fidelity and reduction are  directly in competition with each other, and so it is important to consider what exactly is required of a reduced dataset in order for it serve as a scientifically useful surrogate.

Lossy compressors should be flexible with regard to the structure of the data, generalize to arbitrarily high dimension, and allow control of errors both in the original degrees of freedom and in downstream QoIs.
Compressing data in the same high-dimensional space where it is defined can make more of the data's spatial correlations visible to the compression algorithm, resulting in higher compression ratios.
Similarly, compression algorithms should make use of as much of the data's spatial structure as possible.
Compressing nonuniform or unstructured data as though it were defined on a uniform grid risks obscuring redundancies and patterns in the data, resulting in lower compression ratios.
Another design goal is the control of errors incurred by compression algorithms.
%
%A natural starting point is to bound the error in the `raw' data---i.e., the difference between the original dataset and the reduced dataset output by the compression algorithm.
Often scientists are concerned with the change to the QoIs from the compressed data, hoping to make sure that all of the features in the QoIs are preserved to a high enough accuracy.
The mathematics required to relate errors in the raw data to errors in QoIs is nontrivial, especially for QoIs that are nonlinear and/or obtained by complex post-processing.
Empirical approaches can provide estimates for, but not guaranteed bounds on, QoI errors by extrapolating from previously encountered datasets and QoIs.

Reduction algorithms need to be efficient as well, meaning that they need to use a minimal set of 
computational resources (time to solution, memory, network, computational) and should be able to reduce the time to solution in application workflows. 
Reduction algorithms must further satisfy the following set of requirements: 1) Ability to quantify the uncertainty of errors in the raw-data and the derived QoIs, 2) Ability to be efficient in its use of computational resources, 3) Ability to work with high-dimensional data on structured and unstructured meshes, sets the overarching requirements so that scientists can both trust and efficiently use the communities data reduction algorithms. 

A 2018 survey by Li et al.~\citep{Li:STAR18}
organized data reduction techniques for scientific data into five categories: 
truly lossless, near lossless, lossy, mesh reduction, and derived representations.
Lossless compression includes techniques like 
entropy-based coders (such as Huffman coding~\citep{huffman1952method}, which is
used by bzip2~\citep{seward1996bzip2}, and arithmetic coding~\citep{witten1987arithmetic})
and 
dictionary-based coders (such as LZ77~\citep{ziv1977universal} and LZ78~\citep{ziv1978compression}, which have inspired many variants such as those used in DEFLATE~\citep{deutsch1996deflate}, gzip~\citep{gzip}, and zlib~\citep{zlib}).
That said, lossless compression
often achieves only modest reductions, for example
fpzip achieved a 3.7X reduction on a simulation of a Rayleigh Taylor
instability by the Miranda simulation code~\citep{lindstrom2006fast}.
Near lossless compression refers to rounding errors that occur during
reconstruction from transforms;
their reduction capabilities are often similar to lossless compression.
Lossy compression includes techniques like truncation,
quantization,
predictive coding schemes, 
and transform-based compression schemes.
The Li survey points to many instances of lossy compression packages applied to scientific
data, including MLOC~\citep{gong2012mloc}, fpzip~\citep{diffenderfer2019error}, ISABELA~\citep{lakshminarasimhan2013isabela}, SZ~\citep{Di:2016}, VAPOR~\citep{atmos10090488}, JPEG2000~\citep{skodras2001jpeg}, and zfp~\citep{lindstrom2014fixed}
%{Lindstrom:2014:ZFP}.
Several of these are included in SDRBench~\citep{zhao2020sdrbench},
as well as some additional packages 
that have emerged in recent years:
DCTZ~\citep{zhang2019efficient},
MGARD~\citep{ainsworth2017mgard}, and
TTHRESH~\citep{BLP:19}.
Mesh reduction techniques include decimation (surveyed by Weiss and De Floriani~\citep{weiss2011simplex}), 
multi-resolution techniques,
subsetting (such as with querying with FastBit~\citep{wu2009fastbit}), and temporal sampling (i.e., triggers~\citep{Larsen:ISAV18,8231851,Salloum:2015:EAS:2828612.2828619}).
Derived representation techniques use alternate representations of the data, typically statistical in nature,
but also including approaches like topological features~\citep{DBLP:journals/cgf/HeineLHIFSHG16} and imagery~\citep{ahrens2014image}.
Recent work using machine learning to reduce scientific data~\citep{choi2021neural,glaws2020deep,liu2021high} 
could also be considered a derived representation.

\paragraph{R\&D Necessary for the future}
The provision of realistic numerical bounds is essential if the scientist is to have confidence in applying 
data reduction. In order for existing and future reduction algorithms to be \lq trustworthy\rq,~it will be 
necessary for algorithms to come with some kind of certificate or guarantee on the fidelity of the reduced data 
to the original data. This may take the form of rigorous mathematical bounds on the loss incurred measured in 
a norm that is relevant to the application. More generally, future research in data reduction procedures should 
ideally aim to provide the scientist with the capability to specify a set of application dependent quantities 
of interest which should be preserved to a user-specified tolerance. The reduction procedure should have the 
flexibility to effectively reduce the data whilst maintaining the set of quantities of interest to the level
specified by the user, and providing realistic bounds on the actual loss incurred. Certificates of this type are
essential for the \lq trustworthiness\rq\ of the reduction routines. 

In order for a reduction algorithm to be \lq effective\rq\ it must be capable of providing meaningful levels of
reduction whilst incurring a level of loss that is acceptable to the user. Achieving a balance between these two
competing criteria encapsulates the essential difficulty in developing effective data reduction algorithms, and
constitutes a major challenge for future research in data reduction. Nevertheless, a reduction algorithm is only
effective if it is applicable to the types of data of interest. Many existing reduction algorithms are effective
at reducing structured data such as uniformly spaced data, data specified on tensor product grids etc. However, 
the performance or applicability of the algorithms to more general data formats including unstructured grids,
particle data is less well-understood. Research into understanding whether or not, and how, existing approaches
can be extended to more general types of data will be needed if effective algorithms are to be developed. 

\noindent
[Scott Klasky]
%, klasky@ornl.gov]

\subsection{Dimensional reduction and sparse modeling}
%[Rick Archibald, archibaldrk@ornl.gov]}
Often in physics, a change of basis can greatly simplify the analysis of measured data and promote scientific understanding.  This is the driving force behind dimensional reduction methods, remapping high dimensional data to a compact representation in low dimensional space that preserves information, increases understanding, and reduces analysis and storage costs.  We will briefly review the dominate methods in this field of research, providing insight into these methods and their best uses.
\par
At a high level, dimension reduction methods fall into two major categories, they are either linear or non-linear transformations.  If the data being analyzed is linear, then linear transform methods will provide accurate and robust dimension reductions.  We explain three powerful linear methods, principal components analysis (PCA), linear discriminant analysis (LDA), and 
independent component analysis (ICA).
\par
For the case when the data being analyzed is not linear, then more sophisticated methods must be used. Kernel principal components analysis (KPCA), diffusion maps (DM), and machine learning (ML) are popular choices for discovering non-linear dimension reduction transformations.
\par
One unique driving force in dimension reduction is sparsity in modeling and data.  Sparse phenomena can occur in scientific data, for example a sparse set of vibrational modes in a material.  Sparse spectral techniques are designed to find sparsity and use this to create dimension reduction transformations.  We will discuss Laplacian Eigenmaps and Hessian locally linear embedding (HLLE).
\par
Even in this short review, we have touched on many viable dimension reduction methods.  The key to using these methods is understanding the type of data they were designed to analysis and employing them accordingly.  
\par
PCA has a long history in data analysis \cite{doi:10.1080/14786440109462720}, and performs one type of transformation very effectively.  PCA performs an orthogonal basis transformation where dimensions are ordered to represents as much of the data's variation as possible.  Thus, often most of the data's variation can be represented in a low number of dimensions.
\par
Data variation is just one way of sorting data.  If the data has class information, then this information may provide hints to the best dimension reduction.  LDA uses class information to create dimension reduction, by determining the best transformation that maximize class separation while minimizing the scatter within a class. 
\par
Suppose that your data is the linear combination of independent non-Gaussian signals.  For example, the cocktail party problem is exemplary of this type of data.  Using only the measurements taken from multiple microphones spread across a cocktail party, methods such as ICA \cite{COMON1994287}, can extract the individual speakers as independent components in the muddled recording of the party.  ICA uses maximum likelihood and minimization of mutual information to identify independent components.
\par
For data that is not a linear combination of components, more sophisticated methods must be used.  KPCA is the non-linear reformulation of PCA using a kernel function to construct a complex transform \cite{scholkopf1998nonlinear}.  The key difference between PCA is that Kernel PCA computes the principal eigenvectors of the kernel matrix, rather than those of the covariance matrix.  Thus, the kernel function can be used to expose different relationships in the data.  The application of PCA in the kernel space produces the dimension reduction.
\par
Once common assumption in dimension reduction is that data resides on a manifold in high-dimensional space.  Determining this manifold creates the ultimate representation for dimension reduction.  Diffusion maps (DM) \cite{nadlerdiffusion} uses methods from dynamical systems and statistics to approximate data on manifolds in high-dimensional space.  DM use multiple Markov random walk on the graph of the data, measuring the so-called diffusion distance. Here, data manifolds can be determine by integrating over all paths through the graph creates an isomap of the manifold, where short-circuiting help discover the diffusion path and hence manifold. 
\par
Sparse patterns in high-dimensional data can be identified using sparse spectral methods. These methods setup sparse (generalized) eigen problem that are capable of low dimensional reductions while retaining local structure of the data.
\par
Laplacian Eigenmaps find a low-dimensional data representation and preserving local properties of the manifold \cite{belkin2001laplacian}.  The Laplacian matrix is constructed using a weight function, where the distance between nearest neighbors is used as the input.  Minimization of a cost function based on the graph ensures that points close to each other on the manifold are mapped close to each other in the low-dimensional space, preserving local distances 
\par
Hessian locally linear embedding (HLLE) \cite{donoho2003hessian} is another type of sparse spectral method that learns manifolds in high-dimensional data, maintaining the balance between local geometric information and over-fitting.  Hessian is calculated at every datapoint, and is used to create a localized parameterization of the manifold.  
\par
Machine learning has become a dominate force in the field of dimension reduction and sparse modeling \cite{ray2021various}.  As described in this section, each method carried with it a set of assumptions on the form of data.  Machine learning has the flexibility to learn unique transforms that our outside of the data assumptions in this section.  The limitations to machine learning is centered around that fact that these methods will only perform transformations that they are trained to execute. 

\paragraph*{Acknowledgements}
This work was supported in part by the U.S. Department of Energy, Office of Science, Office of Advanced Scientific Computing Research, Scientific Discovery through Advanced Computing (SciDAC) Program through the FASTMath Institute under Contract No. DE-AC05-00OR22725 at Oak Ridge National Laboratory.

\noindent
[Rick Archibald]

\subsection{ML-enhanced modeling and simulation}
%[Todd Munson, tmunson@mcs.anl.gov]}

Computationally expensive operators in the system of equations being used to simulate the plasma can significantly impact our ability to sufficiently simulate the plasma.  By replacing these expensive operators with less expensive surrogate models, we can improve the overall performance of the simulation.  These surrogate models can be learned by applying machine learning techniques that use a corpus of data to train a neural network model.  The trained model is then used as a replacement for the operator in the simulation.  However, the machine learning surrogate needs to conserve relevant physical quantities, such as mass and energy, for the resulting simulation that includes the learned surrogate model to be stable and meaningful.  These physics-informed machine learning surrogates \cite{raissi2019physics} offer the potential, however, to be much faster to evaluate than the original operator, while providing a sufficient approximation.  As an example, the Fokker-Planck-Landau collision operator has a computational cost that grows at a quadratic rate as the number of species increases and needs to be evaluated many times when forming the right-hand side of the system of equations.  Machine learning surrogates have been successfully developed for this operator \cite{miller2021encoder}.  In other settings, data-driven machine learning models have been developed to estimate closures for plasma fluid models \cite{ma2020machine,maulik2020neural} and fluid turbulence models \cite{kochkov2021machine}, and for coupled simulations.\cite{kruger_machine_2019}
The amount of available data, network architecture, and training methods all have an impact on the quality of the resulting machine learning surrogate models.

An assumption typically made when developing machine learning surrogates is the availability of a large corpus of data.  In this data rich regime, deep neural networks with many parameters and layers can be trained, often resulting in good surrogate models, as the amount of data exceeds the number of parameters.  When only a small amount of training data is available, greater care must be taken in the network architecture and regularization techniques to produce a reasonable surrogate model, especially when the amount of training data available is less than the number of parameters in the surrogate model.  Sparse neural networks that are not fully connected between layers can be beneficial in the data poor regime and hyperparameter optimization methods can be applied to search for a sparse network architecture for the machine learning surrogate model.  Embedded physics knowledge can also reduce the amount of data required to train the surrogate model \cite{raissi2019physics}, as the equations reduce the number of degrees of freedom.

There are two basic approaches for incorporating physics knowledge into the machine learning surrogate models: changing the architecture of the network or changing the training problem to penalize deviations from the physics constraints.  Both approaches have the benefits and limitations.  

Changing the network architecture by adding projection layers guarantees that the machine learning surrogate used in the simulation will conserve the relevant physical quantities.  Whenever a physical quantity that needs to be conserved is added, the network has to be updated and the model retrained.  Moreover, when the training data only approximately conserves the quantities, as is often the case with real and simulated measurements, the trained surrogate model may not satisfy the conservation constraints exactly.  As a premium is placed on the satisfaction of the conservation constraints, the trained surrogate model may not represent the training data as well as one would like.

Adding the deviation from the physics constraints to the loss function used during training can result in a good surrogate model that conserves the physical quantities.  This approach offers flexibility in making a tradeoff between an accurate representation of the training data and the conservation error by adjusting the penalty weights.  Adding new quantities to conserve amounts to adding a new term to the loss function an retraining.  As the conservation constraints are not preserved exactly, one may not be able to completely rely on training machine learning surrogate model in the simulation.  Moreover, choosing the best weights for the terms in the loss function and the formulation of the (scaled) conservation errors can greatly impact the quality of the trained surrogate model.  Methods that dynamically adjust the penalties, using, for example, and augmented Lagrangian formulation, have been developed that can circumvent some of the challenges in determining the best weights.

The surrogate model that is integrated into the simulation may not always be a good approximation to the true operator, particularly under conditions not well represented in the training set or when rare events occur.  By developing metrics to identify when the surrogate model is inadequate, such when the output has a large conservation error, one can store the inputs and the simulation can continue by reverting back to the computationally expensive operator when it is available and computationally tractable and store the correct outputs.  The stored input/output data can be added to the corpus of training data and a retraining strategy employed to produce an updated surrogate model.  A complete workflow where the surrogate model is retrained while the simulation is running, with the improved surrogate model fed back to the simulation, can be pursued.

\paragraph*{Acknowledgements}
T. Munson was supported in part by the U.S. Department of Energy, Office of Science, Office of Advanced Scientific Computing Research, Scientific Discovery through Advanced Computing (SciDAC) Program through the FASTMath Institute under Contract No. DE-AC02-06CH11357 at Argonne National Laboratory.

\noindent
[Todd Munson]
%tmunson@mcs.anl.gov

% Transfer learning, active learning, reinforcement learning, reservoir learning, GANNs.

\subsection{ML Hardware and integration with models}
%[Tom Gibbs, tgibbs@nvidia.com]}
One challenge with achieving performance for simulation models of fusion science problems is sparsity and irregular data formats. This is a problem in multiple domains as well, and the simple explanation is that nature is connected in such a way that the elements that affect one another aren’t contiguous, and the connections aren’t always organized in an orderly fashion. The result is that when the data are arranged in a matrix there are large gaps where the array elements are zero and the columns that are non-zero are often jagged. This in turn presents a problem for digital computers that want to calculate ordered sets of data, and obviously calculating an operation with zero isn’t very efficient. This issue is exacerbated when the natural phenomenon being modeled are inherently noisy or turbulent. In this case the arrays are highly irregular and the methods for modeling them accurately are challenging due to the high level of dimensionality.

The general challenge with these approaches is that most of the numerical algorithms used in science require full 64 bit precision, so two full 64 bit operands must be moved to generate a multiply or divide and with a sparse array these are often multiply or add zero.

There is a large body of work on sparse methods to help alleviate this problem, but they all introduce some level of overhead in their attempt to re-order the matrix or compact the elements. Another related issue is that many of the best in class algorithms use operations that are vector by vector or matrix by vector and these often can’t fully utilize all of the computational infrastructure.

Advanced Machine Learning algorithms have emerged as a mechanism to deal with some of these constraints. The ML algorithms such as Deep Neural Networks, can be classified as Universal Function Approximators. So rather than start with the equations of state that define the natural phenomenon the function approximators are trained from data. There are physics informed methods that will use the equations of state to as part of the model development, but the ML methods don’t directly approximate the equations of state but use them to govern the loss or provide input to the training process. 

In general, this approach has been shown to offer a number of advantages for developing models for complex natural phenomenon in multiple domains. The overarching benefit is that the training of the model and the resulting inference are more efficient as they don’t require full precision and the algorithms used to train and execute the resulting inference can be posed as matrix by matrix operations where a digital computer can be much more efficient.

This allows the hardware to be more efficiently used, or perhaps more efficient hardware to be efficiently used. The training process is generally more expensive but is done far less often than the inference and both are more efficient than classical methods to model the same natural phenomenon. The matrices for the ML methods can be sparse, but they can generally use matrix by matrix operations as well as reduced precision. So, the overall improvement in time to solution and resource consumption is multiple orders of magnitude relative to classical methods. 

The dramatic improvement in time to solution and resource consumption opens the door for the ML methods to be used both for data center as well as experimental use case settings.

\noindent
[Tom Gibbs]
%, tgibbs@nvidia.com]}]

\subsection{Workflow Automation}
%[Ana Gainaru, gainaru@uornl.gov]}
The rate and size of data generated by cutting-edge experimental science facilities and large-scale simulations on current HPC systems is forcing scientists to move toward the creation of autonomous experiments and HPC simulations. However, efficiently moving, storing, and processing large amounts of data away from the point of origin presents an incredible challenge. Machine learning approaches are being used to learn insights from I/O patterns; and in-memory computing, in situ analysis, data staging, and data streaming methods are being explored to transfer data between coupled workflows.  However, many challenges remain to offer scientist the tools they need to efficiently automate their workflows. Modern scientific workflows are often collaborative in nature, consisting of multiple heterogeneous and coupled processes that must dynamically interact and exchange  data on the fly during or after execution. This dynamic nature adds another layer of difficulty in managing these massive datasets. 

Steering experiments in near real time is becoming critical, demanding further automation of the experimental scientific workflow. For example, scientists often run simulations and experiment analysis on different (sometimes geographically distributed) computing resources to simulate different components of the same physical phenomena. These codes need to interact with each other and often they must exchange data with analytical or visualization processes in near real time to help scientists understand the simulation results in a timely fashion. As a result the need to automate these efforts has grown. Research in this field requires moving large amounts of data from the point of origin (e.g., simulations, experiments, instruments) to HPC facilities that can perform reduction, analysis, and visualization.

A workflow management system capable of automating the execution of complex large-scale workflows must provide scientists with the ability to optimize their experiments for maximized acceleration of the scientific process and use the experimental observations efficiently. There are currently several limitations and challenges that the science community faces for constructing resilient, distributed workflows for making NRT decisions. There is limited support for semi-autonomous, resilient execution of a workflow in NRT. Resiliency is constrained to addressing failures through general task restarts, where restarts may not be done in NRT, and policies are not implemented for tasks in order of their priority in a workflow. Support for dynamic control is limited to a set of basic actions that a user may take at runtime. The ability for a user to query or analyze workflow execution and steer a workflow using monitoring data is largely absent.  

Automating workflows as well as coupling experiments with workflows and automating the data movement for real time analysis and visualization is a new research area moved forward by the needs of current large-scale applications. In order to achieve this goal, several projects have made progress in resource allocation across multiple machines, data streaming over large areas, resiliency, security, etc. Once such effort is the National Energy Research Scientific Computing Center (NERSC) Superfacility~\cite{9307775} project that aims to provide an ecosystem of connected facilities and software for the NERSC computing center. Its main focus is on providing a vision for making resource reservations using an API that can be used to connect to the center’s HPC systems. The project does not address resiliency concerns for distributed workflows and do not allow near real time decision making in their process. 

There is currently a long list of workflow management systems and tools used by the scientific community focusing on different aspects of workflows. Examples include Pegasus~\cite{DEELMAN201517}, Kepler~\cite{10.5555/1148437.1148454}, RADICAL-Pilot~\cite{DBLP:journals/corr/MerzkySTJ15}, and others. A common theme across existing workflow management systems is the focus on execution patterns and optimizing computational throughput, dynamic support constrained to task restarts, but almost no support for real time data delivery, monitoring, and workflow steering. The EFFIS~\cite{5452435} framework, initially designed to loosely couple multiple fusion codes running on HPC resources, is a workflow management system that uses a combination of enabling technologies, including ADIOS~\cite{10.1145/1383529.1383533}, Kepler and eSimMon~\cite{barreto2007managing}, a web-based dashboard.  EFFIS is built upon the Cheetah-Savanna~\cite{8943548} suite of workflow tools and provides an API for composing and executing codesign studies for online data analysis on different supercomputers. It supports both the execution of strongly coupled workflows on HPC resources~\cite{suchyta2021effis} and the execution of data streaming from the fusion KSTAR experimental facility to NERSC~\cite{d2017fusion}. EFFIS is being successfully used by applications at Oak Ridge National Laboratory. However, it is only a first step towards providing a workflow infrastructure capable of efficiently coupling complex geographically distributed workflows.

An important aspect of workflow automation is managing distributed resources and dynamically controlling a running workflow. Scheduling schemes for supporting real-time jobs, along with traditional batch jobs on HPC systems, have been evaluated by the community in several studies~\cite{nickolay2021accommodating, Wang2018}. Current approaches include the usage of basic manual intervention and pre-programmed scripts to control a workflow dynamically. Challenges around live monitoring, analysis, and control of running workflows still remain open issues. In addition, runtime control in current solutions is limited to restarting failed tasks. There is little support for more resilient and policy-based execution of a distributed workflow. 

Current workflow system do not provide native support for scientific data management middleware to tune data delivery.  Most workflow systems cannot interact with streaming scientific data management frameworks. They support staging data as files across resources using tools such as GridFTP, but they do not provide low-level tuning of data streams for low-latency data delivery. Data objects are seen as black boxes, and support is provided only to stage and persist them; there are no ways to switch between file- and stream-based options.

\paragraph{R\&D Necessary for the future}

Coupling experiments with workflows containing simulations, surrogate models, analysis and visualization codes requires geographically distributed resources: large-scale systems, edge devices at facilities, and computers at home institutions of the science team members. The science team must be able to discover and provision the resources required to execute their workflows and monitor the data generated by the workflow transparently. Since these workflows can be executed for making near real time decisions, some components are critical to ensure vital information is delivered in a timely fashion. Workflow management systems need to be able to provide rich monitoring and provenance information for a running workflow, an interface to steer the workflow dynamically, a resource management layer for elastic resource provisioning, and a policy-driven design for constructing resilient workflows.

Uninterrupted availability of data needs to be guaranteed. In order to support coupling of experiments with surrogate models across a distributed set of computational resources it is essential to build a workflow infrastructure capable of resiliently executing simulations with analysis and visualization and transferring the information in NRT. Resiliency policies centered around priority-based redundant computations are needed to ensure that a set of surrogate models and analysis services are coupled in near real time fashion in order to make timely decisions and control experiments. 

The ability to dynamically control and steer workflows needs to be provided. Runtime validation of experiments and simulations via dynamically spawning models and analysis tasks is needed for more efficient usage of 
experimental resources. Consequently, this requires computational resources to grow elastically when they are needed. A command-and-control interface is needed so that scientists can make informed decisions on the experiment in near real time. 

It is imperative to have a data streaming and management system capable of moving workflows and data efficiently through the distributed resources. Streaming methods impose several different challenges: the data sources could be many: large-scale experiments, such as the Large Hadron Collider, or the results of large-scale simulations; the data might need to be processed in real time and streamed directly to the data analysis processes completely bypassing the file systems; the data producer and the data consumer are often independent programs running on different nodes or systems geographically distributed. Ultimately, these tools and services will allow the sharing of machine data between the experimental analysis and computational simulations, which will allow scientists to steer their analysis using AI/ML-based surrogate models, helping them better use their time on the experiments. 

\noindent
[Ana Gainaru]

\subsection{Uncertainty Quantification}
%[Habib Najm, hnnajm@sandia.gov]}
% UQ text

Recent decades have seen increasing awareness of the role uncertainty quantification (UQ) can play in science and engineering. UQ encompasses developments in applied mathematics and statistics, including fundamental theory, numerical algorithms, and software tools, for the assessment of uncertainty in models of physical systems and their predictions. It is relevant across the board in the modeling of physical systems, including two essential elements, namely the inverse UQ problem, relevant in learning from experimental, observational, or computational data; and the forward UQ problem involving propagation of uncertainty from inputs to outputs of computational models. A range of other UQ activities build on these fundamental components, including hypothesis testing, model selection and validation, optimal experimental design, as well as robust optimization and control under uncertainty. In the following we present brief highlights of the state of the art and challenges in UQ, focusing primarily on the probabilistic UQ framework.

The primary goal of the inverse UQ problem is model calibration/fitting, or parameter estimation, accounting for data noise/uncertainties and model error, to arrive at learned uncertain model parameters/inputs. The probabilistic framework provides improved conditioning of the often notoriously ill-conditioned inverse problem. Further, specifically in the Bayesian inference context, the use of priors provides for additional regularization that is often indispensable. Nonetheless, the challenges of model complexity, computational cost, and high-dimensionality, have always been a considerable obstacle to the application of statistical inversion in large-scale computational models of physical systems. This is particularly true when, as is often the case, multiple challenges are present simultaneously. The forward UQ problem involves the propagation of uncertainty from model inputs to outputs. While not plagued with ill-conditioning as in the inverse problem, the forward UQ problem is similarly challenged with model complexity, cost, and high-dimensionality.

In order to facilitate inverse and forward UQ in relevant problems, considerable effort has targeted the development of surrogate models that, when fitted to represent the dependence of computational model observables or quantities-of-interest (QoIs) on parameters of interest, can be substituted for the original model. Surrogate models have been built employing a wide array of technologies. One approach, employing expansions in orthogonal basis functions, particularly polynomial chaos (PC)~\citep{Ghanem:1991,LeMaitre:2001b,LeMaitre:2002,Xiu:2002b,Xiu:2002c,Debusschere:2003d,Ghosh:2008} expansions, has been a considerable focus in forward UQ, the result of which is also precisely the surrogate needed for the inverse problem~\citep{Marzouk:2007,Marzouk:2009}. PC constructions have been fitted using generalized sparse-quadrature as well as regression methods, often relying on sparsification via compressive sensing~\citep{Candes:2006b,Candes:2007,Ji:2008,Moore:2012} when high-dimensional. Other surrogate constructions have employed interpolants~\citep{Barthelmann:2000,Xiu:2005,Babuska:2010,Narayan:2014,Stoyanov:2016}, low-rank tensors~\citep{Grasedyck:2013,Oseledets:2013,Gorodetsky:2018}, Gaussian processes~\citep{Williams:1996,Wilson:2011,Bilionis:2013}, and neural networks~\citep{Tripathy:2018,raissi2019physics,Fiorina:2020}. Moreover, multilevel/multifidelity methods have emerged as essential means to facilitate surrogate model constructions in high-cost computational models~\citep{Eldred:2017,Peherstorfer:2018,Fleeter:2020,West:2020}, allowing the use of model computations at varying degrees of fidelity/resolution, and hence cost, to achieve requisite surrogate accuracy at much-reduced cost.

Further, and specifically in the statistical inversion context, advanced Markov chain Monte Carlo (MCMC) methods have been developed to deal with complexity and high-dimensionality of posterior distributions~\citep{Haario:2001,Haario:2006,Solonen:2012,Byrne:2013}. Further, approximate Bayesian computation (ABC) methods have been developed to tackle expensive/intractable models/Likelihoods~\citep{Toni:2009,Peters:2012,Beaumont:2019}. Developments in this area have also pursued the design of reduced representations and distance metrics that address the complexity of model response, particularly in dynamical systems, to provide tractable dynamical observables and Likelihood loss functions that are smooth in parameters of interest while capturing essential dynamical features~\citep{Haario:2015,Craciunescu:2016}. Addressing high-dimensionality in Bayesian inference has also led to advances in the design of MCMC methods for infinite dimensional problems~\citep{Beskos:2017}, and in identifying lower-dimensional subspaces where data is in fact informative~\citep{Cui:2016,Cui:2021} and where MCMC random sampling can be focused. 

These and other developments have been documented in reviews/books~\citep{Ghanem:1991,Najm:2009a,Xiu:2010,Smith:2013,Ghanem:2017}, and deployed in open-source tools~\citep{DAKOTA:web,uqtk:web,Villa:2018}. Resulting capabilities have enabled the use of UQ methods in complex problems of physical relevance, including e.g. transport in porous media~\citep{Ghanem:1998,Ghanem:1998a}, seismic sensing~\citep{Crestel:2018}, fluid dynamics~\citep{LeMaitre:2001b,Lucor:2004b,Lucor:2008}, chemistry~\citep{Najm:2014a,Khalil:2016}, reacting flow~\cite{Hakim:2018}, and materials~\citep{Ganapathysubramanian:2008,Chen:2014}, spanning applications in geophysics, combustion, and climate. 

Despite these achievements, numerous open challenges remain in the practical use of UQ methods. High-dimensionality remains a universal challenge, particularly when combined with model cost and complexity. The identification of a sufficiently low-dimensional subspace of ``important" parameters, e.g. using global sensitivity analysis (GSA), is crucial for facilitating UQ in practical problems. Often, however, high-dimensionality is an inherent, irreducible challenge, e.g. when dealing with models where there is no such lower-dimensional important space. Examples include problems where non-smooth observables/QoIs are of interest, e.g. detailed turbulent motions or material fracture. They also include deep neural network (DNN) models, whose practical utility relies on their expressiveness that comes with the exceedingly large number of weight parameters. The DNN setting is also highly challenged by the remarkable complexity of the loss-surface, and the lack of informed priors. Even where there is a low-dimensional subspace of important parameters, however, GSA is in-itself a challenge with expensive models, particularly when dealing with models having discontinuities/bifurcations. Such artifacts are also generally problematic with surrogate construction. Other challenges include the estimation of probabilities of rare events, and the design of smooth, sufficiently informative, observables in dynamical systems. These challenges render subsequent ``outer loop" UQ activities doubly difficult. This includes e.g. model marginal-likelihood/evidence estimation for model selection purposes, as well as Bayesian optimal experimental design, and optimization/control under uncertainty. 
%\noindent\textit{Acknowledgement}: 
\paragraph*{Acknowledgements}
H.N. Najm was supported by the U.S. Department of Energy, Office of Science, Office of Advanced Scientific Computing Research, Scientific Discovery through Advanced Computing (SciDAC) program.  Sandia National Laboratories is a multimission laboratory managed and operated by National Technology and Engineering Solutions of Sandia, LLC., a wholly owned subsidiary of Honeywell International, Inc., for the U.S. Department of Energy’s National Nuclear Security Administration under contract DE-NA-0003525.

\noindent
[Habib N. Najm]

\subsection{Visualization and Data Understanding}
%[David Pugmire]}
One of the primary challenges facing scientists is extracting understanding from the large amounts of data produced by simulations, experiments, and observational facilities. The use of data across the entire lifetime ranging from real-time to post-hoc analysis is complex and varied, typically requiring a collaborative effort across multiple teams of scientists.
The rapid growth in the rate and size of data generated at these facilities is make a challenging task of gaining understanding even more difficult.

As this complexity has grown, scientists have relied on complex workflows to orchestrate the collection and/or generation data as well as the processing and movement throughout the lifetime of a scientific campaign.
To be truly useful, analysis and visualization tasks must be able to access data in a variety of different ways in these complex workflows. These range from traditional post-hoc visualization where data are accessed from disk to in situ visualization where the data are accessed as it is being generated. Visualization tasks must be able to integrate in a robust manner into these autonomous workflows and be dynamically controlled.

Fusion science requires a number of different techniques for gaining understanding from the data being generated.
These include both 2D and 3D visualization.
In simulatations, because of the nature of fusion science, a large amount of analysis can be done using two dimensional slices through the simulation mesh. These 2D slices can be colored by different quantities in the simulation to describe the science. These slices typically provide a summary view of the fusion around the tokamak.
An example would be the averaged value in the torroidal direction at each point in the plane.
The time evolution of these visualizations can provide valuable insight into behavior of the simulations.
The use of 3D visualization can be used to illustrate the changing complexity of the physics in the torroidal direction. These provide valuable insight into how features in the plasma vary torroidally.
Another important type of visualization is that of the particles used in the simulations. Typically, there is a very large number of particles and only particluar types of particles are of interest. These include particles that become trapped in the plasma or travel to particular regions in the plasma.

One critical aspect of visualization involves derived quantities. These include regions of relatively high energy (called blobs) that can develop in the plasma and move around as the plasma evoloves. A more complex example is the generation of Poincare plots. Poincare plots are used for the analysis of the magnetic field in the plasma.  These are created by advecting a large number of particles around the tokamak and marking the punctures that each particle makes with a plane normal to the torroidal direction. These puncture patterns provide valuable information about the evolution of the plasma.

The visualization of experimental data typically involves 1D and 2D visualizations that evolove over time. The 1D visualizations are typically of time varying curve plots. One common source of 2D data are cameras that operate at very high speeds and are growing in resolution. Feature detection and tracking is one key type of visualization performed for 2D data.

Currently, there are a number of tools that can be used for analysis and visualization.  These include production tools like VisIt~\cite{visit} and ParaView~\cite{paraview} that provide a powerful set of tools for creating visualizations of data.  These tools can be used for post hoc processing as well as in situ processing using LibSim~\cite{libsim} and Catalyst~\cite{catalyst}.
A service-based approach to analysis and visualization~\cite{vaas} builds upon a hardware-portable visualization toolkit (VTKm)~\cite{VTKm}, and a data model (Fides)~\cite{fides} that can be integrated into automated workflows like EFFIS.
The serviced-based approach supports both post-hoc and in situ processing and provides the flexibility for workflow systems to schedule analysis and visualization tasks as they are needed.
To support collaboration between teams of scientists, a web-based dashboard, eSimMon~\cite{esimmon} allows scientists to see in near-real time visualizations of different variables and quantities of interest.

\paragraph{R\&D Necessary for the future}
The emergence of computing ecosystems that couple experiments, simulations and surrogate models, and reliance on streaming data will require significant work for visualization to continue as a critical aid to gaining understanding from data.
This increased complexity will require the use of automated workflows to compose and orchestrate the set of tasks required to do the science.
The resources available to perform the analysis and visualization will dynamically change over the course of a scientific campaign.
Additionally, the resource requirements for visualization will vary as well. Visualization algorithms have different scaling characteristics depending on how they are run.  Recent work has explored cost models for the placement of tasks using different placement strategies~\cite{Kress-2019}~\cite{Kress2020-cost}~\cite{Kress2020-time}.
The rapid increase and size and rate of data are also requiring the use of data reduction techniques. Analysis and visualization tasks must be able to adapt to the uncertainties introduced by data reduction. In order to provide trustable visualization from reduced data, the uncertainity must be conveyed to the scientists. The uncertainity will come from the raw data as well as the algorithm that is producing the visualization. If the uncertainites are too high, the algorithms must be smart enough to request additional data that will lower the errors and/or use different algorithms with higher accuracy.  This of course can require additional resources in order to compute accurate results. As such, there must be an integration with automated workflows in order to ensure that enough data is used with the proper algorithm running on the right amount of resource.
Solutions to these challenges will require visualization to integrate well into the controlling workflows. This includes abstractions for access to data, the ability to be composable and schemas to describe the underlying streamed data. Platform portability will be required for placement across a wide range of computing devices. Performance models will be required so that visualization tasks can be placed on the proper resources with access to the proper accuracy of data. The development and use of smart dashboards, where scientists can see the current status of a simulation or experiment will make it possible to for teams to efficiently collaborate. These dashboards should be customizable by each scientist.
AI can be used in these dashboards to learn the interaction patterns of scientists to ensure that the most relevant visualizations are displayed, features of interest are highlighted and anomalies are highlighted.

\noindent
[David Pugmire]

%Federated systems. streaming data. coupled simulations, coupled experiments.
%Complex workflows.
%Reduced data and uncertainty. how to preserve QoI, especially for complex analysis.
%smart dashboards that allow collaboration.

%\begin{itemize}
%\item Data problem: Growing. particle and field data. Understanding of complexity aided by visualization.
%\item Current techniques: slices, 3d, volume, particle vis. derived quantities (blobs), poincare. Federated makes this more complex. Code coupling and vis. in situ, in transit, dashboards. Tools that can be used. Integration of vis with other technologies (adios, effis, etc)
%\item Future: federated systems and streaming data. Vis Portability. Vis of reduced data. how to preserve QoI, especially in something as complex as poincare analysis.
%  integration into automoated workflows, better dashboards, smart dashboards.
%\end{itemize}

%State of the art:
%\begin{itemize}
%    \item State of the art and what is available
%    \item What is missing for the future
%    \item Do not constrain to what is available to plasmas right now
%    \item Draft due at the end of November
%\end{itemize}

\subsection{ML Control Theory}

% Title page
%\title{ML Control Theory}

% Author list with affiliations
%\author{Ali Mesbah}
%\affiliation{Department of Chemical and Biomolecular Engineering, University of California Berkeley}

%\maketitle

%{\color{red}The first paragraph should be the introduction paragraph, which should be about 200 words (if necessary up to 350 words). In this paragraph, please discuss the background, motivation, importance, research history, etc., of the topic to be discussed in this subsection.} 

Historically, control of plasma processes was based on statistical process control approaches, which are open-loop in nature and are merely suitable for monitoring the process performance. Recent years have witnessed a growing interest in model-based feedback control approaches for confinement fusion reactors and low-temperature plasma processes to cope with intrinsic variabilities of plasmas and exogenous process disturbances. To this end, model predictive control (MPC) \cite{rawlings2017model}, which relies on real-time optimization and is the prime methodology for constrained control, has emerged as a promising advanced control technology for plasma processes (e.g., \cite{maljaars2015control,gidon2017effective}). This stems from the ability of MPC to handle the highly nonlinear and multivariable plasma dynamics and to explicitly account for constraints on process variables, which is crucial for safety-critical plasma applications. However, the conventional MPC paradigm follows strict separation of a design phase, which mainly involves model development and controller tuning using offline data, and a closed-loop implementation phase, during which the controller remains largely intact. Such a controller design strategy can limit the MPC performance for plasma processes whose complex dynamics can span over multiple length- and time-scales. Recent advances in the field of ML, along with enhanced computational, sensing and communication capabilities, have created ample opportunities for safe learning-based control of the hard-to-model behavior of plasma processes at exceedingly fast sampling rates.           

\begin{figure*}[t!]
    \centering
    \includegraphics[width=0.64\textwidth]{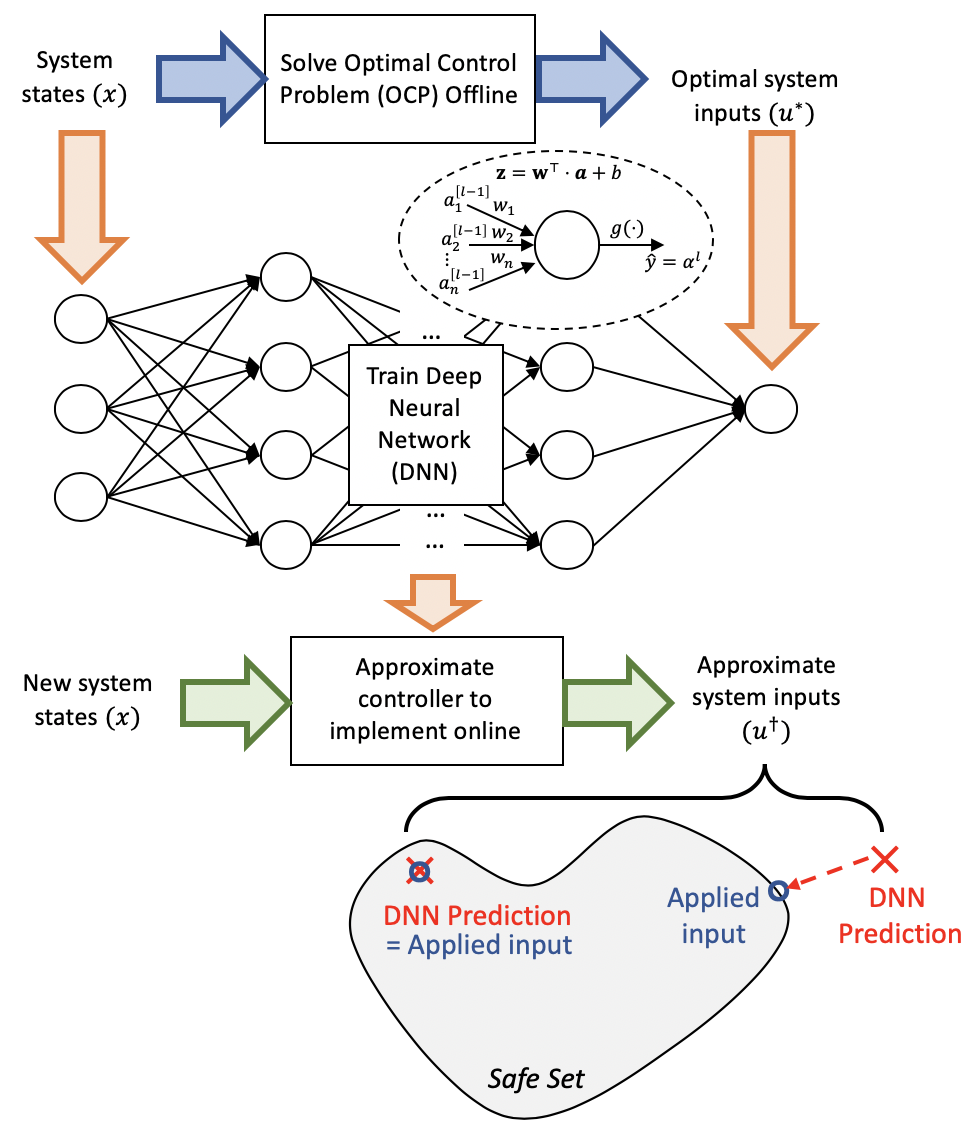}
    \caption{Deep neural networks can effectively approximate optimization-based control laws with a cheap-to-evaluate explicit control law that has low memory requirements. To guarantee satisfaction of safety-critical constraints of plasma processes in the presence of approximation errors and system uncertainties, the control inputs computed by the neural network can be projected onto a safe input set that is constructed using the notion of robust invariant sets. Safe neural network-based controllers can play a pivotal role for control of fast-sampling plasma processes using resource-limited embedded control hardware \cite{bonzanini2020toward}.}
    \label{fig_Mesbah_OCP_Flowchart}
\end{figure*}

Data-driven methods can aid in the design of MPC approaches for plasma processes in two primary ways. \textbf{(i) Learning the system dynamics:} The performance of MPC is heavily dependent on using a suitable and sufficiently accurate model representation of the system dynamics.  %Thus, much of the research in learning-based MPC has focused on automatically improving the model quality \cite{}. 
ML has shown great success for deriving data-driven, multivariable representations of complex system dynamics that are amenable to real-time optimization and control. Data-driven models can embed varying degrees of physics-based knowledge of a process. In the absence of theoretical plasma models, control-oriented models can be readily learned from data that are collected offline \cite{wan2021experiment,gidon2021data}. Alternatively, when theoretical plasma models are available, surrogate modeling, in which dynamic models are trained based on high-fidelity simulation data \cite{dong2021deep,kaptanoglu2021physics}, has proven useful for deriving computationally efficient models suitable for control. Yet, an emerging approach to learning-based MPC is to combine a prior model (data driven or physics-based), which represents our available system knowledge, with a learning-based model that is adapted in real-time \cite{hewing2020learning,lin2020introducing,bonzanini2021learning}. Such a learning-based modeling scheme is particularly useful for capturing the hard-to-model and time-varying nature of the plasma behavior when it cannot be captured \emph{a priori} via offline data or high-fidelity simulation data. To this end, Gaussian process regression has proven especially useful for not only learning the unmodeled system dynamics, but also characterizing the uncertainty of model predictions, which can be incorporated into the MPC design to robustify control actions with respect to uncertainties \cite{bonzanini2021learning}. \textbf{(ii) Learning the MPC law:} Another important research direction in ML for MPC focuses on learning the control law, as opposed to a prediction model. MPC relies on online solution of often a nonlinear optimization problem that can be computationally prohibitive for real-time control of fast sampling systems. This can especially be the case when sophisticated process models are used for MPC, or when the objective is to control the fast process dynamics that would hinge on fast measurement sampling frequencies on the order of KHz to MHz or even possibly faster. ML has proven useful for developing so-called \emph{approximate} MPC approaches that learn a cheap-to-evaluate, explicit expression for the MPC law using data generated from offline solution of an MPC problem \cite{bonzanini2020toward}; see Figure~\ref{fig_Mesbah_OCP_Flowchart}. A variety of function approximators, ranging from polynomials to deep neural networks, have shown promise for approximating optimization-based control laws with surrogates that can be evaluated on fast sampling times. The resulting low-complexity controllers typically exhibit a limited memory footprint, which makes them particularly suitable for implementations via resource-limited (i.e., low power and memory) embedded control systems \cite{lucia2016predictive}. ML can also be used to learn other components of an MPC formulation, such as the control cost function, directly from data, as discussed in \cite{hewing2020learning} and the references therein.   

A largely open area of research in learning-based MPC is how to confer an \emph{active learning} mechanism to a controller to simultaneously explore and exploit the system dynamics towards actively mitigating the model uncertainty. To this end, there has been significant interest in leveraging reinforcement learning \cite{sutton2018reinforcement} and Bayesian optimization \cite{shahriari2015taking} methodologies to design learning-based controllers. These methodologies will allow us to combine learning and feedback policy design into a unified framework that provides a ‘self-optimizing’ feature via systematically balancing learning (i.e., exploration) and feedback control (i.e., exploitation) of an uncertain system \cite{recht2019tour}. Another crucial consideration in learning-based control is to ensure safe learning of the unknown and hard-to-model process behavior. In particular, it is imperative to guarantee safe operation of safety-critical plasma processes despite uncertainties in models and variabilities in the process itself. In general, safety guarantees for learning-based controllers can be established by decoupling  optimization of the control objective function and requirements of constraint satisfaction \cite{hewing2020learning}. Nonetheless, safe learning-based control, particularly for controllers with an active learning mechanism, remains an open and active area of research.

\noindent
[Ali Mesbah]
%, University of California Berkeley]

\section{Basic Plasma Physics and Laboratory Experiments}
%[Coordinator:Wang] 

\subsection{Introduction}
%[Wang, zwang@lanl.gov]}
Recent advances in data science and data methods, together with the continued decline in the cost of computing hardware and data acquisition instruments, are not only reinforcing the traditional roles of experimental and observational data, but also rapidly changing how the data are used in interpretation, prediction, and optimization problems, as illustrated in Table~\ref{DataTb}. We may distinguish small data sets as ones that can be effectively handled manually, and big data sets as ones that are impossible to be processed manually and therefore mandate automated computer techniques. %The data-driven approaches, is sometimes called a new paradigm shift from equation, computations. 

\begin{table*}[!ht]
%% increase table row spacing, adjust to taste
\renewcommand{\arraystretch}{1.1}
% if using array.sty, it might be a good idea to tweak the value of
% \extrarowheight as needed to properly center the text within the cells
\caption{Evolving approaches to and roles of experimental data in Plasma Physics}
\label{DataTb}
\centering
%% Some packages, such as MDW tools, offer better commands for making tables
%% than the plain LaTeX2e tabular which is used here.
\begin{tabular}{|c|c|c|}
\hline
{\bf Application/workflow} & {\bf Small data} & {\bf Big data} \\
\hline
Processing/analysis & manual & computer automation \\
Interpretation & incomplete & self-sufficient \\
Prediction & No/limited & Yes \\
Optimization & No/limited & Yes \\
Control & Maybe & Yes \\
\hline
\end{tabular}
\end{table*}

Interpretation, prediction, optimization and control are too sophisticated for traditional computer programs, which can only repeat the same pre-programmed tasks or workflows without glitches. Machine learning methods such as deep learning bring several important features that are missing from traditional computer programs. Deep learning or multilayer neural networks, which mimic the way the network of neurons in human brain processes information, has the ability to learn without being explicitly programmed. Machine learning process is equivalent to tuning a large number of `weight' parameters associated with neurons. The same neural network architecture can also be re-programmed or `re-trained' for different datasets or multi-tasks. In other words, a neural network developed for material science, biology or even outside natural science can be adopted to solve plasma physics problems.  As another example, neural networks provide a new tool for the fast solution of repetitive nonlinear curve fitting problems encountered in experimental data fitting~\cite{Bishop:1992}. 

Neural networks continue to grow in size and architectural variety from thousands to more than one billion of simple computational units or the `artificial neurons'. In comparison, there are about 100 billion neurons in a human brain. Most parameters of a neural network called weights are determined during the training process. Initial network configurations such as the connections between different neurons must be set manually (a growing number of libraries and pre-designed neural network architectures are now available through open sources such as github). These starting states of a neural network are called hyperparameters, which also include variables like the orders of neuron connections, type of nonlinear truncation functions, the number of layers, with or without loops.

\begin{figure*}[!t]
\centering
\includegraphics[width=0.8\linewidth]{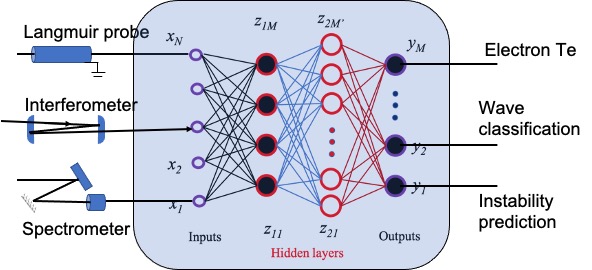}
% where an .eps filename suffix will be assumed under latex, 
% and a .pdf suffix will be assumed for pdflatex; or what has been declared
% via \DeclareGraphicsExtensions.
\caption{A generic approach, which is in-dependent of the hardware details, to construct data-enhanced instrument.  A Langmuir probe, a single-channel interferometer, and one line-of-sight of spectroscopy are shown. The output of such a synthetic instrument, such as electron temperature ($T_e$) measurement, or a binary prediction about the on-set of an instability, is significantly enhanced by the neural network, including noise reduction, frequency retuning. A virtual instrument with multiple inputs are also possible through multi-stream data fusion.}
\label{fig:instrum}
\end{figure*}

Laboratory plasmas also provide rich experimental datasets and data varieties to test and develop machine learning methods, or a plasma-trained `artificial brain', that can potentially benefit other scientific fields. Laboratory plasmas are extremely diverse, ranging from microplasmas produced by short-pulse lasers, to table-top experiments, and to the nearly 30-meter-tall ITER experiment. In spite of the difference in plasma density, temperature, data-driven methods are generic for interpretation, prediction and control problems. A generic approach, which is in-dependent of the hardware details, to construct data-enhanced instrument, is illustrated in Fig.~\ref{fig:instrum}. Below we shall highlight the applications of different data methods in laboratory experiments as illustrative examples. %Furthermore, multiple factors have contribute to the growing popularity of data-driven methods in laboratory plasma experiments. %Data science offers a new
\noindent
\noindent
[Zhehui Wang]
\subsection{Spectroscopy, imaging and tomography}
%[Wang, zwang@lanl.gov]}
Optical, UV, and X-ray spectroscopy are widely used for plasma density ($n_e$), electron (T$_e$), ion (T$_i$), neutral atom ($T_g$) temperature, and impurity measurements. Passive spectroscopy using plasma self-emission is preferable over Langmuir probes for a number of reasons. However, data analysis to retrieve $T_e$ information may be more complicated than a probe measurement. For ion temperature $T_i$, Doppler broadening of ion line width may be used. For $T_e$ measurement,  one common approach is to use intensity ratios  of multiple emission lines, which may be measured using several line-filtered photo-multiplier tubes (PMT)~\cite{Wang:2002}, or a 1D array of photodetectors, or a spectrometer with an imaging camera. For local thermal equilibrium with a temperature $T_e$, if a pair of emission lines originate from the same ground state of an atom or an ion, and the excited states are mostly empty, the line ratios give the value that is proportional to $e^{-\Delta E / kT_e}$. Similar physics-motivated analytical and empirical formula can be derived for other diagnostics including Langmuir probes~\cite{Chalanturnyk:2019}, which usually forms the basis of many instrumentation data interpretation. Neural network can replace such formula and represent much more complicated correlation between a measurement and interpretation. 

Multi-chord spectroscopy, and similarly multi-chord interferometers, reflectometers, can be used to obtain the two-dimensional (2D) profile distribution of the plasma emissions through inversion algorithms. Neural networks have been implemented to reconstruct electron temperature profiles from
multi-energy soft-x-ray arrays and other plasma diagnostics with fast time
resolution~\cite{Clayton:2013}. By training a three-layer fully-connected feedforward neural
network to match fast ($>$10 kHz) X-ray data with Te profiles from Thomson scattering, the
multi-energy soft-x-ray diagnostic can be used to produce $T_e$ profiles with high time resolution. The typical network inputs nodes for soft X-ray signals were up to 20. The number of output nodes for $T_e$ was comparable to the number of inputs. The hidden layer nodes was about 40. A sigmoid activation function, in the form of the
logistic function $f (x) = 1/(1+e^{-x})$ was used to sum up the inputs of each hidden nodes. Adding spectroscopic data as inputs was found to decrease the rms error of the temperature predictions by as much as 50\%. Multi-chord bolometers with the photon radiation absorption range of 2.5 eV–10 keV was reported.

A feedforward fully connected neural network approach has been implemented to measure the electron temperature directly from the EUV/VUV emission spectra (photon wavelength in the range of 50 to 160 nm) of the divertor region of the DIII-D tokamak plasma~\cite{Samuell:2020} [2020]. The plasma temperature is below 100 eV in the region. The best performing neural network had 12 hidden layers of 12 neurons that were sandwiched between a 1000 element input vector (the spectra) and the single output node ($T_e$). Each neuron in the model used an exponential linear unit (ELU) activation function with the exception of the final output neuron which does not have an activation function so that it can take on any value. A Nadam optimizer was used to calculate the changes to the model weights. The Python model construction and training were handled with
the Tensorflow. The full dataset consisted of 1865 input (spectrum time slices) and output pairs, of which 25\% are reserved for evaluation. The rest of the 75\% were further splitted to a 3:1 ratio for training and training assessment.

In addition to tunable and higher emission intensities, laser-based spectroscopy such as Thomson scattering and charged particle spectroscopy such as CHERS can overcome limitation associated with the passive spectroscopy from a plasma emission. Passive spectroscopy gives the line-averaged information along the line-of-sight.  Laser and particle beam techniques can localize the temperature and density in space and time. A three-layer (one hidden layer with eight nodes) neural network approach was used to calculate the electron temperature in Thomson scattering diagnostics, replacing the traditional $\chi^2$ method~\cite{Lee:2016}. One of the main advantages of the neural network was to speed up the data processing time by almost 20 times over $\chi^2$ method. Neural network has also been used to speed up the analysis of collective Thomson scattering (CTS) data~\cite{Berg:2018}. As a result of scattering by fluctuations in the electron density, electric field, magnetic field, and current density, CTS has been used to diagnose ion temperature and fast ion velocity distribution~\cite{Korsholm:2010}. Recovery of $T_i$ and $T_e$ from CTS usually requires time-consuming simulations to produce synthetic spectra from a set of input parameters including $T_e$ and $T_i$. A feedforward artificial neural network with three hidden layers were implemented with SciKit-Learn~\cite{Berg:2018}. The $T_i$ mapping error was less than 5\%.

\noindent
[Zhehui Wang]
\subsection{Sparse measurement and noise}
% In your section on Image analysis, tomography and inverse problem using machine learning and data science, you may include some introduction/background/earlier work, the state-of-the-art, and future perspectives. Regarding the future perspectives, we discussed possibilities such as physics-motivated priors, small experimental data. Another possibility is to weave multiple measuremnts (X-rays with visible light, imaging with single probes, etc) into a `super-instrument'. Including a  figure in your section would be nice.

%\section{Sparse measurements}
The problem of image reconstruction from a small number of measurements or sparse measurement commonly arises in  plasma experiments as well as in computational, medical, and scientific imaging \cite{candes2011compressed, donoho2006compressed, candes2005decoding}. Even by using multi-chord configurations and detector arrays, measurement of a plasma such as through spectroscopy is intrinsically sparse. The number of chords is limited by the real estate and viewing ports around a plasma device. The number of photons recorded is limited by the plasma emissivity, or the laser power in terms of the scattering experiments, and the time duration of signal integration. The detectors have a finite spatial and temporal resolution. The electronics have a finite bandwidth and a finite sampling rate. Even though the state-of-the-art oscilloscopes have now the impressive bandwidth of tens of GHz, it may still be insufficient for example in ultrafast plasma experiments. In short, Shannon's information theory, which requires that the sampling rate or the Nyquist rate must be at least twice the maximum frequency in the signal, can be too restrictive for experiments. Similarly, the mathematical formulation of the inversion algorithms such as the Radon inverse transform also assume a large number of projections, which may not be practical in experiments. In situations when the Nyquist rate are achievable, the volume of data generated may be too large and can result in transmission, storage, and processing challenges. 

The ubiquitous presence of noise can further complicate 2D profile reconstructions from multi-chord line-integrated measurements, for both traditional methods and machine learning~\cite{Kalapanidas:2003}. A related problem in tomography is to use as few as a single 2D projection to reconstruct 3D volumes~\cite{Wolfe:2021}. Noise is probably the hardest type of signal to reproduce based on physics and first principles because of its seemingly random nature. Noise of different origins is present along the full chain of signal generation, propagation, and registration (digitization). For spectroscopy measurements, the limited amount of light can appear as Poisson noise. Background noise is spectral-dependent and experiment-specific, {\it i.e.} the background light for an optical spectroscopy is different from a soft X-ray spectroscopy due to, for example, different geometry of the setup, field of view or the solid angle of the light collection, surface reflectivity, the light path setup that may be susceptible to atmospheric turbulence. Electronic noise, which has different sources by itself, is also not avoidable. Yet noise removal and reduction are needed for any measurements. Noise may be described statistically using a Gaussian model and its variance. Neural network and machine learning for noise classification, denoising, even noise modeling is of growing interest~\cite{Fujii:2017}.

Compressed sensing or compressive sampling principle has emerged as a new framework for data acquisition, detector designs, and signal processing including inversion problems~\cite{ donoho2006compressed, candes2005decoding,Elad:2010}. Compressed sensing spectral imaging system was reported for plasma optical emission spectroscopy~\cite{Usala:2016}.  A single PMT detector and a variable encoding mask (a digital micromirror device) are designed and implemented for measurement of molecular and ion vibrational temperature. In other examples, compressed sensing was used to decompose emission spectra from an extended plasma source such as the Sun~\cite{Cheung:2019}. A combination of compressed sensing and machine learning led to dimensionality reduction so the flow properties such as Reynolds number, pressure and flow field can be obtained from a sparse pressure measurements~\cite{BLK:2013}. More recently, compressed sensing framework is implemented in variational autoencoder and generative adversarial networks~\cite{Bora:2017}.  The method can use 5-10 times
fewer measurements than Lasso for the same accuracy. A canonical imaging system can be represented as 
\begin{equation}
y = Ax + e, 
\end{equation}
where $y\in \mathbb{R}^m$ represents sensor measurements, $x\in \mathbb{R}^n$ represents the unkown image, $A$ represents an imaging operator, and $\eta$ denotes noise in the measurements. 
The problem of reconstructing $x$ from $y$ is underdetermined if $m < n$, and we need to use some prior knowledge about the signal structure. Classical signal priors exploit sparse and low-rank structures in images and videos for their reconstruction \cite{duarte2008single,baraniuk2007compressive, yang2013adaptive,shi2012video,zhao2017video,christopoulos2000jpeg2000,puri1998mpeg,sullivan2004h,wallace1992jpeg,asif2013low,li2017structured,dai2017fully,mota2016adaptive}. However, the natural images exhibits far richer nonlinear structures than sparsity alone. 

A recent trend is to use data-driven methods, mainly based on deep learning and neural networks, to perform image reconstruction. Deep learning-based methods can be broadly divided into the following categories: 
\begin{enumerate}
    \item End-to-end networks that are trained to map the sensor measurements onto the desired images \cite{chen2018fsrnet}. 
    \item Learned neural networks that are used as denoisers plug-and-play priors during the recovery process \cite{venkatakrishnan2013plug}. 
    \item Trained generative networks that are used as priors for natural images~\cite{bora2017compressed}. 
    \item Untrained networks that are learned while performing image reconstruction~\cite{Ulyanov2017DeepIP, heckel2018deep}.
\end{enumerate}

Below we highlight techniques that use \textit{pre-trained} or \textit{untrained} networks within an optimization algorithm in order to leverage the information from both the data-acquisition model and the learned prior. This is a rapidly evolving research area with a number of recent theoretical and practical developments  \cite{Iliadis2016DeepFN,lin2016deep,pan2016hierarchical,santurkar2018generative,srivastava2015unsupervised,shrivastava2017learning}.

{\it Generative Models as Image Priors}
Deep network-based generative models have emerged as useful image priors in recent years. In a nutshell, a deep generative model represents a function $G(\cdot)$ that maps a low-dimensional, latent vector $z$ into a high-dimensional image as $x = G(z)$ \cite{bora2017compressed,hyder2020generative}. 
% Generative adversarial networks (GANs) and variational autoencoders (VAEs) are two popular classes of such generative networks  \cite{goodfellow2014generative,kingma2013auto,hinton2006fast,bengio2013representation}. 
%
The weights of the generative network and the distribution of the latent vectors can either be learned using training images or the generative network can be learned while solving the image recovery problem. 

\begin{figure*}[t]
\centering
\includegraphics[width=1\linewidth]{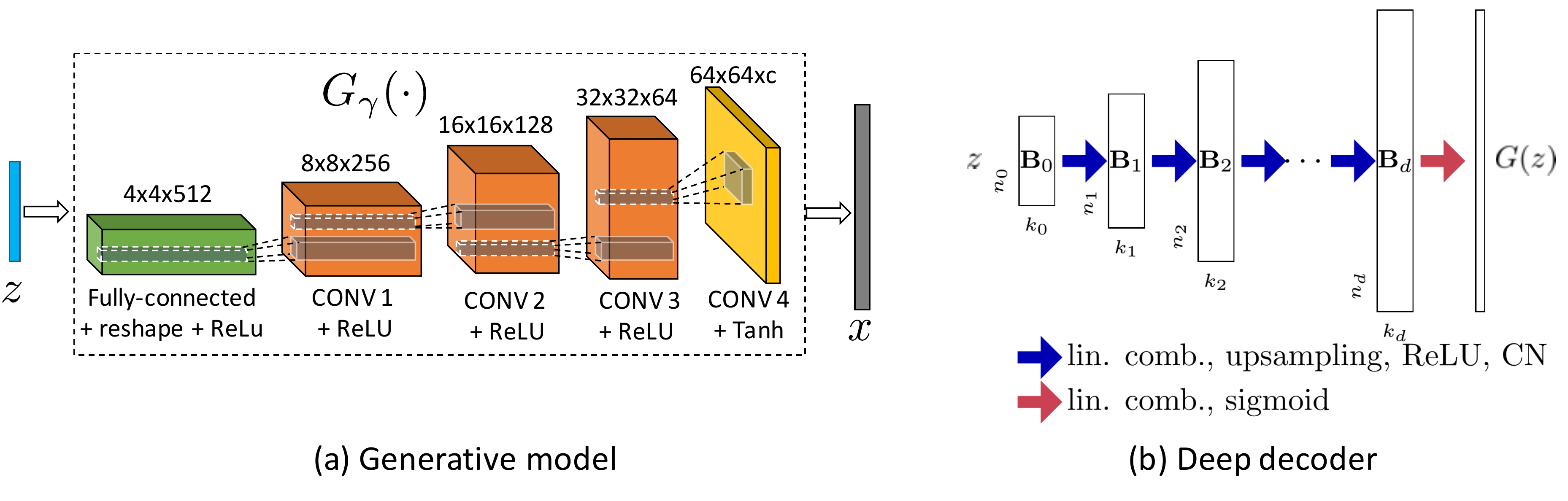}
\caption{Examples of neural networks commonly used as image priors. (a) DCGAN \cite{radford2015unsupervised} architecture that maps a low-dimensional latent vector $z$ into an image as $x = G(z)$. (b) Deep decoder \cite{heckel2018deep} architecture that uses an untrained network as image prior. }\label{fig:generators}
\end{figure*}

Let us denote a generative model as 
\begin{equation}\label{eq:genModel}
    x = G_\gamma(z) \equiv g_{\gamma_L}\circ g_{\gamma_{L-1}} \circ \cdots \circ g_{\gamma_1}(z).
\end{equation}
$G_\gamma(z)$ denotes the overall function for the deep network with $L$ layers that maps a low-dimensional (latent) code $z \in \mathbb{R}^k$ into an image $z \in \mathbb{R}^n$ and $\gamma = \{\gamma_1,\ldots, \gamma_L\}$ represents all the trainable parameters of the deep network. $G_\gamma(\cdot)$ as given in \eqref{eq:genModel} can be viewed as a cascade of $L$ functions $g_{\gamma_l}$ for $l=1,\ldots,L$, each of which represents a mapping between input and output of the respective layer. 
Figure~\ref{fig:generators}(a) illustrates a generative network based on DCGAN architecture that is usually used as an image prior \cite{radford2015unsupervised,hyder2020generative}. 
Some other commonly used generator architectures include U-net \cite{ronneberger2015u} and deep decoder \cite{heckel2018deep} as shown in Figure~\ref{fig:generators}(b). 
To recover an image using generative models as image priors, we can either use a trained or an untrained network. We briefly discuss both approaches below. 

{\it Trained network as an image prior.} 
A number of papers have recently explored the idea of replacing the classical (hand-designed) signal priors with deep generative priors for solving inverse problems \cite{bora2017compressed,hand2016compressed,ulyanov2018deep,van2018compressed}. 
Recovery of an image using trained generative model ($G_\gamma(\cdot)$) can be formulated as the recovery of the latent code ($z$). To learn latent representation of an image with respect to a generator, we often need to solve a nonlinear problem \cite{creswell2018inverting,Bojanowski2018Optimizing, lipton2017precise, zhu2016generative}. Given a pretrained generator $G_\gamma$, measurements $y$, and the measurement operator $A$, we can solve the following optimization problem to recover the low-dimensional latent code: 
\begin{equation}
      \underset{z}{\text{minimize}}\; \|y-A G_\gamma(z)\|_2^2. \label{eq:latentOpt}
\end{equation}
The reconstructed image can be computed as $\hat x = G_\gamma(\hat z)$, where  $\hat z$ denotes the solution of the problem in \eqref{eq:latentOpt}. We can solve \eqref{eq:latentOpt} using a gradient descent-based method that iteratively updates $z$ to minimize the objective function. The gradient of the objective function in \eqref{eq:latentOpt} with respect to $z$ can be computed using backpropagation. This approach is employed in \cite{bora2017compressed,creswell2018inverting,hand2018phase}. An alternative approach is to solve the following (nonlinear) projection based method \cite{shah2018solving,hyder2019alternating}:  
\begin{equation}
      \underset{z,x}{\text{minimize}}\; \|y-A x\|_2^2 \;\; \text{subject to} \;\; x = G_\gamma(z), \label{eq:latentOpt2}
\end{equation}
where we alternately update $x$ via gradient descent and project the estimate onto the range of the generator $G_\gamma(\cdot)$. 
% This projected gradient descent based approach  is described in Algorithm~\ref{chap1:mbir:lbir:alg:gen_pgd}. 

{\it Untrained network as an image prior.} 
Trained networks serve as good image priors, but they require a large number of training samples, which limits their use in settings with limited data. Furthermore, trained generators can only correctly recover images that are close to the training samples. 
In recent years, a number of methods have shown that untrained networks can also be used as image priors \cite{ulyanov2018deep,heckel2018deep}.
Deep image prior method in \cite{ulyanov2018deep} first showed that an over-parameterized network can be trained to generate natural images by early stopping. This observation led to the use of untrained generative models as image priors for solving different inverse problems \cite{ulyanov2018deep,heckel2018deep,van2018compressed,jagatap2019algorithmic,mataev2019deepred,hyder2020generative}. 
A number of theoretical results have also appeared recently that provide conditions under which an untrained network can solve different inverse problems \cite{jagatap2019algorithmic,heckel2020compressive,darestani2020can}. In practice, untrained networks perform almost as good as trained generative networks when the test data lies in the range of the trained generators. Untrained networks perform better than trained networks when the test data do not fall in the range of the trained networks. 

Untrained generative prior is free from limitations as we use random weights to initialize the network and update the weights as we go along. However, it is natural to question the theoretical validity of such priors. 

% Compressive sensing using generative models is first introduced in \cite{bora2017compressed}, which use a trained deep generative network as a prior for image reconstruction from compressive measurements; the reconstruction problem involves optimization over the latent code of the generator. Since the generator is fixed, this approach works well only if the unknown image/video belongs to the range of the generator used. Deep image prior (DIP) is a related method in which an untrained convolutional generative model is used as a prior for solving inverse problems such as inpainting and denoising because of their tendency to generate natural images \cite{Ulyanov2017DeepIP}; the reconstruction problem involves optimization of generator network parameters. Inspired by these observations, a number of methods have been proposed for solving compressive sensing problem by optimizing generator network weights while keeping the latent code fixed at a random value \cite{heckel2018deep,van2018compressed}. Both DIP \cite{Ulyanov2017DeepIP} and deep decoder \cite{heckel2018deep} update the network parameters to generate a given image; therefore, the generator can reconstruct wide range of images.

%\section*{Future Directions}
%% we discussed possibilities such as physics-motivated priors, small experimental data. Another possibility is to weave multiple measuremnts (X-rays with visible light, imaging with single probes, etc) into a `super-instrument'. 

%TBD 

%\bibliography{refs}
%\bibliographystyle{IEEEtran}
%\end{document}

\noindent
[M. Salman Asif and Zhehui Wang]

\subsection{Synthetic instruments and data}
\begin{figure*}[!t]
\centering
\includegraphics[width=0.4\linewidth]{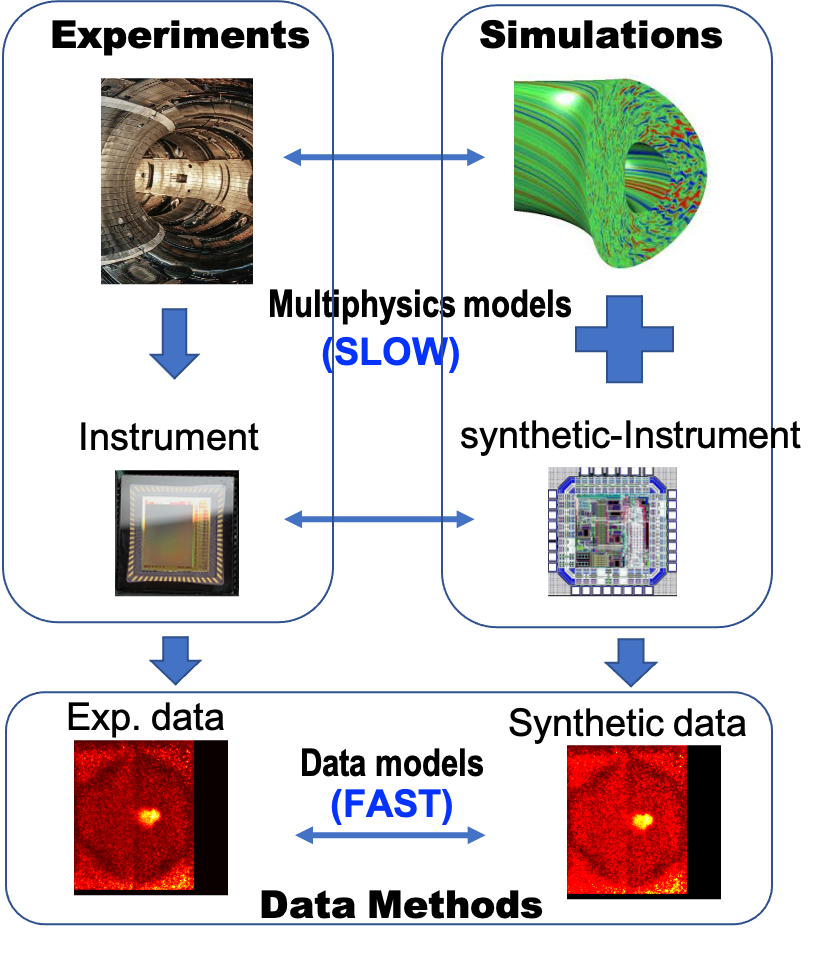}
% where an .eps filename suffix will be assumed under latex, 
% and a .pdf suffix will be assumed for pdflatex; or what has been declared
% via \DeclareGraphicsExtensions.
\caption{Data methods including machine learning motivate development of synthetic instruments, and allow direct and faster interpretation of complex experimental data. Ability to extract information faster from measurements is particularly important for real-time control of plasma experiments.}
\label{fig:syni}
\end{figure*}

Multiphysics simulation tools are now available to design and simulate plasma experiments, up to the full-scale experiments in realistic geometries~\cite{NTP:1994}. Such tools have been adopted for modeling of plasma instruments and data interpretation. The multiphysics model for an instrument is sometimes called a synthetic diagnostic~\cite{Holland:2009,Shi:2016,Kukus:2016}. In parallel to hardware-based instruments for diagnosis of a real plasma, a synthetic diagnostic can be regarded as a numerical instrument for diagnosis of a numerical model of a plasma, as illustrated in Fig.~\ref{fig:syni}. Due to the complexity of the plasmas and instrumental responses, synthetic diagnostics are indispensable for quantitative interpretation of the experimental data from a physical instrument, and for comparison of the experimental data with plasma simulations~\cite{Shi:2016}. As mentioned above, spectroscopy, tomography, interferometry and others such as electron cyclotron emission imaging, millimeter-wave imaging reflectometry are usually line-of-sight or volume integrated and time integrated, while plasma simulations usually give physical quantities such as temperature and density as a function of position and time. The synthetic data generated from a synthetic instrument can be flexibly converted into both the experimental and simulation formats. Another  function  of  synthetic  diagnostics  is  to quantify  uncertainties  and  sensitivities  of  the  instrument to different plasma conditions and noise, with applications in improving instrument design.
In the case of synchrotron emission from runaway electrons, geometric effects are shown to significantly influence the synchrotron spectrum. Simplified emission model that does not include detection physics can lead to incorrect interpretation of the measurements~\cite{Hoppe:2018}. A third function of a synthetic diagnostic is for experimental control and plasma parameter optimization~\cite{Yang:2020}.

In addition to synthetic data generation and `data fusion' between experiments and simulations, synthetic imaging has been proposed to replace hardware or components such as focusing optics in experiment~\cite{Kramer:2004}. Not only that synthetic imaging is simpler, but also that on many occasions, the optics may not be available or difficult to implement. In hard X-ray imaging, for example, the focusing optics is difficult to fabricate due to the small refractive index difference from the vacuum for essentially all materials and the sub-nm X-ray wavelength. For microwave imaging, the wavelengths are several cm, which make the focusing optics very large. There are plenty of examples outside plasma physics. Computational X-ray imaging, including lensless X-ray imaging, have been reported~\cite{Duarte:2019}. A synthetic aperture microwave Imaging  has been used for imaging of laboratory plasmas~\cite{Shevchenko:2012}.

An emerging framework for synethtic data generation is Generative Adversarial Nets (GANs) \cite{Goodfellow2014}. GANs demonstrated that deep learning could discover hierarchical probability distributions of data~\cite{LeCun2015}, which is common for experimental plasma physics and other branches of science. In this framework, generative models are trained in an adversarial process: a discriminative model that learns to determine whether a sample generated by a generative model is from data distribution. The modules that correspond to the generative models and discriminative models are generators and discriminators respectively. Adversarial nets \cite{Goodfellow2014} implemented both generator and discriminator as multilayer perceptrons, and demonstrated its applicability to generating images of datasets such as MNIST \cite{Lecun1998} and CIFAR-10 \cite{Krizhevsky2009}. GANs have the advantage that Markov chains are never needed, only backprop is used to obtain gradients, no inference is needed during learning, and a wide variety of functions can be incorporated into the model.

However, GANs have been known to be unstable to train, and generators often produce nonsensical outputs. Deep convolutional GANs (DCGANs) \cite{radford2015unsupervised} addressed this issue by implementing both generator and discriminator as deep convolutional neural nets. The visualization of the convolutional filters learned by DCGANs empirically showed the connections between the filters and specific objects. This was convincing evidence that DCGANs could learn a hierarchy of representations from object parts. It follows that convolutional GANs is a promising approach to generating images with complex structures.

In addition to the issue of training stability, the unconditioned generative models of GANs can cause difficulties in controlling the modes of data being generated. This is because many interesting problems are more naturally thought of as a probabilistic one-to-many mapping. For example, an image can have multiple tags.  Conditional GANs \cite{Mirza2014} addressed this issue by using conditional probabilistic generative models. This approach allows GANs to be conditioned on class labels, some parts of data, or even data from different domains. Preliminary results of conditional adversarial nets on image tag generations demonstrated the potential of this approach on multi-modal learning.

By following the conditional and convolutional approaches, various GANs were developed for cross-domain image synthesis. Those conditional and convolutional GANs tailored their generators, discriminators and loss functions for specific applications. Image-to-image translation, for instance, is a problem that is involved in many image processing, graphics and vision problems. One of the data-driven image-to-image translation approaches is to learn mappings between paired input and output images by using GANs. For example, in \cite{Isola2017}, a U-Net \cite{ronneberger2015u} based generator was used to learn image-to-image mappings, and a Markovian discriminator called PatchGAN was proposed. This work demonstrated that the proposed GANs could synthesize photos from label maps, reconstruct objects from edge maps, and colorize images. Paired training data is, however, not easy to acquire in practice. CycleGANs \cite{Zhu2017} achieved image-to-image translation on unpaired data by using a cycle-consistency loss function. It has been proven that cycle-consistency is an upper bound of the conditional entropy.% \cite{Li2017}. 
~Qualitative results of CycleGANs were presented on several tasks where paired training data did not exist. For example, collection style transfer, object transfiguration, season transfer, photo enhancement, etc.

In parallel with the studies of conditional and convolutional GANs, unconditional and convolutional GANs were studied in applications that involved intra-domain image synthesis. Image super-resolution (SR), for instance, is about how to recover the finer texture details when images are super-resolved at large upscaling factors. SRGANs \cite{Ledig2017a} employed a deep residual network (ResNet)~\cite{He2016} with skip-connections, SRResNet, as its generator. As the objective of SRGANs was to achieve photo-realistic single image super-resolution, the authors proposed a perceptual loss function which consisted of an adversarial loss and a content loss. They also introduced a mean opinion score (MOS), which evaluated the qualities of reconstructions by humans. They found out that the SRResNet without the adversarial component set a new state of the art on public benchmark datasets when evaluated with the widely used PSNR measure, whereas the SRResNet with the adversarial component, \textit{i.e.} SRGANs, was the best in terms of MOS. More recently, SinGAN \cite{Shaham2019} achieved the unconditional generation of synthetic images by using only one training image. This was achieved by adopting a multiscale approach: the pyramid representation. This work demonstrated that a pyramid of fully convolutional GANs could learn the generative model of the complex structures of a single natural image.

While most existing studies on GANs concentrate on natural images, they have inspired studies of GANs on non-natural data such as medical images. In \cite{Nie2017}, a fully convolutional network (FCN) \cite{Shelhamer2017} was used to learn mappings from magnetic resonance (MR) images to computed tomography (CT). Experimental results showed that this method was accurate and robust for predicting CT images from MR images. Using GANs to accelerate compressed sensing MR imaging (CS-MRI) reconstruction is another example. CS-MRI needs only a small fraction of data to generate full reconstruction. However, this method suffers from long running time due to the extra computational overhead for dictionary training and sparse coding. RefineGAN \cite{Quan2018} built upon ResNet and GANs, with a novel cycle-consistency loss function, so that it shifted the time-consuming process from the reconstruction phase to the training (pre-processing) phase. RefineGAN achieved state-of-the-art CS-MRI reconstructions in terms of running time and image quality.

\begin{figure*}[!t]
\centering
\includegraphics[width=\linewidth]{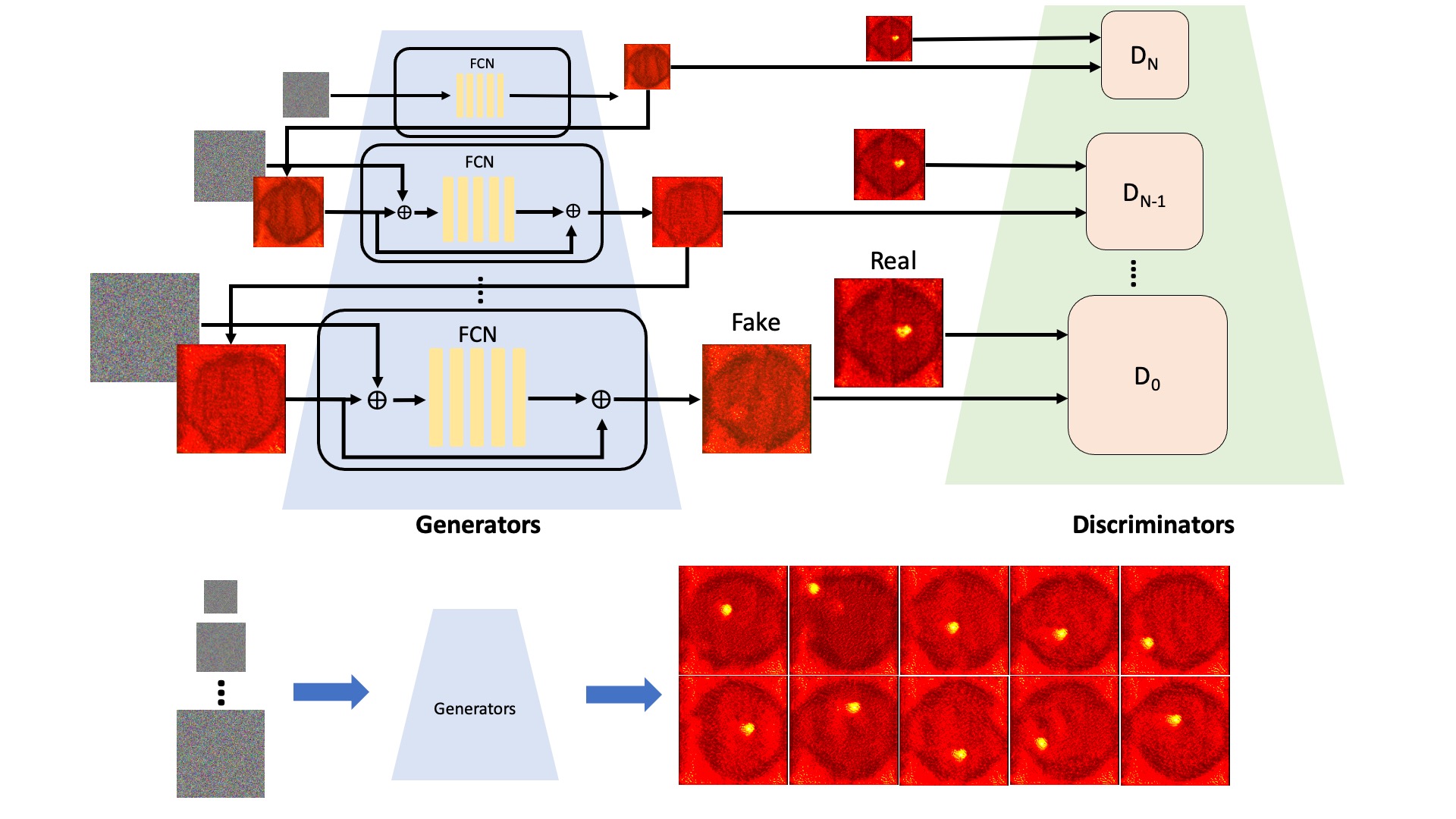}
% where an .eps filename suffix will be assumed under latex, 
% and a .pdf suffix will be assumed for pdflatex; or what has been declared
% via \DeclareGraphicsExtensions.
\caption{A SinGan consists of two pyramids of generators and discriminators at different scales. Each generator is a FCN. $D_i$, where $i = 0,1, …, N$ is the discriminator. Given images with random values, a trained generator of SinGAN can produce a set of synthetic images. The image used in this example is an ICF experimental image~\cite{Wolfe:2021}.}
\label{fig:XH1}
\end{figure*}

Inspired by the previous research and applications of GANs, multiphysics simulation is potentially another area that can use GANs for acceleration. Fig.~\ref{fig:XH1} shows an example of generating synthetic experimental images from a single experimental image by using SinGAN. However, this is just an initial attempt to show the potential of using GANs to accelerate multiphysics simulations. The generation of experimental images is different from the generation of non-physics images in terms of their underlying physical laws. For this reason, physics-informed methods are necessary for the generation of synthetic data that is sensical to actual physical processes. It has been shown that a physics-informed GAN \cite{Yang2020} can approximate the generation of stochastic processes so that it can solve stochastic problems. 

Some of the main bottlenecks in developing and deployment of a synthetic instrument are good physics models for different components of an instrument, the slow process in carrying out multiphysics simulations, esp. for high-fidelity models~\cite{TangCh:2005}. Compared with classical computational methods such as finite difference and finite elements, machine learning method can significantly accelerate the simulation for instrumentation applications. A recent work that combines a convolutional neural network and traditional direct computational method have shown a 40 to 80-fold computational speedups~\cite{Kochkova:2021}. Data methods offer a new way to combine simulations and experimental data~\cite{Humbird:2021}. %$\P$ Experimental data can be divided into point measurements such as probes and line integrated measurements. 

\noindent
[Xinhua Zhang, John L. Kline, Zhehui Wang]
\subsection{Experimental data visualization}
%[Gekelman]
%\section{Experimental data sets and visualization}
The visualization of scientific data is universally accepted as key to understanding complex datasets. A recent article~\cite{Fry:2021} pointed out that the first statistical graph made by Michael Florent van Langren in 1628 of twelve calculations of the distance from Toledo to Rome~\cite{FW:2021}. The large range of distances, what we now call the standard deviation, was meant to convince the Spanish court that better calculations of longitude were necessary as this impacted trade.  Nowadays scientists and much of the general population have no trouble interpreting an $x-y$ , or $x-t$ graph.  This was not so at the dawn of graphics when, for example William Playfair in 1786 used stacks of coins, each stack corresponding to the expenditure of the Royal Navy in a year, to illustrate that the shape of the stack corresponded to line on the graph he created~\cite{FW:2021}.  The problem we now face is in illustrating data which may be inherently more than three dimensional. This subsection highlights examples of experimental plasma data collection and visualization. A comprehensive review paper on the subject would be lengthy, and in all probability obsolete in several years.  In the seventies, plasma data sets of 10 megabytes seemed enormous.  Computers were in their infancy. Commercial software that could draw surfaces with hidden line removal did not exist.  Now terabyte data sets are becoming common.  One cannot comprehend huge lists of numbers and the assiduous use of graphics is key to understanding them.  After all, more than 50 percent of our brain is devoted to processing visual information~\cite{Hagen:2013}. There is a wide variety of commercial and free software, as well as scientific data analysis programs (Python, IDL, Matlab, Mathematica, $\cdots$) which have easy to use graphical routines built into them. We must now avoid drowning the reader of a scientific publication  in a sea of graphs or presenting deceiving graphics.  There are techniques or graphical displays on the horizon for the presentation of multidimensional data.

Fully 3D data is is often calculated in computer simulations ~\cite{Daughton:2011}.  It can also be generated in reproducible, high repetition rate experiments.  We use as an example an experiment involving colliding magnetic flux ropes in a strongly magnetized background plasma ~\cite{Gekelman:2018}.  The flux ropes were kink unstable and designed to collide periodically, at the kink frequency. When the ropes collided magnetic field line reconnection occurred somewhere in the plasma. The process of reconnection results in annihilation of a small portion of the magnetic field. The magnetic energy is converted into heat, flow and waves.  One outstanding question in this experiment, and in general, is where in the large volume of plasma does this this occur?

\begin{figure}[!t]
\centering
\includegraphics[width=3 in]{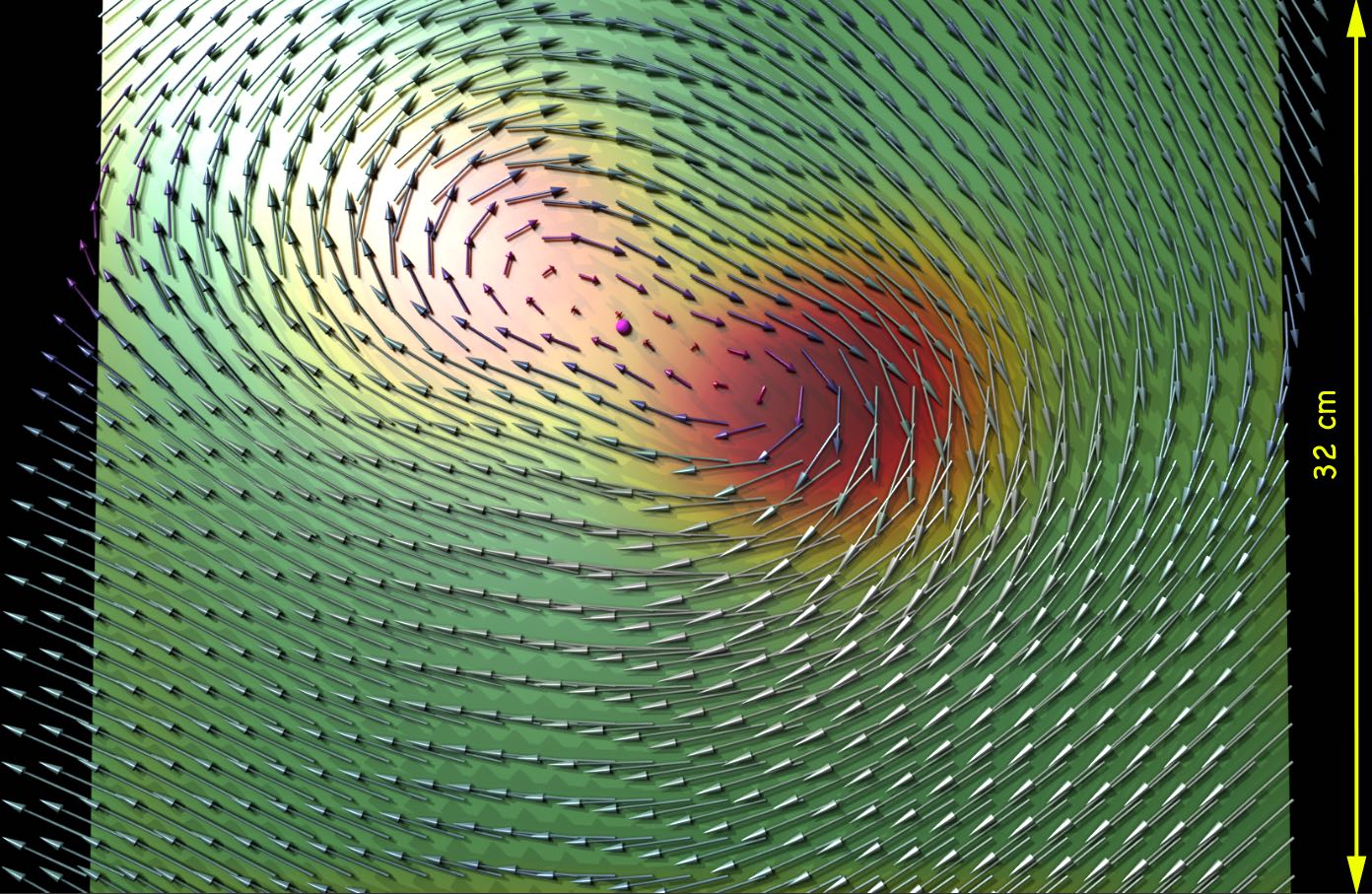}
% where an .eps filename suffix will be assumed under latex, 
% and a .pdf suffix will be assumed for pdflatex; or what has been declared
% via \DeclareGraphicsExtensions.
\caption{Vector plot of the transverse magnetic field at $z = 512$ cm and t = 5.673 ms.  The axial component, $B_z$, is suppressed.   The background colors correspond to the current density on the same plane.  The maximum value is 3.0 A/cm$^2$.  The largest arrow corresponds to a magnetic field of 16 Gauss.}
\label{fig:ucla1}
\end{figure}

\begin{figure*}[!t]
\centering
\includegraphics[width=\linewidth]{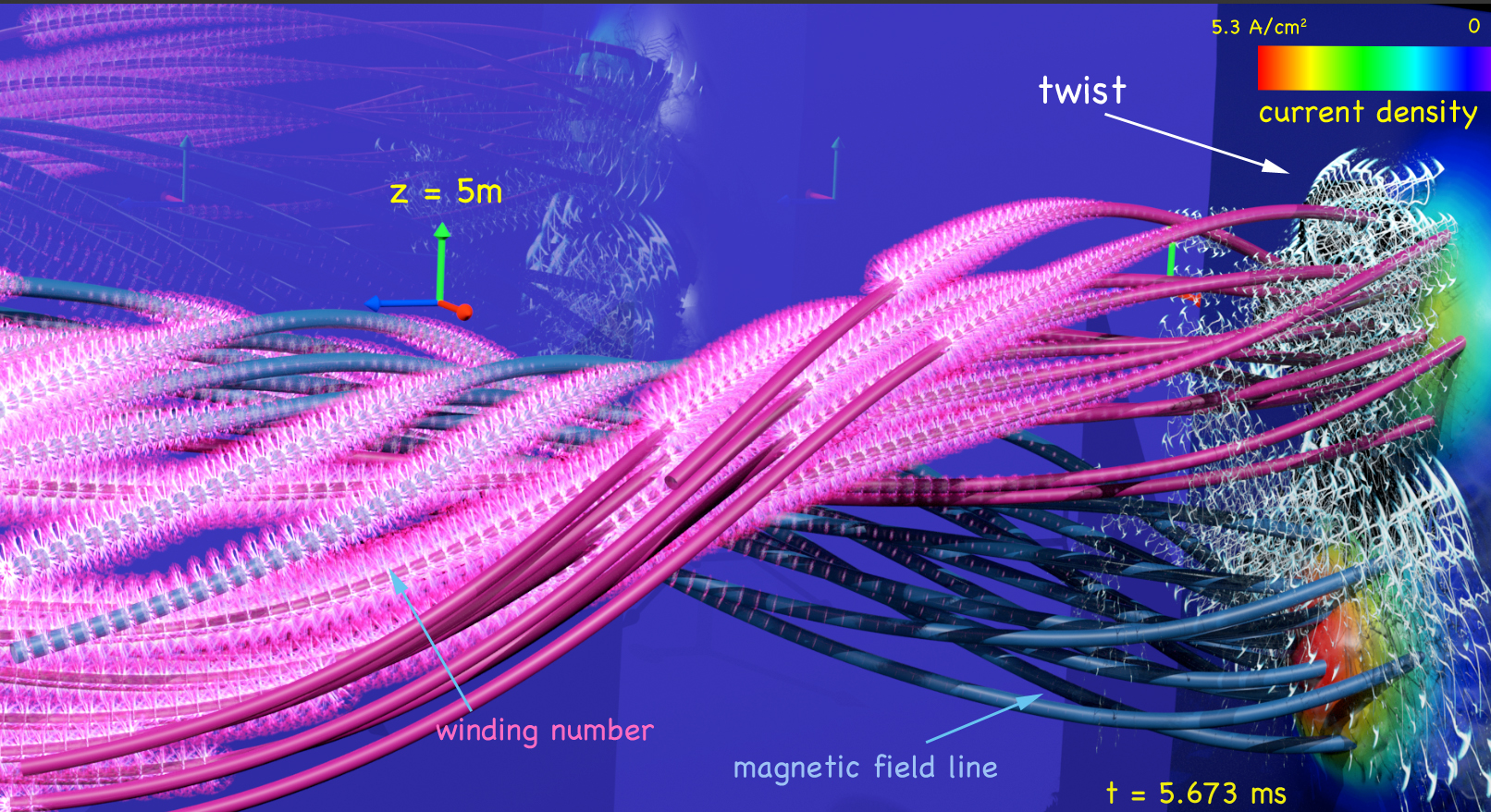}
% where an .eps filename suffix will be assumed under latex, 
% and a .pdf suffix will be assumed for pdflatex; or what has been declared
% via \DeclareGraphicsExtensions.
\caption{Data from an LAPD experiment on magnetic field line reconnection which occurs when two magnetic flux ropes collide.  Isosurfaces of the current in the current channels is shown on the right .   The data plane is at $z=0$. The maximum current density, J, (5.3 Amps/cm$^2$ ) is colored red. A colormap is provided at the top. The data plane at $z = 0$, on the right, is 30 cm on a side, however the axial distance ($z$) spans 9 meters.  The current permeates the volume. The magnetic field, ${\bf B}$, was measured at 48,000 locations. This was used to generate field lines shown as red and blue tubes.   The magnetic twist is depicted as white sparkles and is largest in the first two meters. The winding number ($L$) along the field lines is depicted as red sparkles. The ropes start at $z=0$, the first transverse plane upon which data is acquired is at $z= 64$ cm.}
\label{fig:ucla2}
\end{figure*}

The experiments were carried out in the Large Plasma Device (LAPD) at UCLA ~\cite{Gekelman:2016}. This, coupled to computer controlled probe drives allows the collection of volumetric data sets. Problem required acquisition of a large amount of data. Three axis magnetic pickup coils measured magnetic field from which the vector magnetic field and  plasma currents are derived. Other quantities measured with different sensors were the plasma flow, electron temperature, plasma density, and plasma potential. They were measured at over 42,000 spatial locations and 7,000 timesteps. The  measurement volume was 30 cm on a side in ($\delta x  = \delta y = 0.5$ cm) on 15 planes transverse to the background magnetic field ($B_{0z}$ = 330 G).  The planes were 64 cm apart in $z$, or the axial direction of the cylindrical geometry of the experiment.
One dimensional data in this experiment is not enough to get the true picture of what was occurred. At any given location the magnetic field oscillates at the kink frequency (5.2 kHz) and varies smoothly in the transverse direction. 
Two dimensional data is far more helpful but can sometimes be misleading.  Figure~\ref{fig:ucla1} is a vector map of the transverse ( Bx-By) magnetic field on plane z=512 cm from the start of the ropes and at an instant of time when the flux ropes collide.  The plasma current density is superposed as a color map. A red dot superimposed on the field marks the location at which the temporal data in Fig.~\ref{fig:ucla1}a was acquired. The small transverse field near the center is close to the point of collision. From Fig.~\ref{fig:ucla1} one could guess that the location at which reconnection occurs is somewhere near the red dot, but it is not that simple.  A three dimensional picture, constructed from volumetric data is given in Fig.~\ref{fig:ucla2}. Important topological quantities that shed light on the reconnection location are the quasi-seperatrix layer, magnetic twist (the rotation of  a field line around its neighbors) and the winding number which is a measure of entanglement of field lines~\cite{Gekelman:2020}. Most are displayed in Fig.~\ref{fig:ucla2}. 
The magnetic twist is given by: 
\begin{equation}
T(\vec{r},t)=\int\limits_{\gamma \left( {\vec{r}} \right)}{\frac{\vec{J}\centerdot \vec{B}}{{{B}^{2}}}}ds 
\end{equation}
Where ds is a line element for integration along a fieldline ${\gamma}$.
The twist calculated along the field lines is shown as white sparkles in Fig.~\ref{fig:ucla2}. The winding number is a measure of entanglement of field lines. First one must calculate the winding angle ${\Theta}$.
\begin{equation}
  {\Theta} \left( {{{\vec{r}}}_{0}},\vec{r},z,t \right)=a\tan \left( \frac{{{\gamma }_{y}}\left( \vec{r},z,t \right)-{{{\tilde{\gamma }}}_{y}}\left( {{{\vec{r}}}_{0}},z,t \right)}{{{\gamma }_{x}}\left( \vec{r},z,t \right)-{{{\tilde{\gamma }}}_{x}}\left( {{{\vec{r}}}_{0}},z,t \right)} \right)
\end{equation}

Here ${\gamma}$ is the vector of x,y coordinates for a test field line anchored at ${\vec{r}_0}$, that passes through successive domains D (regions that the field lines pass through) transverse to the background magnetic field. 
There are N ${\vec{r}}$ locations on a plane for all the other field lines $\gamma$, and z the plane in question for which ${\Theta}$ is evaluated.   Once ${\Theta}$ is calculated another test field line is chosen and the calculation is repeated for every ${\gamma}$ in the plane.
To measure the average entanglement of ${\gamma}$ with the rest of the field we integrate ${\Theta}$ over all field lines at positions $\vec{r}$. The winding number L is given by:

\begin{equation}
    L\left( {{{\vec{r}}}_{0}},z,t \right)=\frac{1}{2\pi }\int\limits_{{{D}_{0}}\left( t \right)}{\left[ \Theta \left( {{{\vec{r}}}_{0}},\vec{r},z,t \right)-\Theta \left( {{{\vec{r}}}_{0}},\vec{r},0,t \right) \right]}dA
\end{equation}
The winding number is shown as red sparkles in Fig.~\ref{fig:ucla2}. For most field lines, the winding number begins to grow at about z = 5 meters and it is largest near the axis of the machine. It was established that the reconnection occurred in the region where the twist became small and the winding number large~\cite{Gekelman:2020}. To confirm this one must study what is displayed in Fig.~\ref{fig:ucla2} over many viewing angles.  This is possible with existing software packages.
The upshot of the  analysis is that these topological quantities as well as one not mentioned, the quasi separatrix layer, were used to identify additional three dimensional volumes in which reconnection occurred.   When there was no reconnection these quantities vanished.  To belabor a point, these quantities could not be derived without fully 3D, time dependent data.

Traditional graphics appears in printed scientific publications.  Now many published articles have links which allow downloading of movies of the time development of the data, or fly-arounds to view complex data from many angles. Who knows, one day moving images may be feasible in print publications? Perhaps future software will allow interested viewers to navigate through 3D data in a publishers repository in real time. Televisions with 3D capability have become inexpensive.  While they could be a valuable adjunct to a publication they are hardly used.  The reason may be that there is no standard format for the 3D images/movies between different brands.  The televisions require specialized shutter glasses which are expensive and have short battery lifetimes. Lucrative 3D blockbuster movies have paved the way for the development of sophisticated projectors which can fill giant screens with unsurpassed clarity. The use of 3D in scientific meetings, however, is rare~\cite{Gekelman:2010}.  High quality projectors (necessary for large audiences) are expensive to ship and rent.  They come with a small team of operators and require special screens that reflect light without changing its polarization.  As with television shutter glasses are required for every member of the audience. There is a big push in the gaming world for virtual reality, which necessitate the purchase of clumsy headsets.  One day, these may find a use in scientific visualization.

It is possible to embed holograms in scientific publications as was done for a cover of National Geographic~\cite{NatGeo:1984}.  They are expensive to produce, especially if the image quality is high, but we should not rule out their future use.  There is speculation that images using organic LED’s could be embedded in paper.   This would enable publications to have moving color images. Finally one may look to science fiction to imagine what future visualization systems might be.  Characters in a book by William Gibson~\cite{Gibson:2014} characters have chips implanted in their brain that can make telephone calls and project 3D images in space before them.  The chips are controlled by small movements of their tongues on the upper palate of their mouths.

\noindent
[Walter Gekelman]
\subsection{High-rep rate laser experiments}
%\section{Experimental data sets and visualization}

   The use of high intensity laser pulses as drivers for the next generation of accelerators has received considerable attention over the past decade and demonstrations of multi-GeV electron acceleration~\cite{kmk:1a,kmk:1}, 100 MeV ions~\cite{kmk:2} and energetic positron beams~\cite{kmk:3} have been performed. Beam quality and control is approaching that needed for applications such as x-ray and neutron production as well as for Inertial Fusion Energy (IFE) drivers.  However the main disadvantage of laser sources is the relatively low rep rate and stability of the drivers. For example applications for a laser wakefield accelerator (LWFA) or a laser driven neutron source would be dramatically enhanced if the laser driver rep rate could be increased to 10 Hz or more.  For IFE such rep rates are also necessary.
   
      In a LWFA, the laser pulse drives the relativistic plasma wave via the ponderomotive force, which depends on laser intensity, pulse shape and spectral content. In general, all of these parameters are constantly evolving throughout the acceleration process. Although it is possible to obtain simple expressions for the dependence of electron beams produced by a LWFA with regard to plasma density and laser intensity for an unchanging laser pulse, in reality, the evolution of laser parameters makes analytical treatment less tractable.  Furthermore, there are a large number of input parameters that must be tuned to optimize the accelerator performance. The usual approach to optimization and “machine learning:” is to perform a series of single variable scans in the neighborhood of the expected optimal settings. These scans are challenging, as the input parameters are often coupled and the highly sensitive response of the system can lead to large shot-to-shot variations in output. Moreover, due to the non-linear evolution of the LWFA, altering one input can affect the optimal values of all the other input parameters. Hence, sequential 1D optimizations do not reach the true optimum unless initiated there. A full N-D scan would be prohibitively time consuming for $N > 2$ and so more intelligent search procedures are required~\cite{kmk:4}.  At the University of Michigan we have implemented such optimization using genetic algorithms acting on the actuators of a deformable mirror that controls the laser focal spot characteristics.
      
      \begin{figure*}[!t]
    \centering
\includegraphics[width=1.0\linewidth]{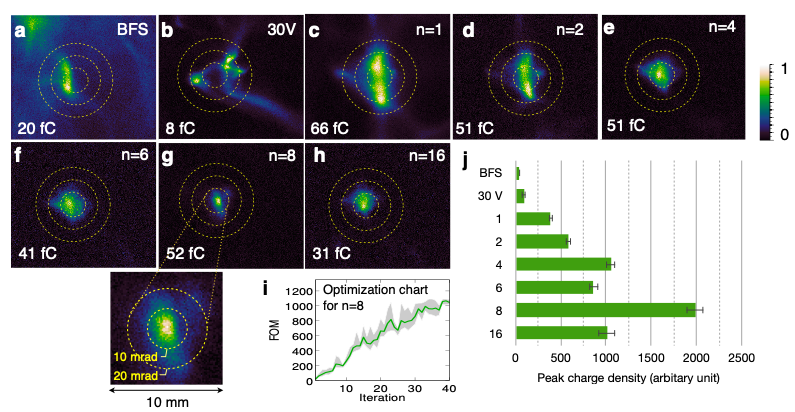}

\caption{Optimization of the electron spatial profile from a Laser Wakefield Accelerator. Electron beam profile image integrated over 50 shots (100 ms exposure time) with a deformable mirror configuration (a) corrected for the best focal spot (BFS) and (b) 30 V on all actuators. (c)-(h) are single-shot electron beam profiles after genetic algorithm optimization using different weighting parameters, n. (i) shows the convergence of the genetic algorithm with n=8. The shaded gray area represents the range of the 10 best children in each iteration and the solid green curve is the average. (j) Comparison of the peak charge density in a single-shot electron image Contours shown are for 20, 40, 60 mrad, centered on the beam centroid.}
\label{fig:kmk1}
\end{figure*}
   
\begin{figure*}[!t]
\centering
\includegraphics[width=0.8\linewidth]{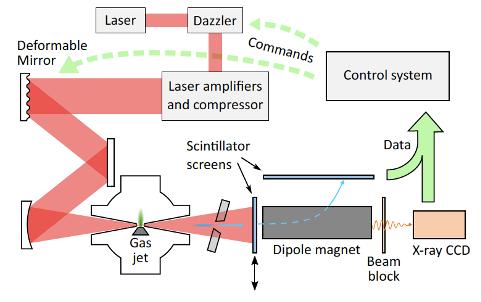}
% where an .eps filename suffix will be assumed under latex, 
% and a .pdf suffix will be assumed for pdflatex; or what has been declared
% via \DeclareGraphicsExtensions.
\caption{A schematic diagram of the experimental setup for machine learning. Pictured are the laser system, Dazzler, deformable mirror, inner chamber, gas jet, and diagnostics}
\label{fig:kmk2}
\end{figure*}

      Machine learning techniques are ideal for these kinds of problem. Consequently it is possible to use genetic algorithms, Bayesian optimization and other methods; using the spatial phase of the laser to optimize a keV electron source (Figures~\ref{fig:kmk1} and~\ref{fig:kmk2}), and subsequently using both spectral and spatial phase to optimize multi-MeV sources~\cite{ShallooNatComm2020}. In these cases, only some of the laser parameters were controlled preventing full optimization of the LWFA which relies on the complex interplay between the laser and the plasma. Further, these optimizations often do not incorporate experimental errors and fluctuations and can be therefore prone to distortion by statistical outliers.  For extension of these techniques to Inertial Confinement Fusion experiments at high rep rate, fully automated laser pulse optimization at high power and energy is needed in addition to control of laser pointing which adds a fluctuating component to the laser pulse. In performing such optimizations, the algorithms will need to build a surrogate model of the parameter space, including the uncertainty arising from the sparsity of the data, fluctuations and measurement variances. 
      
Consequently it is clear that work to develop feedback control of high power high rep rate laser pulses with respect to focal spot shape, temporal pulse shape, spectral control and laser pointing will be required simultaneously. In work up to now the performance of LWFA has been dramatically improved – using deformable mirrors as well as control of the laser pulse shape through the applied phase (Dazzler)~\cite{ShallooNatComm2020}. Extensions of this work to the higher laser energies needed for fusion will enable the use of feedback techniques of the pulses needed for reproducible direct drive implosions at high rep rate. Use of adaptive optics, with genetic algorithms at high rep rate, as well as combining this feedback with the Dazzler and pointing stabilization to optimize all aspects of the laser focal spot for controlling beam properties and reducing instabilities. While optimization and machine learning at 10 Hz works more “slowly” than that at kHz rep rates demonstration of the viability of the technology at these higher energies will be possible in the near term.
	
\paragraph*{Acknowledgement}	
This work was supported by DOE/HEP grant no. DE-SC0016804.

\noindent
[Karl Krushelnick]

\subsection{Charged particle beams}
%[Scheinker, ascheink@lanl.gov]
%%\documentclass[11pt]{article}
%\documentclass[aps,rmp,reprint,amsmath,amssymb,graphicx,longbibliography]{revtex4-1}

%\usepackage[utf8]{inputenc}
%\usepackage[english]{babel}

%%\usepackage{natbib}

%\usepackage{graphicx}
%\usepackage{epstopdf}
%\usepackage{hyperref}

%\begin{document}

%\noindent{\bf Alexander Scheinker}
%\\
%\section{Machine Learning for Particle Accelerators}

%{\it Machine Learning for Particle Accelerators} 
Beam-driven plasma wakefield acceleration (PWFA) can achieve the same energy gain in a single meter, for which a conventional accelerators require several kilometers, but has not yet achieved the same beam quality (in terms of metrics such as energy spread and transverse emittance) as conventional accelerators. PWFA requires extremely intense, high current and sometimes extremely short charged particle bunches with complex beam dynamics and phase space manipulations \cite{ref-FACET,ref-PITZ}. The bunches required for the PWFA process must be extremely short ($\sim$3 fs) to achieve the extremely high peak currents (20 - 200 kA) with bunches having a few nC of current, making them very challenging to control. The PWFA process is extremely sensitive to the detailed longitudinal current profiles of these bunches and requires precise control over these profiles. However, the dynamics of extremely short and intense charged particle beams are difficult to control and quickly/accurately model due to collective effects such as space charge forces and wakefields. Furthermore, diagnostics are extremely limited for such high current, high energy, and short electron bunches. 

For example, the Facility for Advanced Accelerator Experimental Tests (FACET-II) at SLAC National Accelerator Laboratory is being designed to provide custom tailored current profiles for various experiments with bunch lengths as low as (1 $\mu$m or $\sim$3 fs) \cite{ref-FACET-II2,ref-FACET-II}. Another example is the Advanced Proton Driven Plasma Wakefield Acceleration Experiment (AWAKE) which uses transversely focused ($\sim 200 \mu$m), high intensity ($2.5-3.1\times 10^{11}$), high energy (400 GeV) protons from CERN's Super Proton Synchrotron (SPS) accelerator to create a 10-meter long plasma and wakefields into which $\sim18.8$ MeV electron bunches with charge $\sim656$ pC are then injected for acceleration up to energies of 2 GeV\cite{ref_AWAKE_e_ACCEL}.

PWFAs are driven by kilometer long accelerators which are composed of thousands of interacting electromagnetic components including radio frequency (RF) accelerating cavities and magnets. The performance of all of these components is susceptible to drift, e.g. such as thermal drifts. There is also uncertainty in and time variation of the electron distribution coming off of the photo cathode and entering the accelerator. Traditional model-based control and diagnostics approaches are severely limited by such uncertainties and time variation of both the accelerated beam's phase space distribution and the accelerator's components as well as misalignments, thermal cycling, and collective effects such as space charge forces, wakefields, and coherent synchrotron radiation emitted by extremely short high current bunches. Adaptive feedback and machine learning (ML) methods have the potential to aid in developing more advanced controls and diagnostics for complex accelerator facilities.

%\subsubsection{Machine Learning for Static Systems}
{\it Static Systems} For simulation studies or for small accelerators whose properties do not change significantly over time, surrogate models are very useful examples of ML applications in the accelerator community. Neural network-based surrogate models can be trained to quickly map between accelerator parameters and beam properties, providing faster estimates than possible with computationally expensive physics models. Surrogate models can also be used to generate data sets for ML training and for optimization studies \cite{li2018genetic_accel_ML,edelen2020machine_accel_ML,kranjvcevic2021multiobjective_accel_ML,emma2018machine_accel_ML,hanuka2021accurate_accel_ML,zhu2021deep_accel_ML,scheinker2020adaptive_accel_ML}.

An effort has also been made towards developing ML-based accelerator controllers using Bayesian and Gaussian Process (GP) approaches for accelerator tuning \cite{shalloo2020automation,li2019analysis,duris2020bayesian,hao2019reconstruction,li2019bayesian,mcintire2016bayesian}, including various applications at the Large Hadron Collider for optics corrections and detecting faulty beam position monitors \cite{fol2021supervised,fol2019unsupervised,arpaia2021machine,fol2019optics}, and polynomial chaos expansion-based surrogate models for uncertainty quantification \cite{adelmann2019nonintrusive}. Reinforcement learning (RL) tools have also been developed for online accelerator optimization \cite{o2020policy,bruchon2020basic,kain2020sample,hirlaender2020model}. 

%\subsubsection{Limitations of Machine Learning for Time-Varying Systems}
{\it Time-Varying Systems} An open problem in ML is the development of tools for quickly time-varying systems and systems with distribution shifts. If a systems quickly changes with time it is no longer accurately represented by the data that was used to trail the ML model. Therefore the accuracy of the ML methods for accelerators will quickly degrade for systems that change with time, for which previously collected training data is no longer accurate.

%\subsubsection{Transfer Learning for Slowly Changing Systems}

{\it Transfer Learning for Slowly Changing Systems} For systems that change very slowly with time and for which gathering large amounts of new data is feasible without interrupting operations, it is possible to utilize transfer learning techniques in which a network is modified to be accurate for a new data set by taking advantage of some learned feature extraction capabilities and fine-tuning others for the particular problem of interest \cite{goodfellow2016deep}.  

The most common transfer learning technique is re-training. For a particle accelerator a re-training approach may start by using large amounts of simulation-based data to train ML models and then “freeze” most of the weights in the layers that have learned the high-level features of the physical systems for which they were trained, and then fine tune only a few layers, such as input layers that must handle real data rather than simulation-based data as inputs, by using much smaller experimental data sets. Another approach to transfer learning is domain transform in which a much smaller neural network, such as a U-Net approach is developed using a small amount of experimental data and is used as the input layer of our trained NN, the U-Net encodes and decodes data to translate between experimental and simulation domains \cite{zeiler2010deconvolutional}. These transfer learning techniques are not limited to neural networks. For example, they can be applied to GP-based algorithms in which the prior and parameter correlations are first estimated via simulation studies and then fine-tuned with experimental data.

Such transfer learning techniques have been demonstrated to be very successful on a wide range of systems with recent applications including cross-modal implementations \cite{castrejon2016learning}, and both re-training and domain transform were recently demonstrated for mapping electron backscatter diffraction patterns to crystal orientations in which simulation based data was first used and then many orders of magnitude fewer experimental data sets were successfully used for transfer learning to make the networks accurate for experimental data \cite{shen2019convolutional}.

%\subsubsection{Adaptive ML for Time-varying Systems}
{\it Adaptive ML for Time-varying Systems} For most accelerator applications repetitive re-training is not feasible because detailed beam measurements are time-consuming and invasive procedures that interrupt regular operations. Furthermore, for quickly changing systems continuous re-training may be required forever chasing the changes. For such quickly time-varying systems adaptive feedback techniques exist which are model-independent and can automatically compensate for un-modeled disturbances and system changes. Recently, novel adaptive feedback algorithms have been developed which are able to tune large groups of parameters simultaneously based only on noisy scalar measurements with analytic proofs of convergence and analytically known guarantees on parameter update rates, which makes them especially well-suited for particle accelerator problems \cite{scheinker2017model}. 

\begin{figure*}[!ht]
\centering
\includegraphics[width=1.0\linewidth]{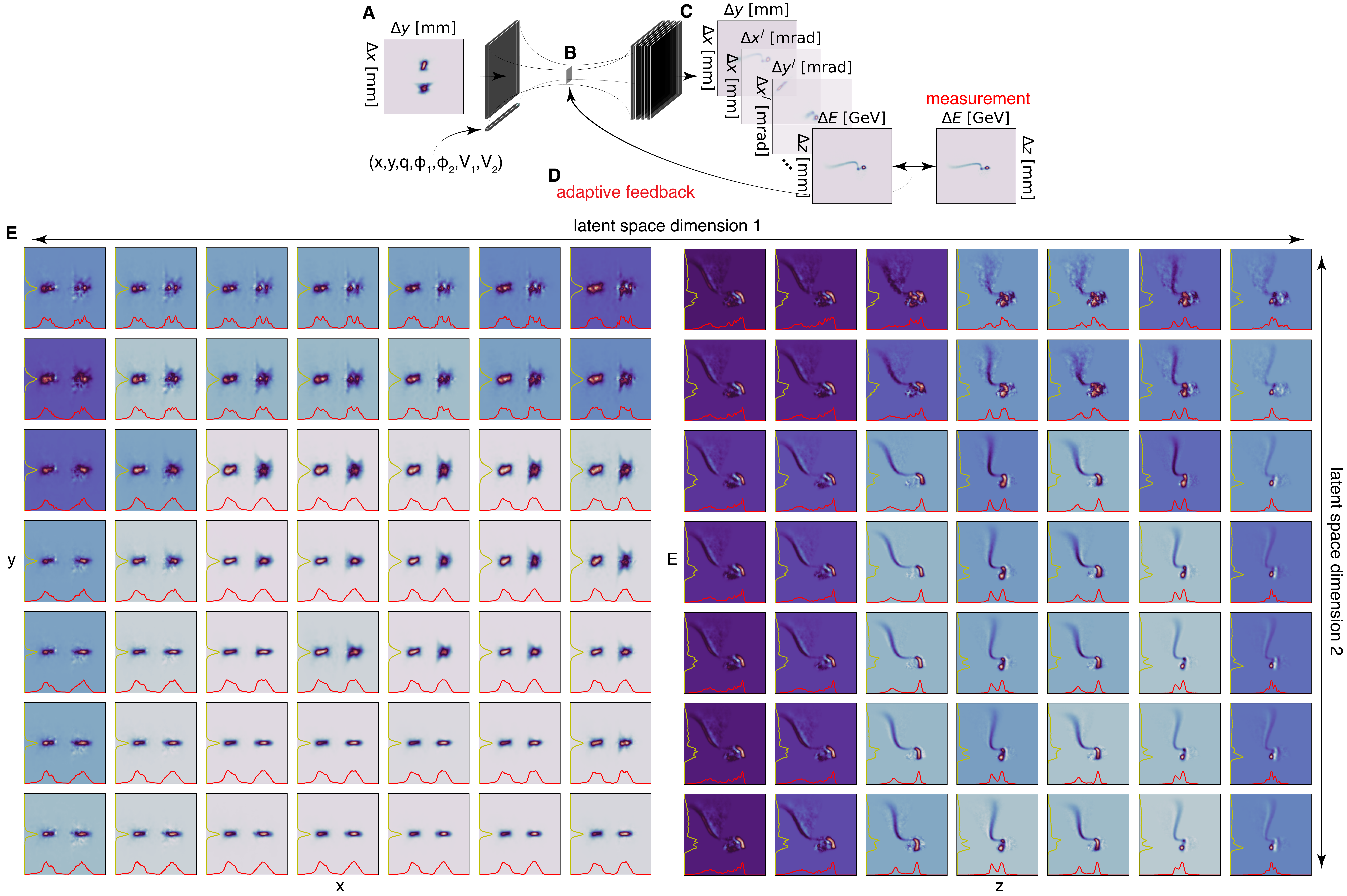}
\caption{
An encoder-decoder convolutional neural network setup is shown which takes an image of an electron beam's $(x,y)$ phase space distribution as an input together with a vector of accelerator parameters (A). The high dimensional inputs are squeezed down to a 2 dimensional latent space (B), from which 75 2D distributions are then generated which are all 15 2D projections of the beam's 6D phase space at 5 different particle accelerator locations (C). Some of the projections, such as the $(z,E)$ longitudinal phase space distributions can be compared to TCAV-based measurements to guide adaptive feedback which takes place in the low dimension latent space to compensate for unknown changes in both the accelerator parameters and in the initial beam distribution (D). The variation of the $(x^/,y^/)$ and $(z,E)$ 2D phase space projections is shown as one moves through the 2D latent space learned by the network and adaptively tuned (E) \cite{scheinker2021adaptive_latentspace}.
}
\label{fig:xpyp_latent_space}
\end{figure*}

Adaptive methods can be applied online in real time for drifting accelerator systems. For example, these methods have now been applied to automatically and quickly maximize the output power of FEL light at both the LCLS and the European XFEL and are able to compensate for un-modeled time variation in real time while optimizing 105 parameters simultaneously \cite{scheinker2019model}. Adaptive methods have also been demonstrated for real-time online multi-objective optimization of the electron beam line at AWAKE at CERN for simultaneous emittance growth minimization and trajectory control \cite{scheinker2020online}. These methods have also been demonstrated at FACET to provide non-invasive longitudinal phase space (LPS) diagnostics that to predict and actively track time-varying TCAV measurements as both accelerator components and initial beam distributions drift with time \cite{scheinker2015adaptive}. Adaptive methods can also be applied for online RL in which optimal feedback control policies are learned directly from data to learn optimal feedback control policies which are parametrized by a set of basis functions whose coefficients are adaptively tuned online \cite{scheinker2021extremum}.

Adaptive methods are usually local feedback-based and can become stuck in local minima. An active area of research is the combination of ML and adaptive feedback in an adaptive ML (AML) approach which combines the robustness of model-independent algorithms with the global learning-power of ML tools such as neural networks. For example, at the Linac Coherent Light Source (LCLS) free electron laser (FEL) at SLAC National Accelerator Laboratory a neural network was combined with adaptive feedback for fast automatic LPS tuning, quickly guiding the system to a neighborhood of the global optimum, and allowing the system to adaptively zoom in on and track the time-varying optimal conditions for fast automatic LPS control of the electron beam \cite{scheinker2018demonstration}. This general AML method has also been utilized for 3D coherent diffraction imaging for accurate reconstructions of 3D electron densities by combining adaptive feedback with 3D convolutional neural networks \cite{scheinker2020adaptive}. 

Novel AML methods are being developed which utilize adaptive feedback to tune the low dimensional latent space of encoder-decoder type convolutional neural networks based on real-time measurements and for online adjustment of inverse models that can provide realistic estimate of the accelerator’s input beam’s phase space distribution based only on downstream diagnostics \cite{scheinker2021adaptive_ML,scheinker2021adaptive}. Such AML tools have the potential to enable truly autonomous accelerator controls and diagnostics so that they can continuously respond to un-modeled changes and disturbances in real time and thereby keep the accelerator performance (beam energy and energy spread, beam loss, phase space quality, etc) at a global optimal, not allowing it to drift as things change with time. 

In a recent example of adaptive latent space tuning a non-invasive diagnostic for the FACET-II beam-line was studied in which a convolutional neural network (CNN) was trained to map inputs of 2D $(x,y)$ electron beam images as well as vectors of 7 accelerator parameters to 75 phase space distributions which were all 15 unique 2D projections of the charged particle beam's 6D phase space at 5 different accelerator locations. The input images were $128\times128$ pixels and so combined with the input vector the total input has a dimensionality of 16391. This high dimensional inputs were reduced down to a 2 dimensional latent space from which the output beam distributions were then generated. By forcing the CNN to generate the large number of phase space projections simultaneously the network was forced to learn correlations between various phase space coordinates. In order to utilize the encoder-decoder as a non-invasive diagnostic, it was then demonstrated that by just comparing the predicted $(z,E)$ projections to their TCAV-based measurements, and adaptively tuning the latent space in order to make them match, all of the other 2D projections of the beam's 6D phase space could be predicted and tracked even as both the input beam and accelerator parameters changed with time \cite{scheinker2021adaptive_latentspace}.  The setup for the adaptive encoder-decoder latent space tuning approach is shown in Figure \ref{fig:xpyp_latent_space}.

%\noindent
%[Alexander Scheinker]

%%\bibliographystyle{te}
%\bibliographystyle{plain}
%\bibliography{A_Scheinker_ML_Plasma_Refs}
%\end{document}

\noindent
[Alexander Scheinker]

\subsection{Control and Optimisation of Plasma Accelerator Experiments}
Plasma accelerators exploit the strong electromagnetic fields supported by plasmas to generate relativistic electron and ion beams. 
In a plasma-based electron accelerator an ultra-short driver, either an intense laser pulse \cite{TajimaPRL1979} or high-current particle beam \cite{ChenPRL1985}, excites a trailing wakefield as it propagates through an underdense plasma. 
Relativistic ion beams can be produced in laser-plasma interactions through use of near-critical or overdense plasma sources \cite{ClarkPRL2000}.
The accelerating fields in these devices can reach hundreds of \si{GV/m}---more than three orders of magnitude higher than available in conventional radio-frequency accelerators---allowing for the production of multi-GeV electron beams over centimetre scale lengths or multi-MeV ion beams in lengths on the order of tens of microns. 

Plasma-based electron accelerators offer a route to drastically reduce the size and cost of brilliant light sources. In this domain, they have demonstrated production of synchrotron-like x-ray beams \cite{KneipNatPhys2010} and FEL gain \cite{WangNature2021}. Further, the technology offers a promising compact alternative to future high-energy colliders based on conventional technology \cite{AlbertNJP2021}.
Compact ion accelerators might find application in medical treatment, material science or ICF technology \cite{AlbertNJP2021}.

However, while the future of plasma-based accelerators is extremely promising, they are not yet devices at a state of technological readiness where they could be used in place of today's radio-frequency accelerators. Some of the critical challenges in making this transition are improving the control and optimisation of the acceleration process and reliably and robustly automating the accelerator operation.

As with any nonlinear system, small changes to the input parameters can constitute a significant shift in the behaviour of the interaction. Plasma accelerators are no exception. In these devices, the relativistic interaction of the intense laser or particle beam with the plasma represents a strongly coupled system that dynamically evolves throughout the acceleration process. Add to this the shot-to-shot fluctuations in driver and plasma source parameters as well as uncertainty and noise in the experimental diagnostics and the task of manually controlling and optimising the multi-dimensional parameter space of these machines becomes onerous.

One route to improving the  performance of plasma accelerators while simultaneously adding automation and advanced diagnostic capability is through the application of machine learning and data science. Here, key experimental controls and diagnostics of the plasma accelerator are given to a machine learning algorithm to exploit their unique capabilities in multi-dimensional optimisation, pattern recognition and predictive analytics. 

\begin{figure*}
	\centering
	\includegraphics[width=\textwidth]{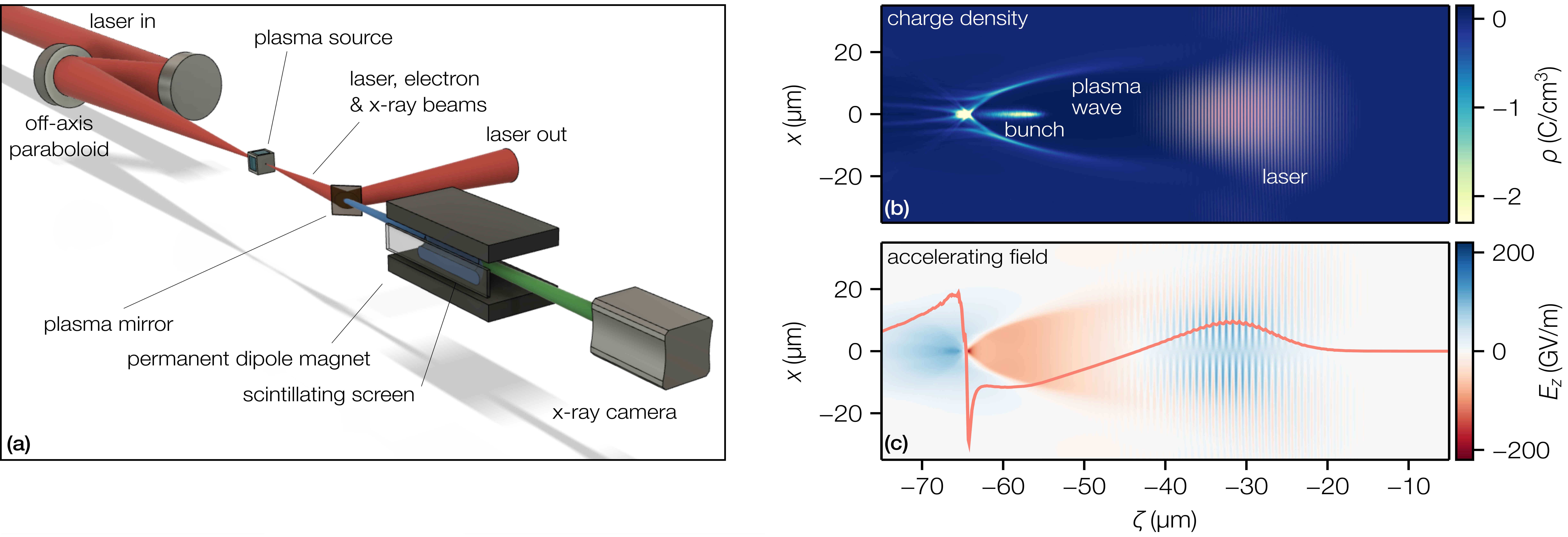}
	\caption{\textbf{(a)} Conceptual layout of a laser-driven plasma accelerator adapted from \cite{ShallooNatComm2020}. \textbf{(b)} Snapshot of a particle-in-cell simulation, performed using FBPIC \cite{LeheCompPhysComms2016}, showing the plasma wave driven in the wake of an intense laser pulse (travelling from left to right) along with an injected electron bunch. \textbf{(c)} The accelerating electric field generated by the separation of charge shown in (b).}
	\label{shalloofig_1}

\end{figure*}

%\section{Current status/state of the art}
{\it State of the art} Plasma based accelerators have recently started to adopt the use of several different supervised machine learning techniques for the control and optimisation of the electron, ion and x-ray beams they produce.

Several key experiments confirmed the fundamental feasibility of applying machine learning techniques for the real-time optimisation of plasma-based acceleration of electrons \cite{HeNatComm2015,StreeterAPL2018,DannPRAB2019,LinOptEx2019} and ions \cite{NayukiRevSci2005,NoamanulHaqNIMA2018}. These experiments utilised genetic algorithms to control specific aspects of the experiment, such as the spatial or spectral phase of the driving laser and in some cases demonstrated optima with order-of-magnitude improvements over manual system optimisation or found significant improvements with unexpected driver properties. 

A key drawback of the genetic algorithm approach was the inability to incorporate experimental uncertainty and shot-to-shot variations in experimental parameters. Recently, Bayesian optimisation based on Gaussian process regression has been explored for the control of plasma accelerators due to its ability to incorporate uncertainty into the optimisation process. This, coupled with the simultaneous tuning of multiple facets of the experimental arrangement, has enabled significant control over the form and parameters of the electron beam phase space \cite{ShallooNatComm2020,JalasPRL2021}. It has additionally allowed for optimisation of specific parameter regimes, such as stable operation, which is of paramount importance for the long-term development of plasma accelerators \cite{JalasPRL2021}.

In addition to the optimisation of the specific experimental outputs, the data generated through long-term operation of these devices can be combined with machine learning and data science techniques to provide insight into the underlying phenomena. 

For example, surrogate models can provide a cheap-to-evaluate, continuous and noise-free abstraction of the complex plasma interaction allowing for an investigation into the underlying parameter dependencies and how they influence the achieved optima.
It has been demonstrated that the Gaussian process models generated during Bayesian optimisation can naturally serve such a purpose \cite{ShallooNatComm2020}. 

Artificial neural networks are also gaining traction as tools for exploring complex experimental datasets. For example, they have found use in explaining and quantifying the influence of drive laser fluctuations on electron beam quality \cite{KirchenPRL2021}. Such knowledge is vital to improving the shot-to-shot stability of these machines.

In a similar fashion, several different supervised learning techniques have been applied in a predictive capacity to compare their performance in determining the charge generated in a laser-plasma accelerator as a function of changes to the laser wavefront \cite{LinPoP2021}.

In the context of plasma-based ion acceleration it has further been shown that surrogate models can replace costly simulations, based on training neural networks with comparably sparse sets of particle-in-cell simulations \cite{DjordjevicPoP2021,DjordjeviPCF2021}.

%\section{Current and future challenges}
{\it Current and Future Challenges} Over the last two decades important proof of principle experiments have shown that plasma based acceleration is a technology that in principle can provide competitive beam parameters for accelerator applications such as brilliant light sources. However, due to limitations of the driver technology and the experimental nature of the setups, the findings of these experiments were often based on a small amount of data or even just single events. 

Today, building on the results of these early experiments, the field is making significant progress in improving the reliability of the acceleration process to allow for stable long term operation \cite{MaierPRX2020,RovigePRAB2020}. Additionally, promising progress has been made in using low-energy high-repetition-rate drivers \cite{GuenotNatPhot2017, SalehiPRX2021} and high-power high-repetition-rate laser drivers are foreseeable in the near future.

This progress in both stability and data availability has been a key enabler for the recent advances in the machine learning and data-driven methods listed above. Consequently, with the current trajectory of the field, machine learning and data-driven research demonstrates great potential but also faces key challenges. These include the aggregation of data at high repetition rate, comprehensive diagnostics of the relevant parameters and lastly the development of algorithms that can handle the large data throughput. 

Therefore, with the transition towards production machines, plasma accelerators will naturally adopt more and more concepts that are currently being established in the field of conventional accelerators \cite{ScheinkerPRL2018,LeemanPRL2019}. This is expected to be especially prevalent in the case of beam-driven plasma accelerators that by their nature operate in very close synergy with conventional machines.

Among these concepts are complex virtual diagnostics \cite{GonzalezNatComm2017, EmmaPRAB2018, HanukaSciRep2021, ConveryPRAB2021} that allow non-invasive measurements of beam properties that would otherwise require destructive diagnostics such as fluorescent screens. For this, machine learning models, typically neural networks, are trained to predict the outcome of an invasive diagnostic from machine parameters that can be measured noninvasively. 

For Bayesian optimisation it has been shown that domain knowledge can be used in physics informed Gaussian processes \cite{DurisPRL2020,HanukaPRAB2021} to increase the speed and robustness of the optimiser. Further, methods for efficient multi-objective optimisation have been explored to find optimal machine states given competing optimisation goals \cite{RousselPRAB2021,EdelenPRAB2020}.

Moreover, reinforcement learning agents \cite{KainPRAB2020} that are either trained on the experiments themselves or on surrogate models resembling these, promise to be a useful tool when confronted with dynamic conditions that tend to be a challenge for other optimisation methods. 

%\section{Concluding Remarks}
{\it Concluding Remarks} Plasma accelerator technology is currently in a transition period, moving from single experiments to study fundamental concepts towards robust machines fit for applications in future light sources, high-energy colliders and beyond. The increase in quality and quantity of data has brought with it a commensurate uptake in machine learning and data science techniques for experimental control, optimisation and data analysis. It is foreseen that in the future, the use of these techniques will rapidly expand. 

Plasma accelerators offer a unique and timely testing ground to translate lessons learned in the control and optimisation of high-repetition-rate \emph{big physics} machines, such as conventional particle accelerators, to the laser-plasma community at large. As such, there is a significant advantage to be gained through close collaboration between members of all facets of laboratory plasma physics research. 

%\bibliography{Shalloo2}% Produces the bibliography via BibTeX.

%\end{document}
%
% ****** End of file apssamp.tex ******

\noindent
[S\"oren Jalas, Manuel Kirchen, and Rob J. Shalloo]

\subsection{Dusty and complex plasmas}

%\section{Dusty and complex plasmas}

Complex plasmas or dusty plasmas consist of nanometer to micrometer sized dust particles immersed in a partially ionized plasma environment~\cite{Mamum:2001}. All plasmas, whether they are in laboratory or natural environment such as the Earth's ionosphere, interplanetary solar wind, the interstellar medium in the Milky way, or intergalactic medium farther away, are dusty to a degree due to the ubiquitous interactions and mixing of plasmas with condensed matter~\cite{Mendis:1994, MerlinoGoree:2004, Wang:2008T}. Supernova or the massive star explosions are a source of dust, or `dust factories', that contribute to the cosmic dust population and have been studied for example by the Spitzer Space Telescope ~\cite{Spitzer}. The discovery of the plasma crystals or Coulomb crystals of dust in low-temperature plasmas in the 1990s by multiple groups was a major milestone in laboratory dusty plasma physics research.  In laboratory plasmas, these micro- and nano-particles usually attain a negative charge due to higher mobility of electrons. The highly charged particles interact with one another electrostatically and exhibit collective behavior such as crystallization, melting, demixing, self-excitation of waves and turbulence, see {\it e.g.} \cite{Piel:2001,fortov05} and references therein. Difference forces including neutral-gas drag force, ion drag force, thermophoretic force, and the Earth's gravity can also affect the dynamics of the individual dust motion and the collective multiple-particle dynamics. Experiments such as PK-3 Plus laboratory on board the International Space
Station (ISS) have been used to isolate the effects of the Earth's gravity~\cite{ThomGer:2008}. Tesla-strong magnetic fields have also been applied in the laboratory to examine the effects of the magnetization~\cite{Thomas:2016}.
The processes of self-organization and phase transition can be observed on the single particle level using laser scattering and imaging cameras such as CCDs. Together with the table-top experimental footprint and modest hardware cost, dusty and complex plasma experiments are highly accessible to data science. 

\begin{figure*}[!htb]
\centering
\includegraphics[width=\linewidth]{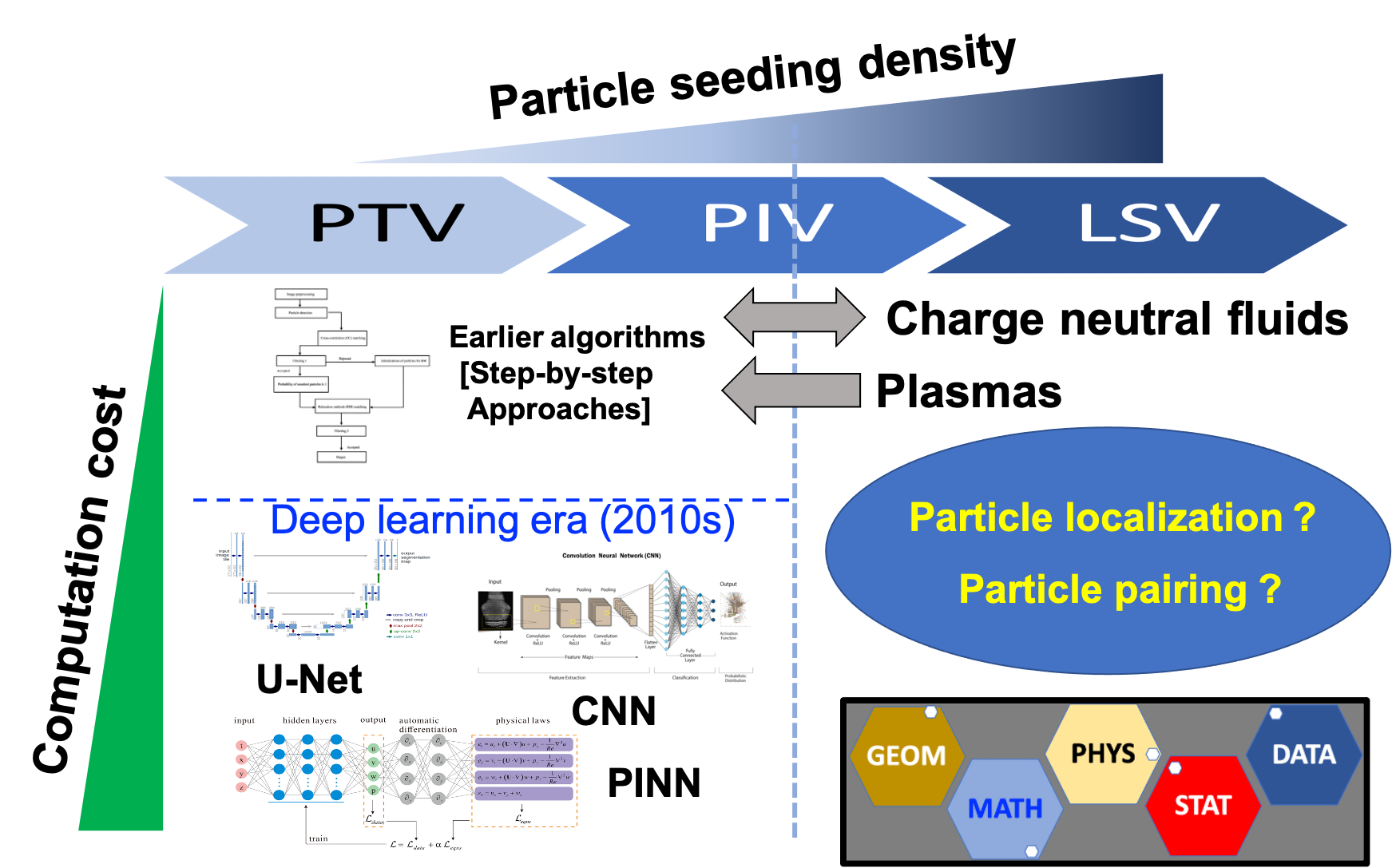}
% where an .eps filename suffix will be assumed under latex, 
% and a .pdf suffix will be assumed for pdflatex; or what has been declared
% via \DeclareGraphicsExtensions.
\caption{As the particle seeding density increases, three particle imaging and tracking methods: particle tracking velocimetry (PTV), particle imaging velocimetry (PIV) and laser speckle velocimetry (LSV) have been developed for charge neutral fluids such as water and gas flows. For plasmas, PTV and PIV are more common. The two central questions for PTV and PIV algorithms to address are how to localize seeding particles from an image, and how to pair up the same particles from different images to form particle trajectory. In the deep learning era (2010s), a growing number of PTV and PIV algorithms such as U-Net, convolutional Neural Networks and physics-informed neural networks (PINN) are being introduced to neural fluid and plasmas. The new algorithms have higher computational cost compared with earlier algorithms which process images step by step and more manually. While most such neural networks are trained by large datasets, they can also take into account of geometry, mathematical, physical, and statistical constraints.}
\label{fig:ptvpiv1}
\end{figure*}

Leveraging the fact that individual dust particles can be detected together with a cloud of dust, tracking individual dust and collective dust motion is an important and unique experimental technique in dusty plasma research. Dust tracking and imaging, see Fig.~\ref{fig:ptvpiv1}, coupled with theory and dust dynamic simulations (a cousin to molecular dynamic simulations), are used to examine a broad range of problems such as the dynamics of dust charging and motion, dust crystal-liquid phase transition, non-thermal and statistical physics, discovery of new phases of dust clusters such as glass phase and supercooled dust liquids, nucleation and dust growth, dust acoustic waves and instabilities, nonlinear physics, formation of 2D and 3D dust structures, and anisotropic dust clusters under microgravity, AC electric field, cryogenic temperature, charged-particle beams and shock wave conditions. For example, electrorheological (ER) complex plasmas can evolve into a string phase when an external AC electric field is applied~\cite{Ivlev:2008}. Fluid demixing and crystallization can be examined with a mixture of two or more types of microcroparticles. Dust Acoustic Waves have been extensively studied theoretically and experimentally~\cite{Rao:1990, Barkan:1995}. Dust acoustic wave turbulence, when coherent dust motion oscillations change to a turbulent state of motion with many harmonic modes, was also reported~\cite{Pramanik:2003}. More recently, through novel multidimensional empirical mode decomposition based on Hilbert-Huang transform, 3D dust acoustic wave turbulence has been decomposed into a zoo of
interacting multiscale acoustic vortices, exhibiting attraction, repulsion, entanglement,
bunching, and synchronization, in the 2 + 1D spatiotemporal space~\cite{Lin:2018}.

Terabyte datasets are available from dusty plasma experiments through particle tracking and imaging~\cite{Wang:2020b}. Dusty plasma movies have been recorded at about 1500 to 5000 frame length, at the rates between 100  -  500 frames s$^{-1}$ and each image size of a few MB per frame~\cite{Ticos:2019}. For an experimental campaign consisting of a few hundred runs, more than 1.5 million movie frames  or more than 1 TB of raw data becomes available~\cite{Ticos:2020}.  Automated particle tracking through machine learning is emerging as a necessary to process the large number of images and to extract the particle trajectories~\cite{Wang:2020b}. Particle tracking and particle imaging velocimetry (PIV) techniques have wider applications than plasma physics. In addition to the traditional probabilistic algorithms, new PTV and PIV algorithms based on U-Net, Convolutional Neural network, and physics-informed machine learning~\cite{Wang:2022} are emerging. Other examples of machine learning applications may be found in the phase transitions in the dust cloud~\cite{Dietz}, the correlation of current-voltage (I-V) characteristics given by a Langmuir probe with the main plasma parameters~\cite{Zing:2021}, to identify the boundary layer between mixed regions of dust particles with different diameters~\cite{Huang:2019}, and the response of a single dust particle levitated in the plasma sheath, to a nonlinear excitation frequency~\cite{Ding:2021}. %[Prediction applications?]

%[Waves and tracking plot]

\noindent
[ Zhehui Wang, Catalin M. Tico\c{s}]
%\red{please spell out the first name of Ticos}

\subsection{Physics and machine learning}
%\subsection{Physics and machine learning}
Prior to the recent introduction of machine learning (ML) models, physics-based hypothesis-driven models are the most powerful tools for natural sciences including plasma physics. ML has now been used in many scientific domains with few exceptions~\cite{WangL:2016, Carrasquilla:2017, Ramprasad:2017, Smith:2017,Butler:2018,Zitnik:2019,Yan:2019,Brunton:2020,LiDe:2021}. ML as a new scientific tool is as generic as traditional physics-based hypothesis-driven methods, and allow broad implementations by different scientific domains and subfields.
Automated data processing through ML has led to the acceleration of every aspect of the scientific activities or `{\it scientific workflows}', from observations and experimental data taking, to hypothesis generation, to model construction, to model execution through computation, and to model validation and predition~\cite{Mjolsness:2001}. 

Some plasma problems parallel their counterparts in other scientific domains, which may justify the use of similar ML algorithms. Understanding plasma waves and instabilities in plasma physics poses similar challenges as in understanding diseases in biology~\cite{Zitnik:2019}. Plasma flow and turbulence, which resemble charge-neutral fluids, are also further enriched in structures due to the electromagnetic interactions~\cite{Heinonen:2020}. New phases of matter, including quantum phases of matter, are expected in high-energy-density plasma experiments due to the extremely high-pressure that can be created~\cite{Hatfield2021}. Plasma-material surface interactions are encountered in both low temperature and high temperature plasmas%~\cite{Kruger:2019}.  
Plasma-material interface engineering poses one of the most significant challenges for both fusion energy and plasma technology applications. The computational complexity are comparable to and may even exceed quantum DFT calculations for materials. A comprehensive physics-based description of this multi-phase system requires integrated approach to plasma physics, material science, and their interactions. The length scales involved range from sub-nm to above 1 m in the largest laboratory plasma apparatus. The temporal scale spans 1 femtosecond to the order of a second. Hundreds of controllable parameters may be needed in search for the best recipe for generating and controlling a plasma, making plasma optimization problems high dimensional. Automation through machine learning is necessary for model reduction, and to accelerate the plasma physics workflows for more accurate predictions, more reliable controls, and more accessible optimization.

One latest trend is to combine machine learning with physics deep learning. Combination of a deep learning architecture and high-dimensional datasets have shown to be more effective than earlier machine learning methods such as support vector machines (SVMs), small multilayer perceptrons (MLPs), random forests and gradient-boosted trees~\cite{KST:2019}. High-dimensional data came from multiple plasma apparatuses and different experimental conditions from about 9000 experiments. Physics consideration guided the selection of more than a dozen features including plasma density, plasma temperature, etc. as the neural network inputs. Physics motivated dimensionless combinations of the
raw measurements were used for input data normalization. 
Reliable predictions with 82\% or better accuracy have been demonstrated on another plasma from the one on which the neural network was trained. Construction of Grad-Shafranov equilibria is usually the first step in understanding and control of magnetically confined plasmas. A five-layer fully-connected deep neural network was reported for solving the Grad–Shafranov equation constrained with measured magnetic
signals in real time ~\cite{Joung:2020}. The computing time was approximately 1 ms on a personal computer, potentially allowing applications in real-time
plasma control. 
An encoder-decoder neural network model of tokamak discharge is developed based on the experimental dataset
alone~\cite{Wan:2021}, without a direct reference to a physics constraint such as the Grad-Shafranov equation.
Electron density, stored energy, and loop voltage were reproduced with close to 90\% fidelity to experimental data from a series of actuator signals using the neural network. 
The method provides an alternative to the physical-driven method for plasma modeling, 
experimental planning and model validation.
Variations of experimental plasma conditions are usually captured by statistical models. The stored energy of a plasma $E_{tot}$, for example, may be a function of input power ($I_p$), plasma geometry ($\Delta$), magnetic field ($B$), ion species ($Z_k$), impurity ($n_i$), etc.  The statistical mean of $E_{tot}$, $\bar{E}_{tot}$, may be given by
\begin{equation}
\bar{E}_{tot} = \sum_j E_j P_j(I_p, \Delta, B, Z_k, n_i, \cdots),
\end{equation} 
where the probability function $P_j$ corresponds to the energy content $E_j$. The statistical variance, $\Delta E_{tot}^2$,  is given by
\begin{equation}
\Delta E_{tot}^2 =\sum_j (E_j - \bar{E}_{tot})^2 P_j(I_p, \Delta, B, Z_k, n_i, \cdots).
\end{equation}
To construct explicit probabilities $P_j$ as a function of $I_p$, $\Delta$ and others present substantial challenges for theory, but important to experiments and controls. ML can be used to obtain implicit correlations between $E_{tot}$ with input power $I_p$, etc. Meanwhile, there may be even features of plasmas that is hard to be captured by explicit physics model~\cite{Gonoskov:2019}. 

Even with the use of physics-motivated quantities and features such as electron temperature, plasma density as neural-network inputs, successful scientific applications of deep learning for feature extraction, pattern recognition, classification, denoising, nonlinear regression, statistical inference can still be perceived as a ‘black-box’ magic~\cite{Hagan:2014, Goodfellow:2016}. One may recognize similarly that modern computer codes are also quite complicated and not necessarily transparent to understanding. Code validation therefore has been an important part of the code development process. This apparent separation of the power of machine learning and artificial intelligence from understanding through the fundamental laws of physics or corollary laws is convenient but not satisfying. The fundamental laws of physics are universally applicable to physics, chemistry, biology, geology, astronomy, and cosmology, to atoms, molecules and bulk materials, to different phases of matter such as gases, fluids, solids, plasmas and Bose-Einstein condensates. The difficulty of {\it ab initio} models  is only that mechanical applications of these laws lead to equations much too complicated to be soluble~\cite{Dirac:1929}.  Other difficulties include incomplete initial and boundary conditions, random noise and errors that may accumulate with time and the number of elementary calculations. Yet another difficulty is that data is sparse. Limited by instrumentation or numerical resolution, data and information sparcity increases as the length scale and time step decrease. These difficulties with the first-principle methods have given rise to corollary or empirical laws such as quasi-linear theory, Kolmogorov turbulence scaling, BBGKY hierarchy, adiabaticity of charged particle motion and many others in plasmas. The corollary laws are approximations to the fundamental laws. They are not intended to be universal and are expected to be broken down. But the corollary laws are effective methods for understanding complex phenomena, and meanwhile are traceable to the fundamental laws. One open question is whether machine learning can be used to derive corollary laws, as a step towards the recovery of the fundamental physics laws behind the data. Another related question is whether such corollary laws, and fundamental laws are as important to machine intelligence as they are to human intelligence. 

Applications of machine learning in physics and its subfields pave the way towards a more satisfactory union between the two; namely interpretable machine learning models based on physics and  vice versa, discovery of new physics aided by machine learning. A theory of artificial intelligence may still be a long way to go~\cite{Agliari:2020}. Interpretation of the machine-learning-based algorithms may lead to even more powerful algorithms for plasma control~\cite{Parsons:2017}. The fundamental laws of physics are incomplete. With the growing evidences for dark matter and dark energy, and the ongoing effort to reconcile general relativity with quantum physics, there are apparently rooms for discovery of fundamental physics through data science. In high-energy particle physics, pattern recognition and machine classification have found applications in data reduction, {\it i.e.}, searching for extremely rare events that may hint at new physics beyond the existing frame work of quantum 
chromodynamics~\cite{Radovic:2018}. Machine learning to recover hidden physics models could be extended to plasma physics~\cite{Rassi:2018, Rackauckas:2020}. 

\begin{figure}[!t]
\centering
\includegraphics[width=2.5in]{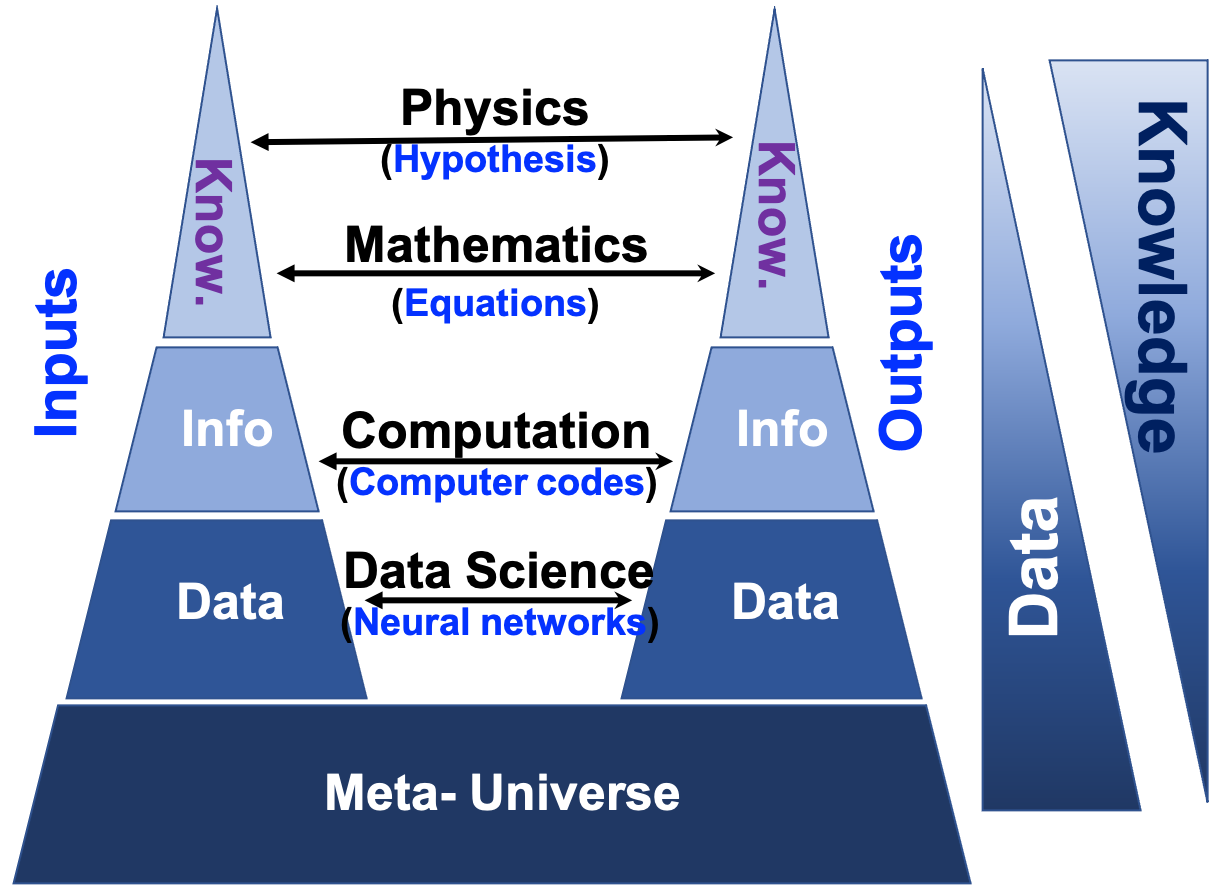}
% where an .eps filename suffix will be assumed under latex, 
% and a .pdf suffix will be assumed for pdflatex; or what has been declared
% via \DeclareGraphicsExtensions.
\caption{The universe is the ultimate source of all scientific data, which collectively may be called `Meta- Universe'. A growing number of methods and tools are used to correlate the data, information and knowledge, shown in hierarchical order as a subset of the Meta- universe.}
\label{fig:digiU}
\end{figure}

Physics can give rise to new concepts in ML and data science, such as physics-enabled and physics-informed machine learning~\cite{Carleo:2019, Karniadakis:2021}. Quantum machine learning is emerging, which could transform both machine learning hardware and software~\cite{Biamonte:2018}. The tensor network structure of quantum mechanics has inspired machine learning methods for classification~\cite{Stoudenmire:2016}. One approach to physics-informed machine learning as discussed above is by using physics-motivated quantities or features as inputs and outputs for machine learning. Therefore, neural networks can be trained to emulate corollary laws such as empirical scaling relations that are widely used in plasma physics. Another approach is to use computer simulations to produce training data for neural networks, which can then be used for nonlinear regression and prediction~\cite{Duraisamy:2018}.  A recent approach to physics-informed machine learning has introduced differential-equation-based loss functions for neural network training. Statistical physics may be used for uncertainty quantification. 

There are also physics concepts that may not be captured by differential equations. One class of such concepts is the principle of symmetry~\cite{Gross:1996}, which includes reflection or mirror symmetry, translational symmetry, and rotational symmetry. Galilean invariance is the hypothetic symmetry for different inertial frames. According to Noether's theorem, symmetry gives rise to conservation laws in physics. Momentum conservation is the consequence of translational symmetry. Energy conservation is derived from time invariant symmetry. Mass conservation are other familiar examples. The probability, probability density of an electron or an ion distribution function, and the intensity of light on a sensor need to be positive. These symmetry, invariants, and the positiveness of many physical quantities may be used to regularize the parameter space of the inputs and outputs of a neural network, or the loss functions. It has already been recognized that image representations by neural network such as CNNs should be invariant due to the translational and rotational symmetries~\cite{Kauderer-Abrams:2017}.  Use kernel-based interpolation to tractably tie parameters, CNN has been generalized to deep symmetry networks~\cite{Gens:2014}. By taking into account of the spherical geometry of an object, Spherical CNNs have been found to be more computationally efficient and accurate for 3D model recognition~\cite{Cohen:2018}. There are also specific symmetries in plasmas related to toroidal geometry of a plasma, periodicity. The concept of collective variables~\cite{Sarra:2021}, when there is no obvious symmetry, might be useful for turbulent plasma feature extraction. Further exploration of these additional physics concepts for machine learning algorithms would become fruitful and rewarding in the near future.

\noindent
[Wenting Li, Zhehui Wang]

\subsection{Challenges and outlook}
Rapid advances in computing hardware, architecture and data acquisition instruments present challenges and opportunities for plasma physics and science at large. One challenge lies in the fact that manual and even semi-manual data mining methods face increasing difficulty in extracting new information and knowledge from the large and multi-dimensional datasets. Data science and machine learning (ML) offer transformative tools for laboratory plasma experiments and physics of plasmas in the big-data era. The classical physics framework, which includes Newton's laws and Maxwell's equations, is the canonical pathway to understand plasmas, guide the designs of plasma experiments and inventions of plasma technologies. Many problems in plasmas rise from the complexity derived from a large number of particles (on the order of 1 mole in some laboratory plasmas), and their interactions with electromagnetic fields and material surfaces. The combination of accumulative computational errors, insufficient knowledge in initial condition, boundary condition and perturbations, and the long computing time even by using the state-of-the-art computers renders the canonical pathway ineffective if possible for reliable predictions and optimization problems in plasmas. There are also NP-hard problems in plasma physics, which may be difficult to both ML and traditional computation. Enabled by heterogenous multi-dimensional data sets including experimental and observational data, simulation data, and other meta-data, data science and machine learning have been successfully or can be used to accelerate all aspects of plasma research or the `scientific data flows'; {\it i.e.}, from observational and experimental data taking, to hypothesis generation, to model construction, to modeling and to model validation. Despite of their practical prowess and simplicity, machine learning methods for plasmas and other scientific domains are not completely understood at this time. Seeking a better union between the established knowledge framework of plasma physics and emerging information science is an exciting new frontier for data-driven plasma physics and laboratory experiments. New results may be anticipated such as in data-driven discovery of new plasma physics, development of scientific machine learning algorithms that will be broadly applicable to problems beyond plasma physics, and quantitative understanding of uncertainties for more effective predictions and optimization, paving the way towards automated plasma knowledge discovery and novel technologies.

\noindent
[Zhehui Wang]

\section{Magnetic Confinement Fusion}
%Coordinator: Benkadda and Chang

\subsection{Introduction}
For the successful realization of the safe, unlimited, and carbon-free magnetically confinement fusion energy, the nonlinear non-local behaviors of $\sim 150$ million $^oC$ plasma in strong magnetic field need to be understood and predicted.  There have been, and will be, a vast amount of experimental and computational data available, which may be used to build surrogate models and digital twins. Since a thermonuclear magnetic fusion device is extremely costly and takes tens of years to build, the digital twins and surrogate models can be highly valued tools for scientific advancement.  Fast surrogate models are also valuable for real-time workflow and control of the on-going long-pulse experiments and improvement of next experiments.

With the rapid advancement of computing power, extreme scale simulations are supporting the magnetic fusion energy research by solving the fundamental equations.  However, the turn-around time for extreme-scale computational study is still too long for near-real-time input to the experimental studies.  Data from such simulations can be used, together with experimental data, to raise the fidelity of the simpler models.  Moreover, AI/ML can be used to replace computationally expensive kernels to accelerate the extreme scale simulations and to enable physics discovery online from the big simulation data and compress the output data without sacrificing the important physics features.

Data driven science in magnetic fusion research is only at the beginning stage.  However, many useful developments have been reported, with some of them already in use in the experiments and modeling. Topics covered here many not be highly comprehensive, but will at least be representative.

\noindent[C.S.Chang]

\subsection{Data-Driven Physics Models}
\label{MCF_models}
Data-driven models have become increasingly popular in the scientific literature in recent years.  One of the basic ideas motivating data-driven modeling is to utilize data from experimental systems (e.g. such as diagnostics, control systems, reactor consumables, and/or maintenance schedules, etc.) as well as simulation models of various fidelity, to derive additional predictive models that are either physically-informed or, at the other extreme, entirely empirical. Physically-informed data-driven models are often either enriched versions of first-principle theoretical physics models (e.g. MHD, gyrokinetcs~\cite{narita:2021}, etc), or they can be extracted models from data that constrain themselves to prescribed physical laws, or conditions.

These techniques are distinct from surrogate model generation for acceleration of multi-physics modelling, which rely on model data for their training sets. This is discussed in section~\ref{sec:surrogates}. 

\begin{figure*}[t!]
  \centering
\includegraphics[width=0.8\linewidth]{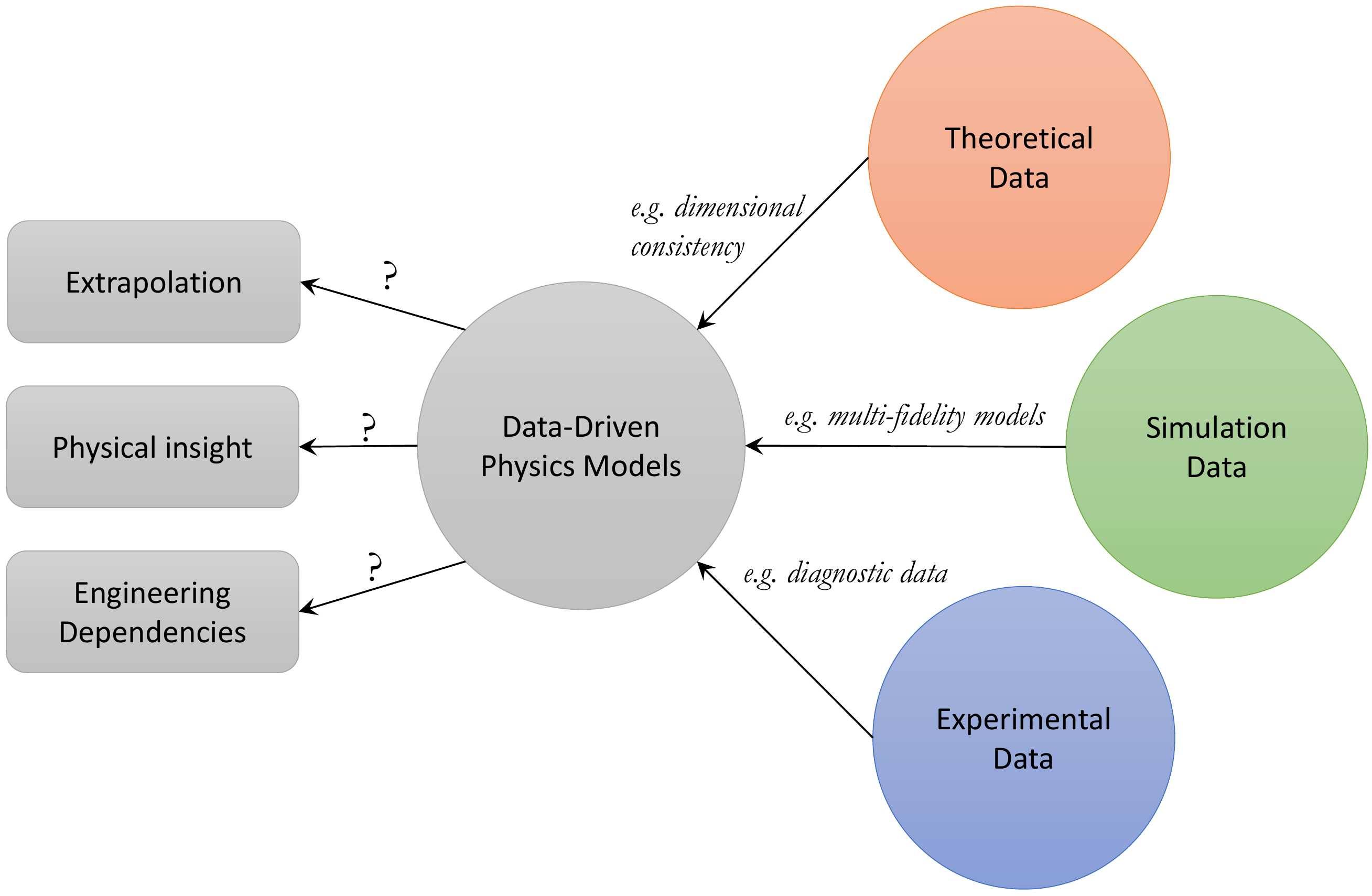} 
  \caption{The relation of data-driven physics models to experimental, simulation, and theoretical data streams.}\label{fig:dd1}
\end{figure*}

Generally a benefit of data-driven modeling is that the ``validation'' of data-driven models against experimental data is, in some sense, baked into the model itself.   In other words, because the experimental data is used to train the model, the model validates against that data naturally, removing many of the concerns regarding whether the observed experimental phenomenon corresponds with (or validates against) the model itself.  The primary concerns that tend to remain include the following open questions: 1) whether these models can extrapolate well to different physics contexts (e.g. different machines or plasma configurations, etc.), 2) how dependent these models become on the underlying engineered hardware that drive some of the physics observed in the experiments (e.g. the specific engineering design and performance impact a specific divertor, cryostat, etc. may have on the resultant model system), and 3) whether these models are too ``blackbox--like'' to extract meaningful physical insight/understanding from.  These common considerations are illustrated in Figure~\ref{fig:dd1}.

While data-driven methods have been utilized in many contexts and for many purposes---such as for identifying error estimates in sophisticated validation studies using traditional physics simulations models \cite{cartier2020posteriori,ISI:000507120000001}, as well as being used in semi-empirical methods \cite{moreau2015combined}, stabilization analysis \cite{olofsson2013subspace}, the development of plasma stability control techniques \cite{goumiri2016modeling}, discharge control systems \cite{treutterer2012management}, deep statistical inference models on experimental data \cite{ISI:000348843100013,ISI:000410776700001,mathews2021uncovering}, as well as feedback control schemes \cite{treutterer2011integrated}---many of these techniques are frequently considered more empirical than physics-based.

As a consequence, efforts have been undertaken to find physics-informed data-driven techniques that are capable of mitigating some of the limitations of these more empirical approaches.  For example, physics-informed neural networks (PINNS) \cite{mathews2021uncovering}, or partial differential equations (PDEs) solved and enriched using Deep Neural Networks (DNNs) \cite{michoski2020solving}, have been recently developed and explored.  These models have generally been used to solve traditional initial-boundary value problems in physics-based PDEs (e.g. multicomponent reactive MHD), but with the added benefits of: 1) significantly improved numerical regularity features, 2) the ability to readily incorporate large data sets into the ``training regime,'' and 3) the ability to simultaneously solve for solutions over an entire parameter sweep (e.g. over not only $(t,x)$ but over $(t,x,\gamma)$, etc.).  The major drawback of using these methods for solving numerical PDEs, however, is: the slower overall run time per forward solve that can render them impractical for high dimensional systems (e.g. gyrokinetics, etc), such as those necessary for understanding the plasma physics that drives magnetic fusion reactors \cite{michoski2020solving}.

Additionally, some data-driven physics models can be applied simultaneously to numerical regimes alongside experimental data, leading to models that are automatically ``discovered'' from within the data \cite{long2018pdenet,de2020pysindy,delahunt2021toolkit}, while remaining consistent with the observed data as well.  Again, these discovered models (discussed more below in section D) can be either largely empirical \cite{long2019pde}, or additionally constrained to be physically consistent with theoretical considerations \cite{PhysRevE.104.015206} or simulation-based considerations (e.g. high fidelity model predictions) \cite{michoski2020solving,chang2021fusion}.   It is generally thought that as the amount of both experimental and simulation data increases, data-driven physics models may become increasingly important for being able to predict and model experimental behaviors while simultaneously connecting the gained insights from these systems to traditional and first principle ways of understanding the plasma physics.

\noindent
[Craig Michoski and Jonathan Citrin]

\subsection{Optimizing experimental workflows with data-driven methods}
\label{MCF_workflow}
%(Erik Olofsson, olofsson@fusion.gat.com)}
%\subsection{Optimizing experimental workflows with data-driven methods (Erik Olofsson, olofsson@fusion.gat.com)}

The experimental campaign planning processes in magnetic confinement fusion are currently not explicitly computer aided or otherwise enhanced with optimization, machine learning, and related machinery. The typical chain of events leading up to experimental scheduling and execution starts with open submission of proposals, followed by expert discussions in topical groups, and finally a selection by committee. It appears highly challenging to formalize this planning process towards a more quantitative exploration-exploitation mechanism but it may be worthwhile attempting to do so. Since the ultimate purpose of the MCF device is to reliably maintain a high-performing MHD-instability-free fusion grade plasma, and several metrics to characterize such plasmas are available, it follows that the campaign planning mechanism could, and also arguably should, somehow consider those metrics algorithmically, in order to optimize the progress towards this purpose.

Explicit human-in-the-loop computer-aided decision support in MCF has been attempted in more focused MCF devices \cite{Baltz:2017} and in ICF optimization enabled by data assimilation \cite{Gaffney_2019}, and also in other process optimizations in experimental physics \cite{Duris:2020} with seemingly excellent results. The integration of such systems into large tokamak user facilities is a novel area which is under-explored. Such systems may require original ideas to effectively allocate experimental resources for multi-user multiple-objective exploration and exploitation.

Practically implementing these types of policies in campaign planning may require a shift of the focus of discussions from what topical areas to prioritize next to what metrics to explore and exploit next, and let sanctioned algorithms automatically generate candidate experiments, which can be further discussed and iterated. Classical experimental design  response surface methods \cite{Myers:2016, Montgomery:2017}, standard Bayesian optimization \cite{Shahriari:2016}, and mechanism design \cite{Nisan:2001, Nisan:2007}, can all be envisioned as part of a toolset to build MCF planning decision support systems. In an abstract sense, any planning system for the experimental workflow is a mechanism that uses past data collected, plus external information including predictive simulation data, to propose where data should be collected next.

Mechanism design (not well known in physical sciences) could even be retrofitted onto existing user facilities planning processes. To introduce the idea, here follows a naive example on optimization of collective valuation. The prototypical optimal social choice mechanism which incentivizes participants to provide truthful inputs is the Vickrey-Clarke-Groves (VCG) mechanism \cite{Nisan:2007}. In the context of collaborative planning on a user facility it could be used as follows. Based on initial community input, management comes up with a shortlist of allocation options which are compliant with resource and contractual constraints and other programmatic boundary conditions. The user facility participants then submit the number of hours they would be willing to work to realize each option. The VCG mechanism selects the option that maximizes the collectively most desirable option (collective eagerness to work on its realization). Crucially, the VCG mechanism uses a formula to charge each participant (extra hours asked to work) such that each participant is best off providing their private true valuation (number of hours actually willing to put up for each option) to the mechanism. Presumably, it also holds that the participants true valuation is positively correlated with their belief in the likelihood of making actual physics progress.

Improvements to programmatic decision support using data-based methods combined with designed value revelation mechanisms is an interesting direction for future research. User facilities are in this sense arenas where groups of tax funded agents compete for access to a machine that can (should) convert their labor into a public good (research output that benefit all, not only the resum\'{e}s of particular individuals) \cite{Kress:2018}. Revised incentive structures and transparent mapping of performance metrics across operational spaces may enhance this public-good aspect.

\noindent
[Erik Olofsson]

\subsection{Diagnostics and Fusion Data Streams}
\label{MCF_diagnostics}
%(Michael Churchill, rchurchi@pppl.gov)}
%o   Interpreting and fusing diagnostic data with ML
%o   RT processing of diagnostic streaming data
In fusion energy plasmas, many disparate diagnostic instruments are simultaneously used in order to cover the multiple physics phenomena covering a range of spatiotemporal scales. In addition, fusion experiments, such as ITER, will run longer pulses, with a goal of eventually running a reactor continuously. The confluence of these facts leads to large, complex datasets with phenomena manifest over long sequences. Fusion scientists have a range of data analysis timescales, from real-time processing for plasma control, to between-shot quick processing of data to give insight to adjustments for next shots, to longer-term deep analysis for science discovery. Diagnostic data analysis has always been fundamental to progress in magnetic confinement fusion energy, and many current and emerging applications of machine learning are aiding scientists in these many tasks making sense of diagnostic data.

Machine learning is being applied to interpreting observed experimental data and extracting from it physical parameters of interest (e.g. electron temperature from line-integrated spectrometer measurements). Traditionally this statistical inference of physics parameters from diagnostic data has been performed under the umbrella of `Integrated Data Analysis' \cite{Dinklage2008}, performing Bayesian analysis leveraging potentially multiple diagnostics. Recent trends are integrating machine learning in the form of neural networks to accelerate the IDA process, which usually either relies on analytic likelihoods, or resorts to slow, sequential MCMC samplers. Neural networks have been trained to do approximate Bayesian inference, replicating a Bayesian model which is used to extract electron temperature from a lithium ion beam emission spectroscopy diagnostic (Li-BES) on the JET tokamak \cite{Pavone2020}. The benefit using a neural network is now the inference of electron temperature (with uncertainties) can be performed in microseconds, versus the tens of minutes typically required for a single experimental time slice, enabling use in between shot or real-time control.

Similar techniques are being applied when the forward model relating physics parameters to observed diagnostic data is a more formal simulator, making the likelihood intractable. Simulation-based inference technique of Neural Posterior Estimation (NPE) use normalizing flow models \cite{Tejero-Cantero2020a} (built with neural networks) to create flexible surrogates, performing the Bayesian inference to infer physics parameters consistent with the simulator, but again producing results is milliseconds. An example application used the fluid plasma and neutral edge transport code UEDGE, which takes in anomalous transport coefficients and produces plasma kinetic profiles of density and temperature. NPE was used to train a normalizing flow model on 10,000 UEDGE simulations, producing a neural network which could then take in profiles of electron/ion density and temperature from diagnostics at the midplane and the outer divertor, and infer the corresponding anomalous transport coefficients, which are consistent with UEDGE \cite{Furia2022}. 

 Various works are using other methods for speeding up and broadening the analysis that can be done with experimental diagnostic data for physics parameter extraction. For example, a simple feed-forward neural network was trained to extract electron temperature from a database of measured spectra from an EUV/VUV spectrometer, based on the measurements of electron temperature from Thomson scattering diagnostics \cite{Samuell2021}. The above works and techniques aim to improve our physical understanding of fusion plasmas by leveraging machine learning to extract physics from experimental diagnsotics. 

Recent trends have focused on various way to accelerate identification of plasma modes or other events directly from diagnostic data, using supervised learning. These applications are for aiding the researcher in identifying items of interest, but also for inclusion of real-time control algorithms. Resevoir computing, a dynamical machine learning model which trains quickly, has been applied successfully to prediction of Alfven eigenmodes in the DIII-D tokamak \cite{jalalvand_alfven_2021}. Neural networks have also been used for very rare and difficult signals to fine, such as solitary bursts before Edge Localized Modes on KSTAR \cite{lee_machine_2021}. Also, convolutional neural networks with dilated convolutions have found utility in working with long sequences for diagnostics with high sampling rate like the Electron Cyclotron Emission imaging (ECEi) diagnostic at DIII-D \cite{Churchill2020}. 

Large-scale data analysis for experimental diagnostics can be accelerated using data science and networking techniques to stream the data from the experiment to large, remote HPC centers. By working with data streams, and leveraging the large HPC compute resources, better and more data analysis can be performed, which can better inform fusion scientists between plasma shots on the best way to optimize the next shot \cite{Churchill2019b}. A demonstration of this used the streaming framework DELTA \cite{kube_near_2022} to stream ECEI diagnostic data from the KSTAR tokamak in Korea to the NERSC HPC center in the USA, and complete spectral analysis of all channel pairs using multiple CPUs on the Cori supercomputer. The entire streaming and analysis completed in 10 minutes, compared to the 10 hours that sequential analysis would take. This opens the door for a range of large-scale parallel analysis, modeling, and simulation to further enhance the information scientists can extract from diagnostic data.  

\noindent
[Michael Churchill]
%o   Interpreting and fusing diagnostic data with ML
%o   RT processing of diagnostic streaming data

%[1]A. Pavone, J. Svensson, S. Kwak, M. Brix, and R. C. Wolf, “Neural network approximated Bayesian inference of edge electron density profiles at JET,” Plasma Physics and Controlled Fusion, Feb. 2020, doi: 10.1088/1361-6587/ab7732.
%[1]A. Jalalvand et al., “Alfvén eigenmode classification based on ECE diagnostics at DIII-D using deep recurrent neural networks,” Nucl. Fusion, vol. 62, no. 2, p. 026007, Dec. 2021, doi: 10.1088/1741-4326/ac3be7.
%[1]J. E. Lee, P. H. Seo, J. G. Bak, and G. S. Yun, “A machine learning approach to identify the universality of solitary perturbations accompanying boundary bursts in magnetized toroidal plasmas,” Sci Rep, vol. 11, no. 1, Art. no. 1, Feb. 2021, doi: 10.1038/s41598-021-83192-2.
%[1]C. M. Samuell, A. G. McLean, C. A. Johnson, F. Glass, and A. E. Jaervinen, “Measuring the electron temperature and identifying plasma detachment using machine learning and spectroscopy,” Review of Scientific Instruments, vol. 92, no. 4, p. 043520, Apr. 2021, doi: 10.1063/5.0034552.

\subsection{Prediction of Tokamak Disruption}
\label{disruption_control}
Disruption, which is an abrupt termination event of tokamak discharge, is one of the biggest issues in fusion energy development \cite{wesson2011oxford}. Magnetic and thermal energy as high as GJ is released in very short time of the order of milliseconds at this event. Consequently, disruption causes harmful damage on tokamak through excessive thermal load on the wall, magnetic force and run-away electrons. Therefore, prediction, avoidance and mitigation of disruption is prerequisite for a tokamak fusion reactor.

Extensive works for disruptions have been done since the early stage of fusion research\cite{boozer2012pop} and intensive works targeting the operation of ITER are being implemented by international collaborating efforts \cite{lehnen2015jnm,indranil2021nf}. Although understanding of physical process of disruption has been deepened by the MHD theory and simulation, prediction capability of disruption still remains limited. Since disruption is highly non-linear dynamics with complex interaction of different physical processes \cite{devaries2011nf}, it is essentially difficult to predict disruption by the framework of time-dependent differential equations defined a-priori. Instead, data-driven approaches based on a-posteriori observation are anticipated to give an induced model reliable for practical use \cite{kates-harbeck2019nature}. In this subsection, development of data-driven models for disruption prediction of tokamak plasmas is reviewed.

Deep-learning algorithm for multi-machine disruption prediction has been proposed and achieved high predicting accuracy across multiple tokamaks \cite{zhu2021nf}. This means device-independent representations of disruptive characteristics have been identified. Simultaneously, this work has shown that non-disruptive property is device dependent and only use of existing tokamaks is still not enough to predict disruption in a new tokamak. It is also noted that synthetic data from numerical simulation do contribute to improvement of prediction capability.

The approaches of interpretable machine learning models, which are contrast methodology of deep learning \cite{zhu2021nf}, neural network \cite{kates-harbeck2019nature} and generative topographic mapping \cite{pau2019nf}, are attracting interests because not only they improve prediction capability but also their resultant expression enables exploration of underlying disruption physics. Physics validation of the model/hypothesis would secure limitation of generalization performance. Also, these approaches have high potential compatibility with actuators for disruption avoidance and mitigation. Radom Forest (RF) algorithm and sparse modeling via Exhaustive Search (ES) and Support Vector Machine (SVM) are referred as examples.

The RF algorithm has been applied to the prediction of disruption is a variety of tokamaks such as DIII-D \cite{rea2019nf}, JET \cite{rea2020fst}, Alcator C-Mod \cite{tinguely2019ppcf}, and EAST \cite{hu2021nf}, and it has been successfully integrated with the real-time plasma control system on DIII-D and EAST. Disruptivity, that is the final probability of disruption, is characterized by the average result of decision trees to classify disruption/non-disruption from training. It should be noted that this approach can quantify the relative contributions of the various input data signals to disruptivity.  Disruptivity is expressed in the decomposition formula of the sum of each feature contribution and bias of the intrinsic value of the sample mean in the classification scheme. Since the decision paths in RF trees provides measures of explainability of input data, effectiveness of new input data is easily assessed. For examples, peaking factors of plasma parameters such as temperature, density and radiation are proved to enable earlier prediction. In other words, selection of input parameters based on hypothesis and physical insight is essential for improvement of a predictor.

\begin{figure*}[t!]
  \centering
   \includegraphics[width=0.7\linewidth]{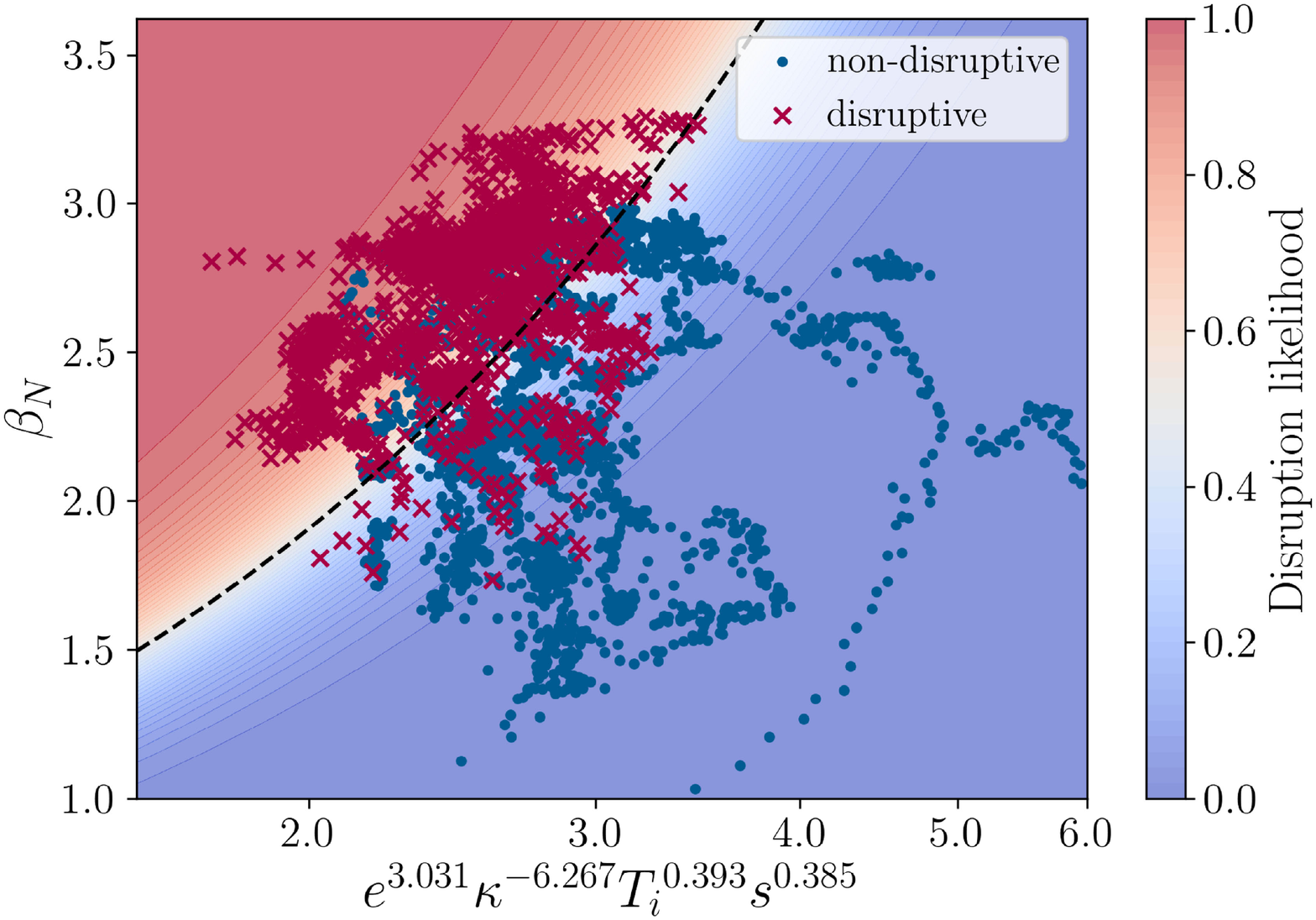} 
  \caption{Contour plot of disruption likelihood. Here, $\beta_N$, $\kappa$, $T_i$, \textit{s} and e are normalized beta, plasma elongation, ion temperature, magnetic shear and Napier’s constant.}
  \label{fig:ptd}
\end{figure*}

Not limited to disruption prediction, the selection of input parameters is an essential issue for machine learning. The ES, which exploits the inherent sparseness in all high dimensional data to extract the maximum amount of information from the data, selects key parameters subject to the SVM classifier for disruption. With regard to high- $\beta$ disruption in JT-60U, four physical parameters have been extracted as key parameters to describe the boundary between the disruptive and the non-disruptive zones \cite{yokoyama2021pfr}. Then it has been found that disruption frequency can be expressed as of the distance from the boundary in multidimensional space. Consequently, the disruption likelihood has been quantified in terms of probability based on this boundary expression. Figure~\ref{fig:ptd} shows the contour plot of the disruption likelihood on the plane of the normalized pressure $\beta_N$ and the function of residual extracted parameters. It is noted that the boundary function is expressed in a power law so as to be compatible with physics discussion. 

Careful deliberation of the expression of the disruptivity/disruption likelihood, which is derived with machine learning, could lead to the elucidation of the underlying physics behind disruptions. Data-driven approach to prediction of tokamak disruption is inevitable for the plasma control system and the device protection system in ITER as well as a next demonstration fusion reactor.

\noindent
[Hiroshi Yamada and Tatsuya Yokoyama]

%(Hiroshi Yamada)

\subsection{Surrogate models of fusion plasma}
\label{MCF_surrogate}
\label{sec:surrogates}
% Still Work in Progress

A challenge in multi-physics simulation of MCF systems~\cite{poli:2018} is to achieve high physics fidelity at a computational burden that is compatible with the desired use-case. This is particularly acute for many-query applications such as sensitivity studies, uncertainty quantification, scenario optimization, and reactor design. Fast simulations can also be applied in control-oriented simulators, where high accuracy is critical for powerful new techniques such as controller design through reinforcement learning~\cite{degrave:2022}.

\begin{figure*}[t!]
  \centering
   \includegraphics[width=0.8\linewidth]{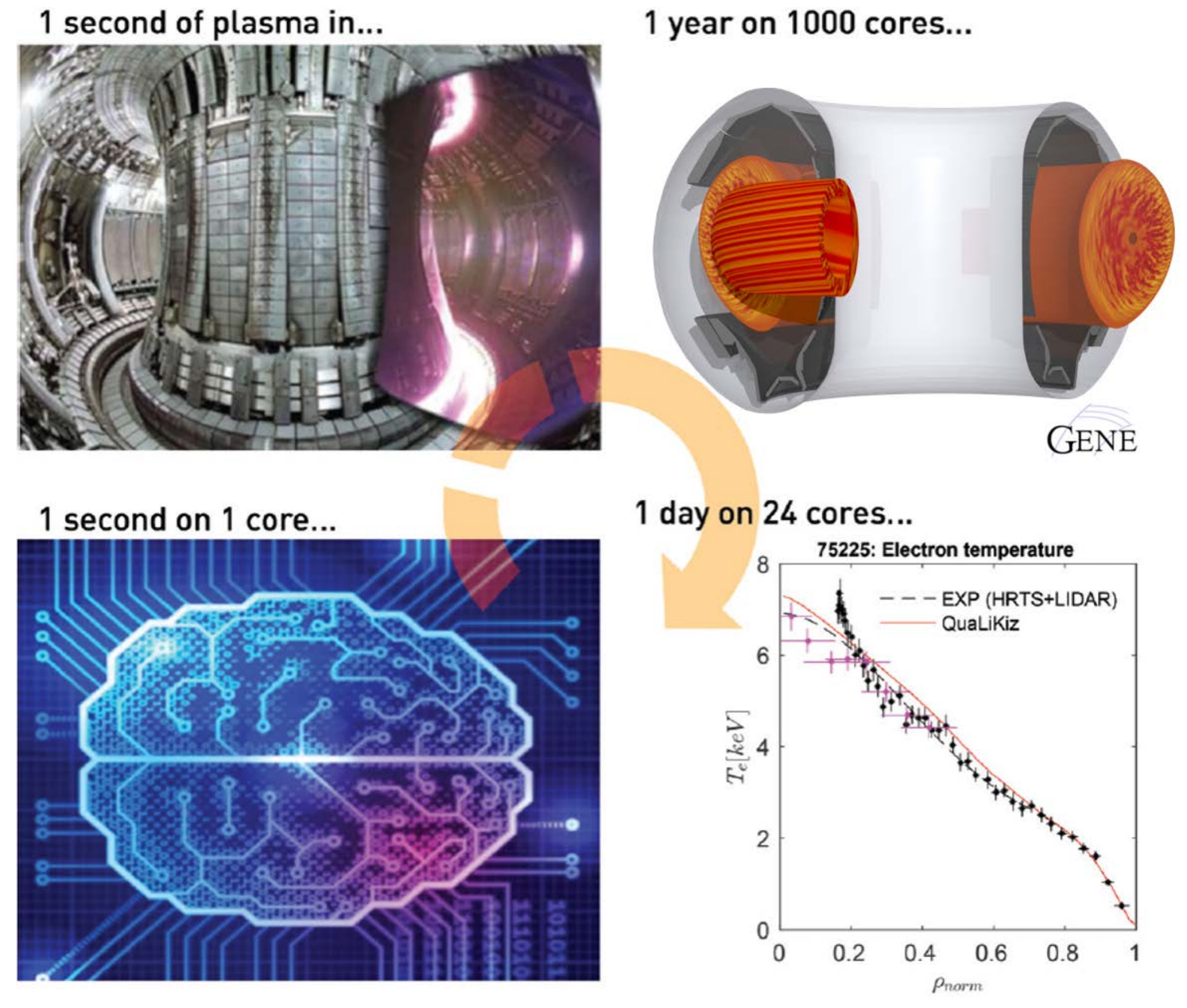} 
  \caption{Hierarchy of models applied towards fast and accurate multi-physics simulation, with the example of tokamak core turbulence. Routinely modelling a tokamak scenario is prohibitively expensive with high-fidelity nonlinear gyrokinetics (upper right panel). However, the high-fidelity model verifies and validates reduced-order-models (lower right plot), which are then applied to generate training sets for ML-surrogates (lower left plot) for fast simulation.}\label{fig:QLKNN}
\end{figure*}

Carrying out regression of the individual physics models that comprise the multi-physics suite, using supervised learning methods, can circumvent the conflicting constraints of model speed and accuracy. The ML-learned surrogate model then provides faster (often by orders of magnitude) multi-physics simulation when applied as drop-in replacements for the original models. The computational cost is relegated to the training set generation phase, facilitated by HPC resources. See Fig.~\ref{fig:QLKNN} for a conceptual overview. In principle, physics models which are too slow for routine direct application in multi-physics simulation can also be incorporated in such a manner, as long as there is sufficient computing resources for generating the required training set. This idea is compelling, since the ML-surrogate has the potential to then be both faster and more accurate than present-day multi-physics modelling capabilities. 

To date, multiple surrogate models have been developed for fast MCF modelling applications, primarily (but not exclusively) applying feed-forward neural network architectures. A non-exhaustive list of examples are summarized below: 
\begin{itemize}
    \item NUBEAM Monte-Carlo neutral beam heating code~\cite{boyer:2019}. Principle component analysis was applied to reduce the dimensionality of the 1D input and output profiles. Extensively validated on DIII-D~\cite{morosohk:2021} and NSTX-U~\cite{boyer:2019}.
    \item Turbulent transport models. The QuaLiKiz-neural-network~\cite{vandeplassche:2020,ho:2021} utilizes prior knowledge of the physical input-output mapping structure for determining physics-informed constraints of network topology and optimization cost functions that improve model fidelity. Applications include JET Tritium ramp-up optimization~\cite{ho:2022} and ITER scenario optimization~\cite{vanmulders:2021}. Similar work was carried out for TGLF~\cite{meneghini:2020} with applications for scenario optimization and control~\cite{morosohk:2020}, as well as the multi-mode model~\cite{morosohk:2021b}. A surrogate of the higher fidelity turbulent transport model GKW has been developed for JT-60U parameters~\cite{narita:2021}.
    \item EPED neural network for pedestal predictions, and core-pedestal coupling workflows~\cite{meneghini:2020}.
    \item 3D MHD equilibrium calculations for stellarator optimization applications~\cite{merlo:2021}
    \item MHD instability calculations, as part of a disruption predictor stack~\cite{piccione:2020}
    \item Surrogate formula for divertor heat-load width built from combined experimental and gyrokinetic simulation data~\cite{chang2021fusion}. 
\end{itemize}        

Further extension of these techniques to incorporate all components of the MCF multi-physics simulation stack provides a pathway towards fast and accurate interpretation of present-day experiments, scenario design and optimization (including inter-shot), and control-oriented modelling. Future devices such as ITER will require the availability of such a Pulse Design Simulator to increase shot efficiency and reduce risks. 

A common challenge in constructing the surrogate models is on the data generation side, particularly for high-fidelity physics models with a higher computational burden. It is critical to establish robust high-volume computation workflows, automated data validation and filtering pipelines, and selective sampling techniques. The neural network outputs also need uncertainty quantification to establish trust zones and flag when the surrogate model is extrapolating. At the simplest level this is achievable by assessing the variance of an ensemble of identically trained models. Ideally, the model UQ should be coupled to an active learning pipeline whereby the model training set can expand when new parameter space is encountered. 

\noindent
[Jonathan Citrin]
%(Jhonathan Citrin)

\subsection{Magnetic Fusion Energy Data Challenges and Solutions}
\label{MCF_data}
%(Brian Sammuli, sammuli@fusion.gat.com)}
%\subsection{MFE Data Challenges and Solutions (Brian Sammuli, sammuli@fusion.gat.com)}

Data access patterns for machine learning (ML) workflows are fundamentally very different than traditional access patterns for magnetic fusion experimental or simulation data. Conventional repositories of experimental data have been designed for small-scale human consumption in the control room and are mostly aimed at simultaneous visualization of small amounts of data gated by the visual/mental response time of the human operator. In significant contrast, ML access patterns are driven by algorithms than can potentially read and use vast amounts of data, requiring substantially more computational resources for data loading and processing. Additionally, issues associated with data curation such as data discovery, cleaning, normalization, and labeling are all critical components of successful fusion ML studies. 

\begin{figure*}[ht]
    \centering
    \includegraphics[width=\linewidth]{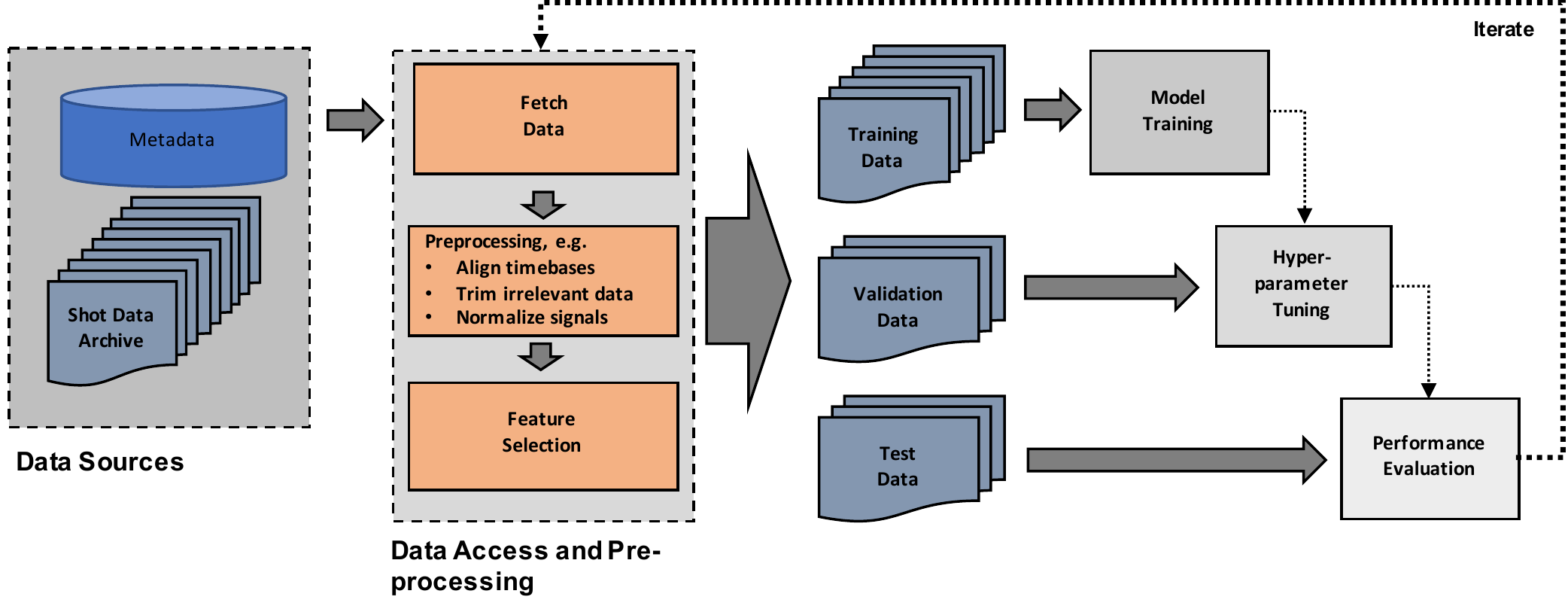}
    \caption{A typical supervised learning workflow for data-driven Magnetic Fusion Energy (MFE) studies. Data exploration, access, and pre-processing are conducted iteratively in conjunction with ML modeling. An FDP would facilitate rapid execution of this loop.}
    \label{fig:data_workflow_diagram}
\end{figure*}

These issues were outlined in the Report of the Workshop on Advancing Fusion with Machine Learning \cite{Humphreys:2019}, which highlighted several limitations of the conventional data repositories; shortcomings that need to be addressed to fully harness the transformational potential that ML could provide in many areas of fusion energy. In particular, the report supports the idea of a community-wide Fusion Data Platform (FDP) targeted at ML research. The core idea of such an FDP is to provide an integrated environment for machine learning and data exploration studies, supported by a common interface.
Data must be staged, and supported with sufficient metadata to support rapid, iterative ML workflows, an example of which is illustrated in Figure  \ref{fig:data_workflow_diagram}. ML studies typically integrate a large number of software tools, so a significant library of tools must also be included to support such workflows. Examples of support tools include data visualization, dimensionality reduction, rapid data space analysis tools, along with the tools needed to actually conduct ML training, testing, and inference.

The \mbox{DIII-D} data archive provides representative examples of both the size and variety of data used by the fusion ML community. It currently consists of $\sim$0.4 petabytes of data accumulated over decades of operation. It contains both raw, unprocessed signal data that is stored in the GA-implemented PTDATA system \cite{mchargptdata}, and processed data (such as equilibrium reconstructions) that is stored in MDSplus \cite{mdsplus}. The data contains a wide array of dimensionalities, ranging from scalars to images, and signals with sample rates spanning multiple orders of magnitude. Historically, the most typical access pattern for this data has been experimental scientists analyzing on the order of 10 ($O(10)$) shots with $O(10)$ signals per shot, with the I/O and processing capabilities of the archive system sized accordingly. The access patterns required for machine learning applications have proven to be significantly more resource intensive. A typical ML study conducted using \mbox{DIII-D} experimental data might be able to take advantage up to the scale of $O(10^5)$ discharges. In recent years, \mbox{DIII-D} has sought to deal with this need for large scale data access by deploying a scaled up data access and processing system. This system includes a complete copy of the DIII-D experimental archives on a BeeGFS parallel file system \cite{beegfs}, along with the TokSearch \cite{SAMMULI201812} framework for parallel data processing, allowing for multiple order of magnitude data processing throughput improvements for typical ML use cases.  

Data discovery relies to a large extent on the ability to perform expressive queries for metadata. For example, a plasma disruption study needs expert-labeled annotations indicating both the time of occurrence and type of disruption. The \mbox{DIII-D} experimental data system is integrated with a Microsoft SQL Server \cite{sqlserver} relational database which records $O(100)$ metadata fields across $O(10)$ tables for each shot, including a schema for recording disruption information. A typical ML application will often gather a preliminary list of shots to process by first querying the relational database. As a simple example, one might be interested in shots from a particular date range. Or, one might have search criteria related to shot length, shot start time, experimental logbook entries, maximum plasma current, etc, all of which can be queried using standard SQL. However, it is worth noting that while the approach taken by \mbox{DIII-D} might effectively utilize one set of tools, there has not been a community-wide effort at standardization, particularly with regard to metadata management, an issue that a dedicated FDP would address. Such standardization would also facilitate increased engagement with domain experts who could more easily provide the annotations needed for classification studies.

Magnetic fusion data is fairly unique in its variety and scope. A single ML study might utilize:

\begin{itemize}
    \item 0-D scalar time series (e.g. magnetics);
    \item 1-D profile time series (e.g. current profile);
    \item 2-D grid data time series (e.g. equilibrium reconstructions);
    \item image time series (e.g. infrared camera data)
\end{itemize}

Each of these items may be stored in a different file format, and each may have one or more associated metadata elements. Such breadth and depth of data underlines the need for a community-wide effort toward standardization, which, given the critical importance of data quality and availability for ML, will have a dramatic impact on the ability of the community to execute data-driven studies.

\noindent
[Brian Sammuli and David P. Schissel]

\subsection{Data Science for extreme scale simulation}
Global nonlinear simulation using fundamental kinetic equations in the whole plasma volume including realistic divertor geometry requires extreme scale simulations.  The soon-to-arrive exascale computers will be great tools, but the size of the filesystem capacity is relatively small compared to the compute node memory.  This brings up the necessity for the online data analysis and data reduction/compression before being written out to the filesystem.  The online data analysis can be done in the simulation codes at every timestep if the analysis routine is well parallelized.  However, there are analysis routines that may not be easily parallelized.  In this case, the simulation data can be offloaded to some analysis nodes using asynchronous RDMA or one-sided MPI data transfer at every timestep.  Thus, the computing does not slow down while the data is analyzed in the analysis nodes.  AI/ML can be used in the analysis nodes for efficient visualization and scientific discovery.  Reduction and compression of the analysis data can also be performed in the analysis nodes.
\begin{figure*}[t!]
  \centering
   \includegraphics[width=1.0\linewidth]{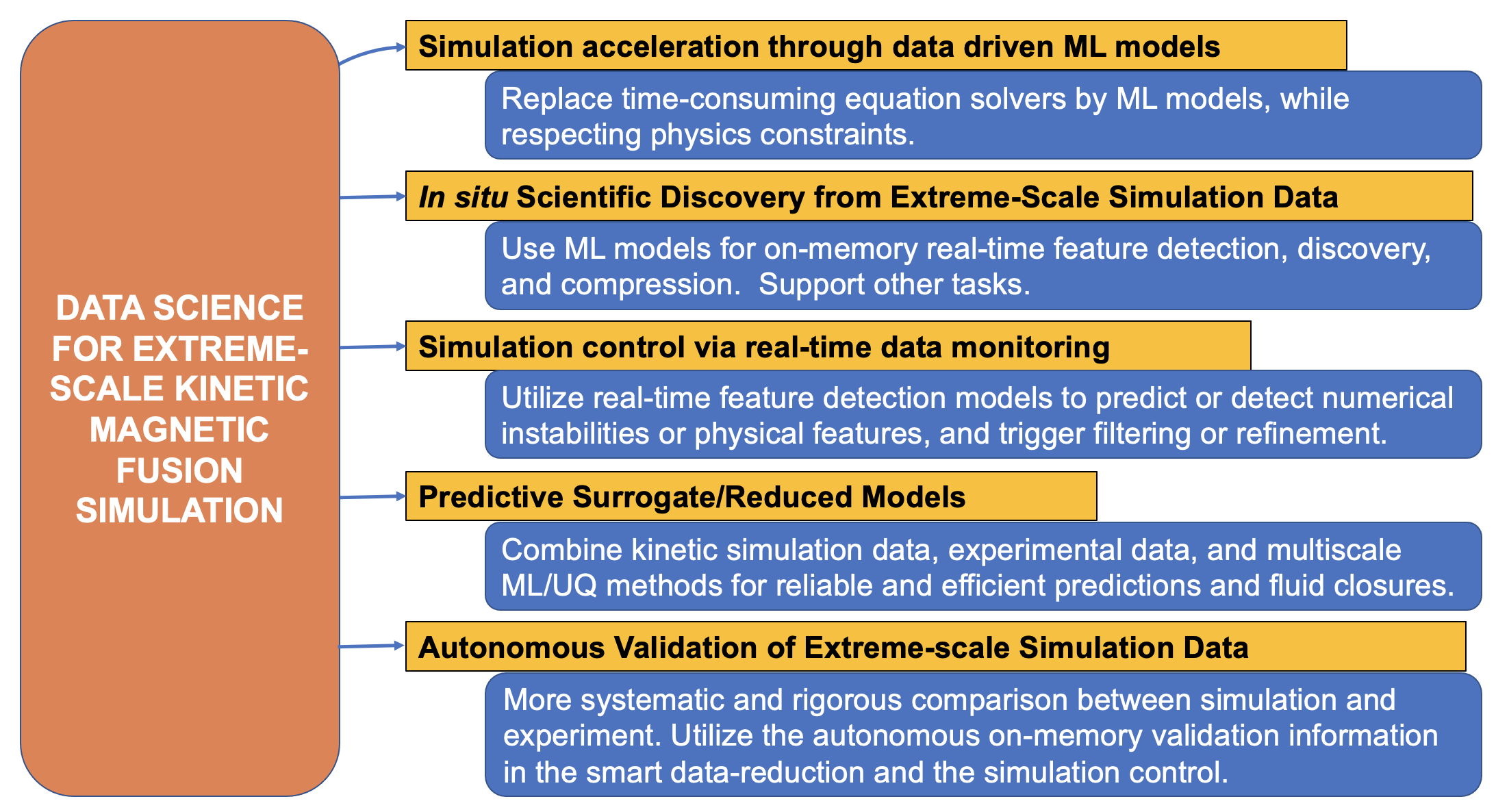}
  \caption{Data science topics for extreme scale kinetic magnetic fusion simulation }\label{fig:simulation}
\end{figure*}

Data driven AI/ML can accelerate the extreme scale simulations by replacing some compute intensive kernels with AI/ML inference routines.  Fokker-Planck collision operation is an example \cite{miller2021encoder}.  Preconditioner and PDE solvers can be good candidates.  However, some difficulty lies in the accuracy and physics property conservation in the data-driven routines: e.g., L2 error, mass conservation, momentum conservation, energy conservation, viscosity conservation, etc. If we aim for 1\% error at the end of 1,000 timesteps simulation, a data-driven routine must have $<10^{-5}$ relative error to avoid accumulation in the possible ``drifting error."  This level of error bound in AI/ML is not easy and requires support from the fundamental AI/ML scientists.

Data driven AI/ML can also perform other functions to help the extreme scale simulations in real time: by detecting and mitigating possible load imbalance, by detecting and suppressing known numerical instabilities, by utlizing UQ techniques to request simulation steering into needed input/output parameter space and to execute autonomous validation tasks using pre-loaded experimental data in the independent data analysis nodes, by combining simulation-experimental data to help construct predictive surrogate models (see subsection~\ref{MCF_models}), etc..  Figure~\ref{fig:simulation} depicts summary of the data science topics for the extreme scale kinetic magnetic fusion simulations.

\noindent
[C.S. Chang]

\subsection{Challenges and outlook}
As for other application science areas, there are numerous challenges in utilizing data-driven sciences in the magnetic confinement fusion research. Besides the challenges and outlook listed in each of the above sub-areas, an important aspect to keep in mind in discussing the challenges and outlook is that magnetic confinement fusion is different from many other scientific projects in that it is an international-scale mission oriented project.  This means that global collaborations among geographically separated large-scale laboratory facilities and between laboratory experiments and high-performance computations are key to the success.

Vast amount of data produced (and stored) in different format at different experimental facilities and by different simulation codes over decades of time span (see Sec.~\ref{MCF_data}) may require building a community-wide federated database and workflow system \cite{federated-data}, which is based on the meta-database management system and which honors individual institution's and code's data format transparently and maps multiple autonomous database systems into a single federated database via wide area network without the need for centralized data mirroring.

\begin{figure*}[t!]
  \centering
   \includegraphics[width=1.0\linewidth]{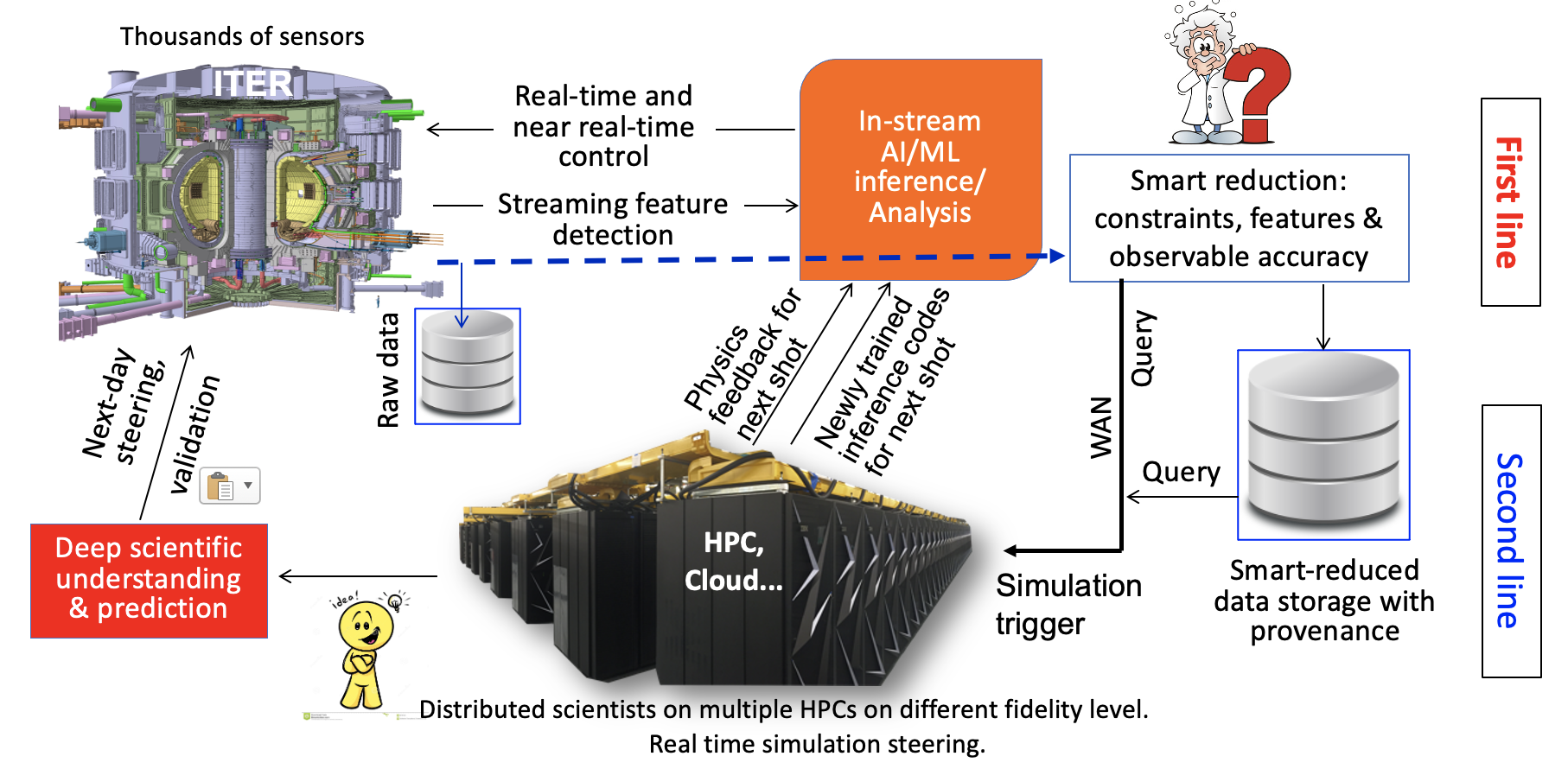}
  \caption{A schematic diagram of the data movement workflow, showing the first line data science region at the experimental site and the second line of data science and HPC studies at remote sites }\label{fig:MCF_workflow}
\end{figure*}

The inference codes can be placed on or nearby the collaborative experiments, such as ITER or future prototype reactors. However, their learning should be performed on remote HPCs, with frequent reinforcement learning for timely update of the inference codes, using streaming data to cope with observational variance. To achieve this, a global management system is needed over wide area network for efficient workflow (see Fig.~\ref{fig:MCF_workflow}).

The continuous accumulation of data to be generated by ITER, or future fusion test reactors, will reach to be enormous (tens of exabytes over the lifetime of ITER experiment).  Historically even on Today's tokamak experiments, once the experimental data hit the permanent storage tape, they are seldom utilized for scientific discovery.  It is a challenge but desirable that the streaming data out of the various experimental diagnostics to be organized according to the features and reduced/compressed without the loss of the features on the way to the permanent storage. In this process, a special request can be sent to the simulation communities, together with the feature-preserved reduced data, for timely study of the observed experimental phenomena and feedback for the design of improved experimental scenarios as mentioned in Sec.~\ref{MCF_diagnostics}. Various AI/ML techniques are expected to be a highly valuable tool in accomplishing this, including the workflow framework building. All the data science techniques discussed in the Fundamental Data Science Section~\ref{Section2} and in this MCF section can be utilized in this workflow framework at various stages.

\noindent
[C.S. Chang]

\section{Inertial confinement fusion and high-energy-density physics} \label{sec:icf_hedp}
%[Coordinator: Spears]}
\subsection{Introduction}
 The field of high energy density physics (HEDP) is typically defined as plasma physics with energy densities $>10^{11}$ J/m$^3$, equivalent to pressures $> 10^6$Bar. HEDP research covers a broad range of systems from strongly-coupled `warm dense' matter, through laboratory astrophysics and inertial confinement fusion, to ultra-intense laser plasma interactions, and more. While these sub-fields probe a zoo of physics phenomena, they are all underpinned by the twin pillars of experimentation and simulation. The difficulties in reaching the conditions of interest in an experiment, collecting high-quality data, and modeling the results means that both pillars rely on the largest experimental and computing resources available worldwide.\par
 We envision data-driven methods as a cross-cutting third pillar that both improves HEDP experiments and simulations, and sits at the interface between the two. Data-driven methods provide an opportunity to efficiently featurize our complex datasets, to reliably combine information from simulations and experiments, and to accelerate the rate at which simulations and experiments can be performed. As a result, HEDP and ICF problems are quickly becoming an important driver of data-driven methods for science. In the remainder of this section, we will describe some aspects of the research in these areas.\par
 
\noindent
[Brian Spears]

\subsection{Representation learning for multimodal data}
%[Anirudh, Kustowski]}
%\documentclass{article}
%\usepackage{graphicx}
%\begin{document}

%\section{Representation learning for multimodal data}

\begin{figure*}
\centering
  \includegraphics[scale=0.5]{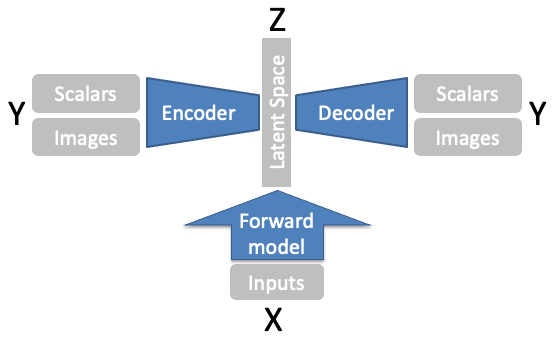}
\caption{Neural network architecture predicting multimodal outputs $Y$ of the simulations. The outputs are first compressed using an autoencoder into a latent space $Z$, and then a forward model is trained to predict these compressed outputs from the inputs $X$.}
\label{fig:MaCC}
\end{figure*}

Predictive models in plasma physics are used to set our expectations about future experiments with varying designs; to explore, optimize, and automate new designs; and to infer important physics parameters that cannot be accurately measured or simulated, thereby allowing for an improved understanding of the experiments. While expensive simulations can generate a variety of diagnostic data types, in many applications, simulations need to be replaced with fast-to-evaluate predictive surrogates, which have traditionally been fitted to only a handful of scalar diagnostic outputs. This approach ignores rich observational and simulated data, such as high spatial and temporal resolution x-ray and neutron images, or neutron yields recorded at multiple azimuths around the burning plasma. These non-scalar detectors are routinely deployed during experiments at nuclear fusion facilities, such as NIF, Omega, and Z, providing additional and more detailed information about processes operating within the plasma. Including these multimodal data can help break degeneracies in the scalar-only models and reduce model uncertainty.

Combining multimodal data poses a challenge as it requires the model to predict thousands of variables in each image or array, and these variables are typically correlated both within and across data modalities. Rather than training the model to predict raw data, ideally, one would like to find a representation of these data in terms of a set of independent variables corresponding to the key physics parameters controlling the experiment. Unfortunately, standard compression techniques cannot detect correlations between different data modalities and collapse an arbitrary combination of data arrays into a set of decorrelated variables. Recent advances in deep learning, however, provide tools for building data-driven representations that both compress and decorrelate multimodal data making them suitable for the inclusion in the predictive models. At the same time, the computing power at the national laboratories has grown to the point, where a sufficiently large number of expensive, radiation hydrodynamics ICF simulations can now be run to train data-hungry deep learning models \cite{nora}.

Equipped with more powerful supercomputers and deep-learning tools, researchers at LLNL have designed a new deep learning architecture to include multimodal data and build more robust predictive surrogates of ICF simulations.\cite{Anirudh_2020} In this architecture, simulation outputs $Y$, consisting of images and scalars, are embedded by an autoencoder into a low-dimensional manifold $Z$ (Figure \ref{fig:MaCC}). The autoencoder consists of two neural networks: an encoder $E: Y \rightarrow Z$ and a decoder $D: Z \rightarrow Y$. To reduce statistical dependencies between the compressed latent variables, a Wasserstein autoencoder is used instead of a standard autoencoder. Adding the adversarial training strategy, in addition to the standard L-2 norm minimization, causes the autoencoder predictions to look like training samples, enforcing consistency with the physics relations built into the simulation. The second part of the architecture is the forward model $F: X \rightarrow Z$ connecting the input design space $X$ with the diagnostic outputs compressed by the autoencoder $Z$. The robustness of this model is improved by imposing a cyclic consistency regularization to penalize predictions that are inconsistent with the pseudo-inverse network, which is trained simultaneously with the forward model.

While the advanced features of this architecture allow the model to predict multimodal simulation outputs nearly perfectly, physicists need to know whether the model also preserves physics relations learned from the simulation. One such relation was investigated by \cite{anirudhneurips}. Using the approximation of the Planck's law, the brightness of images from 4 energy bands was converted into the electron temperature and compared with the ion temperature - one of the scalar diagnostics. These two temperatures are strongly correlated in the simulation outputs. The correlation is very well preserved in the predictions of the neural network model for the validation samples, even though this correlation was not imposed as a constraint during the autoencoder training.

In summary, representation learning enables the inclusion of diverse types of diagnostic data in training of accurate, scalable, predictive surrogates of the simulations. Advanced deep learning techniques allow for building representations that are better at preserving physics relations between predicted diagnostics than standard neural networks. 

%\end{document}
%\bibliography{RepresentationLearning.bib}

%%% Local Variables:
%%% mode: latex
%%% TeX-master: "main"
%%% End:
\label{sec:RepresentationLearning}
\noindent
[Rushil Anirudh and Bogdan Kustowski]  

\subsection{Transfer learning for simulation and experimen}
Standard computer simulations for indirect drive inertial confinement fusion, without platform-specific corrections, often show discrepancy with experiments. In the ICF community, a new approach to calibrating simulations to experimental data has been  shown to create models that can predict the outcome of ICF experiments better than simulations alone. 

This approach leverages a machine learning technique called “transfer learning” to merge simulation data and experimental results into a common model. Transfer learning is when a neural network trained on a large dataset to solve a given task is partially retrained to solve a different, but related task, for which little data is available. For example, a neural network trained on the ImageNet dataset to label random objects (such as cars, trees, cats) can be modified by retraining just a few layers of the neural network to label very specific images, such as the type of aircraft in a photo, which has a significantly smaller training dataset. 

In ICF, transfer learning is used to take simulation-based neural networks and partially retrain them on sparse sets of experimental data, creating a model that is more predictive of experiments than simulation alone. 

Two approaches to transfer learning for ICF have been published in recent years, one which learns a neural network mapping from design input parameters (such as target geometry and laser pulse) to experimental outputs, and one which transforms simulation outputs to experimental outputs via a transfer learned autoencoder. 

\begin{figure*}
  \centering
\includegraphics[width=0.75\textwidth]{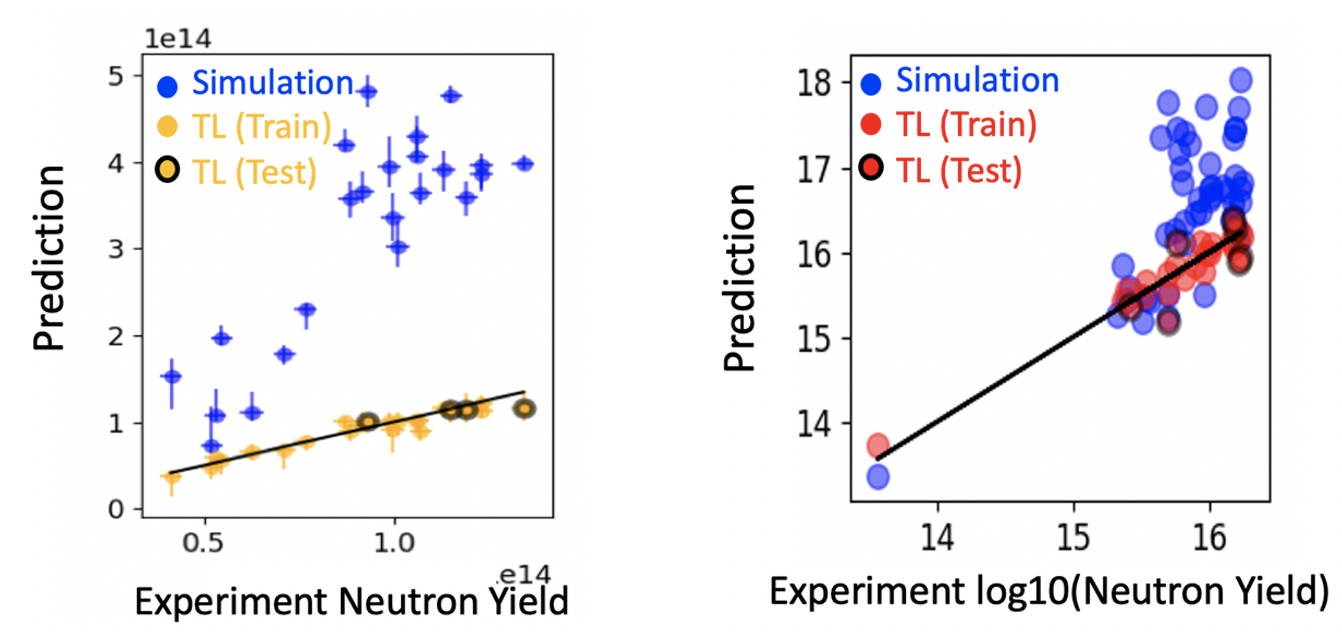} 
\caption{A. Actual versus predicted values of the neutron yield for Omega ICF experiments based on simulations only (blue) and the transfer learned neural network (yellow). B. Actual versus predicted values of the neutron yield for indirect drive ICF experiments from the NIF. The transfer learned autoencoder predictions (blue) are significantly more accurate than the pre-shot simulation predictions (green).
}
\label{fig:tl}
\end{figure*}

The input to output mapping approach was first demonstrated by predicting the outcome of direct drive ICF experiments at the Omega Laser Facility. A neural network trained on 30,000 one-dimensional LILAC ICF simulations was partially retrained on 19 experiments that spanned to same design space as the simulations. The model predicted the subsequent 4 experiments with significantly higher accuracy than the LILAC simulations alone; this is shown in Figure~\ref{fig:tl} A. 

The autoencoder-based transfer learning technique was developed to overcome challenges associated with indirect drive ICF – the expense of integrated hohlraum simulations and the sparsity of indirect drive ICF data. An autoencoder trained on a large database of capsule only simulations learns to encode ICF outputs (such as yield, temperature, density) into a latent space, and decode back to the outputs. The model is transfer learned with pairs of integrated hohlraum pre-shot simulation outputs and corresponding experimental measurements for a database of 50 ICF experiments carried out at the NIF. The resulting model produces an accurate mapping from preshot simulation predictions to expected experimental measurements; resulting predictions from this model are shown in Figure~\ref{fig:tl}B. 

A key benefit of each approach to transfer learning is the ability to immediately update the model after each experiment by retraining the network with the new data. This means the models get more accurate over time, providing a powerful new tool for future design exploration by providing empirically realistic sensitivities to design parameters. Furthermore, the transfer learned models can guide us toward high performing designs more efficiently than simulations alone.

While transfer learning techniques described above predict scalar diagnostic data, multimodal can be incorporated in predictive models using representation learning. Matching additional data types to better inform the model is particularly important in transfer-learned models because they are retrained on only a handful of experimental samples. Machine learning literature, however, does not explain how to apply transfer learning in multimodal architectures with autoencoders, such as the one discussed in Section~\ref{sec:RepresentationLearning}. Multiple retraining options have been tested at LLNL and discussed in~\cite{kustowski2021transfer}. Using synthetic ICF data, the authors demonstrated that retraining the decoder part of the neural network architecture allows for correcting systematic biases in important characteristics of x-ray images, such as the hot spot size, shape, and brightness. Such correction is possible even when only a handful of synthetic experiments is available, as in the case of real ICF experiments. Ongoing research aims to improve this method to handle larger, and more realistic, biases between simulations are real experiments.

Because transfer learning has shown promise at correcting simulated images to match synthetic experiments~\cite{ kustowskifirstpaper}, it will be natural to apply this method to multifidely simulations. An initial model could be trained on a large number of one-dimensional radiation-hydrodynamics simulations, and then elevated to match a smaller number of expensive, two-dimensional simulations, potentially eliminating the need to run thousands of them to train the model from scratch.

%%% Local Variables:
%%% mode: latex
%%% TeX-master: "main"
%%% End:
\label{sec:tl}
\noindent
[Bogdan Kustowski and Kelli Humbird]  

\subsection{Uncertainty quantification and Bayesian inference}
%\section{Uncertainty quantification and Bayesian inference}
%
Quantifying uncertainty presents huge challenges in studies of ICF and HEDP systems that stem from the complexity of both experiments and phyics models. Experimental observations are sparse, difficult to diagnose, and limited in the range of parameter space they can access; as a result, they provide limited information and there is usually some amount of extrapolation to regions where predictions are needed or new physics may be learned. A proper accounting of how much information we have about a system of interest is, fundamentally, a question of uncertainty and this puts uncertainty quantification (UQ) at the forefront of ICF and HEDP research. In recent years data driven methods have been pushing the boundaries of what is possible resulting in more reliable estimates of uncertainty and, hopefully, more predictive computer models.\par
From a data science perspective, experimental datasets are rarely complete enough to make purely experimental-data-driven approaches feasible. Instead, the usual approach is to use the available data to make point checks of physics models (\emph{benchmarking}) or to fit a handful of parameters to observations (\emph{tuning}). The tuned and benchmarked physics model can then be used to make predictions at a new point of interest, with a limited (or no) understanding of the uncertainty in the prediction. Recently, ICF and HEDP researchers have started to formalize the process by treating the physics as a second source of information and to build data-driven models that in are in some way informed by both sources. A variety of approaches have been explored; for example using simplified physics models \cite{Springer_2018,Hsu_2020,Ruby_2020,Ruby_2021}, by using physics considerations to limit the size of the design space \cite{Gopalaswamy_2019,Gopalaswamy_2021}, or by attempting to combine data from disparate but physically related experiments \cite{Nakleh_2021}. Other important efforts aim to pose the benchmarking and tuning of large-scale multiphysics simulations as a Bayesian inference \cite{Gaffney_2019, Lewis_2021}.\par
The Bayesian approach has the advantage that the results automatically capture uncertainties in a statistically consistent manner, while methods that use multiphysics simulations are our best representation of current physics understanding making the results interpretable. However using simulations in a Bayesian inference framework requires huge computational resources since large numbers of simulations runs, each requiring hundreds of cpu-hours to complete, are required. Overcoming this computational barrier has relied on the use of surrogate models \cite{Humbird_2019, Kasim_2021, Anirudh_2020} which aim to replace the simulation with a cheaper approximation. A large set of simulations is run - requiring tens of millions of cpu-hours - and then used to train an approximate interpolator which maps simulation inputs to predicted observables. The key point is that the generation of training data is a massively parallel operation that can leverage leadership-class high performance computing facilities and software tools, while including simulations in the inference directly requires the samples to be run serially.  With a good choice of surrogate \cite{Spears_2018}, a high-fidelity analysis that would be impossible with the simulation itself can be run in a few hours, opening the door for thorough and realistic UQ studies. The process of building surrogate models for large simlations has motivated many of the developments described elsewhere in this paper and has made ICF and HEDP datasets \cite{jag_data} a key driver of developments in scientific machine learning.\par
\begin{figure*}
  \begin{subfigure}[b][][c]{0.35\textwidth}
    \includegraphics[width=\textwidth]{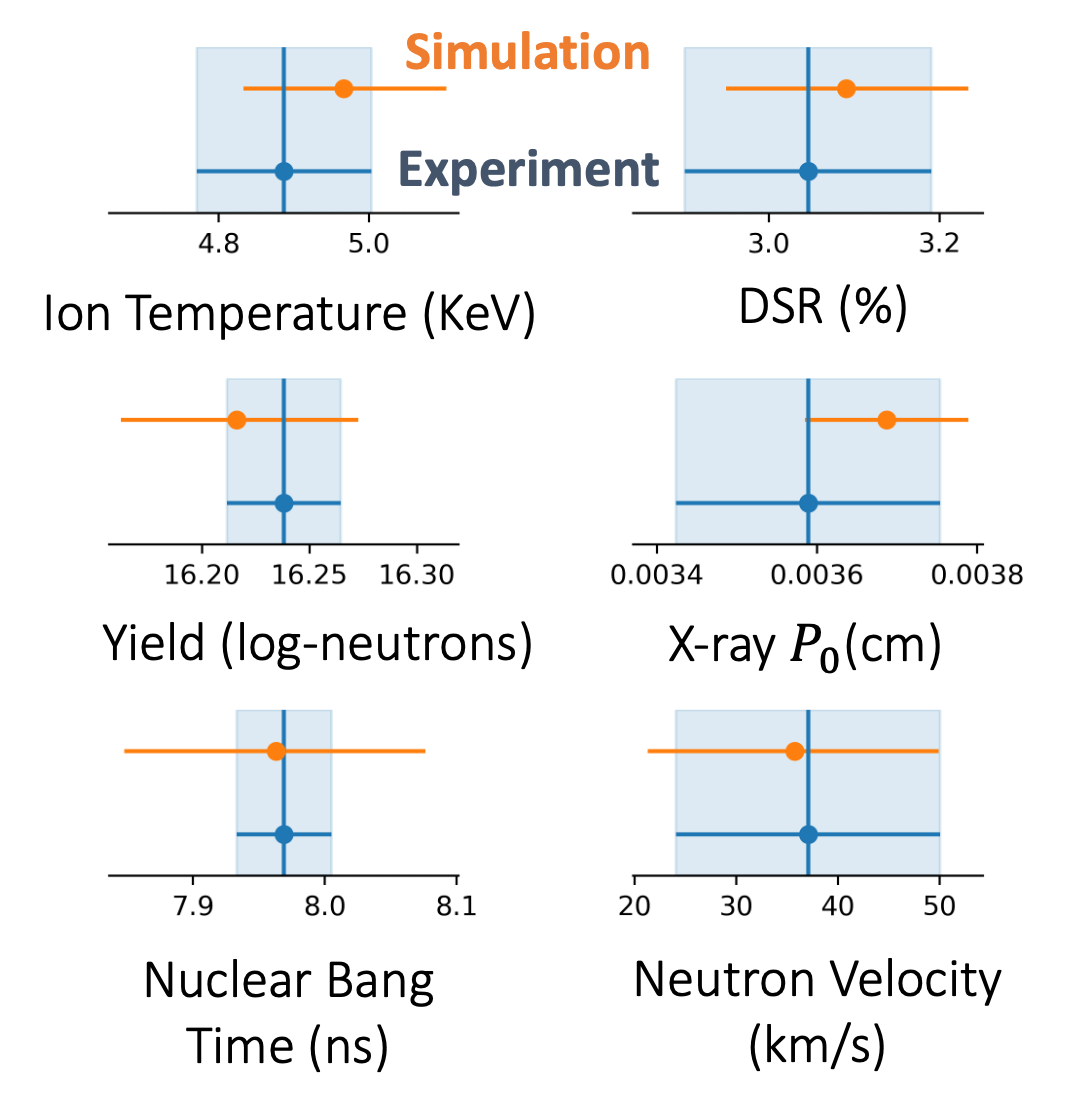}
    \caption{}
    \label{fig:bsps fitted outputs}
  \end{subfigure}
  \hfill
  \begin{subfigure}[b][][c]{0.35\textwidth}
    \includegraphics[width=\textwidth]{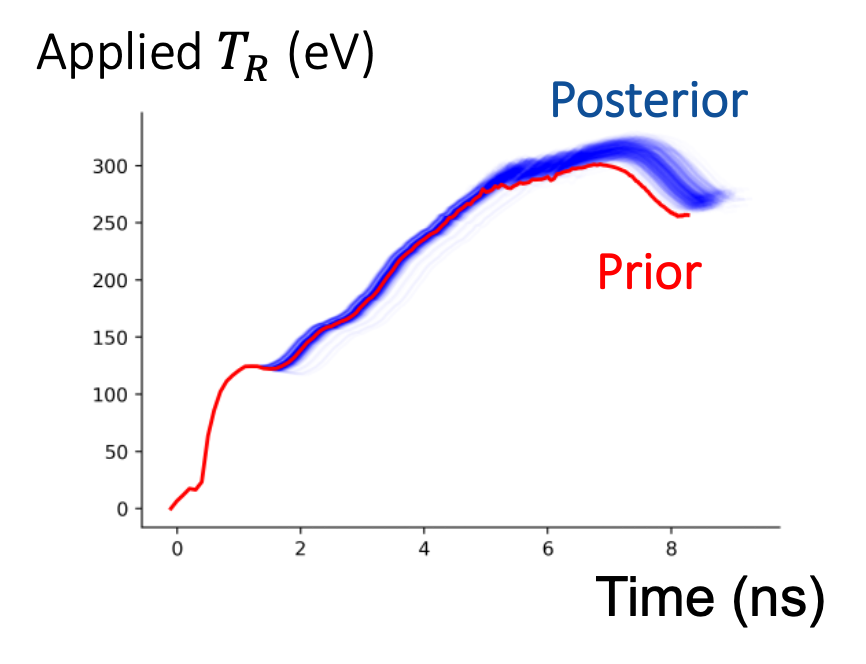}
    \caption{}
    \label{fig:bsps inferred drive}
  \end{subfigure}
  \hfill
  \begin{subfigure}[b][][c]{0.2\textwidth}
    \includegraphics[width=\textwidth]{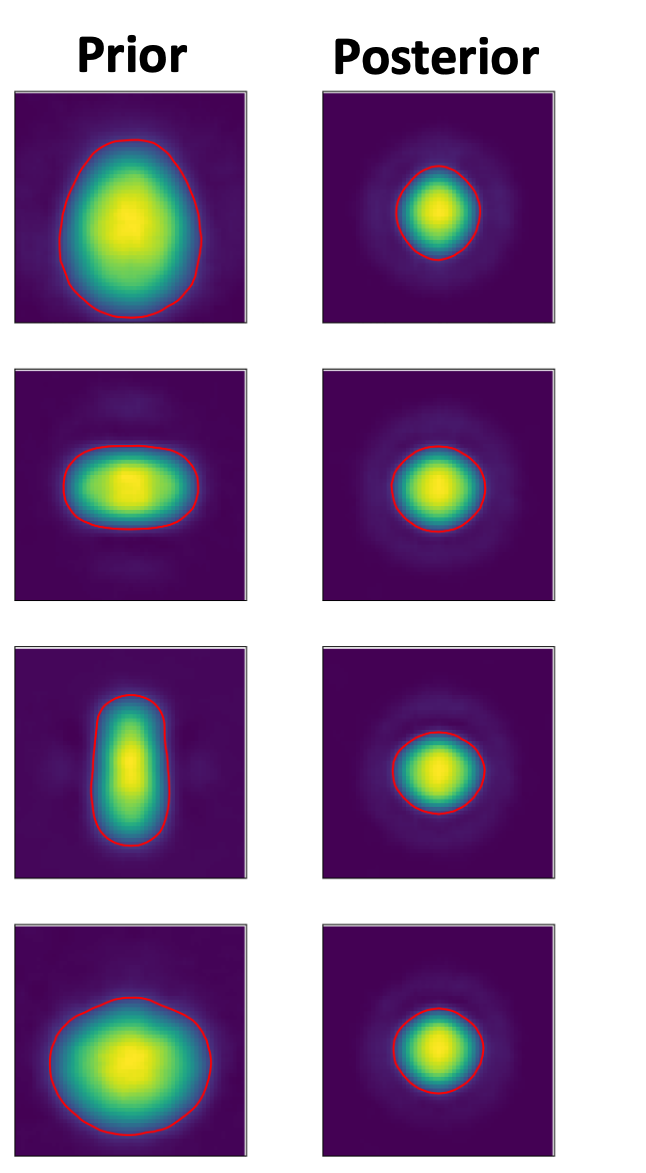}
    \caption{}
    \label{fig:bsps updated images}
  \end{subfigure}

  \caption{Results of a Bayesian inference of inputs to a high-fidelity multiphysics simulation based on experimental data from the National Ignition Facility (shot N180128). Inference was enabled by a deep neural network surrogate trained on 100K expensive ($\sim10$ node-hours / simulation) simulations. Panel (a) shows the quality of match to experimental observables; the Bayesian approach gives an unprecedented quality of fit and provides uncertainties in the match. Panel (b) shows posterior predictive radiation temperature driving the implosion, $T_R$, and demonstrates the inherent interpretability of this approach. Panel (c) shows prior and posterior equatorial X-ray images for the shot, which are enabled by our use of deep neural networks which are highly effective for non-scalar data.}
  \label{fig:bsps}
\end{figure*}

A recent application of the Bayesian approach aimed to interpret results from a series of so-called `BigFoot' ICF implosions at the National Ignition Facility \cite{Baker_2018, Casey_2018} (figure \ref{fig:bsps}). This work used 100,000 2D HYDRA simulations \cite{Marinak_2001}, in a latin hypercube design over 8 input dimensions, to train a novel cycle-consistent deep neural network (DNN) surrogate \cite{Anirudh_2020}. The DNN was trained in an approximate Bayesian manner \cite{Gal_2016}, giving uncertainties in the surrogate prediction which were calibrated by tuning the prior on DNN weights \cite{Anderson_2020}. The trained and calibrated surrogate was used in a Markov chain Monte Carlo inference of probability distributions over the 8-dimensional input space in order to match a set of experimental observables for NIF shot N180128. Comparing the observed quantities with posterior predictive values from the inference (figure \ref{fig:bsps}a) shows a match to multivariate experimental data that would be extremely difficult to achieve with the simulation in the loop, and the Bayesian approach provides a meaningful measure of ther quality of the fit in the form of predicted errorbars. Since the analysis includes high fidelity physics, the fits can be easily intepreted as modifications to radiation drive and degradations (figure \ref{fig:bsps}b). Finally the use of a DNN surrogate allows for the inclusion of non-scalar data like X-ray images (figure \ref{fig:bsps}c) which suggests a path towards future analyses which can use all of the information collected in an experiment (\emph{ie.}, without first projecting non-scalar observations to scalar features).\par

\noindent
[Jim A Gaffney and Jayaraman Thiagarajan]  

\subsection{High-performance computing and simulation acceleration}
The vast amounts of data generated by these simulations and required
for training some of these models can create a substantial demand for
scalable training algorithms and leadership-class HPC resources.  In
developing models for these multimodal data sets we have created new
techniques for composing data-, model-, and ensemble-level parallelism
and working with 100M sample data sets with 1.5B scalar fields and
1.2B images \cite{Jacobs2019}.  Using the LTFB algortithm developed by
Jacobs (\cite{Jacobs2017, Jacobs2019}) enabled the entirety of a
supercomputer like Sierra to be used when training a single model
architecture and was able to produce a single instance of a well
converged model.  Some of the techniques that have been developed are
a coupled, tournament, training algorithm that interwines the training
of a set of model instances to produce a single, best model that has
been trained on a sufficient portion of the training data to
generalize across a held-out tournament and validation data sets.
Additionally, we developed a scalable, in-memory data store and data
ingestion algorithm that is able to fetch a massive, distributed data
set efficiently and use only a single pass over the data for the
entire training regime.  Finally, we have developed methods for both
model- and data-parallel training of each individual instance of the
neural network architecture and optimized it for the IBM Power9 +
Nvidia Volta architecture of the Sierra system.  Building upon these
capabilities enabled us to produce a demonstration on training a
generative molecular model on 1.6B small samples, which was selected
as a finalist for the 2020 Gordon Bell Special Prize for COVID'19
research \cite{Jacobs2020}.  These algorithms have been implemented in
the LBANN scalable deep learning toolkit, which is open-source and is
being optimized for the next generation of leadership-class computing
systems, Fugaku, Frontier, and EL Capitan.

In addition to optimization of deep learning training for HPC systems,
we are also exploring the integration of next generation AI
accelerators and hardware platforms. Specifically, we have integrated
two stream dataflow architectures, the Cerebras CS-1 and SambaNova
SN10-8, into two of our HPC systems, Lassen and Corona, respectively.
Using these systems we have started to evaluate these accelerators may
be able to serve in a Cognitive Simulation workflow, offloading
data-driven, in-the-loop, surrogate models from traditional
GPU-accelerated compute nodes.  

\label{sec:hpc}
\noindent
[Brian van Essen]

\subsection{Design exploration and optimization}
%[Peterson]
%\section{Optimization and Design Exploration}
A key challenge for inertial confinement fusion is the relative lack of experimental data.
Leadership class experimental facilities may only be able to execute a few experiments per week, with single campaigns 
consisting of perhaps dozens of experiments. As such, a major challenge is how to design and optimize an experiment for a desired outcome
(such as high nuclear yield), with only very few opportunities to experimentally test that design. Historically, the community
has heavily leveraged high-fidelity full system numeric simulations to first design experiments in silico. Then only after searching
for a likely-to-be effective design numerically is a candidate design fielded and tested in an experiment. Numerical simulations,
therefore, play a crucial role in the design and optimization of ICF experiments.

However, the digital design of full-system experiments brings with it another set of challenges. First, while ICF drivers and targets
facilitate great flexibility, this flexibility comes with a cost: the
design space is extremely large.
For instance, laser pulses can change their time and space dependent power distribution. An ICF capsule needs to 
define its ablator layer thicknesses and material compositions. And in the case of indirect-drive a hohlraum's material 
and geometry also need to be defined. Furthermore, the tolerances on ICF designs can be very tight, requiring micrometer 
precision. In all, to fully define an ICF experiment can easily require setting a few dozen independent parameters. The setting of these
parameters has historically been done by subject matter experts, who leverage physics knowledge and intuition to smartly find new designs.
A major advance would be to move from this labor-intensive manual process towards automatically discovered and rigorously optimal designs.

Mathematically optimizing functions of several dozen parameters would not be a challenge, except that the simulations are 
expensive. A full indirect-drive coupled hohlraum-capsule simulation can cost a few node-days, and the simplest simulation that 
treats just a capsule with low-fidelity physics models can still take a few core-minutes. Mathematically, this means that the
objective function is very expensive to calculate. Since navigating high-dimensional spaces requires many function evaluations
\footnote{For instance, a simple gradient-based optimization algorithm would need to run at least as many simulations as the size of the search space to calculate a finite-difference approximation of the derivate}, 
mathematical optimization and design exploration for ICF seems to be prohibitively expensive: the search space is too large and the simulations too costly.

However, recent advances in machine learning and computational hardware are beginning to usher in a new era of optimal 
digital design for ICF. Peterson \textit{et al.}~\cite{peterson-zonal-flow} leveraged high-frequency ensemble computing and 
surrogate modeling to discover a digitally-optimized design. The computational workflow to do this was rather complex,
since it had to automatically mange and coordinate the execution and post-processing of several thousand concurrently running
independent HPC simulations. To do so, the authors developed and deployed cloud computing workflow technology 
on the Trinity supercomputer at Los Alamos National Laboratory not to run a large high-fidelity model, but rather to run several 
thousand lower-fidelity models. In all, they were able to execute 60,000 simulations, which spanned a 9-parameter capsule design 
space, enough to adequately train a random forest regression model.
Once trained, the surrogate model was fast enough to embed into a global 
optimization algorithm. The authors also introduced the idea of ``robust design", 
whereby the design parameters themselves could be uncertain (for instance due to finite manufacturability precision or tight engineering tolerances). Instead 
of maximizing the nuclear yield, they maximized the probability that the simulation achieved some threshold yield, given 
variability about the desired target design. After finding a predicted location for a new optimally robust design, the authors then 
double checked the result by running new simulations. Interestingly enough, these new simulations suggested a new kind of 
physics regime for ICF, defined by asymmetric capsule implosions filled with instability-suppressing vortical ``zonal" shear flow. Zonal flows, 
while common in magnetic fusion, had previously been unseen in ICF, and their discovery would not have been possible without an automated 
optimal design framework.

Automated design optimization has also yielded more intuitive results, as in Hatfield \textit{et al.}~\cite{hatfield-genetic}. 
This work avoided the gradient-in-high-spaces problem not with a surrogate model but via a genetic algorithm for use
in ICF capsule optimization. Good
performing simulations had their capsule layer thicknesses and material compositions ``bred" together in an iterative
fashion, with the fittest candidates surviving to breed in subsequent generations. Within a few dozen generations, the best 
design that emerged appeared as a canonical ICF target, with low-density DT gas surrounded by high-density DT ice encased 
in an ablator layer. In this example, an automated optimal design was able to navigate a high-dimensional space and settle on 
a design template not-unlike one that human subject matter experts have learned via decades of study.

While automated design in ICF has shown some early success in being able to discover both intuitive and non-intuitive designs, the solutions 
they discover are inherently limited by the simulator used. That is, if a model disagrees with an experiment the optimal model-based design may 
be of little interest, since it may or not reflect reality. In this case, it could be possible to use techniques such as transfer-learning~\cite{humbird-TL-omega,Humbird:2021} to post-process raw simulation data during the optimization process. That is, the optimization cost-function evaluates not the output of the simulation, but rather the output of a machine learning model that adjusts the simulation output to better match what might occur in an experiment. A similar technique that used statistical linear regression to modify simulation outputs drove an experimental campaign on Omega to record yields~\cite{varchas}.

\begin{table*}[tbp]
\caption{Challenges and opportunities for automated design optimization and exploration for ICF}
\begin{center}
\begin{tabular}{p{0.4\linewidth} | p{0.5\linewidth}}
\hline
\hline
Challenge & Opportunity \\
\hline
Relatively few experiments & Model-based design \\
Costly simulations & Surrogate-enhanced optimization, multi-fidelity optimization \\
High-dimensional design spaces & Bayesian optimization, gradient-free and agent-based optimization \\
Tight engineering tolerances & Stochastic optimization, robust design \\
Complex simulation pipelines & Next-generation hardware; advanced workflow software \\
Simulation-experiment discrepancy & Transfer-learning of simulation data to match experiments \\
\hline
\hline
\end{tabular}
\end{center}
\label{tab:opt-challenges}
\end{table*}%

Given the early numerical and experimental success deploying automated and optimal design exploration, its use for ICF is likely to grow. 
Table~\ref{tab:opt-challenges} summarizes some of the key challenges and opportunities as the field progresses. Surrogate-based and 
gradient-free optimization can be enhanced with Bayesian optimization~\cite{vazirani-bayesopt} techniques that use surrogate model uncertainty to balance exploration 
and exploitation (provided that surrogate models produce uncertainties that increase in unexplored areas). These iterative techniques, in 
contrast to the single-pass or human-in-the-loop surrogate-based optimization, however, become increasingly complex, since simulation post-
processing, surrogate model training and optimization and simulation launching must be automatic. Such heterogeneous, dynamic, high-
frequency computing is less common in a traditional high-performance-computing (HPC) environment than it is in data science. However, the 
melding of AI and scientific computing is a broad trend, and next-generation computer hardware and software will likely see a continued merger 
of machine learning and traditional HPC technologies~\cite{community-workflow}, making the infrastructure needed for automated design more 
common.

%%% Local Variables:
%%% mode: latex
%%% TeX-master: "main"
%%% End:
\label{sec:optimization}
\noindent
[J. Luc Peterson]

\subsection{Self-driving experimental facilities}
%[Ma, Timo Bremer]}
\begin{figure*}
  \centering
\includegraphics[width=\textwidth]{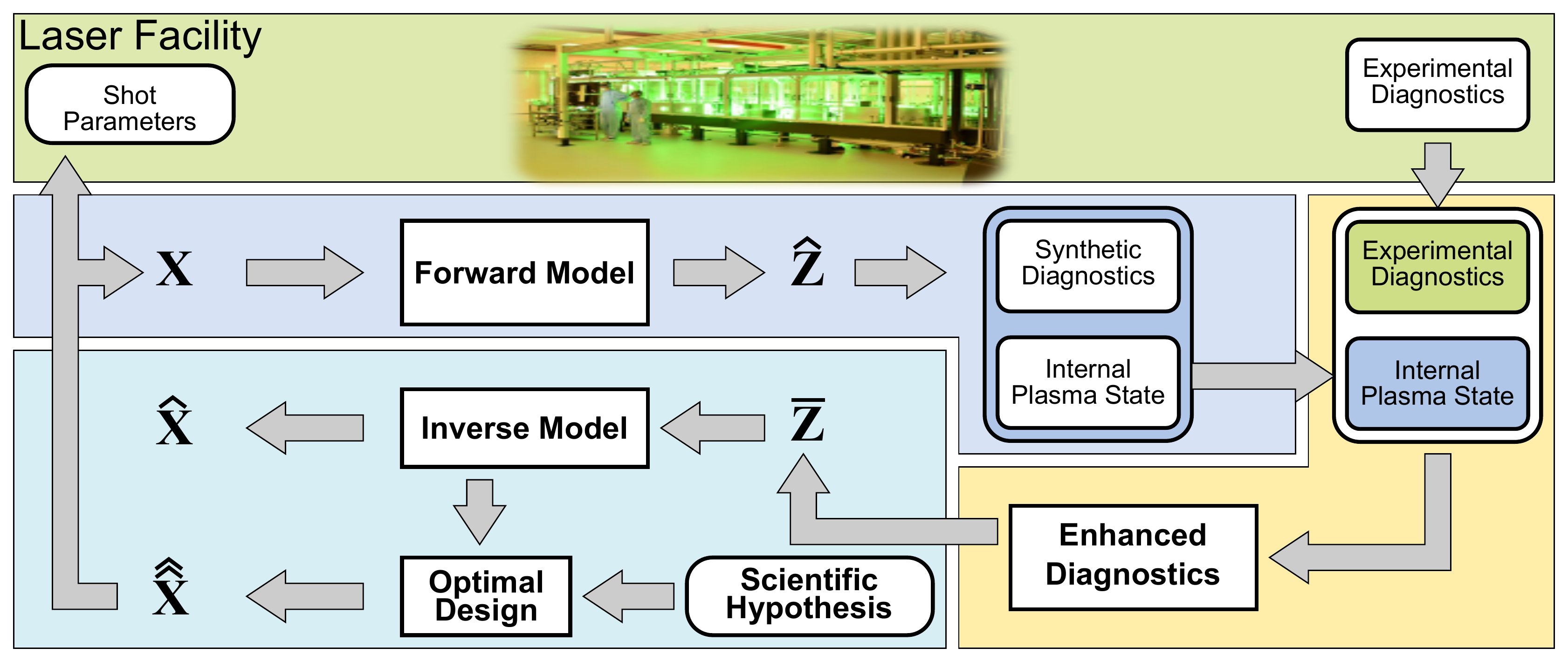}
\caption{A schematic of the envisoned control loop integrating simulation-based models with enhanced diagnostics and optimal design. For each shot taken at the facility (green box) the corresponding forward model produces the expected outcomes in the form of both synthetic diagnostics and predicted internal states. The expected internal state is then combined with the experimental diagnostics (yellow box) and used as the initial condition to estimate a corrected internal state Z. The estimated state is then used directly in the inverse model to potentially improve the surrogate but also informs the next shot by considering the current objective and the results of the previous shot to suggest new shot parameters.}
\label{fig:self_driving}
\end{figure*}

The sections above have introduced various technologies from data representations to design optimizations that address a number of important challenges in plasma science and scientific machine learning in general. Here we show how the combination of these techniques can tackle an even broader challenge to develop self-driving experimental facilities. One of the dominant trends in large scale experiments, manufacturing, and even computational sciences is the rapid increase in automation. Whether it is particle physics, 3D printing, or managing massively parallel workflows, the underlying processes are too complex, and decisions need to be made too quickly for humans to be directly in control. In the context of plasma science we are particularly interested in high repetition laser experiments. The state of the art in laser experiments used to involve one shot per hour or even per day, which provides ample time for an initial analysis and to adjust experimental parameters on-the-fly. Effectively, this created a manual, expert driven optimization loop with each experiment hand-selected and curated. Current systems allow multiple shots a second and soon may reach frequencies of tens or even hundreds of hertz. In this new regime we can no longer optimize individual experiments but need to pre-plan entire sequences or even shot days. This invariably can lead to thousands of experiments being wasted as the preset plan proves less interesting than expected or through mistakes only discovered after the fact. If not addressed these challenges could easily negate much of the benefits the more frequent experiments provide. Instead, combining the various technologies introduced above, we are developing the fully automated and integrated control loop for laser experiments shown in Figure~\ref{fig:self_driving}. The overarching goal is to adjust the various laser controls, i.e., power, pulse shape, etc., denoted as the input parameters $X$ to optimize some scientific objective, such as maximizing electron temperature achieved in the experiment.

To build this system we start with a large ensemble of simulations (Section~\ref{sec:hpc}) designed to mimic a planned experiment as best as possible given the constraints on computing resources and physics knowledge. This results in a large set of outputs representing synthetic diagnostics and internal states of the system (only observable in the simulations). Subsequently, we use the representation learning (Section~\ref{sec:RepresentationLearning}) to entangle all available multimodal output data (of the simulation) into a latent representation ($Z$ in FIgure~\ref{fig:self_driving}), which is then used to build a multimodal forward modal predicting the mapping to the full outputs. Similarly, we build an inverse model and in fact typically these models are linked to ensure internal consistency~\cite{Anirudh_2020}. We then start the experiment using (a set of) inputs initially assumed to provide high quality outputs. Each shot records a set of experimental diagnostics assumed to be a subset of the simulated diagnostics from the simulations. Using manifold projections and ideas from transfer learning (Section~\ref{sec:tl} we then search the data representation for the $Z$ whose corresponding outputs in the forward model best represents the experiments, taking into account experimental noise, distribution shifts, etc. This ultimately leads to set of what we call ``enhanced diagnostics'' which include not only the measures experimental diagnostics but also  unobservable internal state information estimated through the mapping of the forward model. The enhanced diagnostics are then fed into the inverse model providing an estimate of another set of input parameters that represent the inputs that would have resulted in the observed outputs had the forward model be a prefect representation of reality. Using the inverse model as well as the observed differences between the current simulation-driven forward model and the experiment we can then exploit the design optimization techniques of Section~\ref{sec:optimization} to compute a new set of inputs aimed at optimizing the objective. Once connected this chain represents a closed loop optimization approach in which the knowledge encapsulated in a large ensemble of pre-shot simulations is used to autonomously drive high repetition laser experiments. Going forward, the next step is to include self-learning models as well and to use the observed discrepancies in both outputs and estimated inputs to improve both forward and inverse models on the fly. In the limit of sufficient experimental data this will provide the means to incrementally modify the initial simulation based model to create a fully experimentally informed one.

\noindent
[Tammy Ma and Peer-Timo Bremer] 

\subsection{Challenges and outlook}
We have described several elements of ongoing research that aim to make data-driven methods the third pillar of HEDP and ICF research, alongside large-scale experimentation and simulation. While each of these elements is ongoing work, the ultimate aim of the HEDP and ICF community is to tightly couple them into a continuous, iterative process of scientific discovery; high fidelity simulations inform the design of experiments and the resulting data are used to update physics models and propose new experiments at very high throughput. Many of the components of this vision are already in place, and ICF and HEDP research is pushing the remaining pieces forward.
\noindent
[Brian Spears]

\section{Space and astronomical plasmas} 
\subsection{Introduction}
Besides the mysterious dark matter and dark energy, the observable universe is known to consist mostly of plasmas and electromagnetic fields~\cite{Alfven:1986, Ballester:2018}. The mass of the solar system, which hosts an average size star, is dominated by the solar plasma confined to the Sun's gravity. The solar, terrestrial such as auroras and extraterrestrial plasmas such as solar wind, intergalactic clouds are  too large to fit in laboratory experiments. In other words, these natural plasmas would generate orders more data if they were subject to similar measurement schemes in the laboratory. These natural plasmas do share common physical mechanisms and processes such as energy and mass transport on the meter size and smaller scales with laboratory plasmas, which can be probed and measured in a controlled setting. With the recent detection of gravitational waves, a golden age of astrophysics including astrophysical plasma physics has arrived. The growing number of satellite and ground instruments can generate unprecedented amount of observation data from the radio frequency to gamma-ray region of the electromagnetic spectrum, and a lot more will become available through for example the Large Synoptic Survey Telescope (LSST) on the ground and the James Webb telescope in space. Within the solar system, the space instruments can probe the solar, the Earth-bound and the lunar plasmas with unprecedented spatiotemporal resolution through particle detectors, electric probes, magnetic probes and concerted measurements from different satellites.  On the largest spatial and temporal scale of the universe, these data provide information to address open questions and constrain theoretical models regarding the origin, the current state and structures, and the future fate of the universe. On the galactic scale, new phases of matter such as double-pulsar systems~\cite{Lyne:2004} provide unique laboratories and observational data for reconciliation of quantum theory and relativity, and open up new regimes of relativistic and quantum plasmas that only may exist inside a nucleus or matter under extreme pressure~\cite{Bonitz:2019}. On the solar scale,  the data present opportunities for space weather forecasting and protection of the growing number of space assets. On the terrestrial scale, the atmospheric plasmas such as lightning provide opportunities to understand the climate change and other environmental issues. 

The explosive growth of observational data are expected to continue on all length scales from cosmology to terrestrial plasmas. In addition to new windows of observation such as LIGO,  large digital sky surveys across the electromagnetic spectrum are a predominant source of observational data~\cite{Zhang2015b:dt}. For example, between 1997 June and 2001 February the Two Micron All Sky Survey (2MASS) collected 25.4 terabytes (TB) of raw imaging data covering 99.998\% of the celestial sphere in the near-infrared J (1.25 $\mu$m), H (1.65 $\mu$m), and Ks (2.16 $\mu$m) bandpasses~\cite{Skrutskie:2006}. As of 2019, the Infrared Science Archive (IRSA) alone provides access to more than 1 petabyte of data consisting of roughly 1 trillion astronomical measurements, which span wavelengths from 1 micron to 10 millimeters and include all-sky coverage in 24 bands. The IRSA dataset will soon exceed 100 times the data size of the Library of Congress. The Sloan Digital Sky Survey (SDSS) telescope produces 200 GB of data every night. The new LSST telescope captures 6-GB images at 3 GB/s with its 3.2 billion-pixel camera and will generate about 15 TB of raw image data every night. The Cassini mission collected over 600 gigabytes of scientific data from 2004 to 2017~\cite{Azari:2020}. Big data has given rise to the interdisciplinary field of astroinformatics and astrostatistics. The importance of automatic data mining has been recognized by astronomers, cosmologists, astro and space plasma physicists, statisticians and computer scientists alike in recent years~\cite{Longo:2019}, which not surprisingly coincide with the advances in novel neural network structures such as deep learning~\cite{LeCun2015,Goodfellow:2016,Alom:2018}. Even though machine learning and artificial intelligence may not completely replace human intelligence in the foreseeable future, such revolutionary tools may lower the barriers for scientists from other fields and even hobbyists alike to contribute to data analysis and new knowledge mining, through the distributed open-source platforms such as SpaceML~\cite{Koul:2021}.

\noindent
[Zhehui Wang]

\subsection{Space and ground instruments}
\begin{figure*}[!ht]
\centering
\includegraphics[width=0.9\linewidth]{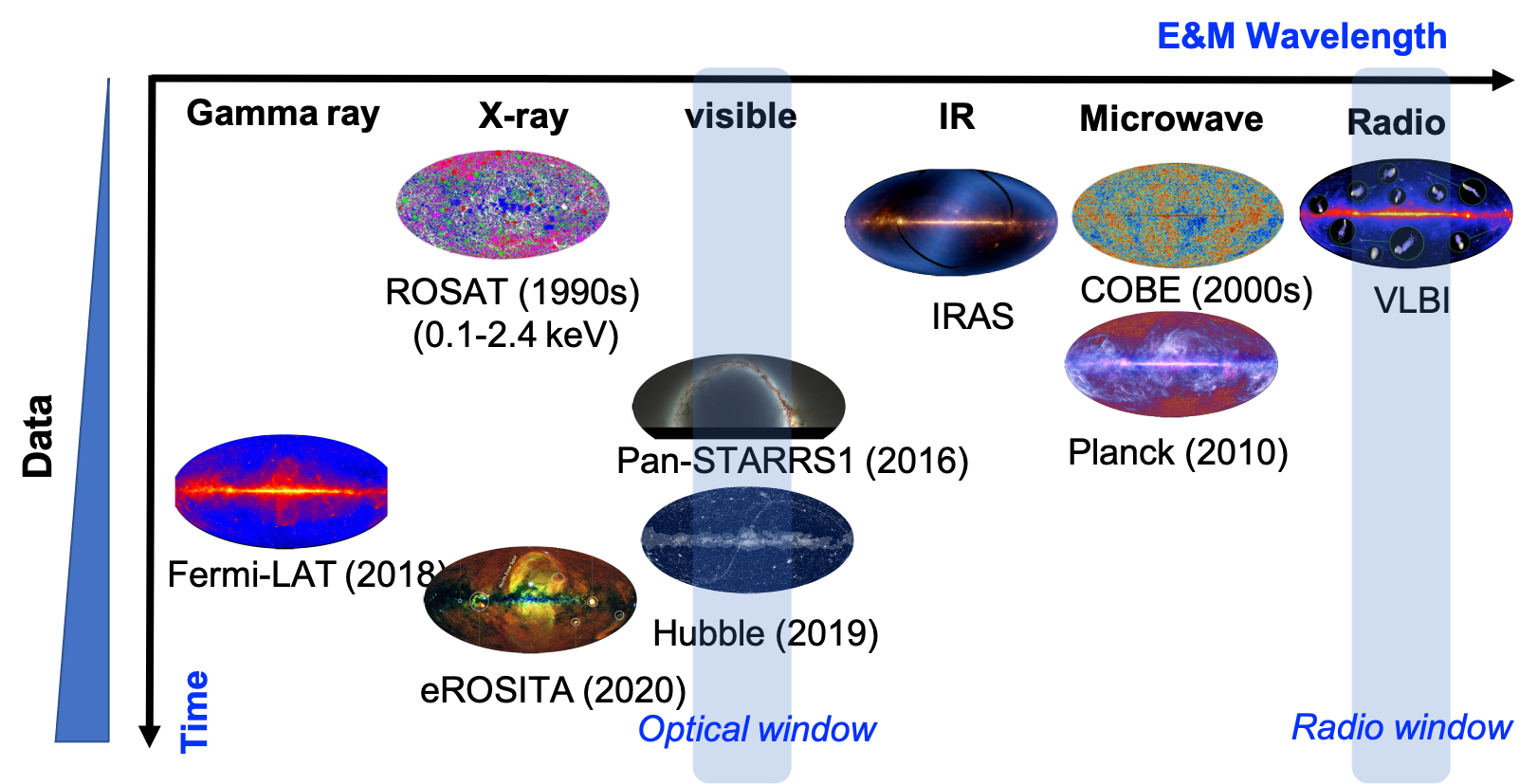}
% where an .eps filename suffix will be assumed under latex, 
% and a .pdf suffix will be assumed for pdflatex; or what has been declared
% via \DeclareGraphicsExtensions.
\caption{Examples of the full-sky surveys of the universe using the electromagnetic (EM) waves. Space-based instruments allow the use of the full EM spectrum. Newer instruments also can produce significantly more data than their predecessors. All these factors combined contribute to the rapid growth in data size and variety.}
\label{fig:InstrumU}
\end{figure*}

%\section{Space and ground instruments}
Firstly, the increasing volume and varieties of observational data from space and astrophysical plasmas are the results of the growing number of ground and satellite-based instruments. Examples of the electromagnetic instruments are summarized in Fig.~\ref{fig:InstrumU}. Ground-based instruments are limited to optical and radio wavelengths due to the absorption of the Earth's atmosphere. Satellite instruments overcome this limitation, and can also stay far away from human-generated background such as lighting. 

%[Figure here, instruments for different scales, space and ground examples.]

Secondly, benefiting from the advances in microelectronics such as CCD and CMOS technologies, which have been characterized by continued reduction of feature size (currently down to nanometers as in the cell phones) or the Moore's law, each instrument has more data capacity due to more pixels or channels, each channel or pixel can have higher data acquisition rate and more data storage. The LSST CCD camera has a pixel size of 10 $\mu$m. Scientific CMOS imagers have been gradually replacing CCD imagers because of their low noise, small pixel format (around 1 microns), and high quantum efficiency (above 90\%)~\cite{Wang:2021}. The microelectronics further allows higher data yield instruments with lower weight, power consumption or more compact size, and therefore a greater number of instruments or channels can fit onto the same payload of a satellite. In addition to continuous improvements in instrumentation hardware, space instruments become more accessible due to the continued decline of the launch cost to the low Earth orbit, from about USD \$100k/kg in the 1980s to \$1-10k/kg in the 2020s. 

Thirdly, advances in detector materials and optics have also given rise to new capabilities in collecting more data, and more efficiently. Astrophonics is a relatively young field that leverages novel photonic components and integration for astronomical instrumentation. Integrated photonic technology is an extension of integrated electronics technology, from processing electrons  to photons. The integrated photonic circuits provide reduced size, weight, and power that is critical for compact instrumentation, especially for space-based systems~\cite{Krainak:2019}. Astrophotonic solutions are already becoming an integral part of existing instruments. Examples include photonic lanterns, complex Bragg gratings, spectrographs and frequency combs, interferometry on-a-chip~\cite{Dinkelaker:2021}. Astrophotonics also enables next generation of large telescopes such as the Extremely Large Telescope (39 m).

Lastly, despite of the advances in instruments and data processing hardware, the sheer volume of data from the space and extra-terrestrial plasmas, which is essentially infinite, requires intelligent data reduction strategies. Traditionally, such strategies come from human intuition, theory and simulations. These established methods and scientific routines are useful in planning a measurement, designing the satellite orbits for the measurement, but are not enough for space-based measurement especially for in-situ measurements. Plasmas within the solar system allow in-situ measurements similar to laboratory experiments. The Parker solar probe has been flying into the Sun's atmosphere since 2018. Equipped with six remote-sensing instruments and four sets of in-situ instruments, the Solar Orbiter spacecraft has been collecting data since 2020. The Parker probe and the Solar Orbiter will not be the last ones of their kinds, since they can generate data that are essential to better understand the solar corona heating, the solar wind acceleration, the 11-year cycle of the solar magnetic activities, and space dust, paving the way towards more reliable space weather forecasting. The Parker Solar Probe is planned for two dozen flybys to the Sun's corona with a temperature up to 1371 $^o$C. Planning an in-situ space plasma measurement ahead of time is like planning a trip, which is difficult due to the indeterministic nature of the space weather, the counterpart of the weather on Earth. 

One emerging trend is to use machine intelligence for onboard data processing and reduction. Machine intelligence has already been routinely used for orbit maneuvers of individual spacecrafts, coordinated positioning of large satellite constellations, satellite communications, rendezvous, sample collection and returns. Onboard classification of images by a 10$\times$10$\times$10 cm$^3$ cubesate in Earth orbit using a random forest classifier was reported~\cite{Thompson:2015}. The classifier was trained  on  the  ground  prior  to  launch  using  test imagery  from  a  high-altitude  balloon  flight. The cubesat used  a non-radiation-hardened  commercial Atmel  AT91SAM9  processor  (210 MHz) that  cost  about  \$40. Another example is $\phi$-sat-1, which has an AI chip to down-select image data before transmitting them down to the Earth. The use of the state-of-the-art machine learning methods such as deep learning onboard has so far been limited by the satellite computation hardware and available power~\cite{Zhang:2018}. 
Deep learning algorithms such as CNN, U-Net are being adapted to fit onboard space applications. An ultralight convolution neural network called CubeSatNet was described for image classification for an eventual implementation on a 1U cubesat~\cite{Maskey:2020}. CubeSatNet had the highest F1 score when compared to trained SVM, DBN and AE models. The trained model, with an accuracy around 90$\%$, was slightly above 100 kB in size and can fit the memory size of an ARM Cortex MCU. A flight-demonstration of various convolutional neural networks using TensorFlow graphs for image processing was described~\cite{Manning:2018}. The constellation of satellites has also given rise to hive learning. 

Using machine intelligence to enhance instrumentation performance and improve data quality does not have to limit to data reduction, including high-dimensional data reduction. Signal degradation by noise or systematics is a common problem, especially for low signal-to-noise scenarios such as exoplanet search by measuring light curves~\cite{Aigrain:2017}. In addition to intrinsic instrument and detector electronic noise, statistical noise from the small flux of photons, external noise or systematics may include instrumentation jitter~\cite{Fergus:2006}, stray star light, and cosmic ray background.  An ensemble of Bayesian neural network called plan-net produced more accurate inferences
than a random forest approach~\cite{Cobb:2019}. The improvements in accuracy and uncertainties led to higher-resolution spectra and physical properties of the atmosphere. Improvements in instrumental resolution can also require more sophisticated models for data interpretation. An unsupervised learning model called ExoGAN~\cite{Zingales:2018}, which combined a generative adversarial network with semantic image inpainting~\cite{Yeh:2017} has reduced data processing time from many hours to minutes or faster, with a factor of several times in speed improvement. ExoGAN could also be retrained for other instruments.   Image inpainting belongs to a class of methods for filling in missing or damaged regions in images. Inpainting can therefore also be used to restore images corrupted by instrument artifacts, remove undesirable objects like bright stars and their halos, and preprocess the Fourier or wavelet transforms~\cite{Pesenson:2009}. Some space instruments may be too large to fit into a launch vehicle. Convolutional neural network has been used to create virtual `super instrument' for monitoring extreme UV solar spectral irradiance~\cite{Szenicer:2019}. The virtual VUV instrument has now been in use as part of a Frontier Development Laboratory project for forecasting ionospheric disturbances, and fill in the missing data from broken sensors.

\noindent
[Zhehui Wang]

\subsection{Space weather prediction}
% Institute for Space Weather Sciences, New Jersey Institute of Technology
Due to the tremendous physical scale, high temperature, strong and dynamic magnetic and velocity fields, and its proximity to Earth, the Sun is regarded as an ideal plasma lab.   In addition, the Sun is the source of space weather (SWx) which is defined by the transients in the space environment traveling from the Sun, through the heliosphere, to Earth. In the recent decade, the difficult task of understanding and predicting violent solar eruptions and their terrestrial impacts has become a strategic national priority, as it affects the life of humans, including communication, transportation, power supplies, national defense, space travel, and more. Its importance is highlighted by the Promoting Research and Observations of Space Weather to Improve the Forecasting of Tomorrow Act (PL 116-181) passed by the US Congress in 2020. Advances of SWx research and forecasting have been made in recent years thanks to a great diversity of
observations from state-of-the-art instrumentation from both ground and space. However, due to increasing spatial and temporal resolutions, researchers are facing tremendous challenges in handling massive amounts of data, especially for operational near-real-time utilization. For example, the flagship solar physics mission, Solar Dynamics Observatory (SDO), produces multiple TBs of data daily. This task becomes more demanding as new facilities probe the rapid dynamics of physical processes at some of the fundamental scales.  
Below are two important areas
of using machine learning (ML) tools to address these challenges,
which can benefit solar and SWx physics significantly.

\begin{figure*}[!ht]
\centering
\includegraphics[width=0.8\linewidth]{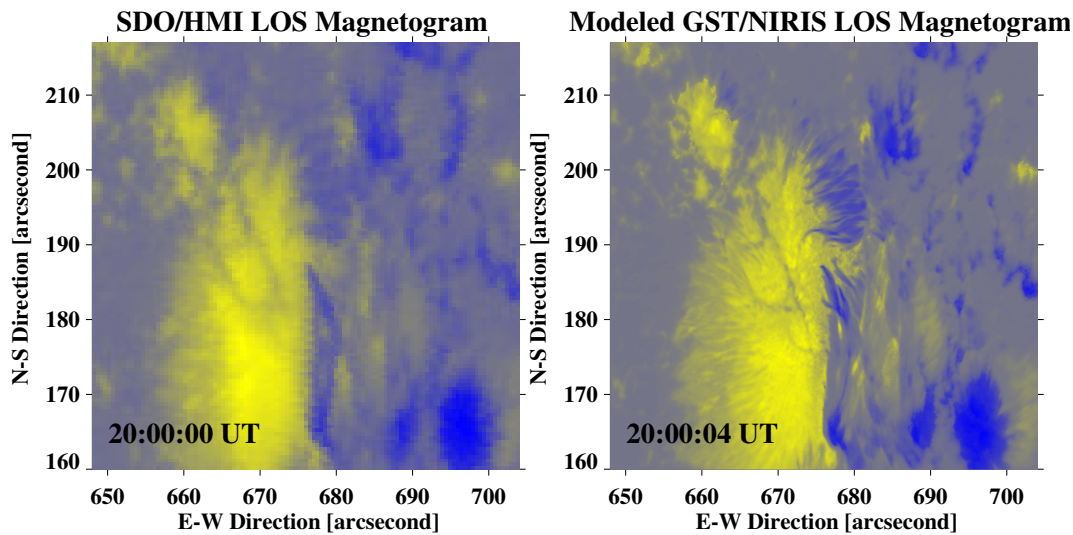}
% where an .eps filename suffix will be assumed under latex, 
% and a .pdf suffix will be assumed for pdflatex; or what has been declared
% via \DeclareGraphicsExtensions.
\caption{Left: SDO/HMI measurements used as independent reference data obtained from the ME Stokes inversion tool developed by the HMI team. Right: The inverted GST/NIRIS LOS magnetic field strengths derived by our CNN model for the same time (20:00 UT on 2015 June 25), and same field of view (FOV). Magnetic structures look similar, while the GST-inverted magnetic map has about five times better spatial resolution (modified from Liu et al. (2020b) \cite{Liu:2020bNJ}).}
\label{fig:NJIT1}
\end{figure*}

%\subsection{Extracting Information Efficiently from Large Volumes of Data in Near Real-time}
{\it Extracting Information Efficiently from Large Volumes of Data in Near Real-time.} A required step of understanding magnetic field evolution prior to the onset of solar eruptions is to derive high-resolution vector magnetic and velocity fields quickly with high precision from spectroscopic observations.  Scientists routinely use standard methods such as the Milne-Eddington (ME) Stokes inversion to deduce the three components of vector magnetic fields, Doppler shifts and other plasma parameters. However, such inversion attempts do not always produce physically meaningful results, especially when Stokes profiles are complicated. Furthermore, the ME inversion for large datasets can be quite time consuming. A Stokes profile can be modeled as waves and a convolutional neural network (CNN) is suited for capturing spatial information of the 
waves \cite{Liu:2020bNJ}. The left panel in Fig.~\ref{fig:NJIT1} presents some results obtained from the CNN model.  The Stokes inversion appears to be quite successful: the ML method is 10 times faster than the ME technique with much reduced 
noise \cite{Liu:2020bNJ}. 
Another example of information extraction using ML is 
SolarUnet \cite{Jiang:2020aNJ} which identifies and tracks solar magnetic flux elements or features. The method consists of a data preprocessing component, a deep learning model implemented as a U-shaped convolutional neural network for fast and accurate image segmentation, and a postprocessing component that prepares tracking results.  This method can be extended to identify and track various other solar and geospace features in large volumes of data.

{\it Predicting Solar Eruptions and SWx Effects Using ML.} The solar and SWx community targets predicting solar eruptions and SWx effects, namely, flares, coronal mass ejections (CMEs), solar energetic particles (SEPs) and geomagnetic storms in near real-time. The predictions use near real-time ML-processed data, some of which are described above. The predictions can be implemented from both empirical and physical aspects, which are complementary. The physical prediction relies on 
advanced physical modeling. For the empirical prediction, ML becomes vitally important.  For example, researchers utilize multiple magnetic parameters for flare prediction, including kernel-based regression analysis \cite{Fu:2008NJ}, ordinal logistic regression combined with support vector machines 
\cite{Song:2009NJ,Yuan:2010NJ,Yuan:2011NJ}, the random forest algorithm \cite{Liu:2017NJ}, ensemble learning methods
\cite{Abduallah:2021NJ}, 
and long short-term memory (LSTM) networks
\cite{Liu:2019NJ,WCT:2020NJ}.
 Liu et al. (2020a) \cite{Liu:2020aNJ} demonstrated the feasibility of using recurrent neural networks (RNNs) to predict CMEs.  In addition, we noted the success of using CNNs in predicting geomagnetic storms \cite{Malanushenko:2020NJ}. This research can be advanced in two directions: (1) applying deep neural network models to perform multi-class prediction including the use of rich spatial-temporal information from ML processed time-series of 2D and 3D images instead of derived magnetic parameters used in the previous studies; (2) adopting a combination of neural networks and statistical methods that innovates on top of off-the-shelf ML algorithms to accommodate the complexity of flaring mechanisms. The second direction will not only benefit SWx prediction, but also introduce novel methodological and theoretical challenges to the foundations of data science.
 
 \noindent
[Haimin Wang, Jason T. L. Wang]

\subsection{Transfer learning to improve historic data}

Modern solar observations provide unprecedented spatial resolution, sensitivity and wavelength coverage.  Solar and SWx research often rely on analysis of large examples of eruptions in the past.  Therefore, it is important to use advanced ML methods to improve these historic data.  Here we present two examples in this direction. 

Kim et al. (2019)~\cite{Kim:2019NJ} generated farside solar magnetograms from STEREO/Extreme UltraViolet Imager (EUVI) 304-Å images using a deep learning model based on conditional generative adversarial networks (cGANs).  This opens an avenue of research to train an ML model using one kind of data and apply it to the other kind through transfer learning. For example, in the past, Halpha, CaK and white-light data are available for over 100 years,  while vector magnetograms are routinely  available  for 10 years.  The method above demonstrates a feasibility of creating vector magnetograms, which are extremely important for SWx research, from historic data.  

\begin{figure*}[!ht]
\centering
\includegraphics[width=0.8\linewidth]{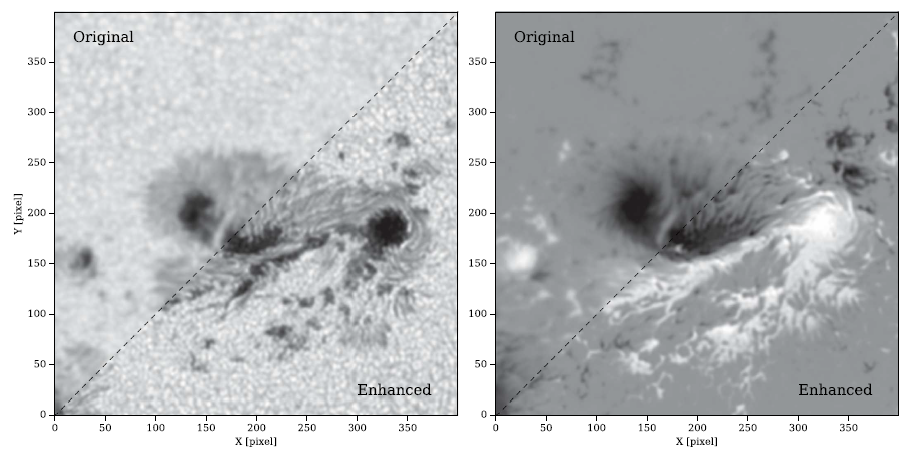}
% where an .eps filename suffix will be assumed under latex, 
% and a .pdf suffix will be assumed for pdflatex; or what has been declared
% via \DeclareGraphicsExtensions.
\caption{Example of a convolutional neural network applied to the intensity (left) and magnetogram (right) of the same region~\cite{Baso:2018NJ};  The FOV is divided into two halves. The upper half shows the original HMI image, without applying the neural network. The lower half shows the enhanced image obtained by applying the neural network to the original image. The original image was re-sampled to have the same scale as the network output. Figure credit: Díaz Baso \& Asensio Ramos, A \& A, vol. 614, p. A5, 2018. reproduced with permission © ESO.
% (need to contact journal to reproduce).
}
\label{fig:NJIT2}
\end{figure*}

The second example is related to resolution improvement of historic data.   The new observations can achieve spatial resolution around 100 km, while historic data had resolution of no better than 1,000 km.  There is a need to improve the resolution of existing data to disclose the dynamic physics of solar active regions.  Such a study has been demonstrated by using the Hinode-HMI/SDO data pairs with a convolutional neural network in Díaz Baso \& Asensio Ramos (2018)~\cite{Baso:2018NJ}, as shown in Fig.~\ref{fig:NJIT2}. Hinode's resolution is 5 times 
better than SDO's. Future work can be extended to improve the resolution even further using observations from large aperture telescopes.

\noindent
[Haimin Wang, Jason T. L. Wang]

\subsection{Surrogate models of fluid closures using machine learning}
% Xueqiao Xu, Lawrence Livermore National Laboratory, Livermore, CA 94550, USA
Many space plasmas can be described by a fundamental kinetic equation for microscopic descriptions or a set of fluid moment equations for macroscopic statistical descriptions. The traditional trade-off by solving a set of fluid equations instead of a kinetic equation is generic accuracy verses practical computability. Direct simulation of physical processes on a kinetic level is still prohibitively expensive.  Any system of moment equations suffers from the “closure problem”: accurately capturing the behavior of an infinite-dimensional kinetic physical system via a few simplified equations. The problem arises when deriving fluid equations through the chains of moment equations for kinetic theories. The resulting lower order moment equations always contain a higher order moment. To truncate the moment hierarchy, a proper closure is thus required to approximate this higher order moment from existing lower order moments for microscopic descriptions, which is conventionally constructed by phenomenological constitutive relations. 

Fluid moment closure hierarchies for kinetic theories are relevant to a wide range of scientific areas of research, including fluid dynamics, plasma physics, neuroscience, radiative transfer equation, and so on. In plasma physics, the widely used Spitzer-Harm closure~\cite{spitzer1953transport}, and similarly, Braginskii closure~\cite{braginskii1965transport} consider a strongly collisional plasma and predict heat flux $q \propto \nabla T$, both of which lack kinetic effects and start to break down when the particle mean-free-path approaches the characteristic length scale (i.e., in weakly collisional regime). Well-known closure models, such as the Landau-fluid closure model (or specifically Hammett and Perkins model~\cite{hammett1990fluid}), can efficiently incorporate certain kinetic effects within fluid models, such as wave-particle resonances. The Landau-fluid closure describes the nonlocal kinetic response of the heat flux to a temperature profile that has significant spatial variations on length scales that are smaller than the microscopic collisional mean free path. Over the years, Landau-fluid closure has been extended to collisional~\cite{chang1992unified,snyder1997HPC,umansky2015modeling}, magnetized plasmas~\cite{guo2012parallel} and with dynamic perturbation~\cite{hunana2018new, wang2019landau}. However, implementing Landau-fluid closures to high performance fluid codes is numerically challenging as they are usually complex functions with both frequency and wave-vectors in Fourier space~\cite{chang1992unified,hunana2018new,wang2019landau}. 

Riding on the rapid development of machine learning (ML)~\cite{LeCun2015}, machine learning moment closures for accurate and efficient fluid moment simulations have made a significant progress recently. The fidelity of the ML surrogate models has been progressively increasing with the aim of reducing the computational cost and capturing the macroscale behavior of the system but use only the microscale model to achieve efficiently integrated multiscale simulations, ranging from learning some complex moment closure functions~\cite{ma2020machine,wang2020ml,Maulik2020ml}, the learned multi-mode (LMM) closure from kinetic simulation data~\cite{Shukla2022LMM}, learning the calculation of the five-fold integral collision operator in the Boltzmann equation~\cite{Xiao2021}, and to learning uniformly accurate surrogate hydrodynamic models for kinetic equations~\cite{Han2019}. The machine learning moment closures has been used for accurate and efficient simulation of polydisperse evaporating sprays~\cite{Scoggins2021,Huang2022}, for the radiative transfer equation~\cite{Huang2021}, and for the moment system of the Boltzmann equation~\cite{Schotthofer2021}. In~\cite{miller2021encoder}, the authors pursue encoder–decoder neural network for solving the nonlinear Fokker–Planck–Landau collision operator in XGC. In~\cite{meneghini2017self}, the surrogate models have been trained for integrated simulations for the calculation of the core turbulent transport fluxes and the pedestal structure.

\begin{figure}[!t]
\centering
\includegraphics[width=3.2in]{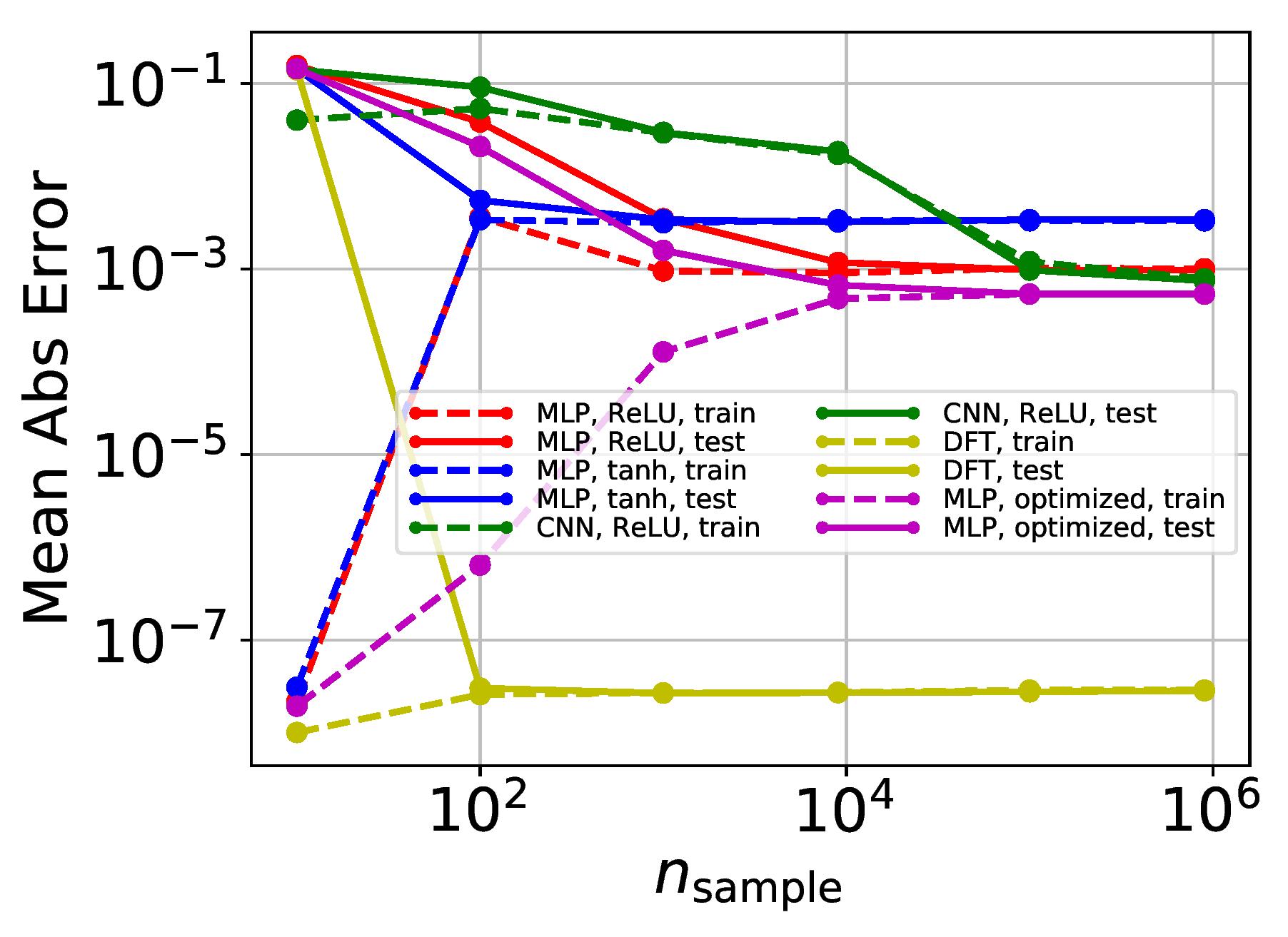}
% where an .eps filename suffix will be assumed under latex, 
% and a .pdf suffix will be assumed for pdflatex; or what has been declared
% via \DeclareGraphicsExtensions.
\caption{Mean-absolute-error versus the number of training samples. Dashed and solid lines denote training and testing error respectively; red, blue, and green lines represent MLP with ReLU (rectified linear unit), MLP with tanh, and CNN with ReLU as the combinations of network and activation  function, respectively. The yellow line represents Discrete Fourier Transform results, while the purple line represents the result of an optimized Bayesian model.~Reproduced from [C. Ma, B. Zhu, X.-Q. Xu, and W. Wang, `Machine Learning surrogate models for Landau fluid closure,' Phys. Plasma., Vol. 27, no. 4, p. 042502 (2020)], with the permission of AIP Publishing.}
\label{fig:sec61}
\end{figure}

Two novel applications of machine learning techniques to Landau-fluid closures in plasma physics were recently published~\cite{ma2020machine,wang2020ml}. In these new studies, the researchers explored how well three different types of neural networks could reproduce the kinetic Landau-fluid closure.  The three networks employed were: multilayer perceptron (MLP), convolutional neural network (CNN), and two-layer discrete Fourier transform (DFT). They found that, with appropriate tuning and optimization, all three types of neural networks were able to accurately predict the closure, while other existing simplified closure models could not yield the same accuracy at equivalent computational speed. Using this new approach, fluid simulations enabled by Deep Learning, with complicated spatio-temporal closure functions predicted by the neural networks, were, for the first time, shown to give the correct Landau damping rate for a wide range of length scales. These results offer a promising pathway to capturing complex phenomena associated with microscopic physics that is still computationally efficient and accurate when applied at the macroscale.

\noindent
[Xueqiao Xu]

\subsection{Magnetic reconnection}
% Giovanni Lapenta, Department of Mathematics, KULeuven, University of Leuven, Belgium
% 
%\documentclass[12]{article}

% Language setting
% Replace `english' with e.g. `spanish' to change the document language
%\usepackage[english]{babel}
%\usepackage{mathptmx}
%\usepackage{baskervald}
%\usepackage[T1]{fontenc}

% Set page size and margins
% Replace `letterpaper' with`a4paper' for UK/EU standard size
%\usepackage[a4paper,top=2cm,bottom=2cm,left=2cm,right=2cm,marginparwidth=1.75cm]{geometry}

% Useful packages
%\usepackage{amsmath}
%\usepackage{graphicx}
%\usepackage[colorlinks=true, allcolors=blue]{hyperref}
%\usepackage{natbib}
%\usepackage{xcolor}
%\title{Methods of machine learning for space sciences}
%\author{Giovanni Lapenta\\
%Department of Mathematics, KULeuven, University of Leuven, Belgium}
%\date{}
\def \bfF {{\bf F}}
 \newcommand{\bfE}{\mathbf{E}}
\newcommand{\bE}{\mathbf{E}}
\newcommand{\bfB}{\mathbf{B}}
 \newcommand{\bfA}{\mathbf{A}}
  \newcommand{\bfS}{\mathbf{S}}
\newcommand{\bB}{\mathbf{B}}
\newcommand{\bfJ}{\mathbf{J}}
\newcommand{\bfV}{\mathbf{V}}
\newcommand{\bJ}{\mathbf{J}}
\newcommand{\bfv}{\mathbf{v}}
\newcommand{\bfu}{\mathbf{u}}
\newcommand{\bfx}{\mathbf{x}}
\newcommand{\bfk}{\mathbf{k}}
\newcommand{\bv}{\mathbf{v}}
\newcommand{\bx}{\mathbf{x}}
\newcommand{\bfa}{\mathbf{a}}
\newcommand{\bfxi}{{\boldsymbol \xi}}
\newcommand{\bfj}{\mathbf{j}}
\newcommand{\rc}{\color{red}}
\newcommand{\bc}{\color{blue}}
\newcommand{\kc}{\color{black}}
\newcommand{\bw}{\mathbf{w}}
\newcommand{\br}{\mathbf{r}}

%\begin{document}
%\maketitle

The amount of observational data and simulated data relative to space plasma physics is growing exponentially: as more computational power becomes available, on the ground or in situ in space, more data is being generated. 
This ever growing data availability is now met with a growing use of machine learning (ML) tools able to consider amounts of data that even a large team of human researchers could not process. 

Generally speaking, ML for data processing can be subdivided into two types. 

First, \emph{supervised ML} tools are designed to replace tedious well known steps of processing with automatic tools. The typical scenario is that of taking a large but manageable set of cases, process them the human way labeling each data set according to our understanding of it. The machine can then learn from this dataset and replace the humans for the task. The archetypal case is that of image recognition, one of the greatest successes in ML. Since 2015, tests have shown ML tools surpass humans in accuracy of image recognition \citep{he2015delving}. In terms of speed there is obviously no contest.  The same techniques can be applied to analyze scientific data transferring human skills to the machine. However, this approach presumes that we already know how to analyze the data and we simply want to transfer this knowledge to a machine.

The second approach is that of \emph{unsupervised ML}. In this case, different methods of ML are applied in ways where the machine learns on its own how to treat the data.  The central idea is to deploy or design a method where the machine arrives at a reduced description of complex dataset and then the human scientist investigates the reduced description attempting to make sense of it in light of our understanding of Nature. The archetypal example is that of classification. The machine can sort all cases into a number of classes, where the number can be preset or can be an unknown of the process itself. At the end, the challenge for the researcher is to understand what the meaning of the different classes is. With this challenge comes the opportunity to discover something new and unexpected. 

The research is progressing at unparalleled speed in both categories of ML tools. Fortunately, the meagerly funded space science community can benefit from the general growth in ML tools developed for other applications. Some specific tools have been developed to make the progress in ML available to the community of space scientists. The aim is to vulgarize the more esoteric aspects of ML and make them accessible to scientists whose background is in space and not in computer science. We mention here the project AIDA that has precisely this goal: \url{www.aida-space.eu}. The AIDA project takes some of the state of the art ML tools and applies them to typical use cases common in space science. Each use case is documented detailing how to use ML tools in a step by step process that is aimed at training non ML experts.

Space science has a peculiar constraint unique to its nature: much of the data generated in space cannot be transferred to the ground due to the limited telemetry. In other applications, the data can always be stored or at least processed, in space the memory onboard  cannot be always transferred in its entirety and only a portion can be downloaded to Earth. This limitation opens the new opportunity for ML deployment in space so that the data can be processed on board and only the outcome of the analysis needs then to be transferred to the ground. This is a pioneering new possibility and great challenges need to be overcome because the processors used in space are much less powerful than those used on the ground due to the intense space radiation environment.  ML tools are highly computing demanding making their deployment in space a great topic of research.

%\section{Example of data analysis for space application: finding reconnection}
{\it Finding reconnection}
Reviewing the explosively growing area of space applications of ML is beyond our scope here and is a futile exercise as many new developments will be published while this manuscript is being processed. We focus instead on a few examples that provide a view of the type of activities that ML can take over. We focus then on only one well known but very difficult task: identifying reconnection regions.  

\emph{Reconnection} \citep{biskamp1996magnetic} is a process that converts magnetic energy into kinetic energy. Its characteristic feature is the breaking and reconnecting of magnetic field lines, giving it its name. Recognizing reconnection is not as simple as 2D cartoons might seem to imply. 

In 2D, one can consider the problem more or less solved in terms of making a definite determination of where reconnection happens: based on the out of plane vector potential, the null points of the in plane magnetic field can be characterized using the Hessian matrix as o-points and x-points giving an unequivocal answer \citep{servidio2009magnetic}. However, there are two problems. First, finding nulls and computing Hessian matrices requires complete spatial information, something we have only in simulations but we do not in experimental data. Often in space we know quantities only at one location (or a handful, in case of multi-spacecraft missions).  Second, Nature is 3D and there is no equally rock solid definition of reconnection in 3D. There are situations where experts might argue endlessly on whether there is or  isn't reconnection.

This provides a unique opportunity to apply ML. Let us then review how different data feeds can be used to find reconnection. 

\begin{figure}
	\includegraphics[width=\linewidth]{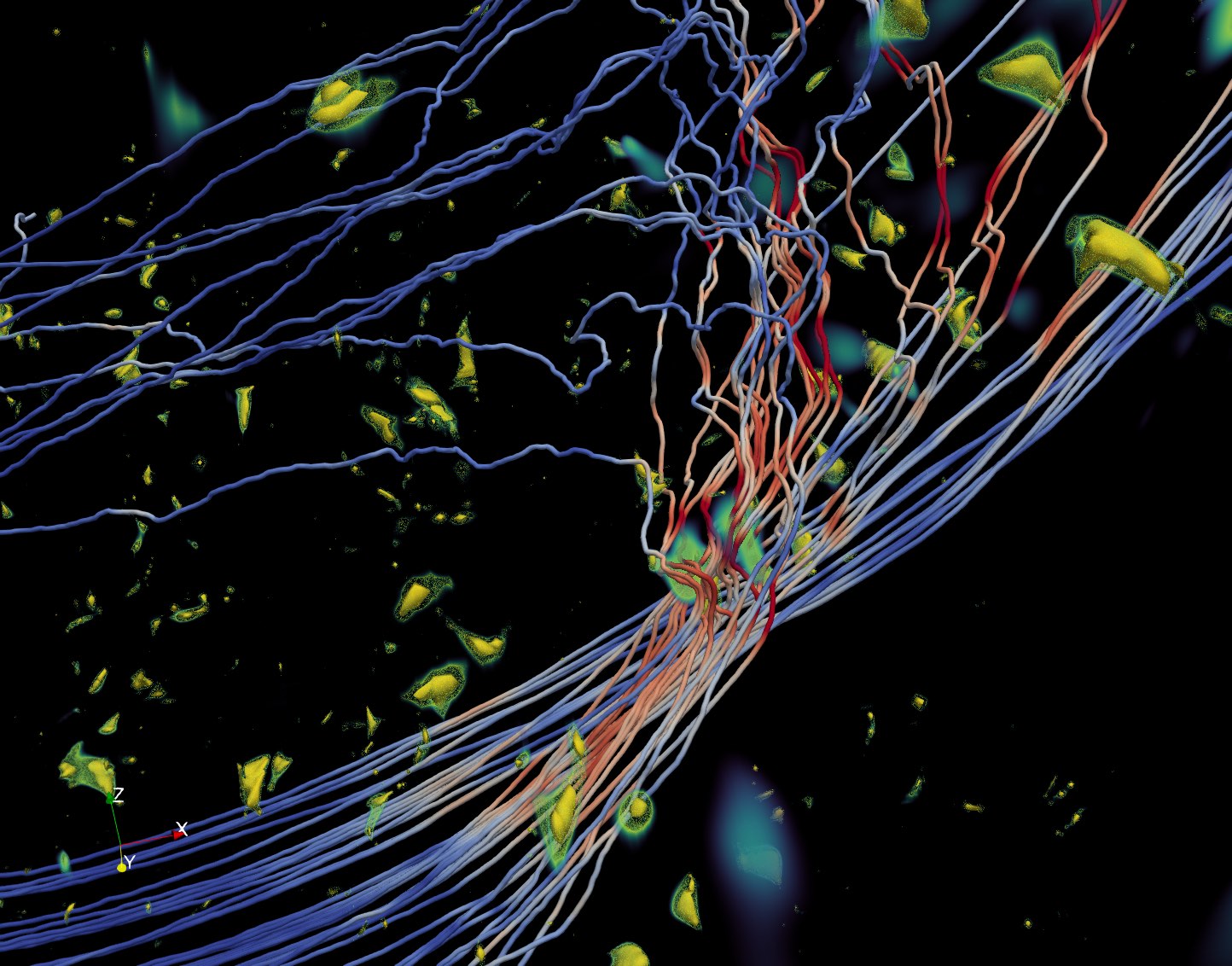}
	\caption{The indicator defined in~\cite{lapenta2021detecting}%\citet{lapenta2021detecting} 
	~identifies many reconnection sites visible in the picture as ghostly yellow-green areas. A group of electron flowlines are shown passing one of these reconnection sites and encountering also others. The flow lines are colored by the intensity of the local electric field that transfers energy between the magnetic field and the electrons, accelerating them and creating a turbulent flow.}
\label{fig:lapenta1}
\end{figure}

Traditionally, reconnection is identified by using a proxy. A review of the different reconnection indicators is provided in~\cite{goldman2016can}.%\citet{goldman2016can}.
~The simplest is finding high speed jets. Of course, many processes can lead to high speed jets and only the expert can combine the analysis of different quantities and arrive at the conclusion that reconnection is really taking place. 
A more recent discovery is that reconnection is associated with peculiar electron velocity distributions that present croissant-shaped features called \emph{crescents} \citep{burch2016electron}. 
An especially convenient way to identify a possible reconnection site is the local measure of the so-called Lorentz  indicator based on computing the speed of a frame  transformation that eliminates the local magnetic field~\citep{lapenta2021detecting}. An example of this indicator is shown in Fig.~\ref{fig:lapenta1} where many 3D reconnection sites are identified in a turbulent region.

This accumulated expertise  provides a great opportunity for creating human-labeled datasets to use as training for a supervised ML tool. However, the intrinsic complexity even for a human to decide what is and  isn't reconnection gives unsupervised ML tools the opportunity for new discoveries. We explore below some methods recently published to identify reconnection with ML. 

%
% QUI
%

%\subsection{Identifying reconnection from velocity distributions}
{\it Identifying reconnection from velocity distributions}
Velocity distribution functions (VDF), $f(v_1,v_2,v_3)$,  are provided as 3D datasets by instruments that measure the count of collected particles in situ. Kinetic simulation can provide a synthetic version of the same information. This information can be in different forms: energy-angles, 3D velocity bins or polynomial expansions (e.g. Hermite and spherical harmonics). The first aspect of this type of data  is the overwhelming size. As an example a typical modern particle in cell simulation produces TB of distribution data, for each time step. At each time step there are millions or billions of such distributions to analyse. A  recent  mission, MMS (Magnetospheric MultiScale) produces in burst mode one distribution every 30ms (though not all can be transferred to the ground due to limited telemetry). A survey of the literature shows that these distributions are rarely used in their full 3D complexity and usually only very few  2D reductions of specific instants are studied. The choice is guided by analysing other quantities that suggest what distribution to study. There is no systematic analysis of all data taken, it is humanly impossible. But not impossible for ML. 

Supervised ML can be trained to recognize features like the crescent using a human-labeled dataset, an application of the widely used image recognition software. However, shapes in VDF are more in the imagination of the viewer than an objective feature. VDF, especially in observed data, are highly noisy and structured. A promising approach is to use unsupervised ML. The complexity of a VDF can be classified using clustering methods. 

The Gaussian Mixture Model (GMM) represents a distribution using a superposition of  overlapping Gaussian distributions \citep{mclachlan2019finite}. With this approach, Ref.~\cite{dupuis2020characterizing}% \citet{dupuis2020characterizing}
~showed that reconnection can be associated with a high number of Gaussian beams, with different classes of distributions capable of identifying the inflow and the outflow region of reconnection. The method automatically determines the number of Gaussian beams using information theory criteria that make the best compromise between efficiency of description (that requires as small a number of beams as possible) and accuracy (that is always higher the more beams are used).  The method also determines the properties (mean  and variance) of each beam. From this ML analysis, physical meaningful quantities can be determined. Especially useful is the determination of the "intrinsic" thermal spread of each beam in the mixture and the "pseudo" thermal speed due, instead, to the relative speed between the different beams in the mixture \citep{goldman2020multibeam}. Ref.~\cite{dupuis2020characterizing}%\citet{dupuis2020characterizing}
~showed that  these physical quantities can be used  to determine the electron and ion diffusion region around a reconnection event.

Another approach to the unsupervised ML analysis of distribution functions uses the subdivision of the VDF in non overlapping beams or arbitrary shape. The k-means method~\citep{macqueen1967some} can be applied for this task leading to the identification of different populations with different physical origin \citep{goldman2021multi}.

%\subsection{Identifying reconnection in spatial data}
{\it Identifying reconnection in spatial data} The quintessential example is the 2D image: in the case of reconnection, this is a 2D view of electromagnetic field component or of a plasma moment. Obviously this type of data can benefit from the methods developed for image processing. 

Convolutional Neural Networks (CNN) \citep{bishop2006pattern} can be trained using expert-labeled images. Ref.~\cite{hu2020identifying}%\citet{hu2020identifying}
~report the example of a dataset of 2000 cases labeled using the expert community via \url{zooniverse.org}.  The project can be accessed via \url{http://aida-space.eu/reconnection} where a tutorial on how to identify reconnection sites is provided. The project is public and unbiased experts  helped with labeling. Once the labeled dataset is available, the CNN can be trained to recognize reconnection.

Unsupervised ML can detect reconnection based on spatial information by using clustering of pixels in the spatial data. Reconnection is identified using physics properties of the resulting classes. Ref.~\cite{sisti2021detecting}%\citet{sisti2021detecting}
~use DBSCAN~\cite{ester1996density}%\citet{ester1996density}
~and k-means \citep{macqueen1967some} to identify current layers with a sufficiently large aspect ratio to flag reconnection.

In principle,  3D datasets and  1D fly-through through datasets can be treated using similar methods and future research will likely investigate this possibility.

\paragraph*{Acknowledgements}
This work has  received funding from the European Union’s Horizon 2020 research and innovation programme under grant agreement No. 776262 (AIDA, \url{www.aida-space.eu}). All examples described above  are available and fully documented in the open source AidaPy software  at: \url{https://gitlab.com/aidaspace/aidapy}. 

%%\section*{Acknowledgments} 
%{\it Acknowledgments} The work described above has  received funding from the European Union’s Horizon 2020 research and innovation programme under grant agreement No. 776262 (AIDA, \url{www.aida-space.eu}). All examples described above  are available and fully documented in the open source AidaPy software  at: \url{https://gitlab.com/aidaspace/aidapy}. 
%\bibliographystyle{apalike}
%\bibliography{sample}

%\end{document}
\noindent
[Giovanni Lapenta]

\subsection{Challenges and outlook}
Besides the mysterious dark matter and dark energy, the observable universe is known to consist mostly of plasmas. The explosive growth of observational and simulation data is expected to continue from cosmological scale to terrestrial size plasmas, which can supply data not accessible to laboratory experiments. The essentially unlimited and heterogeneous data, sophisticated multi-physics models, and lately the universal data mining tools such as deep learning have ushered in the new {\it precision} epoch in cosmology, astrophysics, space and terrestrial science including plasma physics. On the one hand, the wealth of data allows detailed tests of the existing physics-based models, including the underlying fundamental physics such as quantum mechanics and general relativity, and fine-tuning of {\it ad hoc} parameters in some models. On the other end, such data permits systematic searches for new physics motivated by dark matter, dark energy, neutrino mass, high-energy cosmic rays, and quantum information centered around the blackholes.  On the applications front, data science have opened doors to real-time predictions of solar coronal mass ejection, and space weather forecast. Data science has already given rise to new disciplines such as astro-informatics and astro-statistics, it may also provide a generic framework to better integrate plasma-driven physics to the existing models, when plasma effects have so far been left out, for example, of the standard model of cosmology. Data science and machine learning have been successfully or can be used to accelerate all aspects of `scientific data flows' in astronomy and astrophysics; {\it i.e.}, from enhanced instrumentation and data acquisition, to automated feature extraction and classification, to hypothesis generation, to model construction, to modeling and to model validation. Despite of their practical prowess and simplicity, machine learning methods are not fully understood at this time. Seeking a better union between the established knowledge framework of physics-driven models with data-driven models is an exciting new frontier. New results may be anticipated such as in the solving the outstanding problems as mentioned above, development of scientific machine learning algorithms that will be broadly applicable, and quantitative understanding of uncertainties for more effective predictions and optimization, paving the way towards automated space and astro plasma observations, discoveries and novel space technologies.

\noindent
[Zhehui Wang]

\section{Plasma technologies for industrial applications} %[Coordinator:Hamaguchi and Jan Trieschmann]} 
\subsection{Introduction}
Plasma technologies are widely used in industries.~\cite{6Lieberman2005,MakabePetrovic:2016bo}
One of the largest industrial applications of plasmas is plasma processing for semiconductor devices and other related microelectronics devices such as displays and sensors. Especially for the latest and most advanced semiconductor devices, the device dimensions (i.e., typical sizes of transistors) are now approaching the atomic size. Therefore the further miniaturization of a single device is now facing its physical limit and can no longer be expected as a means to pack more devices in a single chip. Instead, the further improvement of device performance must be achieved by other means such as the use of complex three-dimensional device structures and new materials.

Mass production of such complex devices with atomic-scale accuracy poses enormous challenges in their manufacturing technologies. Plasma etching and plasma-enhanced deposition processes~\cite{OehrleinHamaguchi:2018fw} need not only to improve their accuracy in spatial dimensions but also to handle non-conventional materials, such as ferromagnetic metals for magnetoresistive random access memories (MRAMs) and perovskite-type oxides for resistive random access memories (ReRAMs), just to name a few. However, in most cases, the interactions between the newly introduced material surfaces and conventional or newly introduced gaseous species of plasma processing are not well understood, which makes the process development highly challenging and costly.  Furthermore, having a variety of choices for surface materials and process conditions increases the complexity of process development and the exhaustive search for process optimization by experiments becomes prohibitively expensive. One of the possible means to tackle these challenges is to use machine learning (ML) to predict gas-phase and surface reactions of plasma processing, based on the existing knowledge of such systems.

Other technological applications of plasmas that have attracted much attention of the plasma community recently are those for medicine, agriculture, biology, and environmental protection.~\cite{AdamovichBaalrud:2017za,WeltmannKolb:2019ud} 
Although practical applications of these technologies at the industrial level are yet to be seen, some of them are considered to be game-changing innovations. As in plasma applications for semiconductor technologies, gas-phase and surface chemical reactions play critical roles in plasma processes in these fields and the exhaustive search for optimizing their process conditions by experiments can be prohibitively expensive as well. 
This reasoning applies similarly to thin film deposition of hard and functional coatings and plasma assisted catalysis~\cite{bogaerts_2020_2020}. Diagnostics and modelling are crucially challenged by intrinsic multiscale and multiphysics phenomena, including yet to be revealed non-equilibrium plasma-surface chemical reactions. Moreover, the exploration and discovery of novel plasma and solid phase materials systems, e.g., for energy efficient gas conversion and synthesis, are a severe limitation.
Systematic collection of data and the use of data-driven approaches to make full use of such data are expected to enhance the efficiency of process development and promote (or even enable) transitional changes in these fields.

In this section, we present how such data-driven approaches are used in the semiconductor and microelectronics industries in the following three subsections.  From a more academic point of view, examples of data-driven approaches are then presented as new tools to analyze plasma-surface interactions, plasma simulations, plasma chemistry, and plasma medicine in the subsequent subsections. The final subsection briefly summarizes the challenges and outlook in industrial applications of technological plasmas.

\noindent
[Satoshi Hamaguchi and Jan Trieschmann]

\subsection{Data-driven Approaches for Plasma-Assisted Manufacturing in the Semiconductor Industry}
\begin{figure*}[t]
    \centering
    \includegraphics[width=0.8\linewidth]{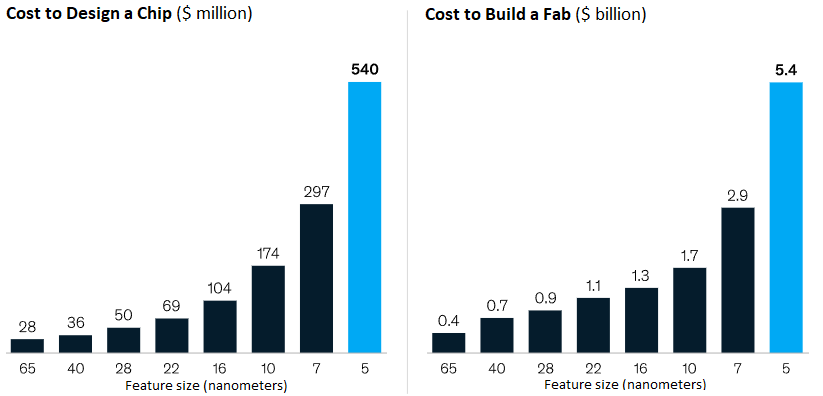}
    \caption{The cost to design state-of-the-art chips and to build the semiconductor fab equipped with latest process modules to fabricate these chips has grown dramatically with the scaling of the nodes \cite{bauer2020}.}
    \label{fig:CostChipsFab}
\end{figure*}

With the explosive growth in data creation, estimated to surpass 180 zettabytes by 2025 due to the increasing popularity of Internet of Things (IoT), there is an unprecedented demand for storage and processing of large volumes of data. Today's data-centric world increasingly relies on semiconductor manufacturing to fabricate chips with integrated circuits that can realize the data storage and computational capabilities required for harnessing data and Artificial Intelligence (AI). Of the hundreds of steps used to fabricate a chip, nearly half use plasma processing. This is because non-equilibrium plasmas offer several benefits over thermal processing, including lower energy barriers to promote surface adsorption, resulting in reduction of high temperature requirement for certain materials; ion acceleration towards the wafer due to sheath physics, resulting in directional behavior; and enhanced surface reactions due to presence of neutrals and ions. These plasma effects will lead to better film uniformity, conformality, and roughness control with atomic layer processing. As the semiconductor industry continues to innovate by building chips with smaller feature size, the cost to design such chips and the cost to equip fabrication facilities with state-of-the-art process tools required for making these chips have increased dramatically (Figure \ref{fig:CostChipsFab}). In addition to the cost, the time taken to complete a chip has increased, as more process steps are required to achieve the desired results \cite{bauer2020}. Thus, it is paramount to accelerate the design and development time of the plasma reactors, optimize the processes used to create the desired features while improving efficiency of engineering staff, and provide adaptive control to mitigate uncertainty at the chamber level, tool level, and fleet level. Smart manufacturing practices and advances in sensing capabilities and product metrology have created unprecedented opportunities for the semiconductor fabrication equipment industry to improve yield, efficiency, and speed to solution using data-driven approaches. 

% \begin{figure}[b!]

\begin{figure*}[t]
    \centering
    \includegraphics[width=0.8\linewidth]{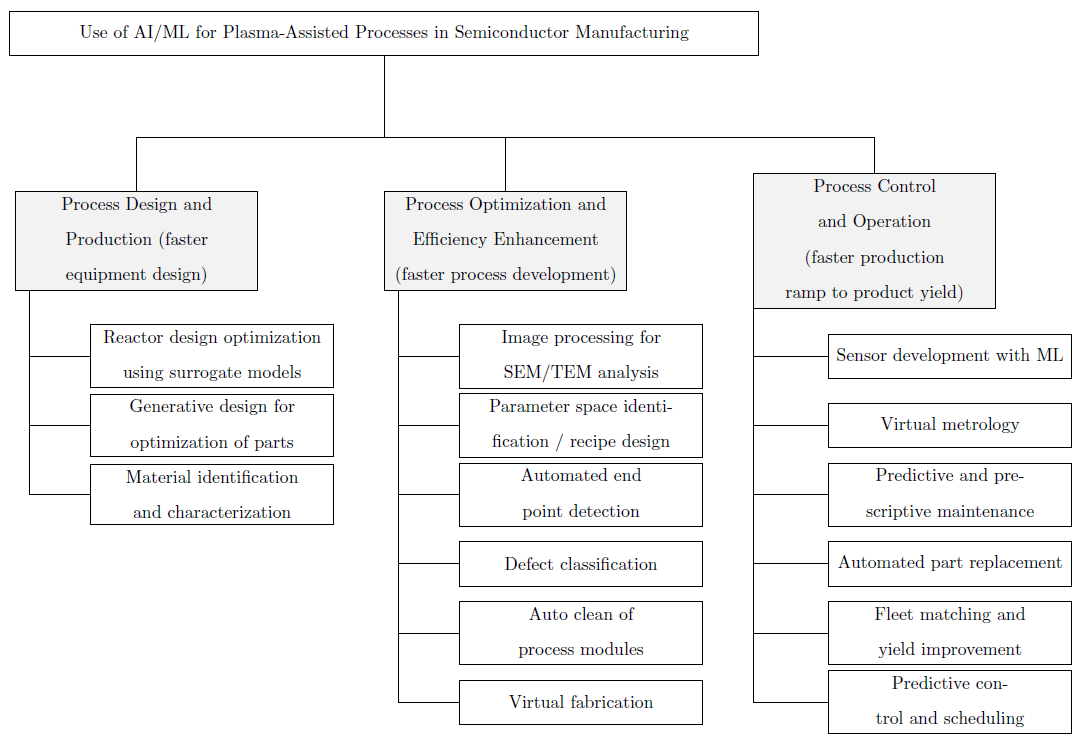}
    \caption{An overview of the applications of AI/ML for the design, development, and operation of plasma-assisted processes for semiconductor manufacturing, towards accelerating the time-to-market of new processes and products for the consumer electronics industry via smart manufacturing practices.}
    \label{fig_AI_diagram}
\end{figure*}

Applications of data-driven approaches for plasma-assisted processes in semiconductor manufacturing can be categorized in three interrelated areas (Figure \ref{fig_AI_diagram}): design and production of plasma processes, optimization of plasma processes and engineering efficiency enhancements, and adaptive process control and operation. As the complexity in shrinking technology nodes has increased, Moore's law has not been followed in recent years, i.e., the cost reduction per bit in case of NAND memory has decreased \cite{fontanajr2017}. To overcome challenges in shrinking technology nodes, equipment makers look to build new processes capable of handling new materials at a much more accelerated timescale in order to meet with the ever challenging demands for new applications. Process design optimization using surrogate models has received increasing attention in the semiconductor industry to facilitate design as well as testing what-if scenarios in a resource-efficient manner. In these approaches, cheap-to-evaluate models based on physics-based simulations are used to construct nonlinear relationships between various design parameters \cite{mesbah2019machine}. Surrogate models are also becoming increasingly important for constructing the digital twin of a system \cite{barricelli2019}, which allows for developing process design and optimization solutions based on fast surrogate evaluations under different operating conditions. For design of parts, especially with additive manufacturing that has gained popularity for quick prototyping and making complex designs possible, generative design approaches have proven useful by combining computational design, simulation, optimization and data visualization. To achieve the most optimal design, an initial design is ``evolved'' under multiple constraints. Such methods can allow process engineers to analyze various trade-offs in the design by determining the Pareto-optimal solution under multiple constraints \cite{oh2019}. Another area of growing importance is material identification and characterization for process design. As new chemistries are introduced in the process reactor and different plasma regimes explored, there is a need for new materials in the system to withstand more challenging conditions, such as corrosion, crack, warpage, and thermal creep. Material informatics create new opportunities to select the correct material for the given application and minimize extensive evaluation cost of materials that may not work in the given conditions. As the industry continues to shrink the technology nodes, equipment makers must constantly add more process knobs to meet the stringent specifications for the layer under consideration, such as deposition or etch rates, uniformity metrics, critical dimensions at desired locations, and other properties (e.g., stress and refractive index). In addition, there are other requirements set at the system level such as defectivity, sustainability, throughput and various cost constraints. Data-driven approaches have shown promise for speeding up process development by optimizing the recipe setpoints for the ideal film, as well as improving efficiency of process engineering given the vast design space for recipe optimization. In order to assess the outcome of a process, automated image analysis capabilities are developed to measure dimensions of interest \cite{darbon2021} and to improve quality of the image \cite{na2021}. Recipe optimization is performed not only based on current data collected, but also prior knowledge developed using machine learning algorithms \cite{suzuki2018, tanaka2018}. To accommodate for upstream film variations and variations in tools, real-time analysis methodologies for end-point detection are developed \cite{chakroun2020}. Data-driven approaches are used to characterize defects automatically by assigning classes to wafer map patters, morphology, and chemical spectra \cite{OLeary2020, batool2021}, as well as detecting and triggering auto-clean routine to improve productivity by minimizing the failures caused by these defects. In addition to process challenges at the unit process level, data-driven approaches can be used for optimizing the entire process flow, allowing engineers to study the sensitivity of a particular layer and build appropriate trade-offs to achieve their desired product \cite{jeong2021}. With proliferation of more sensors in semiconductor manufacturing equipment, new opportunities have also been created for advanced process control, including  operation analytics for online equipment health monitoring to enable predictive and prescriptive maintenance of processing tools \cite{moyne2017}; soft sensing and virtual metrology for enhanced process monitoring and fleet matching for yield improvement; fault detection and classification for timely diagnosis of potential process anomalies \cite{moyne2017}; feedback control strategies such as predictive control and run-to-run control for accommodating process-to-process variability, high product mixes, and process dynamics; and predictive scheduling for improving the overall fab productivity by minimizing idle tool time \cite{morariu2020}.

A fundamental requirement for success of data-driven approaches for the design, optimization and control of plasma-assisted processes in semiconductor manufacturing lies in the interpretability of the data-driven models. As the number of process tuning knobs increases to meet challenging demands for scaling needs in the industry resulting in over 10$^{23}$ possible permutations of recipes, and the continued demand to match system states across a fleet of tools with more than 10$^{100}$ possible states, quantum computing can play a transformative role in the years to come to facilitate AI applications involving complex high-dimensional data, or discrete/combinatorial optimization. To this end, there is a need for further advances in data management, better algorithms, resilience in cyber-physical systems, and innovation in advancing compute and storage of data. Other emerging applications of AI include automated visual inspections of parts, supply chain optimization, and augmenting human capabilities through concepts of extended reality. The field of Industry 4.0 is just beginning for the semiconductor industry and will rapidly grow with the goals of accelerating the time-to-market of new processes and products, as well as the relentless drive for greater productivity and yield.

\paragraph*{Acknowledgements}
The authors would like to thank Kaihan Ashtiani, Michal Danek, Paul Franzen, Sassan Roham, and other colleagues and mentors at Lam Research for valuable discussions over the years to help us better understand the impact digital transformation will have on the semiconductor industry.

\noindent
[Kapil Sawlani and Ali Mesbah]

\subsection{Plasma Information based Virtual Metrology (PI-VM)}
\label{Sec7_Park}
The necessity of a high-value process strategy for the semiconductor- and OLED (Organic Light Emitting Diode) display- manufacturing industry, which requires ultra-fine plasma process technology, is ever-increasing to achieve an increase of device production throughput. 
To manage the process results efficiently in this ultra-fine scale plasma process, an automated control system, such as fault detection and classification (FDC) and advanced process control (APC) logics, is needed. 
It requires the development of a virtual metrology (VM) model, which directs the process control. 
The prediction accuracy of the VM is a crucial component to the performance of the FDC or APC system \cite{1Woong-2014}. 
The VM was developed from classical chemical processes to predict process results based on the statistical analysis of monitored sensor data. 
According to Cheng et al., VM is a method for estimating the manufacturing quality of a process tool based on data sensed from the process tool and without physical metrology operations \cite{2ChengFan2012}.
Therefore, the development of the VM for plasma processes was likewise initiated from statistical approaches. 
Development of the statistically established VM began from correlation analysis of the variables with process results. 
To this end, various machine learning (ML) models are applied to VM modeling \cite{3HeeDuLee1997ONLINEQM,4vanAlbada2007TransformationOA}. 
However, this statistical-method-based VM has shown unsatisfactory prediction accuracy when applied to numerous cases of plasma-aided processes \cite{5Park2015}.

 To develop high-performance VM models, the efficient containment of the ‘good information’ representing parameters – that is, the parameters representing the process plasma state – is needed rather than the direct application of the ML methodologies. 
 These parameters should efficiently mediate between state variables monitored from the sensors and performance variables, and the specificity of a plasma-assisted process mechanism should be considered \cite{1Woong-2014,5Park2015}. 
 Lieberman discussed the importance of the reactions in the plasma volume, sheath, and target surface in terms of the progress of the process reactions, such as etching, deposition, sputtering, and ashing \cite{6Lieberman2005}. 
 These overall reactions are strongly correlated with each other and governed by the properties of the process plasma. 
 Therefore, to develop the VM for plasma-assisted processes, the process plasma information, including parameters representing the reaction properties in the plasma based on the volume-sheath-surface reaction mechanism, is required. 
 To attain this concept of the VM for plasma-assisted processes, new parameters called ‘PI (Plasma Information)’ were introduced that are applicable as powerful variables in 2015. 
 They have been used to predict various process performances such as etch rate, deposition rate, defect particles, etching profile, deposited thin film quality, and spatial uniformity of the processed results. 
 They have been applied to the control and management of the OLED mass production lines last six years \cite{5Park2015,7Park_2018,8Park2019CauseAO,9Park2021,10Park2020,11Kwonma14113005,12Jang2019CharacteristicsOA}.

\begin{figure*}[ht]
    \centering
    \includegraphics[width=0.9\linewidth]{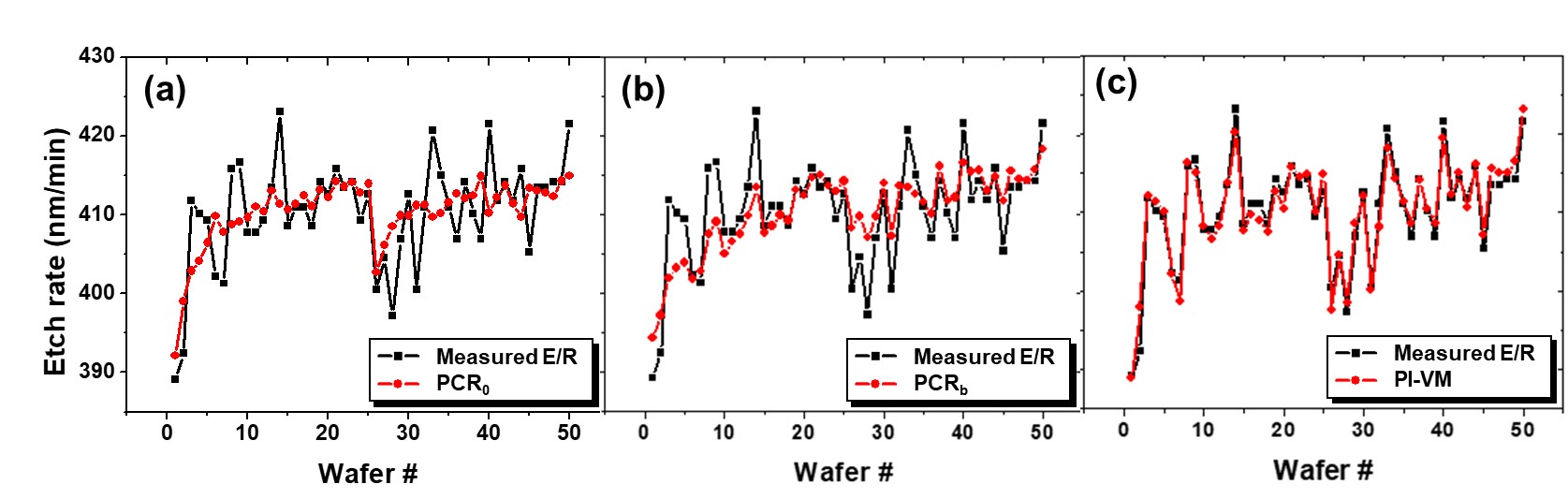}
    \caption{Comparison of the measured etch rate and predicted etch rates of 50 wafers with (a) basic PCR model, PCR\textsubscript{0}, (b) with the adoption of b-factor, PCR\textsubscript{b}, and (c) the fully PI variables adopted PI-VM. }
    \label{fig:CompEtchRates}
\end{figure*}

Figure \ref{fig:CompEtchRates} compares the predicted etch rates for the C\textsubscript{4}F\textsubscript{8} based plasma-assisted silicon oxide etching process with the measured etch rates. 
To test the performance of fundamental ML methodology, principal component regression (PCR) based VM to predict the etch rate was modeled. 
79 equipment engineering system (EES) sensing variables from the power, pressure, gas, chiller, heater, and exhaust system and 1670 parameters from the optical emission spectroscopy (OES) intensities were combined into the PCs and were regressed as shown in Fig.\ref{fig:CompEtchRates} (a). 
The correlation coefficient between the measured etch rate and VM result was R\textsuperscript{2} = 38.8\text{\%}. 
By adopting the PI parameter of b-factor measured by the OES data as PC into the PCR-based VM model (PCR\textsubscript{b}), the correlation coefficient between the measured etch rate and VM result was R\textsuperscript{2} = 57.2\text{\%}, as shown in Fig.\ref{fig:CompEtchRates} (b). 
Here, the b-factor is the shape factor in the generalized form of the electron energy distribution function (EEDF), $f(\varepsilon) \sim exp(-c\varepsilon \textsuperscript{b})$ with the coefficient c and electron energy, $\varepsilon$ \cite{13Park2014CharacteristicsOA}. 
The distribution shape varies from the well-known high-energy tail developed Maxwellian distribution with b=1 to the curtailed Druyvesteyn distribution with b=2 in general \cite{13Park2014CharacteristicsOA,14Gudmundsson_2001}. 
Finally, by adopting pre-sheath potential and surface passivation representing PI parameters synthesized from the monitored OES and EES data, the prediction performance of the VM was enhanced to R\textsuperscript{2}=96.9\text{\%}, as shown in Fig.\ref{fig:CompEtchRates} (c). 
These results imply that selecting the variables according to the reaction mechanisms in the process plasma is important to achieve the performance of the VM for the plasma-assisted process monitoring. 
PI-based VM (PI-VM) modeling, especially includes the characteristics of the EEDFs, can be an efficient method to include the information about the process state into the VM model and is useful to obtain high-performance of the VM applicable to the real field \cite{5Park2015}.
   
Developed PI-VM algorithms were applied to the mass production line of the OLED display manufacturing to solve four kinds of problems that occurred in the real field: The defect particle caused process fault prediction \cite{7Park_2018}, root cause analysis of the high-aspect-ratio contact (HARC) etching process faults \cite{8Park2019CauseAO}, the management of the mass production discontinuities with a proper application of the in-situ dry cleaning (ISD) \cite{9Park2021}, and the micro-uniformity problems in the process results \cite{10Park2020}. 
These PI-VM models optimized for each issue have shown enough prediction accuracy to apply for the long-periodic mass production running. 
Therefore, by applying the PI-VM models to the control of the OLED display manufacturing processes, overall production yields were relevantly progressed in the last six years.
Especially, the mass production management referring to the discontinuity qualifying PI-VM and effective application of ISDs, yield loss was successfully suppressed by about 25\text{\%} for 42 process chambers in the fab.  

\paragraph*{Acknowledgements}
This work is technically supported by the director Jae-Ho Yang, the vice presidents Jaehyung Lee and Jeonggen Yoo, and the executive vice president Insoo Cho of Samsung Display Co., Ltd. The authors would like to thank their support sincerely.

\noindent
[Seolhye Park, Jaegu Seong, and Gon-Ho Kim]

\subsection{Data management in manufacturing}
%Shinagawa
%\paragraph*{Importance of plasma domain knowledge as it relates to data quality in advanced process control applications in semiconductor manufacturing}

Semiconductor device technology is now far below the 10 nm critical dimension in manufacturing with its sights set on 2 nm.  
Successful device scaling, historically driven by lithographic patterning, is now driven by plasma etch. The tight process control afforded by modern plasma sources has enabled scaling.  
Moving past 3 nm requires even tighter levels of control. 
Tighter control translates to atomistic control. 
“Smart manufacturing” initiative is a means to meeting this end. 
Its purpose is to enable adaptability in plasma process tools \cite{Semi} facilitating the reliable and accurate advanced process control (APC) systems delivering nanometer-scale precision.
APC in the form of run-to-run control enables continuous process tuning. 
Process output parameters are monitored by metrology tools potentially including virtual metrology (VM) models. 
Adjustments to process tuning knobs are dictated by control models’ responses to measured deviations between the process outputs and control limits \cite{Moyne2001}. 
The control model itself is a barrier to achieving accuracy and reliability. 
The relationship between control parameters and the surface processes to be controlled is complex even for nominally simple plasma processes. 
Numerical plasma models require significant computational resources, making them difficult to use in a control context directly. 
The accuracy and reliability of theory-based numerical models are also an issue. Despite decades of progress, numerical models are difficult to validate. 
Artificial intelligence (AI)/machine learning (ML) technologies enable high accuracy VM model prediction by capturing variations originated from complex plasma and surface reaction phenomena without reliance on physical assumptions \cite{Himmel1993,Hung2007,Lynn2010inproceedings}. 
While AI/ML algorithms are attractive options for implementing high accuracy VM models, there are some disadvantages. 
They require a large number of training data sets and lack the inference capability needed to link the predicted variations to their root causes. 
Plasma diagnostics paired with appropriate sensor technologies can reduce the advanced data processing load while maintaining or even improving model accuracy.  
This is done through direct extraction of variables that should correlate with target metrics via theory or model \cite{Shinagawa2016patent,Lynn2011thesisPhD,Kwon2021,Kim2021_IEEE}, a methodology termed “data quality improvement.”  
Data quality improvement relies heavily on the selection of appropriate in-situ sensors, which in turn requires specific plasma domain knowledge:

\begin{itemize}
    \item 	Type 1 domain knowledge – what to measure: plasma variables such as the ion flux, neutral flux, and deposition rate defined by phenomenological surface reaction models or interpretation of post-process profile formation using theoretical mechanisms \cite{Osano2008_AVS,Cooperberg2002_JVST,Gottscho1992_JVST}.
    \item   Type 2 domain knowledge – how to measure: non-invasive in-situ sensors to measure plasma variables that are derived from type 1 domain knowledge.
\end{itemize}

\begin{figure*}[ht]
\begin{center}
\includegraphics[width=0.8\linewidth]{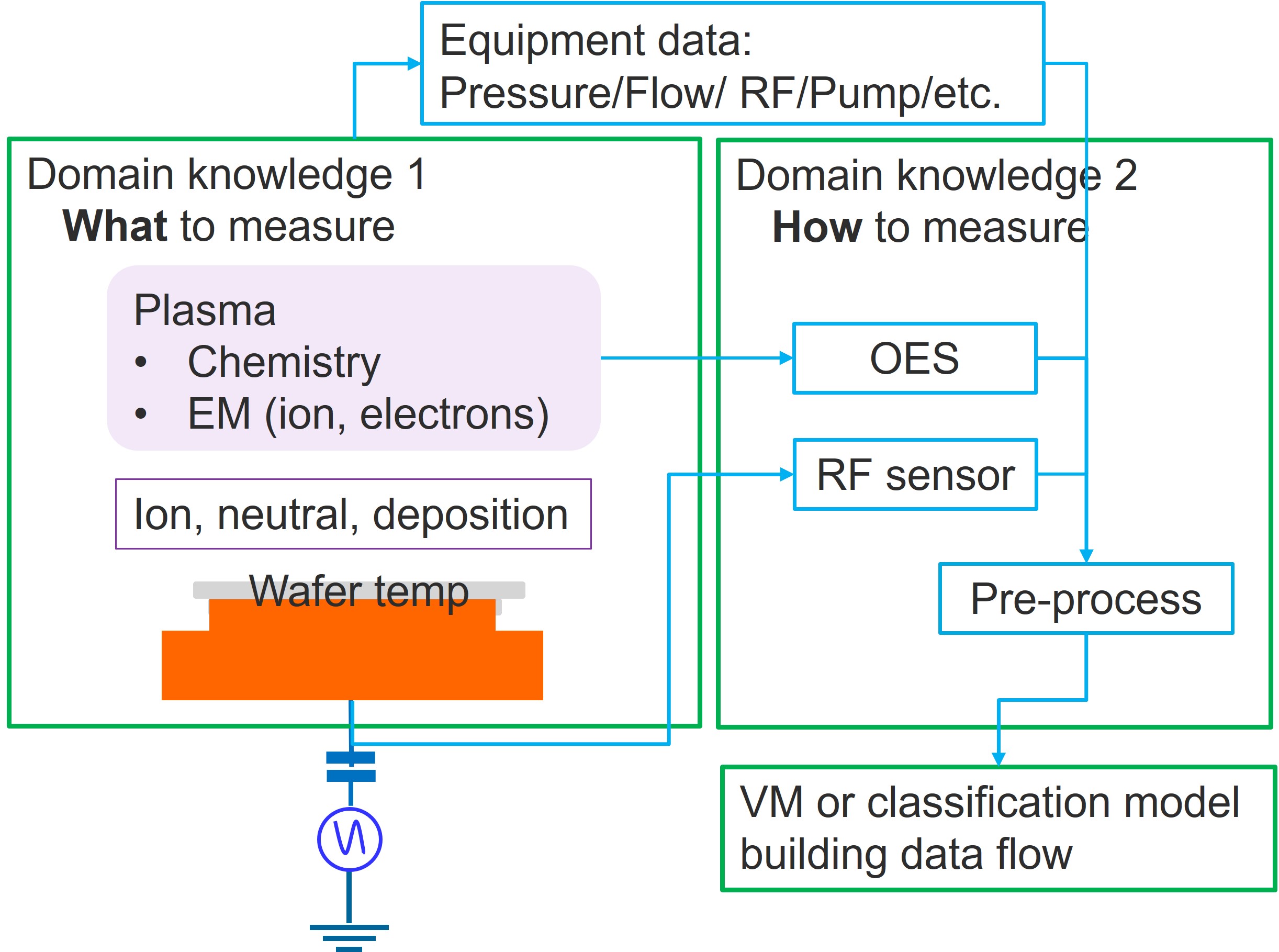}
\caption{Schematic of a generic plasma process tool and in-situ sensors.}
\label{dataMfig1}
\end{center}
\end{figure*}

\begin{figure*}[bht]
\begin{center}
\includegraphics[width=0.9\linewidth]{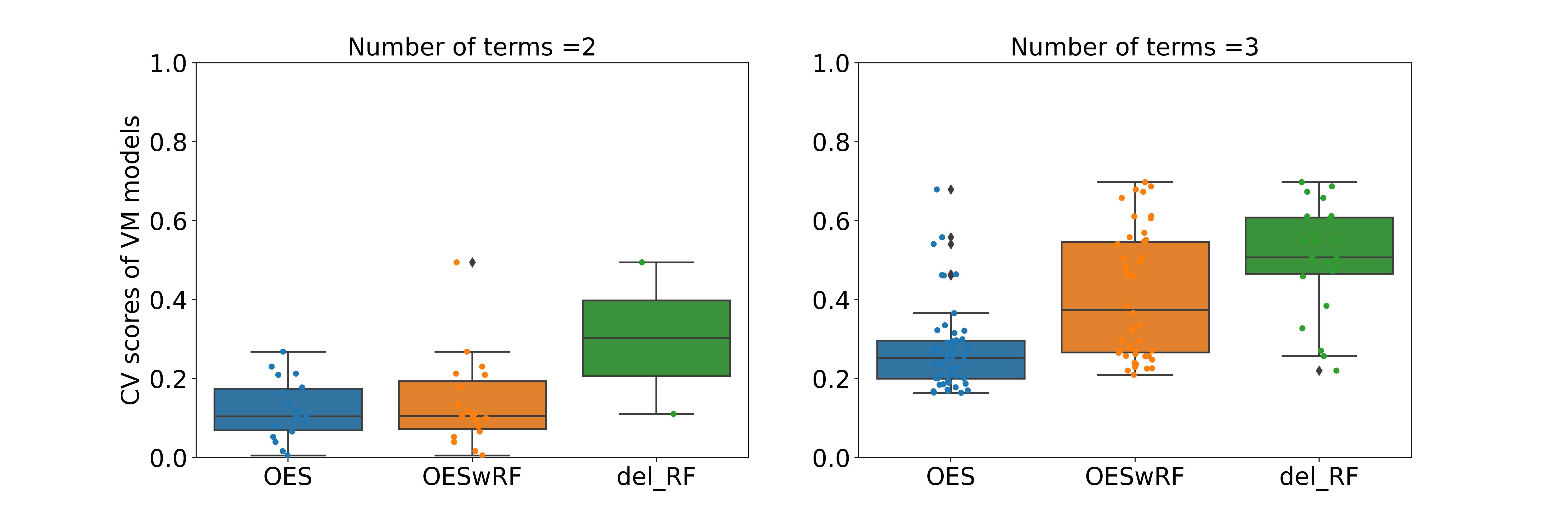}
\caption{CV scores of the VM models built with OES only (blue) and OES with RF sensor (orange) data sets. The VM models that include RF sensor data were grouped into del\_RF (green). Each data point represents the CV score of VM models generated from the exhaustive least square regressions. The CV scores of VM models were significantly improved when RF sensor data were included, indicating the improvement of data quality with RF sensor data.}
%CV score vs. data groups for each number of features.}
\label{dataMfig2}
\end{center}
\end{figure*}

Pre-processing measured data with the interpretive functions afforded by type 1 and 2 domain knowledge is key. Pre-processing involves not only conversion of raw data into plasma variables but also is important for errors removal \cite{Shinagawa2018patent}. 
Optical emission spectroscopy (OES) provides a good example of useful pre-processing. 
OES intensities most often vary during production runs due to varied transmittance through the view window caused by film deposits. 
The intensity variation is independent of plasma condition, hence registered as an error. 
One way to reduce error is to normalize the OES spectra by the OES intensity at chosen wavelength \cite{Tsutsui2019VirtualMM}. 
Both sensor and pre-processing method selection (i.e., data quality) can be evaluated by benchmarking VM model performance with versus without the studied sensor data added to other default sensor data sets. 
The following example illustrates one such evaluation. 
Thermal oxide (TOX) flat wafer etching rates were varied by installing various combinations of new and used chamber parts. 
OES and RF sensor data were collected during the etching of TOX wafers. 
VM models were constructed to predict TOX etching rates using exhaustive least square regression with pre-processed OES and RF sesor data. The number of terms in the VM models was limited below 4 to enhance the sensitivity to data quality. 
The impact of RF sensor data on VM model performance was evaluated using cross-validation (CV) scores calculated as an average of R2 values from each fold of the five-fold cross-validation.  
Figure \ref{dataMfig2} shows CV scores of all VM models generated from exhaustive least square regression with two data sets – OES only (OES) and OES with RF sensor data (OESwRF). 
del{\_}RF represents models that include RF sensor data i.e.) del{\_}RF $=$ OESwRF - OES.  As can be seen, significantly improved high-performance VM models were generated with RF sensor data. 
The results illustrate overall data quality improvement by adding RF sensor data with pre-processing. 

Successful development and deployment of APC to meet the tight control limits demanded by sub-10 nm technology plasma processes require AI/ML to be augmented by improved data quality. 
Data quality improvement with domain-knowledge-aided pre-processing was illustrated in this paper for the simple example of TOX etch. 
The availability of non-invasive in-situ sensors for plasma and surface parameters is an issue. Therefore, the concerted development of these sensors will be an area of emphasis for the industry. 
An area of particular importance for sensor development is drift-free molecular species measurement during production runs. 
The ubiquitousness of pulsing in plasma processing poses additional challenges and opportunities. Faster data rates are needed for in-situ sensors to be able to characterize the important aspects of complex pulse trains.
   
\paragraph*{Acknowledgements}
The authors would like to thank Alok Ranjan and Hiromasa Mochiki for their support and valuable feedback.

\noindent
[Jun Shinagawa and Peter Ventzek]

\subsection{Data-driven analysis and multi-scale modeling of plasma-surface interactions}
\label{Sec7_Triesschmann}

% Title page
%\title{VIII.E. Data-driven analysis and multi-scale modeling of plasma-surface interactions}

% Author list with affiliations
%\author{Jan Trieschmann}
%\affiliation{Electrodynamics and Physical Electronics Group, Brandenburg University of Technology Cottbus-Senftenberg, Siemens-Halske-Ring 14, 03046 Cottbus, Germany}
%\affiliation{Theoretical Electrical Engineering, Faculty of Engineering, Kiel University, 24143 Kiel, Germany}

%\maketitle

%{\color{red}The first paragraph should be the introduction paragraph, which should be about 200 words (if necessary up to 350 words). In this paragraph, please discuss the background, motivation, importance, research history, etc., of the topic to be discussed in this subsection.}

%{\color {red} The main body paragraphs should be about 400 word long (if necessary up to 800 word long.)  In the main body, please present the definition of the problems, a simple outline of the methods to solve them (with proper references on technical issues of the methods), and remaining challenges.}

%{\color{red} The final paragraph should give concluding remarks; should be about 100 words (if necessary up to 250 words)}

The majority of technological (and fusion) plasmas is subject to interactions with bounding surfaces. It is essential for plasma processing, but typically considered inevitable in fusion devices with harsh plasma environmental conditions. The role of plasma-surface interaction (PSI) is generally bi-directional: (1) Particles from the plasma volume may cause modification of surface material (e.g., etching/deposition, chemical reactions, structure and phase transition). (2) The surface may influence the plasma volume through particles emanating from the walls due to related physical phenomena (e.g., sputtering, chemical reactions, secondary electron emission). This feedback implies that PSI cannot be considered independent, but consistently coupled. It requires a bi-directional relation following (1) and (2) between plasma and surface conditions at multiple time and length scales.

Several data-driven approaches have taken PSI into account macroscopically for plasma process control. They used plasma information based virtual metrology for plasma etching with experimental data sources \cite{park_micro-range_2021}, as well as model predictive control for atmospheric pressure plasma dose delivery \cite{gidon_predictive_2019} or reactive magnetron sputtering close to mode transition \cite{woelfel_control-oriented_2021}. In contrast, theoretical multi-scale analyses of technological plasmas have been restricted to classical modeling and simulation (e.g., combining molecular dynamics, binary collision approximation, and kinetic Monte Carlo models at the atomic level \cite{ito_triple_2018}; or unidirectional coupling the reactor scale to the feature scale in complex capacitive radio frequency plasmas \cite{denpoh_multiscale_2020}; list not exhaustive).

So far the focus has been on route (1) toward the surface. The physical complexity and the computational expenses of atomic level PSI models restrict return route (2) toward the plasma. If considered, the latter is often reduced to simple analytical approximations. This may be a severe limitation when complex surface chemical dynamics need to be captured accurately (e.g., plasma-enhanced catalysis or atomic layer deposition/etching) \cite{bogaerts_2020_2020,national_academies_of_sciences_engineering_and_medicine_plasma_2021}. Rigorous treatment of PSI is moreover required if emission from the surfaces may significantly influence the plasma discharge itself. Data-driven PSI models may capture these dynamics at a non-prohibitive computational effort.

\begin{figure*}[ht]
\begin{center}
\includegraphics[width=0.7\linewidth]{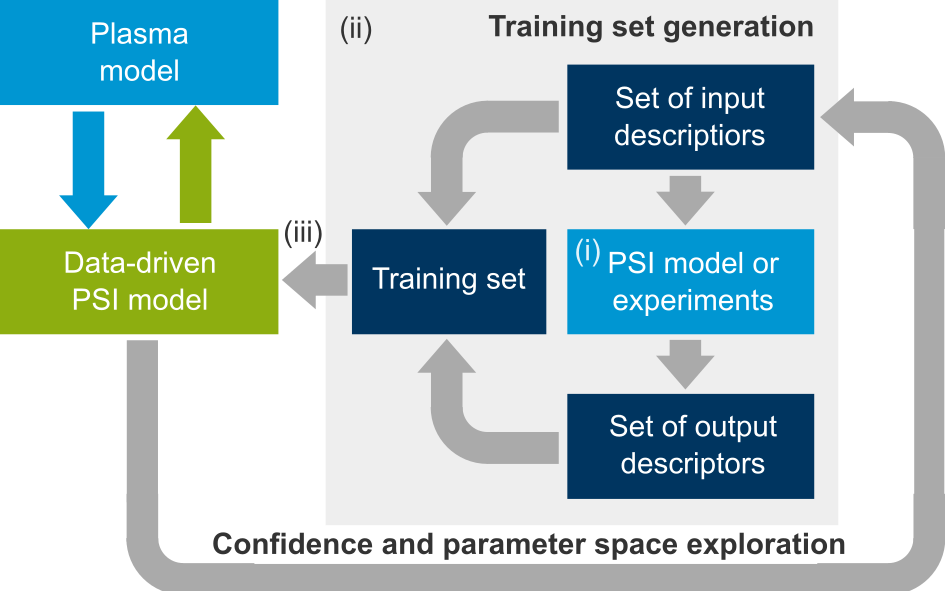}
\caption{Schematic of a generic data-driven PSI model.}
\label{Fig1_JT}
\end{center}
\end{figure*}

The procedure of establishing corresponding data-driven PSI models may differ in detail, but a rather generic scheme is outlined in Fig.~\ref{Fig1_JT} as follows: (i) Data retrieval from measurements or simulations. (ii) Feature selection through identification of reliable physical descriptors. (iii) Establishing of a regression relating descriptors (model inputs) to targets (model outputs), possibly with uncertainties (systematic or statistical). Each step is indispensable and could require several iterations, depending on the utilized procedure.

While a manifold of surface interaction phenomena may be considered, data-driven approaches to PSI have focused on the analysis of sputtering due to energetic particle impingement (e.g., ions, fast neutrals, photons). While its fundamental nature may seem simple, it poses a non-trivial problem due to the nonlinear dynamics of the collision cascade in the solid subsequent to interaction. In the absence of a widely applicable analytical description from first principles, data-driven approaches have been suggested to establish generalized relations inferred from the data \cite{kruger_machine_2019,kino_characterization_2021,preuss_bayesian_2019,gergs_efficient_2021-1}.

(i) The amount of data accessible for data-driven PSI modeling of sputtering varies significantly. For instance, a well-defined data set of experimental sputtering yields for different ion-solid combinations is publicly available \cite{yamamura_energy_1996} and has been successfully used \cite{kino_characterization_2021}. These are limited to integral information, however, eliminating the details of the flux and energy distributions emanating from the surface. Energy and angle resolved data from Monte Carlo simulations (with binary collision approximation) provide a compromise between computational costs and physical fidelity \cite{kruger_machine_2019,gergs_efficient_2021-1}. Accurate physical simulation data at the atomic level (e.g., molecular dynamics) are typically sparse and may require data augmentation, because the computational cost to obtain large data sets imposes a significant challenge.

(ii) The process of defining independent features depends on the requirement of physical interpretability. Given a set of possibly correlated physical variables, a subset of descriptive physical parameters has been devised by hierarchical clustering and corresponding descriptor analysis for sputtering yield regression \cite{kino_characterization_2021}. In contrast, the concept of variational autoencoder artificial neural networks has been applied to provide a descriptive set of latent parameters at the cost of a complicated physical interpretability \cite{gergs_efficient_2021-1}. Uncertainty quantification of physical descriptors using Bayesian analysis has devised confidence bounds in inference of the sputtering yield, suggesting a more accurate surface binding energy \cite{preuss_bayesian_2019}.

\begin{figure}[ht]
\begin{center}
\includegraphics[width=0.8\linewidth]{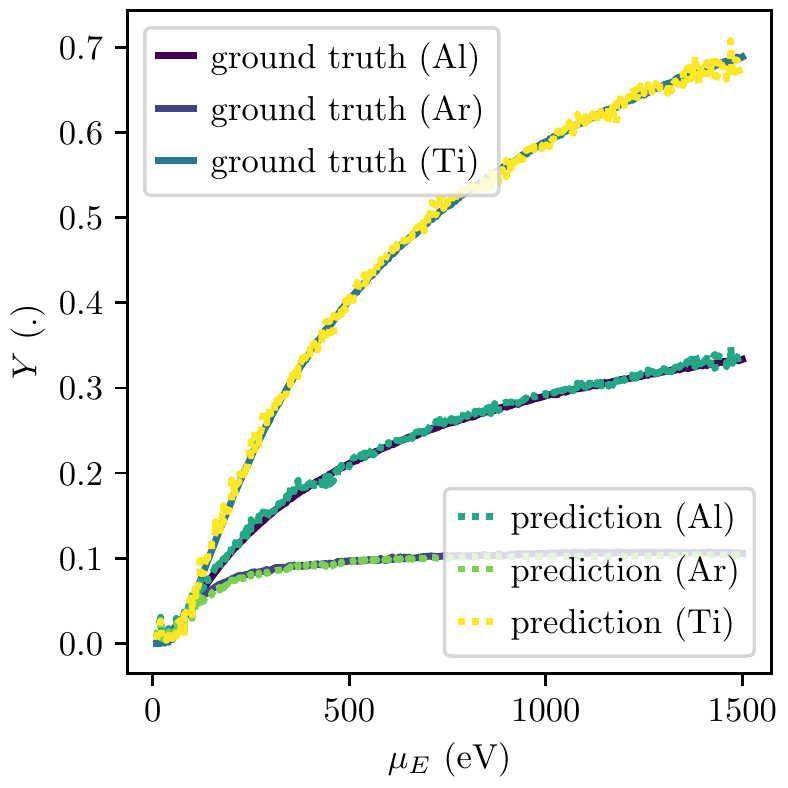}
\caption{Yield per sputtered species (Al, Ar, Ti) as a function of the mean ion energy for an Al$_x$Ti$_{1-x}$ surface with initial stoichiometry $x=0.3$. Ground truth compared to artificial neural network predictions.
Reproduced with the permission of the American Vacuum Society, from
Ref.~\citenum{gergs_efficient_2021-1}.
%``Efficient plasma-surface interaction surrogate model for sputtering processes based on autoencoder neural networks'', accepted for publication in the Journal of Vacuum Science and Technology B [\href{https://doi.org/10.1116/6.0001485}{Link}]. \cite{gergs_efficient_2021-1}
}
\label{Fig2_JT}
\end{center}
\end{figure}

(iii) The ultimate goal is the design of a PSI regression task. While kernel ridge regression was successfully applied for the inference of sputtering yields as a function of the incident particle properties \cite{kino_characterization_2021}, Gaussian process regression has proven capable of simultaneously providing sputtering yields and corresponding uncertainty bounds \cite{ikuse_gpr-based_2021}. Finally, the capability to capture the complex non-linear relation between incoming ion energy distributions and outgoing energy and angular distributions of sputtered particles using artificial neural networks has been demonstrated. It facilitates detailed PSI evaluation during plasma simulation run-time (cf. Fig.~\ref{Fig2_JT}) at tremendously reduced computational cost \cite{kruger_machine_2019,gergs_efficient_2021-1}.

The outlined steps focus on reported approaches to data-driven PSI modeling of sputtering. An extension of similar procedures to other PSI mechanisms like plasma-induced electron emission or surface chemical reactions is due. For instance, the complex transient interplay between reactive plasma and surface dynamics inherent to plasma catalysis or atmospheric pressure plasma in contact with surfaces/liquids may only be resolved with data-driven PSI models. In this context, data-driven chemical reaction pathway analysis \cite{ulissi_address_2017} and active/transfer learning strategies for computationally costly atomic scale simulations \cite{diaw_multiscale_2020} should be considered. Data-driven PSI models may ultimately permit a continuous and high fidelity physical description of technological plasmas, providing guidelines for future research and exploration.

%\section*{Acknowledgements}
\paragraph*{Acknowledgements}
J. Trieschmann acknowledges valuable input from Tobias Gergs, in particular Fig.~\ref{Fig2_JT}, helpful discussions with Borislav Borislavov, and continuous support from Thomas Mussenbrock. Funded by the Deutsche Forschungsgemeinschaft (DFG, German Research Foundation) -- Project-ID 138690629 -- TRR 87 and -- Project-ID 434434223 -- SFB 1461. Financial support from Tokyo Electron Technology Solutions Ltd. is acknowledged.

\noindent
[Jan Trieschmann]

%\nocite{*}

%\bibliographystyle{ieeetr}
%\bibliography{DDPS_PSI}% Produces the bibliography via BibTeX.

%
% ****** End of file apssamp.tex ******

\subsection{Neural-network potentials (NNPs) for the analysis of plasma-surface interactions with molecular dynamics (MD) simulations} 
\label{Sec7_Chen}

The surfaces of a fusion reactor will inevitably be exposed to harsh environmental conditions. 
Besides neutron fluxes, material erosion and fuel retention will limit their lifetime, especially in the divertor region. 
Experimental investigations at these conditions are difficult to impossible. 
Therefore, theoretical materials science is increasingly playing a role to quantify the plasma-surface interactions. 
On the atomic level, two techniques play a major role, molecular dynamics simulations (MD) \cite{1Alder1959StudiesIM} where the many-body system is studied in detail by modelling its time evolution and the binary collision approximation (BCA) \cite{2Robinson1991ComputerSS} theory where the 
%approximation of binary collisions between a projectile ion (or atom) and a surface atom is made. 
path of a projectile ion or atom is determined by a sequence of binary collisions.
In BCA, scattering integrals are normally calculated by Monte-Carlo methods to average over angular and energetic distributions and the collision cascades are then derived. 
The assumption of binary collisions works best at projectile energies from keV and up but not at lower energies where many-body effects are important. 
In MD, the total potential energy surface (PES) is the key ingredient. 
It contains all the information about the system and the trajectories of all atoms under consideration are derived from it. 
Molecular dynamics simulations have only recently been applied to systems where bond breaking and bond formation happen since in this case, analytical potential energy expressions are difficult to derive. 
Such events are, however, happening all the time in sputtering processes.
From humble beginnings like the Sutton Chen potential \cite{3SuttonChen1990} quite successful interaction models like the bond-order potentials \cite{4Tersoff1988} were devised. 
They are analytical expressions that can be evaluated quickly on the computer and especially the latter were used in the investigation of several plasma-facing materials \cite{6Lasa2014ModellingOW,7SAFI2015805}.
%However, they need a lot of work to be derived 
However, their construction is demanding in terms of human effort
and their mathematical form is sometimes not flexible enough. 
About 15 years ago, with the increased employment of machine-learning, techniques were developed to construct the potential energy hypersurface (PES) nonparametrically with neural networks \cite{8Behler2007GeneralizedNR} or Gaussian approximation potentials (GAP) \cite{9Bernstein2019DeNE}. 
Both methods allow for the necessary flexibility and, being parametrized via quantum chemical calculations, can model plasma surface interactions (PSIs) accurately 
%by MD.
for subsequent use in MD simulations.
In the next paragraph, we give an example of typical NN-potential-based MD modelling.

Finding a suitable NN-based PES can be divided into three independent subproblems: (a) Converting the cartesian coordinates of the atoms into descriptors that can be input to the NN. (b) Finding an optimal NN architecture and (c) training the NN. Subsequently, the MD simulations produce statistically meaningful sample directories that are 
%analysed.
analyzed with respect to sputtering yields and many other material properties.

\begin{figure*}[ht]
    \centering
    \includegraphics[width=0.9\linewidth]{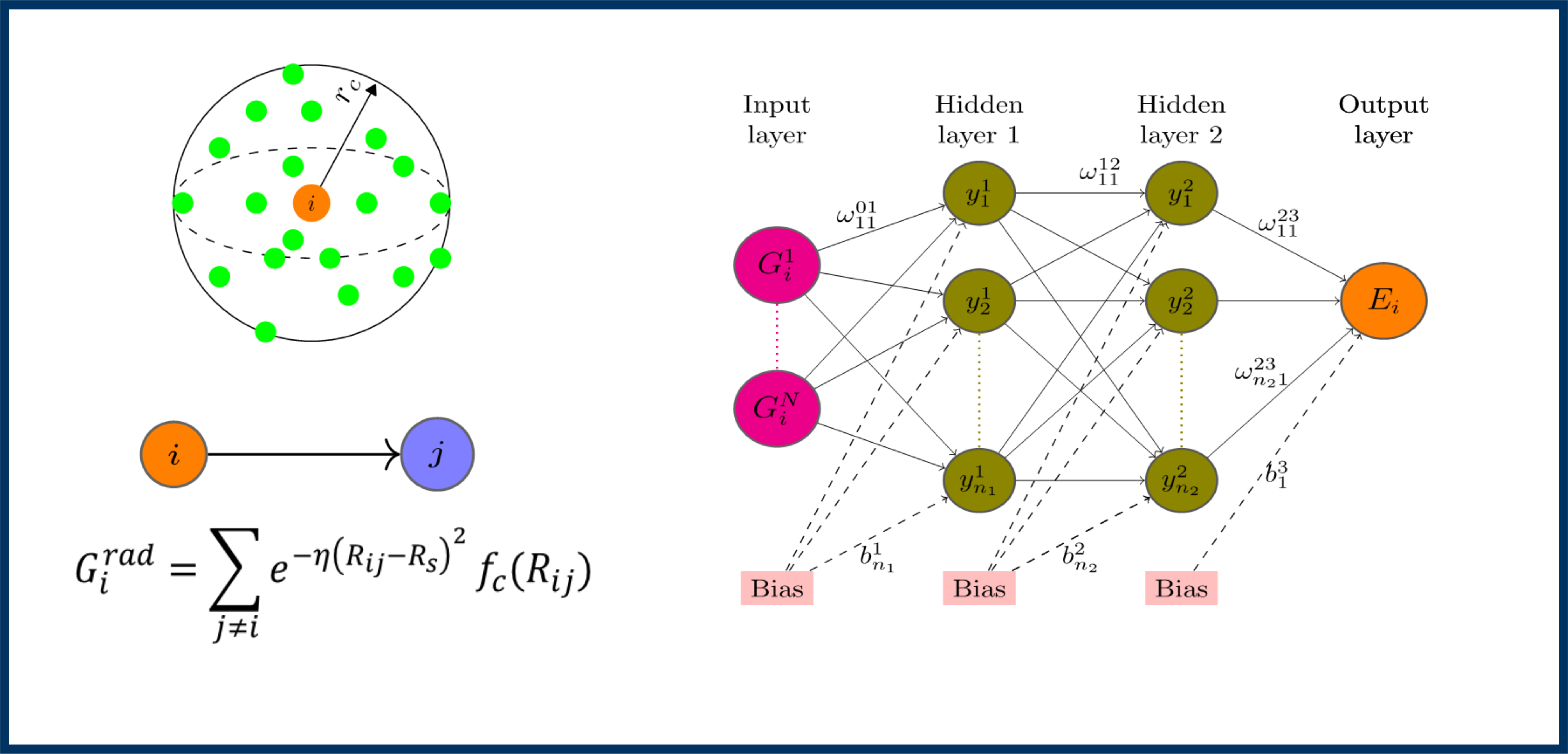}
    \caption{Left side from top to bottom: An atom i and its surrounding atoms, one of the corresponding radial vectors and one radial symmetry function G\textsuperscript{rad}. Right side: Two G\textsuperscript{rad} functions are input and the energy of atom i is an output of an NN with two hidden layers.}
    \label{fig:ChenFig1}
\end{figure*}

The conversion of cartesian coordinates into symmetry-adapted descriptors (a) is necessary because the energy of an atom stems from its environment and must be independent of translation, rotation and the permutation of like atoms. 
In the Behler method, the descriptors are radial and angular basis functions and their coefficients are calculated by projecting the atomic environment onto them. 
The invariant coefficients are input to the NN. Optimal descriptors are at least as important as having an optimal NN (or GAP) architecture and are an area of ongoing research \cite{10Shapeev2016MomentTP}. 
The left side of Figure \ref{fig:ChenFig1} shows as an example how the weight of a radial symmetry function G\textsuperscript{rad} is derived from all neighbours j of atom i. 
This is done flexibly for several Gaussian functions with varying exponents and midpoints to construct the radial density.

 The NN (b) itself can have various architectures but is often a simple feed-forward NN with as many input neurons as basis functions, two hidden layers of the same size and one output layer delivering the energy for one atom. The atomic energies are then summed up.

Training (c) is the process of finding the best NN weights and offsets and is similar to other applications of NNs. 
From simple backpropagation over Marquardt-Levenberg fitting to Kalman filters many techniques are used. 
%It is done performing quantum chemical MD simulations, normally of the density-functional type, and then feeding the coordinates, energies and forces into the network. 
The training data are symmetry adapted atomic coordinates and associated energies and forces from trajectories derived by direct quantum chemical MD simulations.
Sometimes the potential energy and the forces are divided into one part that is treated with simple analytical expressions and the NN takes care of the rest. 
This is advisable for charged systems where simple electrostatic interactions make up the largest part.

The process of network training is normally iterative. A trained NN is used in MD runs. 
Some MD configurations are checked by quantum chemical methods if the NN energies and forces are accurate up to a threshold. 
If they are not, such configurations are used in retraining. 
After a few cycles, a NN–based PES is obtained that is accurate within the limits of the parameter space.

\begin{figure*}[ht]
    \centering
    \includegraphics[width=0.9\linewidth]{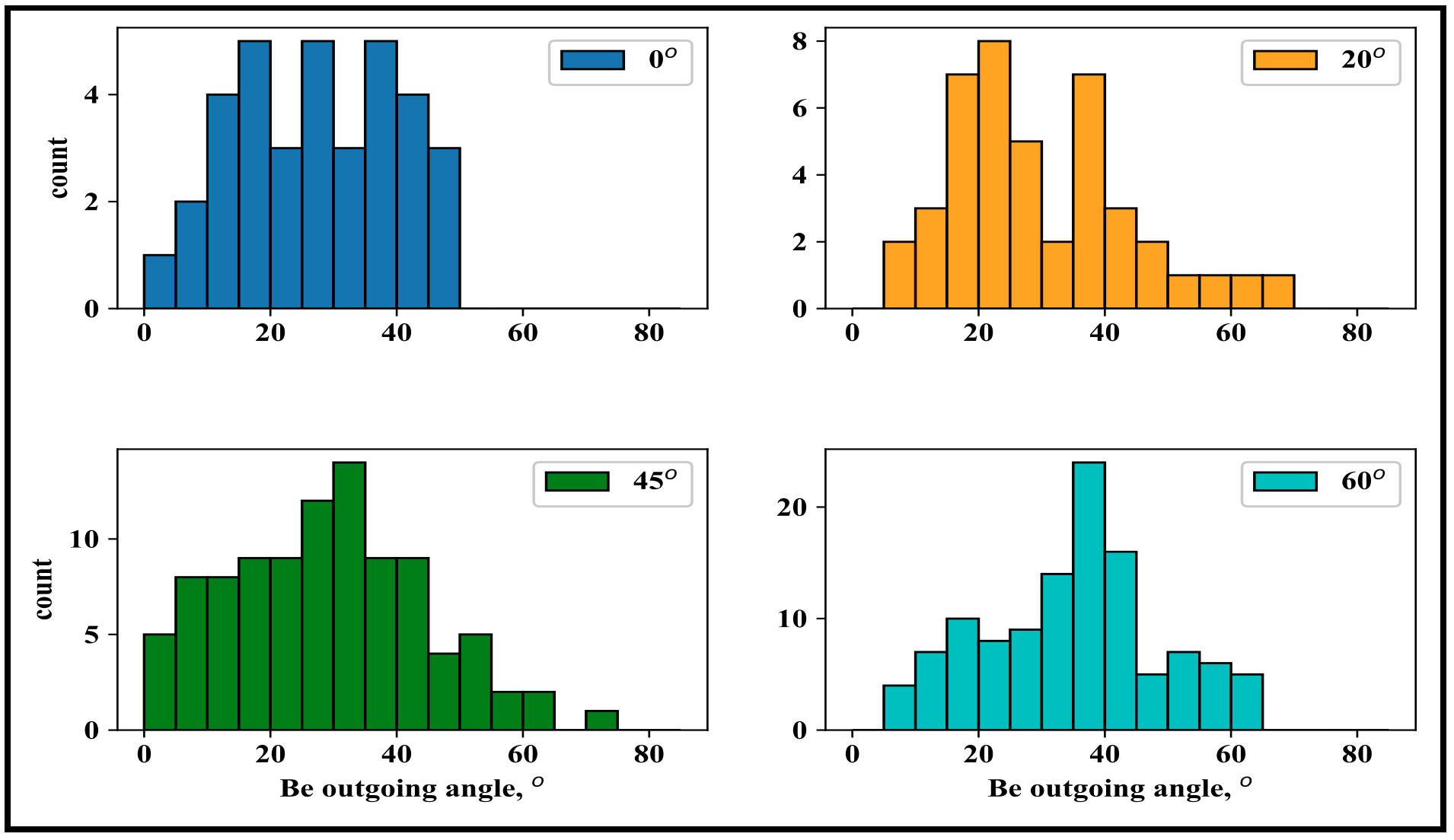}
    \caption{Effect of the angles (0\textdegree, 20\textdegree,45\textdegree, and 60\textdegree to the surface normal) of deuterium atoms incoming with 100 eV on the angular distributions with which Be atoms are sputtered \cite{11Chen2020article}.}
    \label{fig:ChenFig2}
\end{figure*}

Then production runs can be performed like in conventional MD simulations. 
For calculating sputtering yields where energy and angle of the incoming particle are variable, for each energy/angle combination about 5000 trajectories are necessary to achieve a good statistic. 
Figure \ref{fig:ChenFig2} shows results from sputtering simulations of a Be\textsubscript{2}W surface. 
The trajectories of MD runs with different angles of the incoming deuterium atoms are analysed to obtain density distributions (histograms) of the angles with which Be atoms are sputtered away \cite{11Chen2020article}. 
Similar studies have been performed also for other surfaces as well \cite{12Chen2020article}.

In reality, 
%more than two parameters
more than two environmental parameters 
are important, such as the surface temperature, atomistic surface details and so on. 
Then, unfortunately, the limits of MD are quickly 
%reached. It 
reached due to finite computational resources. MD
is also not practical in the MeV range since the integration of the equations of motion would require too small a timestep. 
MD with ML-based PES is, however, by now an established technique that is increasingly used to study PSI–relevant processes such as sputtering, retention, diffusion, bubble formation and diffusion.

 Computational materials science is now becoming a useful tool and the modelling of plasma-surface interactions by means of molecular dynamics simulations is evolving rapidly. 
 It is recognized now that the optimal descriptors are of utmost importance, even more so than the mathematical shape of the potential energy function. 
 At the same time, one becomes aware that the automatization of the training/simulation/improvement cycle is necessary. 
 This is not trivial and ‘active learning’ \cite{13Wiki} methods can be used to achieve this goal. 
 It is quite possible that the methods described here will be soon available in computer codes in a more standardized fashion and will be used even more. 
 At the same time improvements and new algorithms are published in short intervals, indicating that machine learning is far from mature, even as a tool for representing complex potential energy surfaces.

% \begin{figure}[ht]
% \begin{center}
% \includegraphics[width=0.5\linewidth]{Fig1}
% \caption{At most two figures are allowed in a single subsection. If the author does not own the copyright of the figure, he/she must obtain permission to use it in this article and list proper references. }
% \label{Fig1}
% \end{center}
% \end{figure}

% \begin{figure}[ht]
% \begin{center}
% \includegraphics[width=0.5\linewidth]{Fig2}
% \caption{Second figure. }
% \label{Fig2}
% \end{center}
% \end{figure}

\paragraph*{Acknowledgements}
This work has partly been carried out within the framework of the EUROfusion Consortium 
%and has received funding from the Euratom research and training program 2014-2018 and 2019-2020 under grant agreement No 633053.
funded by the European Union via the Euratom Research and Training Programme (Grant Agreement No 101052200 — EUROfusion). Views and opinions expressed do not necessarily reflect those of the European Union or the European Commission.

\noindent
[Lei Chen and Michael Probst]

\subsection{ML-based numerical simulation of low-temperature plasmas}
% ***********
%% See the REVTeX 4 README file
% It also requires running BibTeX. The commands are as follows:
%
%
%\documentclass[
%aps,
%amsmath,amssymb,
%reprint,
%preprint,
%]{revtex4-2} 

%\preprint{APS/123-QED}

%\usepackage{graphicx}% Include figure files
%\usepackage{xcolor}% Color
%\usepackage{dcolumn}% Align table columns on decimal point
%\usepackage{bm}% bold math

%\begin{document}

% Title page
%\title{VII. C. ML-based numerical simulation of low-temperature plasmas}

% Author list with affiliations
%\author{Satoru Kawaguchi}
%\affiliation{Division of Information and Electronic Engineering, Muroran Institute of Technology, Muroran, %Hokkaido, 050-8585, Japan}

%\maketitle

Charged-particle transport plays a key role in generating and maintaining low-temperature plasmas. The Boltzmann equation (BE) provides the basis for elucidating charged-particle transport in plasmas. However, since the BE is the transport equation in phase space, it has still been limited to simulate the spatio-temporal development of the charged-particle transport from solving the BE numerically. This is due to the curse of dimensionality, exponential growth of computational cost with respect to the dimension. Such difficulty clearly appears in three-dimensional (3D) and higher-dimensional simulations using mesh-based methods, such as finite difference methods. Physics-informed neural network (PINN) has attracted attention to solve the partial differential equation (PDE). In the PINN approach, the latent solution of the PDE is represented by artificial neural network (ANN), and the ANN is trained to respect both the PDE, often describing the law of physics, and boundary conditions. When a function is represented by ANN, partial derivatives of the function with respect to its variables can be calculated analytically by taking advantage of automatic differentiation; therefore, the PINN approach enables us to solve the PDE without generating grids and meshes and would allow us to tackle high-dimensional problems. The PINN approach was proposed by Raissi et al.\cite{raissi2019physics} They demonstrated this approach to solve one-dimensional (1D) Burgers’ equation and Shrödinger equation with Dirichlet boundary conditions. The PINN approach has been applied for solving a wide range of problems with various boundary conditions and constraints. Kawaguchi et al. \cite{Kawaguchi2020} employed this approach for solving the BE for two-dimensional (2D) equilibrium electron velocity distribution function (EVDF) in Reid's ramp model gas and Ar under dc uniform electric fields with normalization constraint of the EVDF. Rao et al. \cite{Rao2020} simulated incompressible laminar flows with Dirichlet and Neumann boundary conditions. Zobeiry and Humfeld \cite{Zobeiry2021} applied the PINN approach to solving 1D and 2D heat transfer equations with convection boundary conditions. A comprehensive review of the PINN is available in reference \cite{Karniadakis:2021}. In this subsection, the procedure for solving the PDE through the PINN approach is presented. The BE for 3D equilibrium EVDF under crossed dc uniform electric and magnetic fields in a boundary free space is chosen as an example of the PDE \cite{kawaguchi2022}. Such EVDF would be important to analyze the electron transport properties in magnetized plasmas, which are employed in material processing.

\begin{figure*}[ht]
\begin{center}
\includegraphics[width=0.8\linewidth]{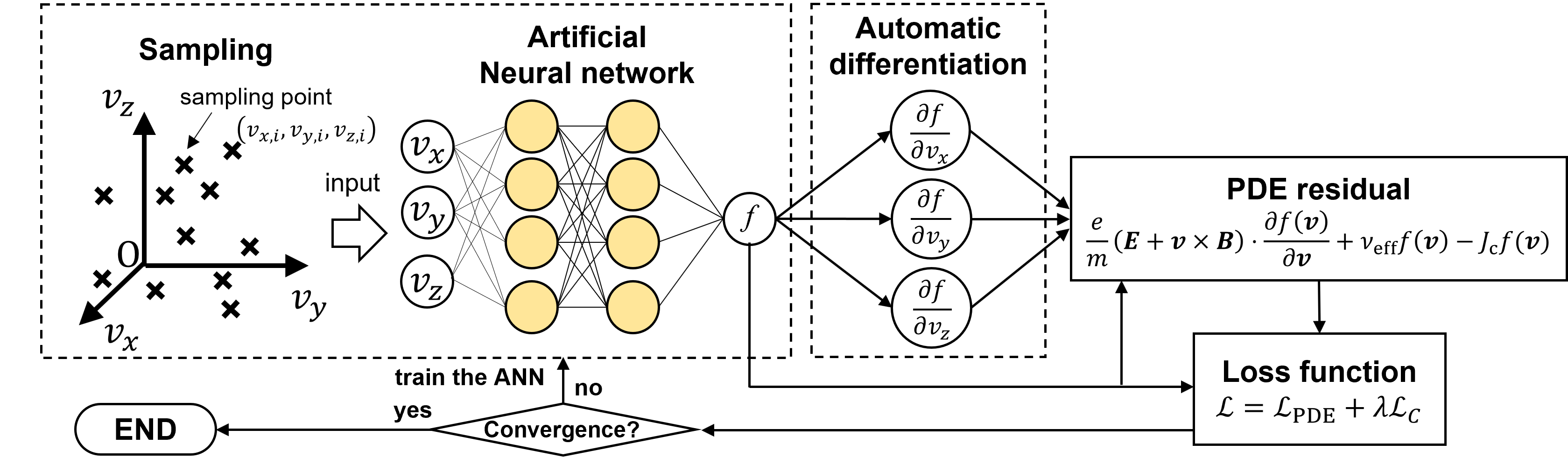}
\caption{Schematic diagram of the PINN approach for solving the PDE.}
\label{fig:diagram_PINN}
\end{center}
\end{figure*}

\begin{figure*}[ht]
\begin{center}
\includegraphics[width=0.8\linewidth]{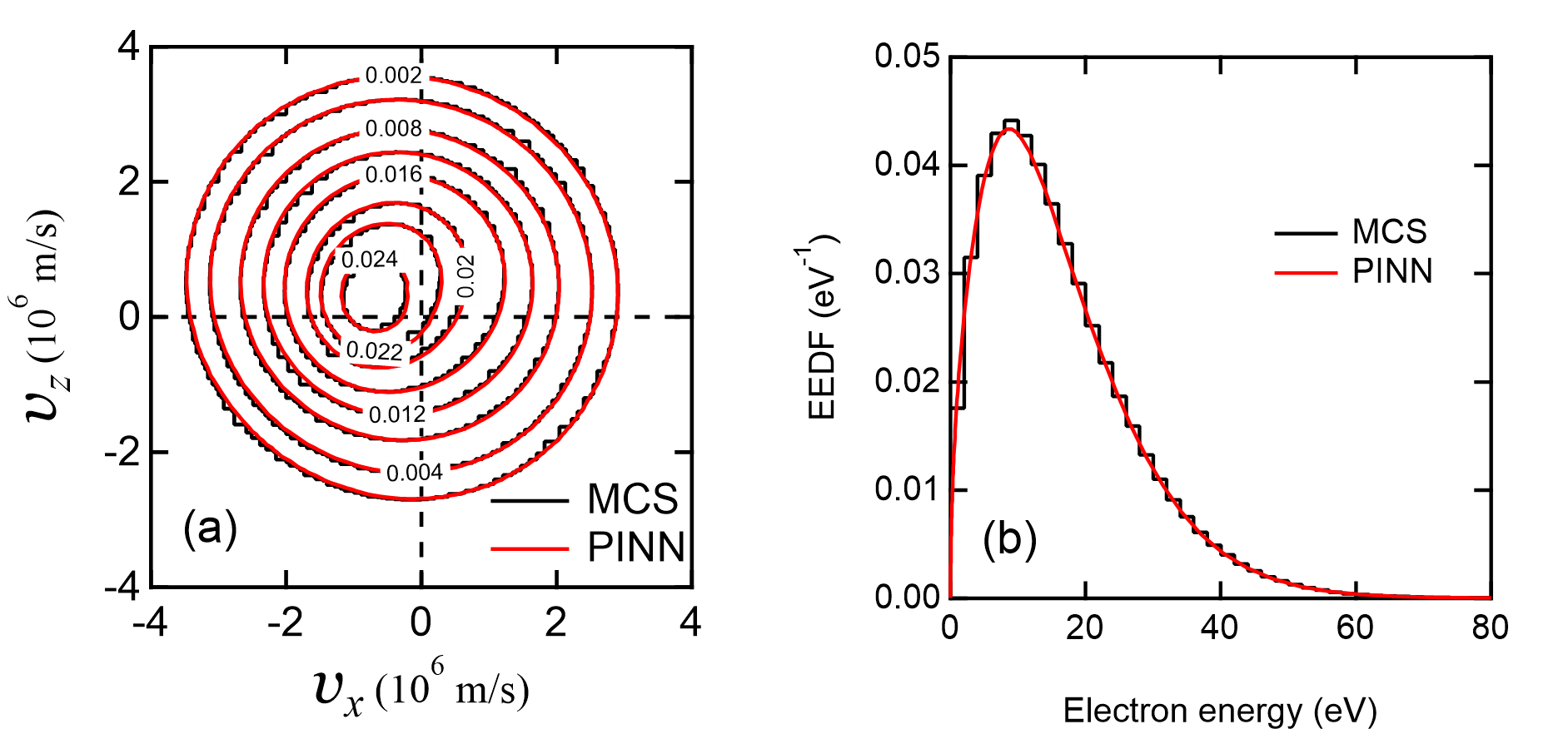}
\caption{(a) Contour plot of the EVDF projected into a $v_x-v_z$ plane. (b) Electron energy distribution function as a function of the electron energy. The strength of the reduced electric field $E/N$ and that of the reduced magnetic flux density $B/N$ are set to 2000 Td (1 Td = $10^{-21}\ \mathrm{Vm^2}$) and 2000 Hx (1 Hx = $10^{-27}\  \mathrm{Tm^3}$), respectively. Here, $N = 3.535\times10^{22} \mathrm{m^{-3}}$ denotes the number density of gas molecules.}
\label{fig:EVDF_EEDF}
\end{center}
\end{figure*}

The equilibrium EVDF $f(\bm{v})$ under dc uniform electric and magnetic fields is governed by
\begin{equation}
    \label{eqn:BEq}
    \frac{e}{m}(\bm{E} + \bm{v}\times\bm{B})\cdot
    \frac{\partial f(\bm{v})}{\partial\bm{v}} 
    + \nu_{eff}f(\bm{v}) - J_{c}f(\bm{v}) = 0,
\end{equation}
where $\bm{E} = (0, 0, -E)$ is the electric field, $\bm{B} = (0, -B, 0)$ is the magnetic field, $e$ is the electron charge, $m$ is the electron mass, $\bm{v} = (v_x, v_y, v_z)$ is the electron velocity, $\nu_{eff}$ is the effective ionization collision frequency, and $J_{c}f(\bm{v})$ is a collision term. Here, $\mathrm{SF_6}$ is chosen as ambient gas, and collisions between an electron and a gas molecule for elastic, excitation, electron attachment, and ionization are considered in the collision term. Figure \ref{fig:diagram_PINN} shows the schematic diagram for solving the Eq. \ref{eqn:BEq} using the PINN approach. The latent solution of Eq. \ref{eqn:BEq} is represented by ANN. How well the ANN respects the PDE and boundary conditions is measured by a loss function
\begin{equation}
    \mathcal{L} = \mathcal{L}_{PDE} + \lambda \mathcal{L}_{C},
\end{equation}
where $\mathcal{L}_{PDE}$ and $\mathcal{L}_{C}$ represent the residual of the PDE and boundary conditions, respectively, and $\lambda$ is the parameter controlling $\mathcal{L}_{C}$. The term $\mathcal{L}_{PDE}$ is given by
\begin{equation}
    \mathcal{L}_{PDE} = \frac{1}{N_{PDE}}\sum_{i=1}^{N_{PDE}}|R(\bm{v}_{i})|^2,
\end{equation}
where $R$ is the residual of the PDE, namely the left-hand side of Eq. \ref{eqn:BEq} and $\bm{v}_{i} = (v_{x,i}, v_{y,i}, v_{z,i})$ denotes a point sampled on the domain of the solution. The partial derivatives of $f(\bm{v})$ with respect to $v_x$, $v_y$, and $v_z$ are calculated by using automatic differentiation. If the Dirichlet boundary condition were applied, $\mathcal{L}_{C}$ could be described by
\begin{equation}
    \mathcal{L}_{C} = \frac{1}{N_{C}}\sum_{j=1}^{N_{C}}|f(\bm{v}_{j})-\hat{f}(\bm{v}_{j})|^2,
\end{equation}
where $\bm{v}_j$ is sampled on the boundary of the domain and $\hat{f}(\bm{v}_{j})$ is a given value at $\bm{v}_j$. In the present calculation, the normalization constraint $\int_{-\infty}^{\infty}f(\bm{v})d\bm{v} = 1$ is applied, and the term $\lambda\mathcal{L}_{C}$ is truncated. Instead, the collision term on $\mathcal{L}_{PDE}$ is calculated by using normalized EVDF. The ANN has weight and bias parameters, and they are optimized to minimize the value of $\mathcal{L}$ by gradient descent based method, such as Adam \cite{kingma2017adam}, until the value of $\mathcal{L}$ reaches a minimum. There is flexibility in how to sample points. We can simply sample points by using presudorandom numbers. The Latin hypercube sampling and quasi-random numbers are used to sample points uniformly. Adaptive sampling method in which the distribution of the sampling points is improved by considering that of $\mathcal{L}$ is proposed \cite{Lu2021}. Scaling the ANN input is important, and they should be distributed on $[-1, 1]$. The appropriate architecture of the ANN would vary with the problem to be solved and is tuned empirically by users at present. Designing an effective ANN architecture for solving the PDE accurately has been investigated \cite{Wang2021}. Figure \ref{fig:EVDF_EEDF} shows the EVDF projected into a $v_x-v_z$ plane and the electron energy distribution function (EEDF) calculated from the EVDF. The EVDF and EEDF calculated from the Monte Carlo simulation (MCS) are also shown as reference data. The PINN can successfully reproduce the MCS results. In this calculation, the EVDF is represented by feedforward ANN having 41700 parameters. The EVDF in the same condition was also calculated using the mesh-based method \cite{Sugawara2019} and was stored in a 3D array the size of which is $10000\times45\times750$. Given that the precision of floating points employed in the calculations is the same, the PINN allows us to represent the 3D EVDF properly with approximately 0.01\% of the memory capacity required in the mesh-based method.

A physics-informed neural network (PINN) provides a novel mesh-free approach to solve the partial differential equations, allowing us to deal with high-dimensional problems. For the electron Boltzmann equation, it is confirmed that the PINN approach can significantly reduce the memory capacity required for representing the EVDF properly compared to the mesh-based method. The PINN approach has been applied to various problems regarding fluid dynamics, heat transfer, electromagnetics, and so forth. Combining PINNs for various scientific disciplines would enable us to represent multiphysics systems and would contribute to advances in plasma simulation. In this case, constituent neural networks would be trained not so much to minimize their loss functions as to minimize the loss function for the system, for example, the sum of the loss functions for each neural network.

\noindent
[Satoru Kawaguchi]

\subsection{Reduction of chemical reaction models}
\label{Sec7_vanDijk}
\paragraph*{\bf Introduction}
The number of species that can be formed in plasmas can be considerable.
Dozens of electronically excited states may need to be considered to correctly
predict the rates of ionization, recombination and radiative processes, even when
the plasma is created in an atomic gas such as argon \cite{Bogaerts1998} or mercury
vapor \cite{Dij2001/1}.
When the plasma is created in a mixture of molecular gases, the complexity further
increases, especially when a rise of the gas temperature results in the onset of
a multitude of non-electronic reactions. Among the many contemporary technologically
relevant examples are plasmas in methane (36 species, 367 reactions) \cite{deBie-Verheyde-2011}, 
air (84 species, 1880 reactions) \cite{14_Gaens_2013}
and 
in carbon-dioxide (72 species, 5732 reactions) \cite{Koelman-Heijkers-2017}.
Incorporating such chemistries in full into a space- and time-resolved computer
simulation may be tempting, but is at present hardly feasible.
Therefore, an analysis and, when possible, a reduction of such
plasma-chemistries is called for, and that task has been accomplished
even for rather complicated chemistries, see for example
Refs. \cite{deBie-Verheyde-2011, Bogearts-deBie-2017}.
And although computers have gotten exponentially faster for the past decades,
Gustafson's law suggests that the problems we try to solve with them
continuously get bigger as well \cite{Gustafson-1988}. Therefore the need
for more systematic and automated methods grows and it is no surprise 
that {\em plasma chemistry reduction} continues to be a subject of
great interest.

Like any modeling effort, an attempt to achieve a chemistry reduction should
start with a precise statement of the scope of the model and the observables
that the model aims to reproduce. If these observables are not influenced by
a particular minority species, that species may be removed from the species list.
But in another study, that minority species may be among the key observables,
for example because, in spite of its small abundance, it is responsible for
degradation of the plasma device.
Also, the relevant time scales must be
part of a model specification. A plasma reactor model may target the plasma
behavior on a timescale of milliseconds and in such case it may be desirable
to eliminate the nanosecond timescales from the model. But in a model of
a Laser Induced Fluorescence experiment, these smallest time scales are
the relevant ones, and the long-term dynamics of the plasma can be disregarded
\cite{VanDerHeijden2000}.

This section discusses a number of methods that have been considered for
plasma-chemical reduction in the past. Furthermore, recent works that are related
to the subject will be summarized. Special attention will be paid to the
suitability of methods that originate from adjacent fields of science, such
as combustion engineering for plasma-chemical reduction.

%Main paragraph 400 words, up to 800.
%
%{\color {red} The main body paragraphs should be about 400 word
%long (if necessary up to 800 word long.)  In the main body, please
%present the definition of the problems, a simple outline of the
%methods to solve them (with proper references on technical issues
%of the methods), and remaining challenges.}

\paragraph*{\bf Timescale-based Reduction Schemes}
The chemical composition of a plasma can be characterized by the particle
densities $n_i$ of the components $i$. The temporal and spatial variations of
these components can be calculated from a set of balance equations that are
given by
\begin{equation}
	\frac{\partial{n_i}}{\partial{t}} + \nabla\cdot\vec{\Gamma}_i = S_i,
\end{equation}
where $\vec{\Gamma}_i$ and $S_i$ are the particle flux density and the volumetric production
rate of particles of type $i$. Depending on the transport coefficients, the electric field and
on the
reaction scheme that underlies the sources and sinks that end up in $S_i$, the density $n_i$
may be affected by transport, or may follow from {\em chemical equilibrium}, which is to say
that $S_i\approx 0$.

The {\em Quasi Steady State Approximation} (QSSA), which amounts to setting $S_i=0$
for (near-)equilibrium species, has been around since the early 1900s
\cite{Bodenstein-1913,Tomlin-Turanyi-1997}. In the 1960s, Bates, Kingston and McWhirter
\cite{Bat1962} used the QSSA for the excited states in atomic plasma.
If the source terms for these states are only due to radiative and electron-impact processes,
these are {\em linear} in the densities of those species and the authors demonstrated
that this allows the elimination of the excited state densities from the system of transport
equations, in combination with a correction of the rate coefficients for ionization and
recombination for the remaining atom and ion ground state. These corrections account
for indirect or {\em ladder-like} processes.
The result is an important tool for chemistry reduction, since the number of atomic states that
is considered in the transport model is reduced from dozens to only two, without sacrificing
the physical validity of the model.

A generalization and a more explicit algebraic perspective on this procedure
were provided in \cite{Dij2001/1}. When we bundle the sources and densities of the atomic and
ion states in column vectors
$\bvec{S} = [ \cdots S_i \cdots ]^T$ and $\bvec{n} = [ \cdots n_i \cdots ]^T$, one can
write $\bvec{S}=\bmat{M}\bvec{n}$, where the matrix $\bmat{M}$ depends on the electron
temperature (through the rate coefficients), the electron density and on the opacities
of the plasma for resonant radiation.
When the `non-local` densities are placed at the top of these vectors, the reduced system
can be partitioned in transport-sensitive $(t)$ and local ($l$) blocks,
\begin{equation}
  \left[\begin{matrix}
    \bvec{S}_t \\
    \bvec{0}
  \end{matrix}\right]
	=
  \left[\begin{matrix}
    \bmat{M}_{tt} & \bmat{M}_{lt} \\
    \bmat{M}_{tl} & \bmat{M}_{ll}
  \end{matrix}\right]
  \left[\begin{matrix}
    \bvec{n}_t \\
    \bvec{n}_l
  \end{matrix}\right].
\end{equation}
Solving the second block of equations for $\bvec{n}_l$ and substituting the result in the
first block yields
\begin{equation}
	\bvec{n}_l = -\bmat{M}_{ll}^{-1}\bmat{M}_{tl}\bvec{n}_t,
	\quad\quad
	\bvec{S}_t = (\bmat{M}_{tt} - \bmat{M}_{lt}\bmat{M}_{ll}^{-1}\bmat{M}_{tl})\bvec{n}_t.
\end{equation}
The first equation expresses the densities of the local states in terms of those that
are affected by transport. The second equation expresses the sources of the transport-sensitive
levels in terms of their densities. The {\em effective} coefficient matrix contains the
direct processes ($\bmat{M}_{tt}$) and a correction for the indirect or ladder-like processes
that involve the states that no longer need to be modeled explicitly.

This elaboration demonstrates the technique that underlies many chemical reduction
schemes. It shows that the {\em locality} of species
densities can be used to replace differential equations with algebraic
ones. It also shows that these species may still influence the kinetics of the remaining
species via indirect processes.

A drawback of the QSSA method is that accurate error
estimates can only be obtained by running the solution both with and without
QSSA, and comparing the results \cite{Tomlin-Turanyi-1997,Maas-Pope-1992}. 
A detailed overview of more recent methods for analysis of chemistries that do sot
suffer from this problem can be found in \cite{Tomlin-Turanyi-1997,Maas-2020}. For
the reduction of chemistries based on timescales, a few classes of techniques are
available, many of which find a root in combustion engineering.
One of the earliest numerical approaches is the Computational Singular
Perturbation (CSP), first described in 1985 by Lam
\cite{Lam-1985,Lam-Goussis-1988,Lam-Goussis-1994,Lam-2013}. The goal of this family of
methods is to automate the process of simplifying systems of differential
equations like the ones encountered in chemical reaction systems, a task that up
till then was executed manually. Variations include Linear CSP (LCSP), Non-linear CSP 
(NCSP), and CSP without eigenvalue decomposition \cite{Zhao-Lam-2019}.

\begin{figure*}
  \includegraphics[width=0.6\linewidth]{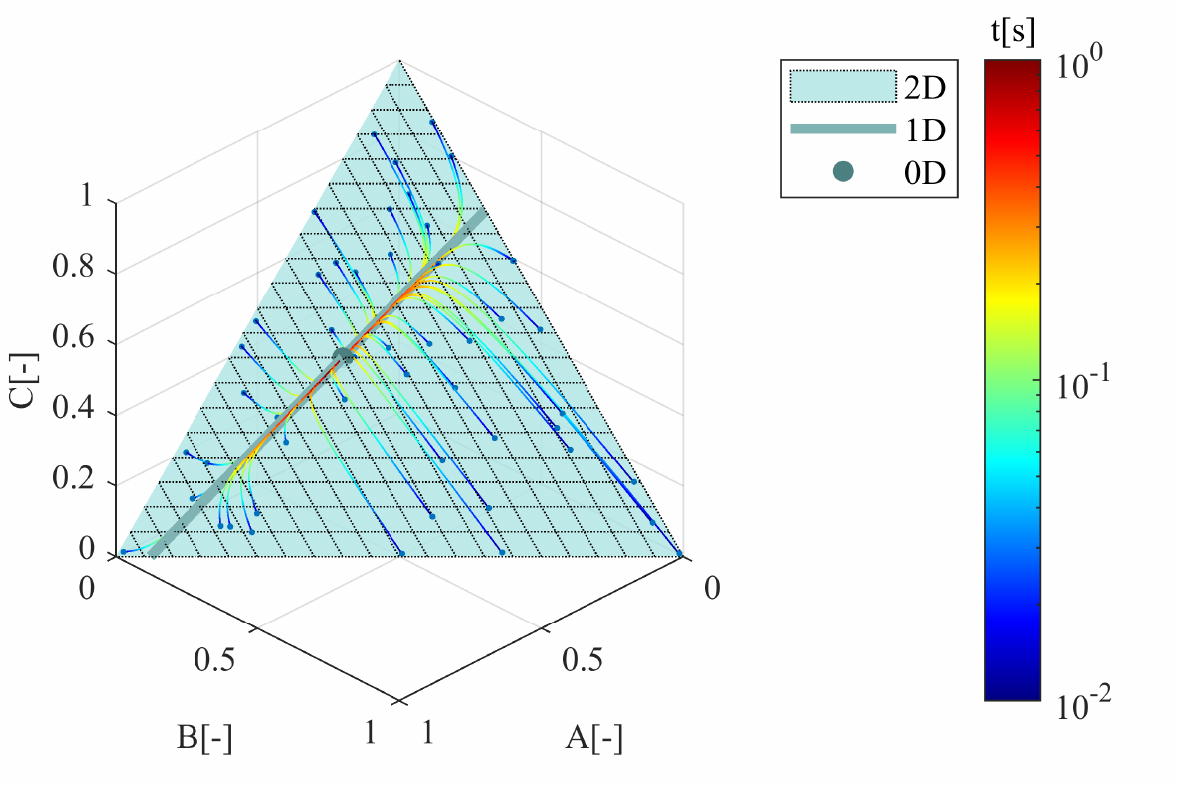}
  \caption{0D, 1D and 2D manifolds for the chemistry described in
    \cite{Maas-2020}. The time evolution of random initial compositions on the
    2D manifold is shown, showing that the compositions first converge onto the
    1D manifold, before settling onto the 0D manifold. Image recreated from
    \cite{Maas-2020}.}
  \label{fig:maas_manifold}
\end{figure*}

In 1992 the Intrinsic Low-Dimensional Manifold (ILDM) family of methods was
pioneered by Maas \& Pope \cite{Maas-Pope-1992,Maas-Pope-1992-2}. This family of
methods recognizes that the time scales involved in the chemical reactions in a
mixture often span multiple orders of magnitude. The fastest equilibration
processes attract the systems towards a low-dimensional subspace in phase-space,
the so-called low-dimensional manifold. This effect is demonstrated in figure
\ref{fig:maas_manifold} for the imaginary chemistry from \cite{Maas-2020},
consisting of species A, B and C. The reaction space of this chemistry is
confined to a two-dimensional manifold, described by $\mathrm{A} + \mathrm{B}
+ \mathrm{C} = 1$. It can be observed that any random initial composition on
this surface quickly converges onto a one-dimensional manifold, before
eventually settling at the equilibrium composition, a zero-dimensional manifold.
Various methods of finding such manifolds for arbitrarily complex chemistries
exist, including Flamelet Generated Manifolds (FGM)
\cite{vanOijen-Donini-2016}, Trajectory Generated Manifolds (TGM)
\cite{Pope-Maas-1993} and ILDM assisted by In-Situ Adaptive Tabulation (ISAT)
\cite{Pope-1997,Ding-Readshaw-2021}. 

Applications of CSP or ILDM to plasma chemistry are still scarce, an example can be
found in \cite{Rehman-Kemaneci-2016}. The reason may be a lack of awareness within
the community of such reduction methods, or the fact that it simply takes more time
for techniques to transfer to a different field of science. Another reason is
that in plasmas more parameters come into play (electron energy, opacities),
and that often their gradients are not co-aligned, frustrating methods that rely
on quasi-one-dimensional behaviour such as FGM \cite{vanOijen-Donini-2016}.

\paragraph*{\bf Recent Developments, Outlook}
Various innovative strategies have been proposed in the past five years. As an
example, Principal Component Analysis (PCA) has been applied to plasmas for
the first time \cite{Peerenboom-Parente-2015,Bellemans-Magin-2017}. Also the
method of Pathway Analysis (PWA) \cite{Lehman-2004} has seen renewed interest
\cite{Lehman-2004,Markosyan-Luque-2014}, and has been applied applied to
large plasma chemistries, see for example Refs.
\cite{Koelman-Yordanova-2019,Kruszelnicki-Lietz-2019}.
More recently, graph theory and machine learning are being used to extract
information from complex chemistries
\cite{Murakami-Sakai-2020,Hanicinec-Mohr-2021}. While an ultimate solution to
the problem of plasma-chemical reduction is not yet in sight, these
developments bear great promise for the future of the field.

%{\color{red} The final paragraph should give concluding remarks;
%should be about 100 words (if necessary up to 250 words). In
%this short conclusion and summary, please emphasize the future
%direction of this topic. In addition, before submitting your
%manuscript, please check for typos by e.g., free software such as
%https://app.grammarly.com/}

\paragraph*{Acknowledgements} %This activity 
The work by J. van Dijk and R. H. S. Bud{\'e} is co-funded by PPS-contribution Research and Innovation of the Ministry of Economic Affairs and Climate Policy (The Netherlands), and ASML.

\noindent
[Jan van Dijk and Rick H. S. Bud{\'e}]

\subsection{Biological data and plasma medicine}
\label{Sec7_Wende}
%\paragraph*{Data driven plasma science – from the biomedical point of view}

In 2003, when Stoffels and colleagues first reported on the non-lethal manipulation of mammalian cells by a non-equilibrium plasma (“plasma needle”), a new chapter of plasma physics began \cite{Stoffels2003}. 
Besides the widely accepted technical application of plasma processes, and the inactivation of prokaryotic bacteria reported since the mid 1990ies \cite{Laroussi1996}, the report highlighted a new facet of plasma and sparked a surge of research projects all around the globe. 
For the last almost 20 years, a number of breakthroughs have been made and non-equilibrium atmospheric pressure plasmas – which are, for the sake of biomedical and clinical researchers, often simply called “cold plasmas” or “gas plasmas”, have found their way into the clinics and ambulant care with a number of certified medical devices in the market. 
The number of publications on plasma medicine rose from less than five in 2003/2004 to more than 800 per year (2020, Google Scholar).
In the beginning, the new interdisciplinary studies were published in journals with a traditional engineering or physical scope. 
While these journals still publish data on biomedical plasma research, journals with a broader scope and readership beyond the plasma research community are increasingly targeted. 
Among these, numerous medical or interdisciplinary journals dominate. 
With the increasing impact of the research on foreign communities, clinicians, funding agencies, and the public, increasing awareness of the validity, interchangeability, and reproducibility of results can be felt in the community. 
Adherence to the FAIR data use policies (see also Sec. VIII E)\cite{Wilkinson2016},   international approaches to define a universal plasma dose, or actions on standardization, are representative for this “coming-of-age” time of the research field. 
Naturally, this affects all aspects of the topic, but the larger variance of biomedical experiments and the resulting data, and medical safety aspects accelerated the correspondent efforts. 
When surveying current publications on biomedical aspects of cold plasma, the use of bioinformatics tools has become the normal case \cite{Clemen2021GasPT,Nasri2021,Wenske2021}. 
Currently, when proteomics (proteins) or lipidomics (lipids) data are a central piece of the paper, most journals desire the upload of these data into public repositories to ensure their long-term persistence and preservation. 
A number of dedicated databases have evolved, e.g. the members of the ProteomXchange consortium \url{http://www.proteomexchange.org/} such as MassIVE, PeptideAtlas or PRIDE for proteomics data, Metabolomics Workbench \url{https://www.metabolomicsworkbench.org/} for small molecules including lipids, or the Genome Sequence Archive \url{https://ngdc.cncb.ac.cn/gsa/} for genomic information. 
The major benefit for any research community is the long-term conservation of the data independent of individual working groups, the possibility to share the data with colleagues to allow additional data analysis approaches and the increase of reliability and reproducibility as defined by the FAIR Guiding Principles for scientific data management and stewardship that were introduced in 2016 \url{https://www.go-fair.org/fair-principles/}\cite{Wilkinson2016}. 
In the plasma science community with a special focus on plasma medicine, a dedicated repository INPTDAT has been established \url{https://www.inptdat.de/}, adhering to the FAIR principles, as discussed in Sec. VIII E. 

\begin{figure*}[ht]
\begin{center}
\includegraphics[width=0.95\linewidth]{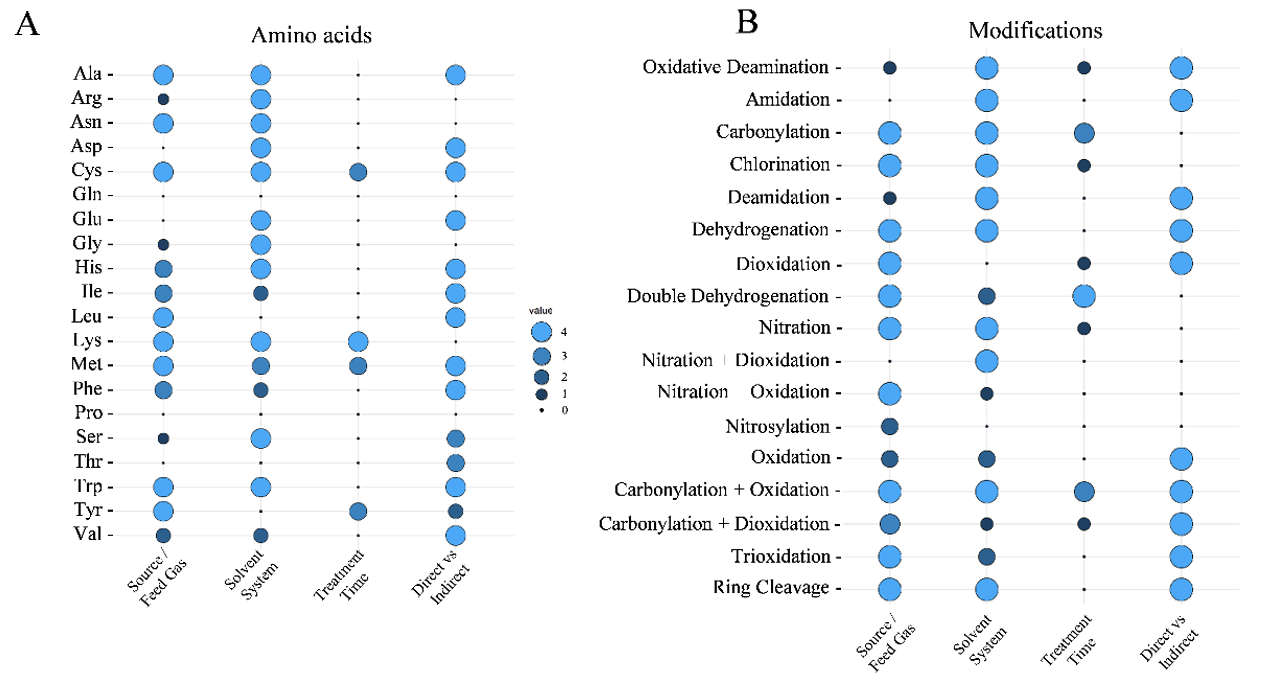}
\caption{Impact of plasma source/gas phase composition, solvent treatment time, and treatment mode on the extent on amino acid modification (A) or type of modification introduced by a plasma treatment (B). A large circle indicates a strong correlation. For example: the modification oxidative deamination (replacement of nitrogen by oxygen) is influenced by the solvent type and the treatment mode (direct), but to a minor extent only to the plasma source or the treatment time (B). Reprinted from Wenske et al., J. Appl. Phys, 129 (2021). Copyright 2021 Author(s), licensed under a Creative Commons Attribution (CC BY) License \cite{Wenske2021}.}
\label{biofig1}
\end{center}
\end{figure*}

To understand the impact of cold physical plasma in biological systems, K. Wende et al. have deployed methods like high-content imaging, flow cytometry, transcriptomics, and proteomics in a number of {\em in vitro} and {\em in vivo} models, e.g. \cite{Clemen2021GasPT,Schmidt2015NonthermalPA,Bekeschus2017,Bekeschus2018,ZOCHER2019101416,Schmidt2017}. 
To analyse the significant amount of raw data, softwares like GeneSpring (Agilent), Kaluza (BeckmanCoulter), Tibco Spotfire (Tibco Software), Byonic (ProteinMetrics), or Proteome Discoverer (Thermo) are used. 
The bottom line of all studies presented here is the major role that reactive oxygen and nitrogen species occupy to trigger the observed events. 
Since cellular signalling of both pro- and eukaryotic cells uses the same reactive species, there is a “common language” between the gas-phase phenomena of plasmas and biological systems. 
However, due to long distances between the generation and the assumed place of action, a direct contribution by short-lived species such as singlet and atomic oxygen, or peroxynitrite is questionable. 
For this reason, we pursue the hypothesis that the short-lived species chemically modify biomolecules in close vicinity to the point of impact. 
Subsequently, either the chemical energy of the reactive species is preserved – e.g. as a radical or peroxide, or the modified molecule is perceived as a signal molecule, or its functionality is changed significantly. 
In the first steps of validity testing, it was observed, that a MHz-driven dielectric barrier argon jet (kINPen, neoplas Germany) has a significant impact on cysteine and tyrosine. 
The reaction products reflected the gas phase composition and the reactive species formed, permitting its use to compare plasma sources and conditions and to infer on plasma liquid chemistry and gas-liquid interphase chemistry \cite{Wende2020,Bruno2020,Bruno2019}. 
The concept was extended using artificial peptides, providing a more complex chemical environment and a greater variety of chemical structures to be attacked by the plasma-generated species \cite{Wenske2021,WenskeSeb2020}. 
Again, this approach involved high-resolution mass spectrometry and the use of an advanced software solution to filter the raw data for relevant information on oxidative post-translational protein modifications (oxPTMs, Byonic, ProteinMetrics, Palo Alto, USA ). 
As a result, the introduction of 17 different oxPTMs was determined along with four main targets: cysteine, methionine, tryptophane, and tyrosine. 
For example,  in the two decapeptides Ala-Asp-Gln-Gly-His-Leu-Lys-Ser-Trp-Tyr and Ala-Cys-Glu-Gly-Lyl-Ile-leu-Lys-Tyr-Val the modification nitration (+44.98 m/z, +N + 2O –H) is introduced in dependence on gas-phase composition (Ar $>>$ Ar/O\textsubscript{2}), and plasma source (kINPen $>>$ COST jet), and solvent system (H\textsubscript{2}O $>>$ PBS). 
Since an aromatic structure and an acidic pH promote nitration, it is most prominent in tyrosine and water as a solvent. 
In figure \ref{biofig1}, the role of the investigated conditions on the extent of amino acid modifications (Fig. \ref{biofig1}A) or on the type of observed modification (Fig. \ref{biofig1}B) is visualized after statistical analysis by the Software package R. 
The data allow insight on the likelihood that a certain amino acid is modified by a plasma treatment when a specific condition is met and how a certain modification can be triggered by the choice of condition (model) or can be expected in an {\em in vivo} setting.

A prominent example is the occurrence of dioxidations (+31.98 m/z, + 2O) that is strictly linked to a direct plasma treatment plus suitable gas phase composition (oxygen admix), setting the stage for singlet oxygen as the underlying reactive species. 
For further analysis and details see Wenske et al. 2021 \cite{Wenske2021}. 

\begin{figure*}[bht]
\begin{center}
\includegraphics[width=0.8\linewidth]{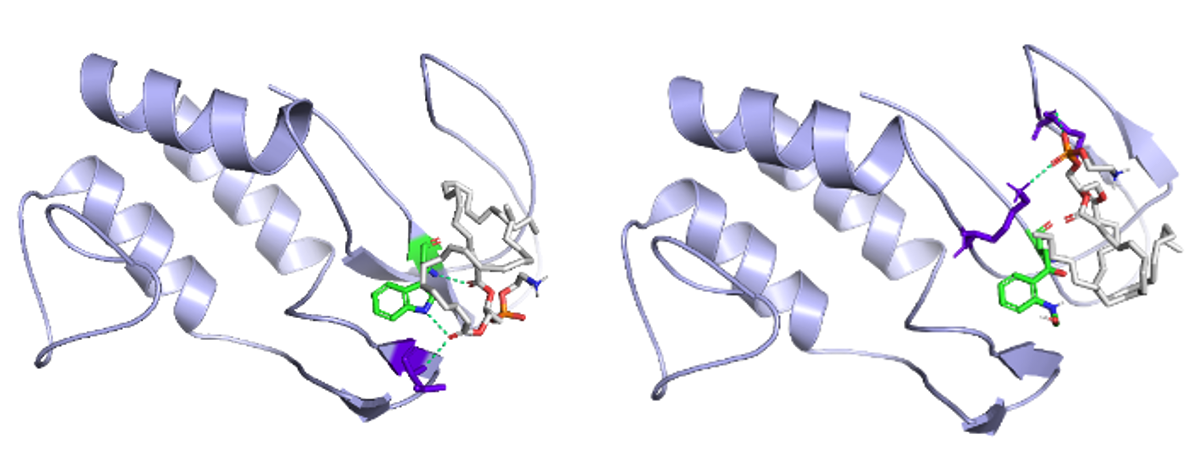}
\caption{Impact of argon plasma jet (kINPen) on phospholipase A2 secondary structure. Control (left) and after direct plasma treatment (right). The residue tryptophan 128 is dioxidized, yielding a structural change and inhibiting enzyme function. Reprinted from Nasri et al., European Chemistry Journal, 27 (2021). Copyright 2021 Author(s), licensed under a Creative Commons Attribution (CC BY) License \cite{Nasri2021}.}
\label{biofig2}
\end{center}
\end{figure*}

%To prove the impact of plasma-driven oxPTMs on protein function, the enzyme phospholipase A\textsubscript{2} is a relevant player in inflammatory processes via supply of unsaturated fatty acids for enzymatic oxidation yielding signalling molecules. 
The impact of plasma-driven oxPTMs on protein function was shown for a number of proteins. One example is the enzyme phospholipase A2 which is a relevant player in inflammatory processes by supplying unsaturated fatty acids as precursors for signaling molecules. 
A necessary step in the cleavage of membrane lipids (phosphatidylcholins) is the docking of the proteins C-terminus to the membranes polar head groups. 
A plasma treatment by the kINPen disrupts the docking and subsequently enzyme activity, strongly suggesting that the biomedical application of cold plasma may utilize the (in-) activation of proteins to achieve effectivity. 
Via high-resolution mass spectrometry/bioinformatics and molecular dynamics simulation (GROMACS[56] program package (version 5.0) OPLS-AA/L all-atom force field), the amino acid residue tryptophan 128 was identified to be the target of plasma-derived singlet oxygen dioxidation, yielding a ring-open kynurenine derivative that subsequently distorted the secondary structure of the C-terminal $\beta$-sheets of PLA\textsubscript{2} (Fig. \ref{biofig2}) \cite{Nasri2021}. 
In a similar manner, it was shown by Clemen et al. in an animal model, that protein oxidation triggers a more strict response of the immune system, opening the avenue to plasma-driven cancer vaccination \cite{Clemen2021GasPT}.

In conclusion, the hypothesis that plasma-derived reactive species modify biomolecules that subsequently modulate physiological processes has to be accepted: oxPTMs are introduced not only in model peptides but also in also full proteins, changing their perception and role. 

\paragraph*{Acknowledgements}
The work of Kristian Wende is funded by the German Ministry of Education and Research (BMBF), grant. numbers 03Z22DN12 and 03Z22D511.

\noindent
[Kristian Wende]

\subsection{Challenges and outlook}
Plasma processing involves complex physical and chemical systems in nonthermal equilibrium conditions. In addition, spatial and time scales involved in those systems vary widely from the atomic scales to the manufacturing tool scales. For example, in a typical plasma processing tool, macroscopic parameters such as gas compositions, gas pressure, and applied power to the plasma source are used as control nobs to form nano-meter scale complex device structures on a wafer surface. The conventional first-principles-based approaches to analyzing plasma processing systems, i.e., numerical solutions to the fundamental physics equations describing the systems, are in general not free from input parameters; they typically require fundamental data such as reaction rates in the gas phase and on surfaces.  Furthermore, such approaches are, even if available, typically time-consuming and often accumulate errors arising from inaccurate input parameters in their analyses. Therefore, although such analyses are undoubtedly important for a better understanding of the nature of plasma processing, more quantitatively reliable and timely analyses are also required for practical applications such as plasma system control and new process development. 

Data-driven approaches may offer solutions to such requirements.  For example, a large amount of numerical simulation data and/or measurement data of experimental/manufacturing systems may be used to create machine-learned regression models or surrogate models to predict system characteristics such as etch rates, sputtering yields, and interatomic forces, as discussed in Secs.~\ref{Sec7_Park}, \ref{Sec7_Triesschmann}, and \ref{Sec7_Chen}. Reduction of the dimensions of extremely large data sets to make the data more tractable by computation is also another challenge, as discussed in Sec.~\ref{Sec7_vanDijk} for chemical reactions in plasmas.

Although a large amount of data may be obtained from individual plasma processing tools and their processed material surfaces, what remains a challenge in plasma technologies is the shortage (or sometimes lack) of fundamental data on elementary processes that can be applied to any processing tools, such as chemical reaction rates of specific surface materials with specific incident gaseous species that characterize the plasma surface interaction. Of course, it is unrealistic to expect to obtain such data for all possible combinations of surfaces and gaseous species exhaustively. However, it is desirable to establish new techniques for high-throughput screening to obtain fundamental chemical reaction data associated with desired plasma processing efficiently. In general physics of plasmas is better understood than their chemistry, so such chemical data combined with the conventional first-principles-based approaches as well as the latest data-driven approaches would allow far more accurate analyses of plasma processing and drive faster and more cost-effective development of new processes and better plasma control techniques.

\noindent
[Satoshi Hamaguchi]

\section{Plasma and Related Database}
\subsection{Introduction}
In the study of any of the different plasmas discussed in this review a common challenge is to obtain a thorough understanding of the physical and chemical properties of plasmas. In order to determine such properties, it is essential to assemble authoritative databases that allow the design, diagnostics and monitoring of the plasma. The plasma community has been active in assembling such databases which include:

\begin{itemize}
    \item Atomic and Molecular databases detailing both spectroscopic data (commonly used as plasma diagnostics to identify key plasma species) and collisional data characterizing electron, ion and photon interactions with those atomic and molecular species within the plasma and knowledge of both the cross sections and reaction rates for such collisions, both in the gas phase and on the surfaces of the plasma reactor;
    \item Material databases which provide data on the properties used in the design and operation of plasma systems with databases for fusion reactors being amongst the most extensive;  
    \item Plasma Chemistry databases which provide access to complete and validated data for plasma modelling with pre-assembled and validated chemistry sets;
    \item Low temperature plasma databases which have been amongst the most common databases since these have been constructed to support specific industrial plasmas such as those used in the semiconductor, lighting and medical industries.
\end{itemize}

However, the compilation of such databases remains a major challenge and the necessary coordinated infrastructure and funding to build and sustain them has often been lacking. This in turn challenges the broader scientific community to recognize that their fields also rely upon the compilation and access to relevant databases and that a united research community must then confront the funders of research (government and industrial), specifying that scientific and technological progress is based upon a strong fundamental bedrock and that if this is neglected then the scientific and technological advances they require will not occur and their investment will not be rewarded.

This section reviews the current status of the different databases and gives indications as to present data deficits. Core to all databases are the criteria for data selection: whether the database then recommends data sets or leaves the user to select data is an important parameter. In particular, recommended data sets allow individual models to be cross correlated. Methods and community practice in establishing recommended datasets will also be presented.

\noindent
[Nigel J. Mason]

\subsection{Atomic and molecular database}
Atomic and molecular (AM) processes are elementary processes in plasmas and 
important to understand microscopic behavior of plasmas and radiative processes
in plasmas. Radiative and collisional processes of atoms and molecules govern 
the energy balance of plasmas. It is also useful to use emissions from atoms and
molecules for spectroscopic diagnostics, for example, 
to know impurity behavior in fusion plasmas and 
plasma properties such as electron temperature and density.
AM data such as wavelengths and transition probabilities of 
emission lines or collision cross sections are important fundamental data 
to describe atomic and molecular processes. AM databases
compile and store such important data since the 1970s
and provide them for users in various research fields \cite{murakami2012}.
In recent years many databases are available through the internet,
and there have been some attempts to provide such data more conveniently 
for users. 
As a new attempt, databases are used to train machine learning methods, for example, 
to estimate a set of electron-impact cross sections from swarm transport data
\cite{stokes2020}.

There are two kinds of AM databases available; one has evaluated data and 
the other has original data. 

The former databases contain evaluated one value (or one data set) 
for one process, eg., one wavelength for one specific transition and one
set of ionization cross-sections as a function of collision energy 
for a specific atom. 
Data evaluation is done by organizers of the database in various ways.
Accuracy of data is carefully examined experimentally, by checking the method
of the study, or by comparing with other data, and one value or one data set is
selected and stored in the database.
NIST Atomic Spectra Database \cite{NIST_ASD} is this type of database
for atomic wavelengths, transition probabilities, energy levels, and ionization 
potentials.
The atomic database in CHIANTI \cite{CHIANTI} for spectroscopic diagnostics 
for solar physics is also this type. 
IAEA ALLADIN database contains evaluated data of cross-sections and rate
coefficients for electron collisions,
photon collisions and heavy particle collisions \cite{IAEA_Alladin}, but several data sets
evaluated by different research groups are stored for one process.

The second type of database contains many data for one process obtained 
by various theoretical or experimental studies. 
All data or data set have their references on their origins and 
users can track the data source. 
Users can compare several data sets for one process such as
ionization cross sections for a specific atom, and can evaluate and select
data by themselves.
NIFS atomic and molecular numerical database is this type of database for
collision cross-sections and rate coefficients for ionization, excitation, 
recombination and charge exchange processes of atoms and molecules \cite{NIFS_DB}.
Users can compare experimental and theoretical data for one process 
with a graphic output of the database.
Open ADAS \cite{OpenADAS} is also this type of database for data set relevant
for spectroscopic diagnostics of fuson and solar plasmas. Various theoretical
data sets are stored for fundamental data such as a set of energy levels and
electron impact excitation effective collision strengths. 
Derived data calculated with ADAS software package are also available,
such as photo emissivity coefficients for emission line intensities of 
an atomic ion.
Databases that provide one set of calculated data for one
process is also categorized as this second type, such as
opacity databases \cite{TOPbase, NIST-opacity}.

There are some attempts to access various AM databases from one website.
LXCat, the Plasma Data Exchange Project \cite{LXcat} is the project to collect 
AM data from various databases for low temperature plasmas 
and to provide them to users from one website.
Databases on electron scattering cross sections, differential scattering
cross sections, and swarm transport data as well as online Boltzmann equation
solvers are available.
VAMDC, the Virtual Atomic and Molecular Data Center \cite{VAMDC} is also the project
to access various databases from one website and to provide data with the
same XML format. Currently 46 databases on spectral lines,
opacities and collision cross sections of atoms and molecules are connected
to VAMDC, including NIST ASD, CHIANTI, and NIFS databases. 
The XML schema, XSAMS, was developed under the collaboration coordinated
by IAEA Atomic and Molecular Data Unit.

\begin{figure*}[ht]
\begin{center}
\includegraphics[width=1.0\linewidth]{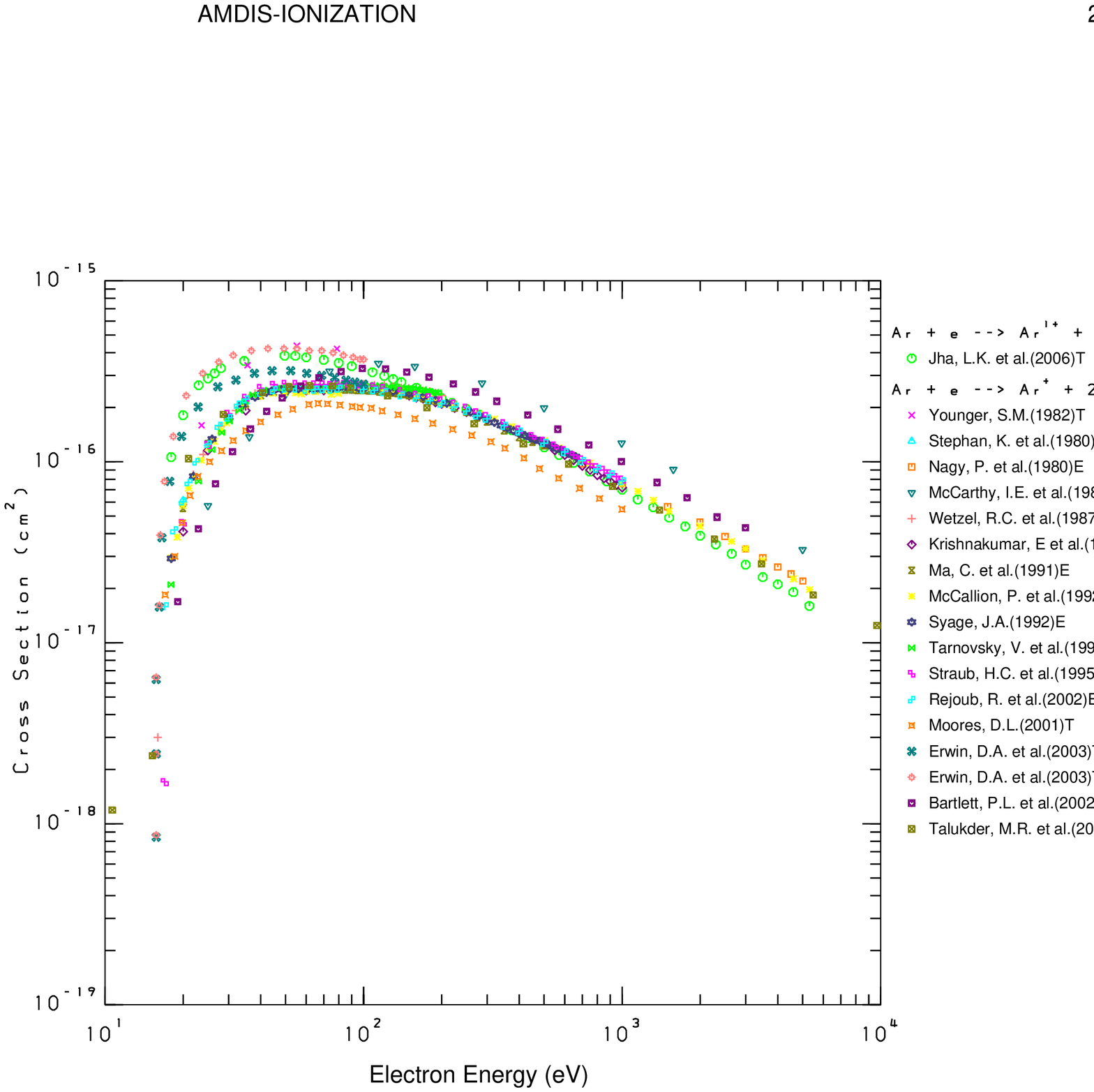}
\caption{An example of electron-impact ionization cross section of Ar atom, 
taken from NIFS database.T or E at the end of each legend indicates theoretical or experimental data. }
\label{Fig1}
\end{center}
\end{figure*}

%{\color{red} The final paragraph should give concluding remarks; should be about 100 words (if necessary up to 250 words).
% https://app.grammarly.com/} 

Current atomic and molecular numerical databases have been developed and
maintained to be available for communities with big efforts by researchers 
on atomic physics and various plasma physics for many years.
Databases on such fundamental data are useful for various applications.
Data needs from communities give motivation to studies for atomic physicists
and the help and efforts of data providers are largely appreciated.
Continuous efforts to maintain these databases must be supported by communities.

\paragraph*{Acknowledgements}
IM acknowledges many atomic physicists supporting the AM database activities
and users from communities.
This work is partly supported by
the Japan Society of the Promotion of Science (JSPS) Core-to-Core Program JPJSCCA2019002. 

\noindent
[Izumi Murakami]

%\nocite{*}

%
% ****** End of file apssamp.tex ******

\subsection{Materials database}
In most industrial plasmas, as well as in fusion plasmas, the plasma is ‘contained’ and therefore plasma surface interactions are important in determining the operation and characterisation of the plasma. Many plasmas are specifically designed to interact with surfaces, for example atmospheric plasmas are being used to sterilise surfaces in medicine \cite{1_Sakudo_2019, 2_Simoncicova_2019} which requires understanding  both of the ‘sterilising agents’ in the plasma (ions, UV photons, radicals) and the properties of the surfaces. Indeed, medical applications are a good example of the myriad of materials with which a plasma may interact – metals, plastics, ceramics and glass.  Plasma treatment is recognised as a valuable method for treating surfaces and may be scaled up for large scale manufacturing, for example introducing hydrophobic properties in materials \cite{3_Zille_2015}. Plasmas may ‘activate’ processes on surfaces or even activate drugs \cite{4_Laroussi_2018, 5_Lingge_2021}. Plasma waste remediation and waste treatment \cite{6_Sanito_2021} requires a detailed knowledge of plasma surface interactions including with (and in) liquids and may be used even for radioactive waste \cite{7_Prado_2020}. However, to date there are no databases that focus on plasma interactions with such materials and there have been few studies to explore in detail the physico-chemical changes induced by plasmas across such a range of materials. Rather, publications are scattered and often present a limited data set for one plasma and one material, making cross comparison difficult. 

In contrast, the fusion community has developed a detailed materials database since the materials used in plasma confinement chambers and the plasma-wall interactions are pivotal to the operation and sustaining of a fusion plasma. Accordingly, the fusion plasma community has developed and maintained databases that detail and analyse the properties of relevant materials and their critical parameters for fusion environments. In Europe this work has been performed under the EUROfusion programme with the data recorded in EUROfusion database and handbook \cite{8_Gorley_2020}. The database has established protocols to obtain the raw data, introduce screening procedures and data storage to ensure quality and thence acceptance (and adoption) by the international community. Similarly, the International Atomic Energy Authority (IAEA) has compiled data and published reviews for many years often resulting from IAEA Coordinated Research Projects (CRPs) – for example, the recent CRP on Plasma-Wall Interaction for Irradiated Tungsten and Tungsten Alloys in Fusion Devices \cite{9_IAEA_project}.  Such reviews are commonly published in the IAEA’s journal series Atomic and Plasma-Material Interaction Data for Fusion (APID) with 18 volumes from  1991- 2019 \cite{10_IAEA_search}. Unfortunately, not all this data is yet available on-line but IAEA has a large repository of databases: https://amdis.iaea.org/databases/.

Newer resources for nuclear fusion energy research hosted by the IAEA focus on atomistic modelling of candidate materials for fusion reactors: molecular dynamics simulations of collision cascades (CascadesDB \cite{11_CascadesDB}) and DFT simulations of radiation-induced defect structures (DefectDB \cite{12_DeFecTdb}). These have been developed and are maintained with the active support of the fusion materials modelling community and, in the case of CascadesDB, provide powerful visualization and data exploration tools \cite{13_CascadesDB} and allow downloads in multiple data formats (XML, JSON, plain text).

A further database, under development, HCDB \cite{14_hcdb}, hosts a heterogeneous collection of data in a hierarchical format, combining the structure of a relational database whilst providing some of the schemaless flexibility of NoSQL database technologies. For example, experimental results from a round-robin comparison exercise on deuterium retention in standardised steel samples may be stored in the same database as literature values for hydrogen diffusion coefficients in different materials without the need to construct new databases for each of these applications.

The IAEA’s Atomic and Molecular Bibliographic Data System AMBDAS \cite{15_AMBDAS} includes data on surface processes including chemical reactions, desorption, reflection, secondary electron emission, sputtering, trapping (and detrapping) and atomic and molecular processes on the surface such as neutralisation, ionisation and dissociation.  The ALADDIN database \cite{16_ALADDIN} has both atomic and molecular and particle-surface data and together these two online databases provide the most detailed and accessible materials data albeit with focus on fusion community and the materials used in fusion reactors.

\noindent
%\red{[Who are the authors? H. K. Chung? His/her name is missing in the author's list]}
\noindent
[Nigel J. Mason and Christian Hill]

\subsection{Plasma Chemistry databases}
% (J Tennyson and QDB)
Plasmas are strong sources of chemistry both in their treatment of surfaces and the (often complex) chemistry within the plasma leading to creation of reactive species that in turn provide the main resource for the action of the plasma. It is therefore important that the chemistry of the plasma is understood if the plasma properties are to be characterised and, through this, natural plasma phenomena such as aurorae unravelled. In the development of industrial plasmas such chemistry should be both derived and modelled if the plasma’s functionality is to be tuned and optimised for plasma usage. Thus the assembly of plasma chemical databases is an important part of future plasma development. 

Tennyson {\it et al.} \cite{11_Tennyson_2017} defined three criteria for developing a chemistry inclusive plasma model: (1) The chemistry should be \emph{complete}, that is contain all the important reactions for the given plasma; (2) It should be \emph{consistent}, that is the reactions should not be unbalanced, thus resulting in the plasma composition being driven away from the true composition; and (3) the plasma chemistry should be \emph{correct}. This  last criterion is difficult to demonstrated on purely theoretical grounds alone and therefore requires validation by  experimental measurements made in plasmas.

Addressing the first criterion, for a given plasma composition, there are sets of species that are present in the plasma and a set of processes, generally called reactions, that will link the species or different states of the species. This reaction set is described as the ‘chemistry’ for that plasma. However, assembling plasma chemistries is far from straightforward since even for relatively simple systems such as a microwave molecular nitrogen plasma some 15 species are necessary to characterise the plasma including: the seven lowest vibrationally excited states of the nitrogen molecule in the ground state $N_2(X^1\Sigma{}^+_g)\nu{} = 0$ to 6, the metastable molecule $N_2(A^3\Sigma{}^+_u)$, the ground state atom $N(^4S)$,  two metastable atoms $N(^2D)$ and $N(^2P)$ and five ionic species $N$, $N^+$, $N_2^+$, $N_3^+$  $N_4^+$  \cite{12_Klute_2021}. With these 15 species more than 100 ‘reactions’ may be necessary to define the inherent plasma chemistry, most of which have never been measured. For even the simplest industrial plasmas, such as those used in etching, the number of reactions taking place may be more than a thousand making it unfeasible to make a ‘complete’ model. It is therefore necessary to determine the ‘critical’ or most important reactions to characterise and describe the physical and chemical properties of the plasma. However, since several important reactions remain completely uncharacterized (e.g. those involving molecular radicals), it is possible that models will neglect key processes due to the unavailability of such data.

This lack of data is a therefore a challenge in meeting the second criteria that the data set should be ‘consistent’ since some reaction pathways may be indeterminate or even unknown such that production and destruction routes for important reactants may not be complete. For example, in atmospheric pressure plasmas the role of water (humidity) may be an important criterion and explain differences in day-to-day operations. In atmospheric pressure plasmas many ions are ‘solvated’ and thus their chemical properties altered by their attachment to one or more water molecules whilst during the plasma operation such clusters may be fragmented releasing reactive ions into the plasma once again.  If such cluster chemistry is not accounted for, the true composition and density of reactive species (e.g. OH radicals) will not be accurate resulting in the modelled plasma composition being different from measurements.

The final criterion that the modelled plasma chemistry should be shown to be correct requires some modelled parameters to be measured. Selection of such parameters is not trivial, for example the number density of some species may rely upon spectroscopic measurements. While spectroscopy may be used to identify species, deriving number densities by spectroscopic measurements is difficult since excited species are populated both by direct excitation and by ‘cascade’ as higher excited atomic/molecular states decay into the lower state and such cascade cross sections are largely unknown. Such cascade processes are responsible for more that 80\% of the formation of metastable species in many plasmas. 

Despite these challenges and limitations plasma chemistry databases have been assembled for different research fields. One of the most complete is the KIDA database \cite{13_Wakelam_2012} a database for astrochemical (interstellar medium and planetary atmospheres) studies that contains over 700 species and up to 10,000 reactions tuned to low temperature environment of space. The data has been assembled into several ‘networks’ for specific conditions (e.g. distinctive planetary atmospheres): https://kida.astrochem-tools.org/networks.html. This database provides references to all included reactions whilst commenting on their validity (making corrections where necessary) and where there are several alternative values may make recommendations as to the values to use.

The Quantemol chemistry database (QDB) \cite{11_Tennyson_2017} is a commercial database that contains chemistry data for industrial plasma modelling from pre-assembled and validated chemistry sets allowing users to assemble their own unique database for their specific plasma. It has about 50 pre-assembled datasets used in common plasma etching processes incorporating electron, heavy particle, photon collision cross sections and atomic and molecular species reaction rates. It also hosts some data for surface processes split into two categories: data for plasma simulations such as sticking coefficients for atomic oxygen, atomic fluorine, fluorocarbons, and silane radicals; and data for surface mechanisms such as specific etches, where the it provides a set of individual reactions with their associated probabilities.

\begin{widetext}

\begin{table}[!ht]
%% increase table row spacing, adjust to taste
\renewcommand{\arraystretch}{1.1}
% if using array.sty, it might be a good idea to tweak the value of
% \extrarowheight as needed to properly center the text within the cells
\caption{Actively maintained databases containing electron-molecule collision cross sections and other data to importance for plasma modelling applicitions}
\label{tab:DBs}
\centering
%% Some packages, such as MDW tools, offer better commands for making tables
%% than the plain LaTeX2e tabular which is used here.
\begin{tabular}{llll}
\hline\hline
Database&
Electron-collision data&
Target field&
Other data \\
\hline
LXCat \cite{LXcat}&
Excitation processes&
Plasma physics&
Atomic cross sections\\
QDB \cite{11_Tennyson_2017}&
Excitation processes&
Technological plasmas&
Chemical reaction rates\\
NIFS \cite{NIFS}&
Excitation processes& 
Fusion&
Chemical reaction rates\\
NFRI \cite{NFRI}&
Excitation processes & 
Fusion&
Chemical reaction rates\\
ALADDIN \cite{ALADDIN}&
Excitation processes& Fusion&
Chemical reaction rates\\
Phys4Entry \cite{jt628}&
Vibrational excitation&
Atmospheric re-entry&
Heavy particle inelastic
cross sections.\\
BASECOL \cite{jt547}&
Rotational excitation&
Astrophysics&
Heavy particle inelastic
cross sections.\\
KIDA \cite{13_Wakelam_2012}&
Dissociative recombination&
Astrophysics&
Chemical reaction rates\\
UfDA \cite{UfDA}&
Dissociative recombination&
Astrophysics&
Chemical reaction rates\\
IDEADB &
Dissociative electron
attachement\\
\hline
\end{tabular}
\end{table}

\end{widetext}

Such chemical databases are expected to increase in coming years as the chemistry induced by plasmas is utilised in more applications, including medical processes \cite{14_Gaens_2013,15_Dvorska_2020} and waste treatment \cite{16_Magureanu_2018,17_Magureanu_2021}.

\noindent
[Jonathan Tennyson]

\subsection{Low temperature plasma database}

% Title page
%\title{Low temperature plasma database}

% Author list with affiliations
%\author{Markus M. Becker}
%\affiliation{Leibniz Institute for Plasma Science and Technology (INP), Felix-Hausdorff-Str. 2, 17489 Greifswald, Germany}
%\author{Second Author's Name}
%\affiliation{Second Author's affiliation and address}

%\maketitle

%{\color{red}The first paragraph should be the introduction paragraph, which should be about 200 words (if necessary up to 350 words). In this paragraph, please discuss the background, motivation, importance, research history, etc., of the topic to be discussed in this subsection.}
%
In the field of low temperature plasma science, central databases providing fundamental data for the analysis and interpretation of measurement results, theoretical modeling and simulations have been used and maintained since many years. These include, for example, the NIST atomic spectra database~\cite{Ralchenko2005}, LXCat~\cite{Pitchford-2017-ID4155} for electron and ion scattering cross sections, swarm parameters, reaction rates, energy distribution functions, etc.,  and Quantemol-DB~\cite{11_Tennyson_2017}
for plasma species, reactions, and chemistries. 
However, the results of application-oriented research in the area of low temperature plasmas are mainly published in traditional journal publications and poorly structured and often not accessible in digital form for direct re-use.
This not only suspends the continuous life cycle of research data, but also inhibits technology transfer, since comprehensive data sets for comparison and validation studies are often lacking. 
In particular the application of  artificial intelligence/machine learning methods to data-driven science and technology requires large data sets in well-defined formats.
Data must be shared with machine-readable metadata containing information on how the data can be accessed, how they can inter-operate with applications or work flows for analysis, storage and processing, and in which context they can be re-used. 
Initiatives in many research fields are  underway to develop or advance systems and standards for documentation and sharing of research data to meet these requirements and to make it easier to find such  data, make it interoperable and re-usable in accordance with the FAIR data principles~\cite{Wilkinson2016,fair4fusion,Fabry2021,Chen2019}.
%Ohno-Machado2017
Furthermore, funding agencies and publishers are starting to issue policies requiring researchers to preserve and share the research data collected during the course of a research grant or presented in a paper. 
Both the practical needs and formal requirements have motivated work on providing a central database for research data in low temperature plasma science.

%{\color {red} The main body paragraphs should be about 400 word long (if necessary up to 800 word long.)  In the main body, please present the definition of the problems, a simple outline of the methods to solve them (with proper references on technical issues of the methods), and remaining challenges.}
%
In general, three options are available for publishing research data in digital form: First, institutional repositories, which are operated by universities or individual research institutions and accommodate data from all disciplines,
%(e.g.\ The York Research Database~\cite{yorkdb}, DataSpace at Princeton University~\cite{princetondb}); 
second, subject-specific repositories for collecting research data from a specific research area,
%(e.g.\ Durham High Energy Physics Database (HEPData)~\cite{Maguire_2017}, NOMAD Repository for materials data~\cite{nomadPaper}); 
and third generic repositories which are open to all types of data from any source, such as Figshare
%~\cite{figshare} 
or Zenodo.
%~\cite{zenodo}. 
Each option has its own advantages and disadvantages. Institutional solutions, for example, can be linked easily to local data management and quality assurance processes. Generic repositories generally impose no restrictions or quality criteria on the data, making them particularly easy for individual researchers to use. Subject-specific databases have the advantage over the former that the data can be documented and stored according to appropriate  metadata standards and data models. This aspect is particularly important in the context of data-driven research where data should be findable and re-usable by automated processes. 
Many research communities with large-scale experiments and mostly homogeneous data already have established solutions, e.g. high-energy physics and astrophysics~\cite{Hatfield2021,Garofalo2016}.
\begin{figure*}[t]\centering%
        \includegraphics[width=0.7\linewidth]{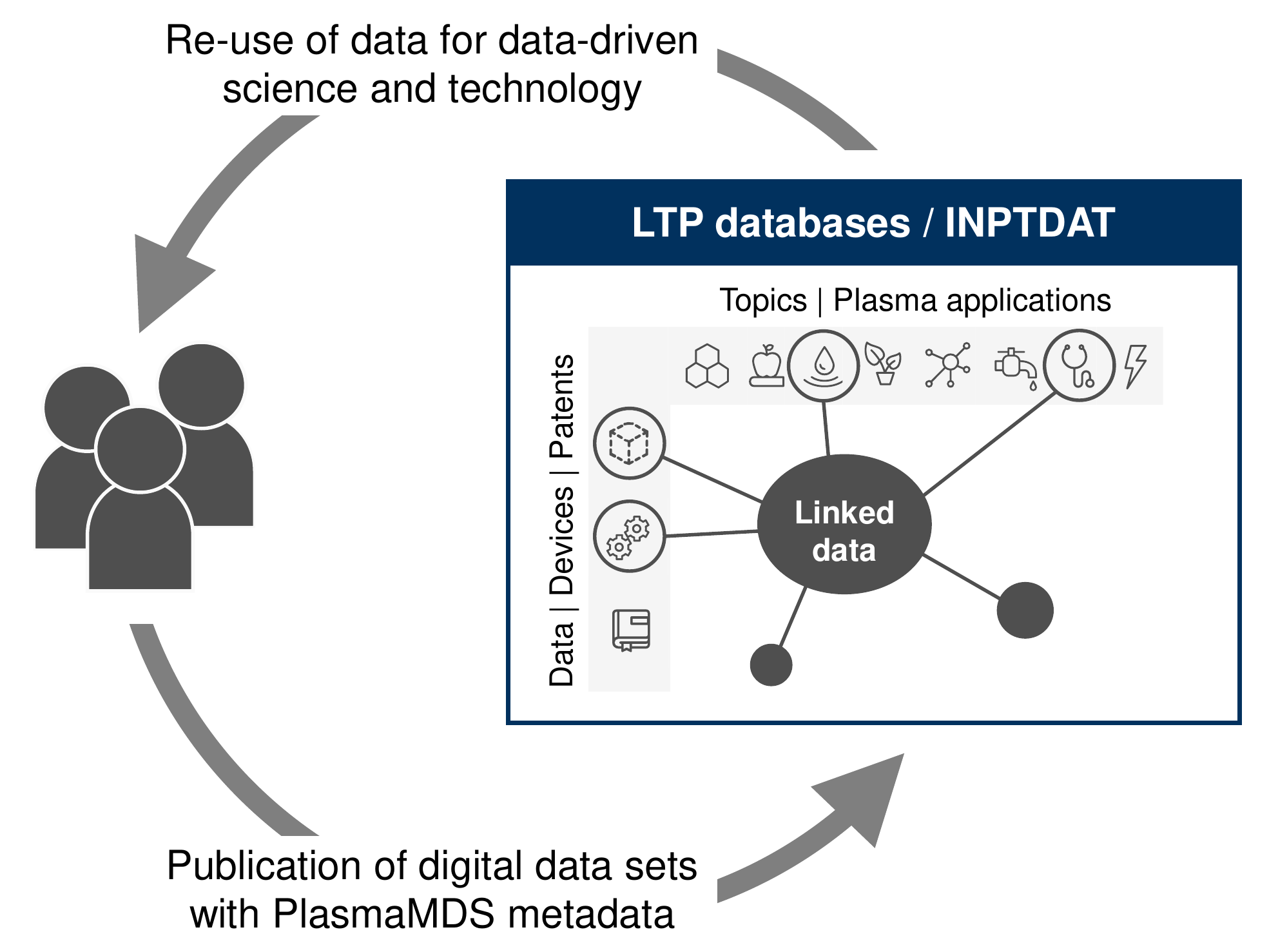}%
        \caption{Concept of a data life cycle supporting data-driven science and technology in low temperature plasma science by means of INPTDAT and the plasma metadata schema PlasmaMDS.}%
        \label{fig:LTPdatabase}%
\end{figure*}
Research in low temperature plasma science, however, is often characterized by small-scale table-top experiments involving diverse methods and devices.
Furthermore, application oriented research in the field of plasma science often involves researchers from other disciplines, like electrical engineering, biology and medicine. 
As a result, research data is extremely heterogeneous and convenient infrastructures are needed to manage and link these data in the sense of making them available for data-driven research. 
The data platform INPTDAT and PlasmaMDS, a metadata schema (MDS) for the uniform description of data in the field of applied plasma science have recently been developed to address this challenge~\cite{Franke2020}.
As illustrated in figure~\ref{fig:LTPdatabase},
the concept underlying these developments is that data obtained in the course of research in a specific subject area by means of a specific experiment and involving specific devices are assigned by the data producers to the respective topic, to a concrete application if applicable, as well as to the experiment, devices and substrates used. 
In this way, a graph of linked data and further information, e.g. from patents and device descriptions is created, and research data available for specific applications, devices, and/or substrates can be found and re-used immediately.
This is particularly beneficial if similar experiments or devices are used in different subject areas and for various applications. An example from the field of plasma technology is a plasma source being used both in plasma surface technology for the functionalization of materials and in plasma medicine for biomedical applications. 
Up to now, the data and knowledge gained in the respective fields (plasma surface technology and plasma medicine) have only rarely been brought together and re-used in an interdisciplinary manner. 
The concept implemented by INPTDAT and PlasmaMDS supports cross-domain re-use of research data by making the data directly accessible for machines and scientists from different fields via  linking with topics, applications, methods and devices.
If this approach is further developed and established in the following years according to the needs, and a community consensus on sharing and documenting research data is reached, a basis for the broad application of data-driven research, development and technology transfer can be achieved.
In this endeavor, data does not necessarily have to be collected in a central location, but can remain with the data providers and will be linked via uniform metadata descriptions and a common metadata catalog. 
The research department Plasmas with Complex Interactions at Ruhr University Bochum has already adapted this approach and, following the example of INPTDAT, set up its own data repository implementing PlasmaMDS~\cite{rdpcidat}. 
With publicly shared and collaboratively developed software and standards, a basis for further dissemination has been provided~\cite{plasma-mds-github}.

%{\color{red} The final paragraph should give concluding remarks; should be about 100 words (if necessary up to 250 words). In this short conclusion and summary, please emphasize the future direction of this topic. In addition, before submitting your manuscript, please check for typos by e.g., free software such as https://app.grammarly.com/} 
%
In conclusion, widespread re-use of data for data-driven research and technology transfer in low temperature plasma science requires that more data are provided and described in appropriate formats. The open data platform INPTDAT and the plasma metadata schema PlasmaMDS are only the first steps in this direction. 
Further work is currently being carried out on semantic cross-linking of data by means of knowledge graphs~\cite{Becker-gec2020}, whereby the participation of the community in developing common terminologies, schemas and ontologies for the extremely diverse requirements in different applications of low temperature plasma science and technology will be important in the future.

\paragraph*{Acknowledgements}
M. M. Becker acknowledges funding by the German Federal Ministry of Education and Research (BMBF) under the grant marks 16FDM005 and 16QK03A.

\noindent
[Markus M. Becker]

%\nocite{*}

%\bibliographystyle{ieeetr}
%\bibliography{DDPS}% Produces the bibliography via BibTeX.

%\end{document}
%
% ****** End of file apssamp.tex ******

%\subsection{Magnetic fusion databases [Schissel, schissel@fusion.gat.com; Sammuli]}
%text here

\subsection{Selections of recommended  data}
%(  J Tennyson and Mason)
The compilation of data in itself is valuable but when confronted by multiple data sets for the same cross section or reaction how is the user to select one set over another? This is a major challenge for user community members, who often do not have a detailed knowledge of the methods by which such data is collected and thus cannot easily distinguish between the myriad of data presented to them. When should they use experimental data, when should they use theoretical data? Is the data collected or calculated by one methodology more reliable than that of another? Is newer data necessarily more reliable that older data? These questions are often asked by the user and modelling communities and some data providers (such as QDB) offer a service to provide recommended data sets having the expertise to analyse the data and determine recommended and self-consistent data sets. However more broadly how are recommended data sets derived and is it necessary?

To answer the question of whether there is a need for recommended datasets it is only necessary to consider the use of spectroscopy to determine the number density of excited species in a plasma. The cross sections used for a specific spectral emission may be used to determine the number density of the emitting species; if different cross sectional data is used to calibrate different instruments viewing the plasma then the same observational data will be ‘translated’ into different number densities. Accordingly for projects such as JET plasma, and in future ITER, it is recognised that agreed cross sections for key diagnostics should be agreed \cite{21_IAEA_2016}.  

Similarly, many of the discrepancies between different models may be due to the use of different cross sections and reaction rates rather than different physical and chemical processes included in each model. Unravelling the data used in different models and the influence of the choice of that data has attracted the attention of both data compilers and users in recent years with discussions of the methodology to provide ‘recommended data’ being held in several meetings, for example those chaired by the IAEA and by data centres such as VAMDC (https://vamdc.org) and VESPA (http://www.europlanet-vespa.eu). Some broad guidelines in recommending data sets have emerged from such meetings:
\begin{enumerate}
    \item All recommended data should have been previously published and therefore have been subject to peer review; \item Estimates of uncertainties in the data should be provided. This is standard for experimental data but has been less common in theoretical data. However, recently publishers have required a discussion of uncertainties in theoretical/computational data \cite{22_Chung_2016};
    \item It is preferable for recommended data to be in datasets rather than individual processes. For example, consider electron scattering cross section data: individual cross sections may be recommended from different sources but the summation of these individual cross sections should be consistent with the recommended total cross section. Similarly, integrated differential cross sections should be consistent with the recommended integral cross section, summed partial ionisation cross sections consistent the recommended total ionisation cross section, and momentum transfer cross sections with recommended elastic and inelastic cross sections. 
\end{enumerate}

These guidelines demonstrate that providing recommended datasets is a challenging exercise and requires wide knowledge of the methods by which such data is generated and often the researchers involved. Experimental data are often prone to systematic effects that are known to the community, for example the community may know the energy and angular ranges over which data has been demonstrated to be reliable and ranges in which systematic effects may lead to larger uncertainties.  Extrapolation of data over angular ranges to obtain an integral cross section may be known to be problematic in some systems (e.g. electron scattering from targets with dipole moments may show strong forward scattering in regions where experimental errors are large). Some theoretical methods may also be known to be more accurate over some particular energy range.  These limitations are not always clear to the general user looking at published data but are known by the community. Therefore, it is the community with its expertise that is best suited to provide recommended datasets. However, with a few exceptions (e.g. Nuclear Data Section, International Atomic Energy Agency (IAEA) and the atmospheric community with its HITRAN database), there are few institutional structures to compile and recommend datasets in part due to lack of funding for such activities.

International organisations such as the IAEA are able to provide a stable and long-term platform for database resources serving particular communities. As computing infrastructure, including cloud computing facilities, become cheaper and more available, this has enabled such institutions to collect and serve a wider variety of data. For example, the well-established ALADDIN database of evaluated plasma collisional data at the IAEA is now supplemented by a larger database of unevaluated data, CollisionDB \cite{23_CollisionDB}, which accepts (with provenance) data from all published sources and provides a searchable interface enabling such data to be compared, aggregated and assessed.

Recent European Union activities, leading to creation of VAMDC and VESPA are encouraging but ensuring the sustainability of such efforts remains a challenge. Hence most data compilations are due to the efforts and enthusiasm of individuals such as the KIDA and LXCat astrochemistry and plasma databases with data for individual targets are published by small academic consortia and often result from a specific need they have identified for other research. Several initiatives have tried to provide a longer term approach to multiple targets for example initiatives to develop recommended datasets for electron scattering from molecular targets used in semiconductor industry by Christophorou and Olthoff at NIST \cite{24_Christophorou_2002} and, more recently, teams led by Mi-Young Song and Jung-Sik Yoon at the Plasma Technology Research Center, Korea Institute of Fusion Energy, Korea both focused on low temperature plasmas \cite{25_Song_2015}. If such recommended data sets are to be updated and new ones compiled in future, much greater emphasis and funding support must be given to such activities and the next generation of researchers convinced on the need to participate and lead such initiatives.

\noindent
[Jonathan Tennyson and Nigel J. Mason]

\subsection{Challenges and outlook}
The variety of plasma based systems has, and will continue, to expand from the study of astrochemistry and planetary atmospheres to the use of atmospheric plasmas for waste treatment and medicine. The need to redesign basic industrial plasmas for semiconductor processing using feedstock gases that comply with environmental protection (e.g. low global warming and ozone depletion potentials) has been recognised since the Kyoto protocol (designed in 1997 and entered into force in 2005) but at the recent COP26 meeting it was recognised that the targets set for 2020 had not been met and with current global uncertainties, few are optimistic of new targets being met. The design of new plasma treatment systems and their optimisation both in energy and net emissions are likely to become even more important whilst the need to accelerate the development of commercial nuclear fusion as an alternative to fossil fuels will place new emphasis on knowledge of atomic and molecular collisions, spectroscopy and, crucially, surface interactions in such plasmas.  

The collection, compilation and preparation of recommend data sets for plasma studies therefore remains one of the most important, yet also the most challenging aspects of modern plasma research. The increasing development of ‘virtual factories’ and the concept of a ‘digital twin’ \cite{31_Yildiz_2020, 32_Cho_2022} in which a plasma processing plant and procedure is modelled prior to construction places increasing emphasis on the quality and quantity of the input data used in such models. However, despite the recognition for the need to collect and compile such data, the community is small and in many cases, such as atomic and molecular data, the this community is steadily declining in numbers as other areas of science and technology attract more funding. This is a dangerous trend since the production of such data underpins all aspects of plasma technology, from the provision of diagnostics for characterising the plasma to the design of plasma itself for specific applications.

The amount of data required is already far in excess of that practical to assemble by experiment, with many targets being unsuitable for experimental research (short lived radicals, highly reactive species, species that are obtained only from highly toxic precursors) hence the majority of the data must be evaluated by theoretical calculations with the limited experimental data being used to benchmark such calculations. Whilst semi-empirical methods may be attractive to users and commercial packages such as Quantemol are available, they should be used with caution and the user is advised to co-operate with a more experienced user and take advantage of expert advice where offered (as in case of Quantemol \cite{33_Quantemol}).  

Recently there have been some attempts to use machine learning \cite{34_Butler_2018, 35_Zhong_2019} to derive data sets with a machine learning based method being to construct a model for predicting total ionisation cross sections $Q_{ion}$ of large molecules without the high cost of \emph{ab initio} calculations. The model is learned from the data composed of the calculated $Q_{ion}$ of the small molecules with fewer constituent atoms and the electron numbers of the corresponding molecules in a training set by a support vector machine (SVM) \cite{33_Quantemol}. Initial results are in broad agreement with experimental and semi-classical calculations so may be valid for higher energies, but whether they are robust enough for lower energies where the structural properties of the target are important and ‘resonances’ are formed is an open question and it is such low energy interactions that are most relevant in  the myriad of low temperature industrial plasmas.    

In conclusion, the need for collection, compilation of fundamental data underpinning the operation of plasmas is widely recognised by the community and there have been several attempts to address the challenge of providing such data to user communities with the creation of several international databases. However, this work remains poorly supported and too often relies of the efforts of a few active individuals which is not sustainable. A long-term strategy for the maintenance and review of databases is required and should be instilled in the training of the next generation of researchers.

\noindent
[Nigel J. Mason]

\section{Summary}
In this review article, the latest studies and their results in data-driven plasma science are summarized for various applications ranging from basic plasma physics to nuclear fusion, space and astronomical plasmas, and industrial plasmas. In addition, we presented a review on fundamental data science that serves as the basis for all analytical techniques used in different plasma applications and databases that serve as vital resources for the wide scientific community. 
It is seen that many common techniques and ideas are used for different applications. From a large amount of observational or computational data, some important features are extracted by regression or classification techniques, and such features are used to control plasma dynamics or to predict certain properties of the system involving plasmas. A large amount of data is also used to construct surrogate models for the systems of interest and such models can be used as alternatives to the corresponding first-principle-based computation of the system equations.  While the first-principles computation of a model system continues to be important for a better understanding of the underlying mechanisms of the system, there are many other important uses of such computation. One of such important uses is the prediction of system behavior. A surrogate model that requires only short or instantaneous computation time can be used to predict the system behaviors in real-time.  The reduction of large-scale computation is one of the goals that data-driven plasmas science attempts to achieve. 

Shortage or lack of experimental data is one of the most important challenges in this field. Probably this problem is more application-specific and what kind of data should be collected and in what way depend strongly on the system of interest. Fast and systematic ways of obtaining useful data, such as high-throughput screening, will continue to be sought after in this field with specially designed experimental systems. Design of experiments with Bayesian inference, for example, is also widely used for such purposes. 

Although we attempted to cover an extensive range of examples of data-driven analyses in plasma science, what is presented in this review article is by no means exhaustive. Unfortunately, many important studies are still missing in this article. Furthermore, the field is rapidly developing and, within several years, some of the results written here may become obsolete. This is why we named this review article ``2022 Review’' with the year of publication. We hope to update this review article with more extensive examples of the latest important developments as the field progresses. 

\noindent
[Satoshi Hamaguchi]

%\vspace{24pt}

%\section*{Acknowledgements}
%text here

%\nocite{*}

\section*{Acknowledgements}
Acknowledgements by the subsection authors are given at the end of each subsection, if any. 
S. Benkadda and S. Hamaguchi acknowledges support by the CNRS International Research Project (IRP) FJ-IPL.
S. Hamaguchi acknowledges support by Japan Society of Promotion of Science (JSPS) Grants-in-Aid for Scientific Research (S) 15H05736 and (A) 21H04453, JSPS Core-to-Core Program No. JPJSCCA2019002, and Osaka University International Joint Research Promotion Programs. This work was also supported in part by the U.S. Department of Energy through the Los Alamos National Laboratory. Los Alamos National Laboratory is operated by Triad National Security, LLC, for the National Nuclear Security Administration of U.S. Department of Energy (Contract No. 89233218CNA000001). 
Z. Wang wishes to thank Christoph R\"ath (Institut f{\"u}r Materialphysik im Weltraum, Deutsches Zentrum f{\"u}r Luft- und Raumfahrt, Germany), Chengkun Huang (Los Alamos National Laboratory, USA), Ghanshyam Pilania (Los Alamos National Laboratory, USA), Platon Karpov (Department of Astronomy $\&$ Astrophysics, University of California, Santa Cruz, USA) for stimulating discussions.
This work (section \ref{sec:icf_hedp}) was performed under the auspices of the U.S. Department of Energy by Lawrence Livermore National Laboratory under Contract DE-AC52-07NA27344.
C.S. Chang and R.M. Churchill acknowledge support from US DOE Office of Fusion Energy Science and Office of Advanced Computing Research under contract DE-AC02-09CH11466 to Princeton University on behalf of Princeton Plasma Physics Laboratory, SciDAC Partnership Center for High-performance Boundary Plasma Simulation, and Theory Department.

\section*{Conflicts of interest}
The authors declare that there are no conflicts of interest.

\section*{Authors Contributions}
S. Benkadda,  C.S. Chang, S. Hamaguchi, N. J. Mason, B. Spears, and Z. Wang acted as the section conveners of this review article. Especially, C.S.Chang hosted the manuscript website and organized and edited Secs. II and IV, Z. Wang Secs. III and VI, B. Spears Sec.V, S. Hamaguchi Sec.VII, and N. J. Mason Sec. VIII.  The author(s) of each subsection is (are) listed at the end of the subsection. S. Hamaguchi and Z. Wang also checked and edited the final version of the article.  

\section*{Data Availability}
The data that support the findings of this study are available from the corresponding author upon reasonable request.

\newpage

\bibliographystyle{ieeetr}
\bibliography{DDPS,
section_2/UQ,
section_2/ML_Sim,
section_3/sec3_mainS.bib,
section_3/AScheinker.bib,
section_3/Shalloo2.bib,
section_4/DDPS_4,
section_4/DDPS_MFE, 
section_4/ddps_roadmap_2022.bib,
section_6/Lapenta.bib,
section_6/ICF_and_HEDP/optimization.bib,
section_6/ICF_and_HEDP/uq_ref.bib,
section_6/ICF_and_HEDP/hpc_ref.bib,
section_7/Hamaguchi_7.bib,
section_7/Chen_NNP,
section_7/Park_VM, 
section_7/Mesbah_VIIB_bib.bib,
section_7/DDPS_tech, section_8/Murakami.bib,
section_8/Chung_IAEA.bib,
section_8/Tennyson_QDB.bib,
section_8/Tennyson_Mason.bib,
section_8/Mason.bib}% Produces the bibliography via BibTeX.

\end{document}